\documentclass[]{aa}
\usepackage{xcolor}
\usepackage{array}
\newcolumntype{P}[1]{>{\centering\arraybackslash}p{#1}}
\usepackage{subfig} 
\usepackage{graphicx}
\usepackage{txfonts}

\usepackage{upgreek}
\usepackage{geometry}
\usepackage{multirow}
\usepackage{amsfonts}
\usepackage{amsmath}
\usepackage{amssymb}
\usepackage{booktabs}
\usepackage{siunitx}
\usepackage{color, colortbl}
\usepackage[flushleft]{threeparttable}
\usepackage{natbib}

\usepackage{threeparttable}
\definecolor{Gray}{gray}{0.9}

\def\HeII{{\ion{He}{II}}}

\def\OIII{{[\ion{O}{III}]}}
\def\OII{{[\ion{O}{II}]}}
\def\OI{{[\ion{O}{I}]}}

\def\SII{{[\ion{S}{II}]}}
\def\NII{{[\ion{N}{II}]}}

\newcommand{\kms}{$\,$km$\,$s$^{-1}$}
\newcommand{\ergs}{$\,$erg$\,$s$^{-1}$}

\begin{document}

\title{AGN-driven outflows and the AGN feedback efficiency in young radio galaxies}
\titlerunning{CRG AGN-driven outflows}
\author{F. Santoro \inst{1,2}\fnmsep\thanks{email: santoro@mpia.de},
             C. Tadhunter\inst{2},
              D. Baron \inst{3},
             R. Morganti\inst{4,5},  
			 J. Holt \inst{6,7}	}
				
\institute{Max Plank Institute for Astronomy, Königstuhl 17, 69117, Heidelberg, Germany \and Department of Physics and Astronomy, University of Sheffield, Sheffield S3 7RH, United Kingdom.\and School of Physics and Astronomy, Tel-Aviv University, Tel Aviv 69978, Israel. \and ASTRON, the Netherlands Institute for Radio Astronomy, PO 2, 7990 AA, Dwingeloo, The Netherlands.\and Kapteyn Astronomical Institute, University of Groningen, PO 800, 9700 AV, Groningen, The Netherlands.\and Anton Pannekoek Institute, University of Amsterdam, Postbus 94249, 1090 GE, Amsterdam, The Netherlands. \and Netherlands Research School for Astronomy, Science Park 904, 1098 XH, Amsterdam, The Netherlands.}

\date{Received 03/08/2020; accepted 20/09/2020}
 
\abstract {Active galactic nuclei (AGN) feedback operated by the expansion of radio jets can play a crucial role in driving gaseous outflows on galaxy scales. Galaxies hosting young radio AGN, whose jets are in the first phases of expansion through the surrounding interstellar medium (ISM), are the ideal targets to probe the energetic significance of this mechanism. 
In this paper, we characterise the warm ionised gas outflows in a sample of nine young radio sources from the 2Jy sample, combining X-shooter spectroscopy and Hubble Space Telescope (HST) imaging data. 
We find that the warm outflows have similar radial extents ($\sim$0.06 -- 2\,kpc) as radio sources, consistent with the idea that `jet mode' AGN feedback is the dominant driver of the outflows detected in young radio galaxies.
Exploiting the broad spectral coverage of the X-shooter data, we have used the ratios of trans-auroral emission lines of \SII\ and \OII\ to estimate the electron densities, finding that most of the outflows have gas densities ($\log( n_e~cm^{-3})\sim 3-4.8$), which we speculate could be the result of compression by jet-induced shocks. 
Combining our estimates of the emission-line luminosities, radii, and densities, we find that the kinetic powers of the warm outflows are a relatively small fraction of the energies available from the accretion of material onto the central supermassive black hole (SMBH), reflecting AGN feedback efficiencies below 1$\%$ in most cases.
Overall, the warm outflows detected in our sample are strikingly similar to those found in nearby ultraluminous infrared galaxies (ULIRGs), but more energetic and with a high feedback efficiencies on average than the general population of nearby AGN of similar bolometric luminosity; this is likely to reflect a high degree of coupling between the jets and the near-nuclear ISM in the early stages of radio source evolution.

}


\maketitle
%
\section{Introduction}

The feedback effect of outflows driven by active galactic nuclei (AGN) is now routinely incorporated into models of galaxy evolution, and has been used to explain the relative dearth of high mass galaxies \citep{Benson2003,Bower2006}, as well as the correlations between black hole mass and host galaxy properties \citep{Silk1998,Fabian1999,DiMatteo2005}. However, the AGN feedback effect is likely to be complex, involving a range of physical mechanisms on different spatial scales.

In `jet mode' feedback  (sometimes labelled `maintenance mode') relativistic jets excavate cavities in the large-scale hot (>$10^7$K) interstellar medium (ISM) of the host galaxies, galaxy groups, or  clusters of galaxies on scales of 10s of kpc, and also drive shocks into the hot gas, thus preventing the gas from cooling to form stars \citep{Best2005,McNamara2012}. This feedback mode is often associated with radio-loud AGN in which the central super-massive black hole (SMBH) are accreting at a relatively low rate, thus leading to a radiatively inefficient accretion flow. However, this mode is also likely to be important in the (rarer) subset of radio-loud AGN that are accreting at higher rates and harbour radiatively efficient accretion disks.

On the other hand, in `quasar mode’ feedback the outflows driven by radiatively-efficient AGN act to heat and expel the pre-existing cooler gas in the host galaxies that would otherwise form stars. The range of radial scales over which this form of feedback operates is currently controversial, with estimates ranging from 10s of pc to $>$10\,kpc \citep{Greene2012,Harrison2012,Liu2013,Harrison2014,Husemann2016,Villar-Mart2016,Tadhunter2018,Revalski2018,Fischer2018,Baron19a}. Although this feedback mode is often linked to winds  driven by the radiation pressure of the central AGN \citep{King2015}, relativistic jets may play a significant role, even in cases in which the radio luminosity is relatively modest ($L_{1.4~\rm{GHz}}<10^{24}$\,W Hz$^{-1}$). Indeed, there is growing evidence that radio jets can have a broader impact than considered so far, and may provide the dominant outflow driving mechanism for AGN over a wide range of radio powers. 

This is based both on statistical studies of large samples of SDSS-selected AGN \citep[e.g.][]{Mullaney2013,Comerford2020} and on a growing number of observations of individual objects in which the outflows show detailed morphological associations with the radio lobes on kpc scales (e.g. IC~5063: \citealt{Morganti2007,Tadhunter2014,Morganti2015}; SDSS J165315.06+232943.0:\citealt{Villar-Mart2017}; NGC~613: \citealt{Audibert2019}; 3C~273: \citealt{Husemann2019a}; HE~1353-1917: \citealt{Husemann2019b}; ESO~428-G14: \citealt{May2018}; NGC~5929: \citealt{Riffel2014}; NGC~1386: \citealp{Rodriguez-Ardila2017}; and the targets in the sample of \citealt{Jarvis2019}).

Recent numerical simulations have also demonstrated that, despite their highly collimated nature, the relativistic jets of radio-loud AGN can inflate extensive bubbles of outflowing gas as they fight their way through the dense and inhomogeneous ISM in the central regions of galaxies \citep{Wagner2013,Mukherjee2016,Mukherjee2018}. This process is particularly important in the first phase of expansion, when the jets are just born or still young (i.e. $< 10^6$ yr). In this way, the outflows driven by the jets on kpc scales can be as broad and extensive as those driven by the radiation pressure of the central AGN. However, we do not yet fully understand how this feedback mechanism -- jets acting on the cooler phases of the ISM -- works in detail; there also remain considerable uncertainties about the masses and kinetic powers of the resulting jet-induced outflows, and the extent to which they can truly affect the evolution of the host galaxies.

Representing a high-radio-power population of AGN in which the nascent radio jets are starting to expand through the central regions of the host galaxies, Gigaherz Peaked Sources (GPS: with diameters D < 1 kpc) and Compact Steep Spectrum (CSS: D < 15kpc) sources \citep{O'Dea1998} are key objects for testing models of jet-induced feedback on kpc scales. There is now clear evidence from both spectral ageing and source expansion studies that CSS/GPS sources are genuinely young rather than merely ``frustrated'' by their interaction with the dense circum-nuclear gas \citep{Owsianik1998,1999A&A...345..769M,Tschager2000,2009AN....330..193G,An2012}.

Optical imaging and spectroscopy observations of CSS/GPS sources have demonstrated that their emission-line regions tend to be aligned with, and on similar scales to, the radio structures \citep{deVries1997,Axon2000,Labiano2008,Batcheldor2007,Santoro2018} – reminiscent of the ``alignment effect’’ observed on larger spatial scales in high-$z$ radio galaxies \citep{McCarthy1987M,Best1996}. Their optical spectra often show strong emission lines disturbed kinematics that are usually more extreme, in terms of line widths and velocity shifts, than those associated with samples of extended radio sources with similar redshifts and radio powers \citep{Gelderman1994,Holt2008,Holt2009,Shih2013,Molyneux2019}. However, although these observations support the idea that the jets in CSS/GPS sources are interacting strongly with the cooler phases of the ISM  in the host galaxies, the masses and energetic significance of the resulting outflows have yet to be accurately quantified in most objects. This is because it has proved challenging to quantify key properties of the outflow regions such as their densities, spatial extents and degree of dust extinction \citep[e.g. see][]{Harrison2018}.

In particular, estimates of electron density are key to precisely quantifying the warm outflows. In the optical band, it is common practice to measure electron densities using the \OII\ 3729/3726 or \SII\ 6717/6731  line ratios (we will refer to these line ratios as the `classical \OII\ and \SII' line ratios). However, due to the relatively low critical densities of the transitions involved, these ratios become insensitive at density above  $n_e\sim$10$^{3.5}$ cm$^{-3}$. Thus, alternative methods are required to determine whether the warm outflows contain high density components. 
The first studies which go in this direction -- for example, using trans-auroral \SII\ and \OII\ ratios  
-- are finding electron densities that are up to two orders of magnitude higher than typically estimated or assumed in studies of warm outflows, demonstrating that components of the outflowing gas have higher densities than the global ISM of the AGN host galaxies \citep{Holt2011,Rose2018,Santoro2018,Spence2018,Baron19b,Davies2020}. This has important implications for estimates of key outflow parameters such as the mass outflow rates, kinetic powers, and AGN feedback efficiencies. 

Here we use deep X-shooter/VLT observations, supplemented by Hubble Space Telescope (HST) imaging observations, to study AGN feedback and its efficiency in driving warm ionised gas outflows in a sample of 9 compact radio sources selected from the southern 2Jy sample \citep{tadhunter98,dicken09}. In particular, we take advantage of the broad wavelength coverage of the X-shooter observations to probe the presence of high density outflowing gas via a technique, pioneered by \cite{Holt2011} and \cite{Santoro2018} for compact radio sources, which uses the \OII(3726+3729)/(7319+7330) and the \SII(4069+4076)/(6717+6731) line ratios (we will refer to these as the `trans-auroral \OII\ and \SII\ line ratios' or, in short, `tr\OII' and `tr\SII') and investigate how this affects  the derived outflow and AGN feedback properties.

In Sec.~\ref{Observations and data reduction} we describe the sample selection, the observations and the data reduction strategy. In Sec.~\ref{Data analysis} we discuss the modelling of the nuclear spectra of our targets, including stellar continuum and gas emission lines. In Sec.~\ref{resuts} we describe the criteria adopted to identify gas outflows and determine their kinematical properties (Sec.~\ref{gaskinematics}), spatial extents (Sec.~\ref{radii}), gas densities and levels of dust extinction (Sec.~\ref{densities}). In Sec.~\ref{outProperties} we calculate the mass outflow rates, the outflow kinetic powers and the AGN feedback efficiencies for our compact radio sources and compare them with those of other AGN samples available in the literature in Sec.~\ref{discussion}. Finally, in Sec.~\ref{conclusion} we discuss our findings, focusing on how  more accurate estimates of the outflow densities affect the way we quantify the AGN feedback efficiency, and on the relative importance of jet-induced feedback in the near-nuclear regions of AGN host galaxies.

Throughout this paper we adopt a flat $\Lambda$CDM cosmology with H$_{0}$=70 \kms Mpc$^{-1}$, $\Omega_0$ = 0.28 and $\Omega_{\lambdaup}$ = 0.72.

\section{Observations and data reduction}\label{Observations and data reduction}

\subsection{The Sample}
\label{sec:sample}

\begin{table*}
\centering 
\resizebox{\textwidth}{!}{
\begin{tabular} { llcccccc } 
\toprule             
\noalign{\smallskip}
 Radio ID & Infrared ID & Type  & Redshift      & Angular- & Radio power     & Radio    & Radio     \\
          &             &       & (Literature)  & to-liner & log~P$_{\rm 5GHz}$ & PA &  D$_L$ \\
          &             &       &               & kpc arcsec$^{-1}$  & [W Hz$^{-1}$]   & [degrees] & [kpc] \\ 
\toprule
PKS~0023--26 & J002549.10-260213.0 & CSS	 &0.322 &4.73  & 27.43   & --34 & 3.06  	  \\
PKS~0252--71 & J025246.13-710435.7 & CSS	 &0.566 &6.60  & 27.55 &   7 & 0.95   \\
PKS~1151--34 & J115421.74-350529.5 & CSS	 &0.258 &4.05  &  26.98 & 72 & 0.37  \\
PKS~1306--09 & J130839.21-095031.2 & CSS	 &0.464 &5.94  & 27.39 &  --41 &  2.21  \\
PKS~1549--79 & J15565889-7914042  & CF 	 &0.152 &2.67  & 27.00 & 90 & 0.37   \\
PKS~1814--63 & J181934.98-634548.0 & CSS	 &0.063 &1.22  & 26.54 & --20 & 0.30  \\
PKS~1934--63 & J193925.01-634245.0 & GPS	 &0.183 &3.11  & 27.31 &   89 &   0.13 \\
PKS~2135--209 & J213749.96-204231.9 & CSS &0.635 &6.96  & 27.58 & 52  & 1.16\\
PKS~2314+03 (3C~459) & J231635.21+040517.6  & CC     &0.220 &3.59  & 27.65 &   95 &  0.71  \\
\bottomrule
\end{tabular}
}
\caption{The main properties of the radio sources in our sample, namely their radio ID (col~1), infrared ID (col~2), radio classification (col~3), redshift (col~4), angular-to-linear conversion factor (col~5), luminosity at 5GHz (col~6), as well as radio source position angle (col~7) and linear diameter (col~8). The infrared ID refers to the Spitzer Space Telescope Source List (SSTSL2) for all the galaxies apart from PKS~1549--79 for which we report the 2MASS ID. The redshifts
for the sources are based on emission-line measurements presented in \citet{tadhunter98}, with the exception of PKS~1549--79 for which we have
used the redshift estimate for the low-ionisation lines from \citet{Tadhunter2001}. The radio classifications of the sources (CC = Compact Core, CF = Compact Flat spectrum, CSS = Compact Steep Spectrum, GPS = Gigahertz-Peaked Spectrum) have been taken from literature \citep[see][]{Holt2008, 2002A&A...392..841T}. The radio luminosities have been taken from \cite{Holt2008} and \cite{Morganti1993}, while the radio sources position angles remaining sources are from \cite{2002A&A...392..841T}, with the exception of PKS~1549--79 and PKS~2314+03 whose radio position angles are from \cite{Holt2008}.  In most cases, the radio diameters represent the distances between the two brightest radio components -- usually the two radio lobes the radio source -- and are taken from \cite{2002A&A...392..841T}. However, in the case of the flat-spectrum core source PKS~1549--79, which has a highly asymmetric core-jet radio structure \citep{Holt2006,Oosterloo2019}, the number given here is the full diameter of the source from \citet{Oosterloo2019}. Moreover, in the case of PKS~2314+03 we give the diameter of the steep-spectrum compact core component from \cite{Thomasson2003}, whereas the larger-scale FRII radio source associated with this object is much more extended ($D_L \sim29$\,kpc).
}
\label{Table_general}
\end{table*}

Our sample selection has been performed starting from the southern 2Jy sample, with steep radio spectra and redshifts in the range  $0.05 < z < 0.7$, as described in \citet{tadhunter98} and \citet{dicken09}. The southern 2Jy sample is a complete sample of radio galaxies that have been extensively studied across the electromagnetic spectrum thanks to observations in the optical \citep{Tadhunter1993,tadhunter98}, the IR \citep{Dicken2008,Inskip2010}, the radio \citep{Morganti1993,Morganti1997,Morganti1999,Venturi2000} and the X-ray \citep{Ming2014} bands.

We selected sources which are in the crucial phase of a nascent radio jet starting to expand through the near-nuclear ISM. According to this criterion, the objects chosen for the current study (see Table~\ref{Table_general}) comprise a complete sub-sample of all 7 CSS and GPS sources ($D < 15$\,kpc) in the 
southern 2Jy sample, with the addition of PKS~2314+03 (3C~459) and PKS~1549--79. We added PKS~2314+03 because, despite being an extended ($D\sim29$\,kpc) FRII radio source, its compact radio core has a steep spectrum and shows similarities with CSS/GPS sources \citep[see][]{Thomasson2003}. On the other hand, PKS~1549--79 was included since, although it is a highly asymmetric radio source with a bright, flat-spectrum core, there is evidence that its radio structures are intrinsically compact, rather than just appearing compact as a result of its jets pointing close to our line of sight \citep[see discussion in][]{Holt2006,Oosterloo2019}. In Table~\ref{Table_general} we list the final targets that are part of our sample and their main radio properties. 

All of the sources in our sample have high resolution VLBI radio observations \citep{2002A&A...392..841T,Thomasson2003,Oosterloo2019}, and  previous optical, mm and radio observations have provided evidence for AGN-driven outflows in multiple gas phases in many of the objects \citep[see e.g. the work by][]{Holt2006,Holt2008,Holt2009,Morganti2005,Santoro2018,Oosterloo2019}.
\cite{Santoro2018} carried out a detailed study of the warm ionised gas for PKS 1934-63 which serves as a pilot for the current paper. For this reason this source has also been included in our sample, and we redirect the reader to the \cite{Santoro2018} paper for the details on the data analysis process.

\begin{table*}
\centering
\begin{threeparttable}
\begin{tabular} { lcccccc} 
\toprule            
\noalign{\smallskip}
Object &   Program & UVB  & VIS & NIR    &   Seeing [arcsec] & Slit PA [degrees] \\    
(1) &   (2) & (3)  & (4) & (5)    &  (6) & (7) \\               
\toprule
PKS~0023--26	&	 087.B-0614(A)   &   6$\times$900  sec     &   12$\times$450  sec       &  18$\times$300  sec      & 1.23$\pm$0.02 & 0\\
PKS~0252--71	&	   087.B-0614(A)    &   4$\times$900  sec     &   ~8$\times$450  sec       &  12$\times$300  sec       & 1.12$\pm$0.02	 & 25\\
PKS~1151--34	&    087.B-0614(A)   &   4$\times$900  sec     &   ~8$\times$450  sec       &  12$\times$300  sec       & 0.65$\pm$0.01 & 12\\
PKS~1306--09	&	 087.B-0614(A)    &   6$\times$900  sec     &   12$\times$450  sec       &  18$\times$300  sec      & 0.70$\pm$0.02 & 18\\
PKS~1549--79*	&  060.A-9412(A)     &   4$\times$675  sec     &   ~8$\times$338  sec       &  24$\times$225  sec       & $\sim$0.8 & 26\\
PKS~1814--63	&	 087.B-0614(A)     &   5$\times$900  sec     &   10$\times$450   sec      &  15$\times$300  sec      & 1.34$\pm$0.01 & 20\\
PKS~1934--63	&	 087.B-0614(A)     &   5$\times$900   sec    &   10$\times$450  sec       &  15$\times$300  sec        & 0.97$\pm$0.06 & 104 \\
PKS~2135--209 & 089.B-0695(A)     &   7$\times$900   sec    &   14$\times$450  sec       &  21$\times$300  sec      & 0.99$\pm$0.03 & 44 \\
PKS~2314+03  &	087.B-0614(A)     &   4$\times$450  sec   &   ~8$\times$225 sec      &  12$\times$150 sec        &  1.09$\pm$0.01  & 5\\
\bottomrule
\end{tabular}

\begin{tablenotes}
\footnotesize
\item[*] For PKS~1549--79 we have no formal estimate of the seeing due to the lack of acquisition images. 
\end{tablenotes}
\end{threeparttable}

\caption{Table reporting the details of the X-shooter observations: name of target (1), ESO program number (2), number and duration of exposures in the UVB (3), VIS (4) and NIR (5) arm, estimated seeing in arcseconds (6) and slit position angle measured from North to East in degrees (7).}
\label{Table_obs}
\end{table*}

\subsection{Observations}\label{sec:observations}

\subsubsection{X-shooter observations}\label{sec:X-shooter-observations}

We carried out an observational campaign for the full sample using X-shooter at the VLT/UT2 in SLIT mode. The observing program and the period of execution are reported in Table~\ref{Table_obs}, together with the exposure times for the visual (VIS), the ultraviolet-blue (UVB) and the near-IR (NIR) arm.
Sky subtraction was facilitated by nodding the source within the slit for most galaxies of the sample. However,  for PKS~1151--34 and PKS~1814--63 the slit was nodded to a separate sky aperture due to the presence of a nearby companion galaxy along the slit and extended starlight on the scale of the slit respectively.
The instrument SLIT mode was used with a 1.6$\times$11 arcsec slit for the UVB arm, a 1.5$\times$11 arcsec slit for the VIS arm, and a 1.5$\times$11 arcsec slit for the NIR arm. The resulting long-slit spectra have pixel sizes of 0.16, 0.16 and 0.20 arcsec in the spatial direction for the UVB, VIS and NIR arms respectively.
In Table~\ref{Table_obs} we also report the slit position angle and the estimated seeing of the observations for each galaxy. Note that in most
cases the slit was not aligned with the radio axis (see Tables~\ref{Table_general} and \ref{Table_obs}).

The seeing full-width at half maximum (FWHM) was estimated by using acquisition images of our targets taken during the observing time. For each galaxy, we selected a few stars in the acquisition images, extracted their spatial profiles (using a mock slit with the same width used for the actual observations) and fit them with Gaussian functions. The seeing, as reported in Table~\ref{Table_obs} is the average FWHM of the best-fit Gaussian functions; uncertainties have been derived as the standard error of the DIMM seeing values recorded during the observations period by the observatory. For PKS~1549--79 we were not able to retrieve any acquisition image and thus lack an estimate of the seeing FWHM. Therefore, as a reference a we report that the average DIMM seeing recorded during the observations, but this is likely to underestimate the true seeing for this object, which was observed at a high air mass.

Data reduction of the X-shooter data was performed using ESO REFLEX and following the same approach used in \cite{Santoro2018}. This includes a standard data reduction (e.g. bias subtraction, flat fielding and flux calibration). In addition, second-order calibrations were applied to remove hot and bad pixels and improve the sky subtraction. 
We derived the average uncertainty in the wavelength calibration and the average instrumental spectral resolution by fitting sky emission lines and  measuring their line centres and FWHM, respectively. We find that the wavelength calibration uncertainty is 20~\kms, 5~\kms\ and 3~\kms, while the instrumental spectral resolution is 90~\kms, 60~\kms\ and 90~\kms\ for the UVB, VIS, and NIR arms, respectively.  The relative flux calibration accuracy was estimated to be between 5 and 10$\%$ taking into account flux variations due to calibration with different standard stars. 

We extracted the nuclear X-shooter spectra of our galaxies by setting an aperture with diameter equal to three times the estimated seeing as reported in Table~\ref{Table_obs}. 

\subsubsection{HST observations}\label{sec:observationsHST}

In order to determine the extents of their warm outflows, three of the sources in our sample --- PKS~0023--26, PKS~1306--09, and PKS~1549--79 --- were observed using the High Resolution Camera (HRC) or Wide Field Camera (WFC) of the Advanced Camera for Surveys (ACS) mounted on the HST. The reduction and analysis of the ACS/HRC \OIII\ imaging observations for PKS~1549--79 is presented in \citet{Batcheldor2007}, and further analysis and discussion on the relationship between the \OIII\ and radio structures in this source is presented in \citet{Oosterloo2019}. Here we present new ACS/WFC observations for PKS~0023--26 and PKS~1306--09, which were taken under HST programme GO12579 (PI J. Holt). 

\begin{figure}[]
\begin{minipage}[t]{0.5\textwidth}
\includegraphics[width=\textwidth]{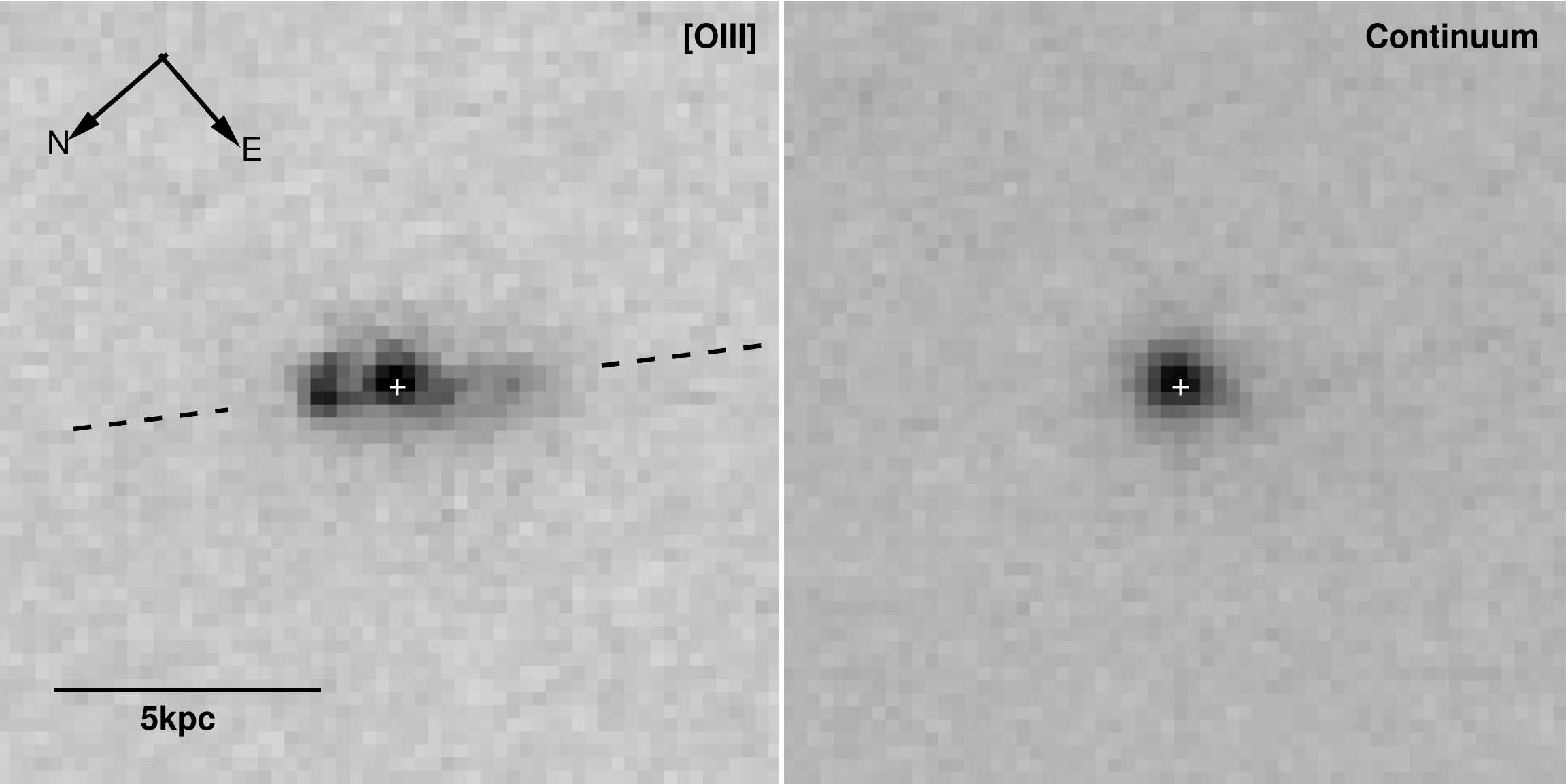}
\end{minipage}
\begin{minipage}[t]{0.5\textwidth}
\includegraphics[width=\textwidth]{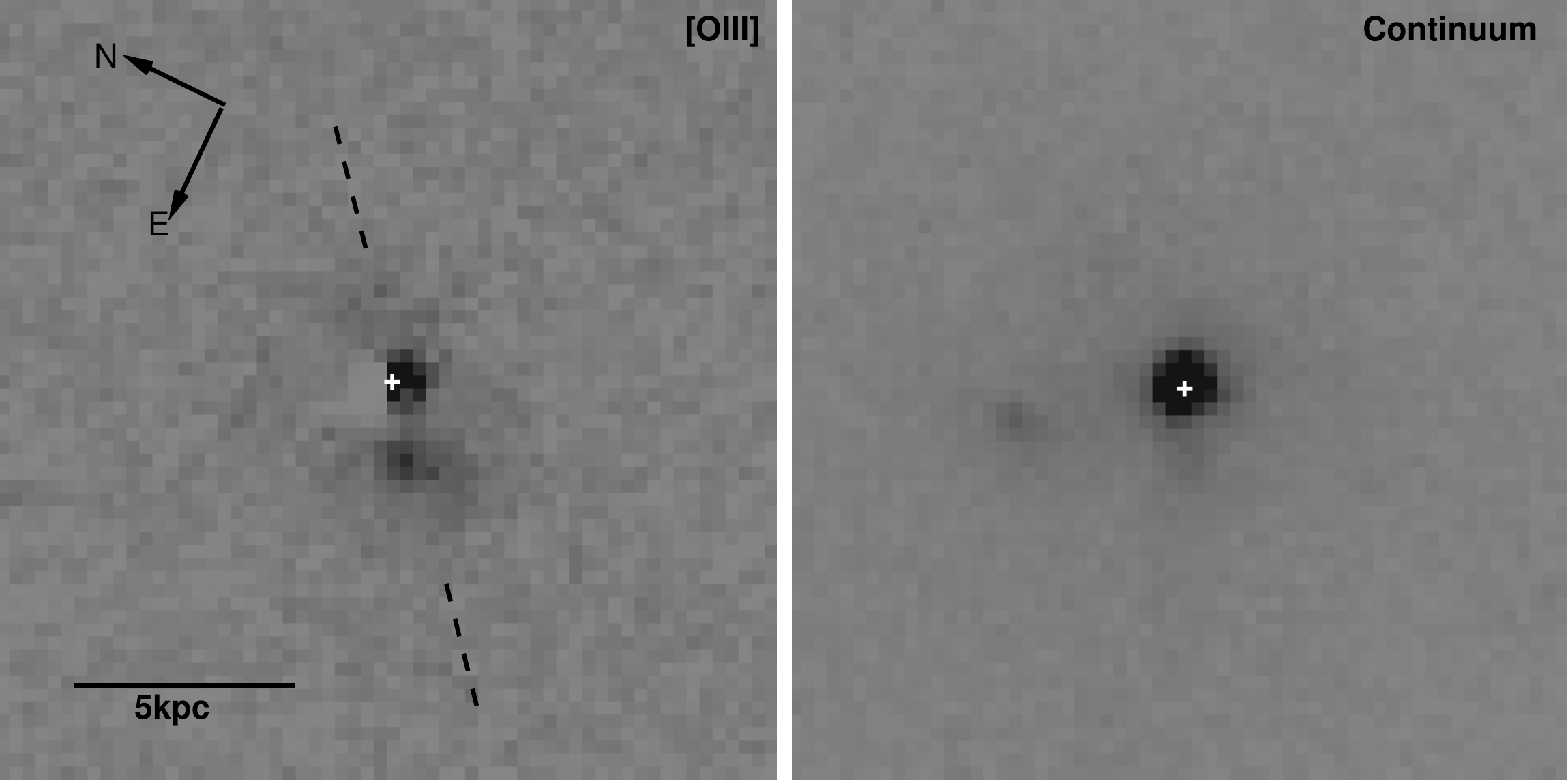}
\end{minipage}
\caption{HST ACS/WFC images of PKS~0023--26 (top) and PKS~1306--09 (bottom). The left-hand panels show the narrow-band \OIII\ images, whereas the right-hand panels show the intermediate-band continuum images. In each case, the  dashed line shows the direction of the radio axis. Note that in both cases, the brightest extended \OIII\ structures are closely aligned in angle with the radio axes (within $\sim$10 degrees).}
\label{HSTfigure}
\end{figure}

The new HST observations are detailed in Table~\ref{hst}. They were taken using the ramp filters in ACS/WFC, with narrow-band filters ($\Delta \lambda \sim 140$\,\AA) centred on the wavelengths of the redshifted \OIII$\lambda$5007\AA\ features, and medium-band filters ($\Delta \lambda \sim 580$\,\AA) centred on adjacent continuum regions, in order to facilitate continuum subtraction. Observations in each filter comprised four separate exposures taken in a box dither pattern. The data were reduced using the standard CALACS pipeline \citep{Pavlovski2005} including de-striping and charge transfer effect (CTE) corrections. Following image registration, the continuum images were subtracted from the emission-line images to create line-free \OIII\ emission line images; these are compared with the stellar continuum images in Figure \ref{HSTfigure}.

\begin{table}
\centering 
\begin{tabular} { lccc } 
\toprule
Object         & Filter  & Central & Exposure \\
               &         & wavelength    & time (total) \\
               &        & [\AA]         & [s] \\

\toprule
PKS~0023--26   & FR656N   & 6582          & 660      \\
         	   & FR647M   & 6068          & 420      \\
PKS~1306--09   & FR716N   & 7303          & 688       \\
               & FR647M   & 6751          & 420      \\
\bottomrule
\end{tabular}
\caption{Details of the HST ACS/WFC observations of PKS~0023--26 and PKS~1306--09.}
\label{hst}
\end{table}

\section{Data analysis}\label{Data analysis}

\subsection{Stellar continuum modelling and redshift determination}
\label{sec:stellarpop}

To properly investigate the physical and kinematic properties of the warm ionised gas in the galaxies of our sample, it is crucial to subtract the contribution of the starlight from their nuclear spectra and determine an accurate systemic velocity.  
To perform this task, the nuclear spectra of the galaxies were fitted using pPXF \citep{Capp2004,Capp2017}  in combination with a set of stellar models from \cite{2003MNRAS.344.1000B} with solar metallicity and ages of 0.005, 0.025, 0.1, 0.29, 0.64, 1.4, 2.5, 5, 11 and 12.3 Gyr. All the emission lines associated with the warm ionised gas were masked out during the fitting procedure. The nuclear spectrum of each galaxy was first de-reddened taking into account Galactic extinction and then fitted using the redshift reported in Table~\ref{Table_general} as a first guess for the systemic velocity. 
The E(B-V) values for the Galactic extinction towards the direction of each galaxy were taken from the NASA/IPAC infrared science archive\footnote{https://irsa.ipac.caltech.edu/applications/DUST/} and the reddening correction was performed using the \cite{Cardelli1989} reddening law. In Table  \ref{Table_redshift} we report the E(B-V) values adopted to correct for Galactic extinction, and the redshifts derived from the starlight fitting procedure described above.

By using this procedure, we were able to recover and subtract the stellar continuum in the nuclear spectra of seven out of the nine sources of our sample. However, two sources --- PKS~1814--63 and PKS~1151--34 --- required a separate treatment. In the case of PKS~1814--63, the light from a bright, nearby star close (in projection) to the galaxy nucleus contaminates the nuclear spectrum. After considering different stellar spectra taken from the online X-shooter spectral library\footnote{http://xsl.u-strasbg.fr/}, we obtained an optimal fit of the nuclear spectrum continuum features by running pPXF and using a combination the stellar template spectrum of a G5 star at 
zero redshift (i.e. HD 8724),  and a single stellar model from \cite{2003MNRAS.344.1000B} with age 11 Gyr and solar metallicity at the redshift of the galaxy.
On the other hand, the presence of prominent emission lines  from the Broad Line Region (BLR) and continuum from accretion disk of the AGN in the nuclear spectrum of PKS~1151--34 (classified as a type 1.5 Seyfert galaxy by \citealt{Veron-Cetty2006}) prevents us from performing a careful subtraction of the starlight across the full spectrum. However, we were still able to estimate the redshift of this source by limiting the starlight fitting to the wavelength range between about 3400 and 4000 \AA\, where the stellar features of the Ca~H and K absorption doublet are prominent and there is less contamination due to AGN light.

The best models for the starlight continuum of our targets are shown in Appendix~\ref{AppendixFits} together with the continuum-subtracted spectra that have been used to perform the analysis described in the following sections.

\begin{table}
\centering 

\begin{tabular} {lcl} 
\toprule
 Object & E(B-V)$_{Gal}$ & z$_{pPXF}$  \\  
\toprule
PKS~0023--26~	 	 &  0.014		 &	 0.32188$\pm$0.00004 \\ 
PKS~0252--71~  &   0.031 		  &  0.56443$\pm$0.00003\\ 
PKS~1151--34~	 	 & 		0.071	 &	 0.2579$\pm$0.0001\\ 
PKS~1306--09~	 	 & 	0.040		 & 0.46692$\pm$0.00007\\ 
PKS~1549--79~	 	 &  	0.175			& 	0.15256$\pm$0.00005\\ 
PKS~1814--63~	  &  	0.074				&  0.06373$\pm$0.00002\\ 
PKS~1934--63~	 	 & 		0.073		 &  0.18255$\pm$0.00003\\ 
PKS~2135--209  	 & 	0.293	& 0.63607$\pm$0.00005\\ 
PKS~2314+03   	 &   0.056			& 0.21986$\pm$0.00004 \\  
\bottomrule
\end{tabular}

\caption{Radio sources in the sample and their galactic E(B-V) values (col 2) taken from the NASA/IPAC infrared science archive, and stellar absorption-line redshifts (col 3) derived from our stellar continuum fit. }
\label{Table_redshift}

\end{table}

\subsection{Emission line modelling}\label{sec:kinmodels}

\begin{figure*}[]
{\centering

\begin{minipage}[t]{0.4\textwidth}
\includegraphics[width=\textwidth, height=0.21\textheight]{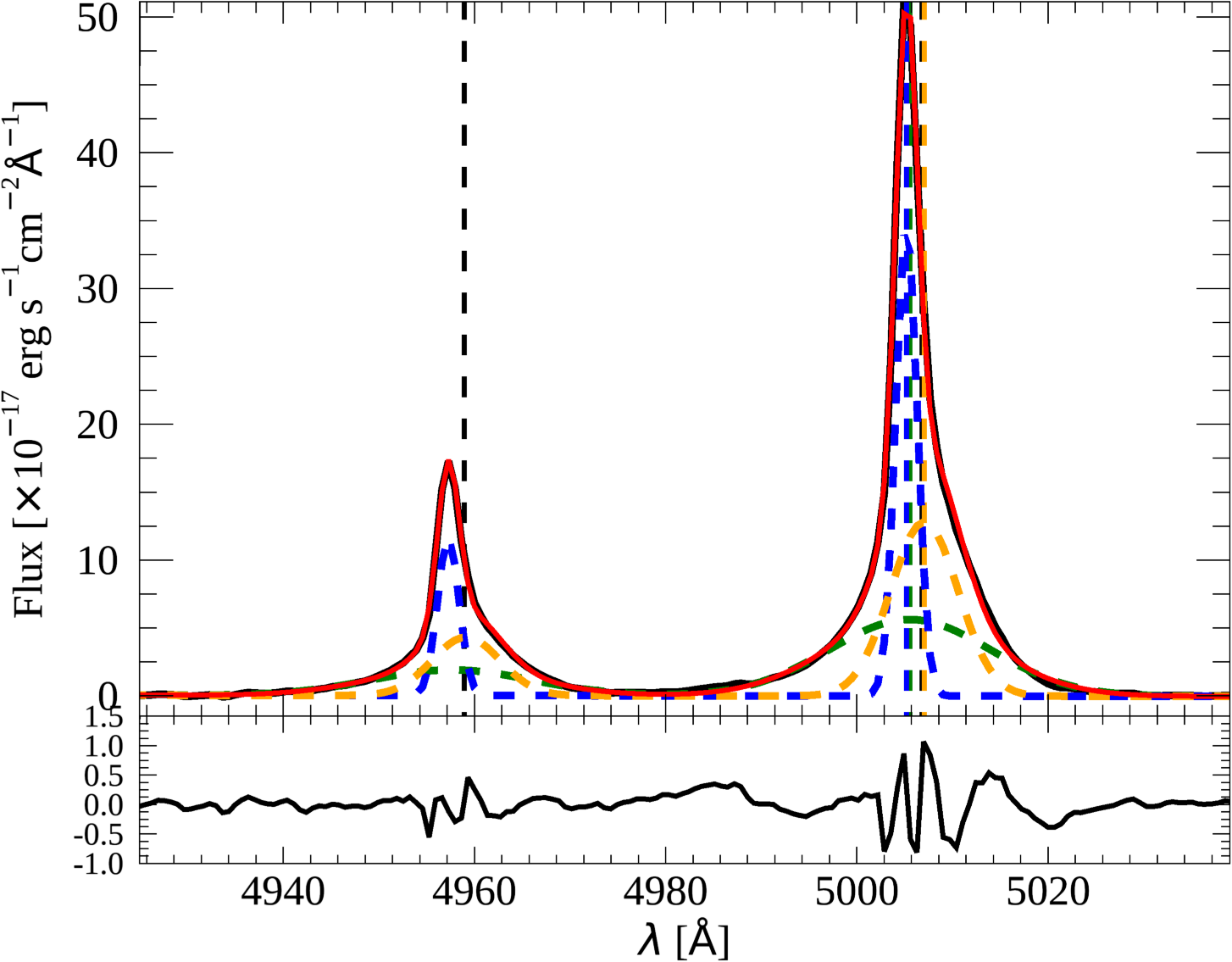}
\end{minipage}
\begin{minipage}[t]{0.4\textwidth}
\includegraphics[width=\textwidth, height=0.21\textheight]{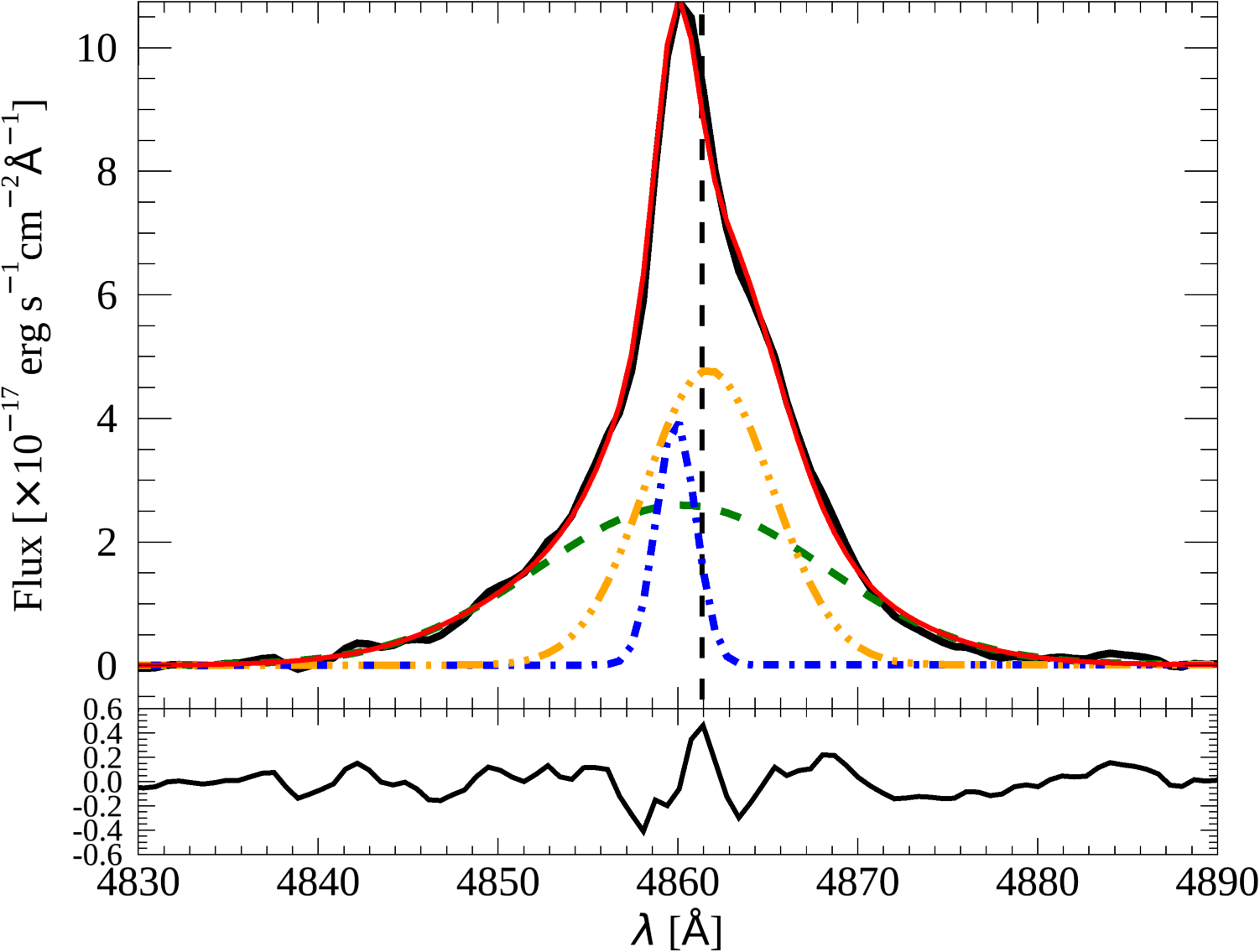}
\end{minipage}

\begin{minipage}[b]{0.4\textwidth}
\includegraphics[width=\textwidth, height=0.21\textheight]{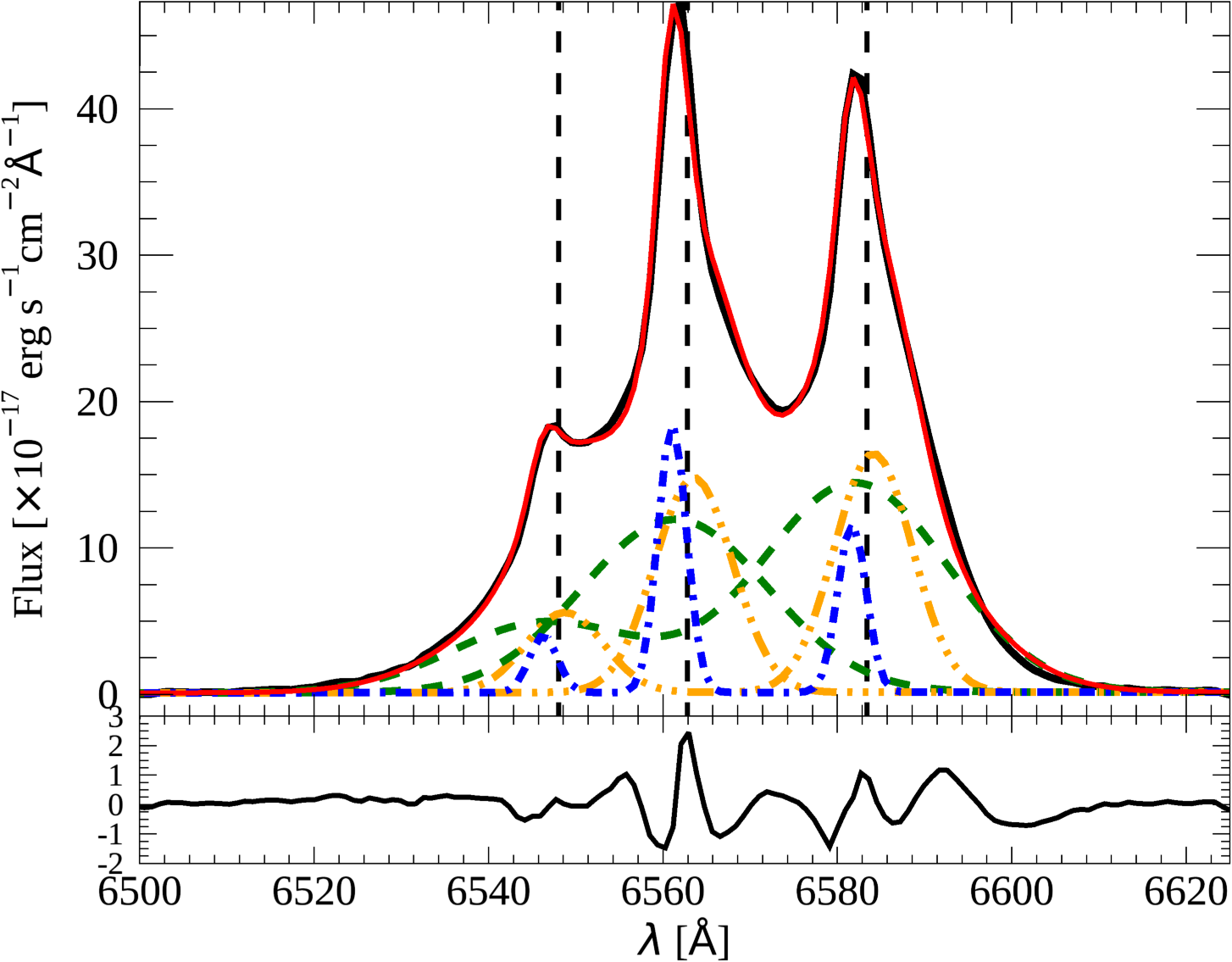}
\end{minipage}

\begin{minipage}[t]{0.4\textwidth}
\includegraphics[width=\textwidth, height=0.21\textheight]{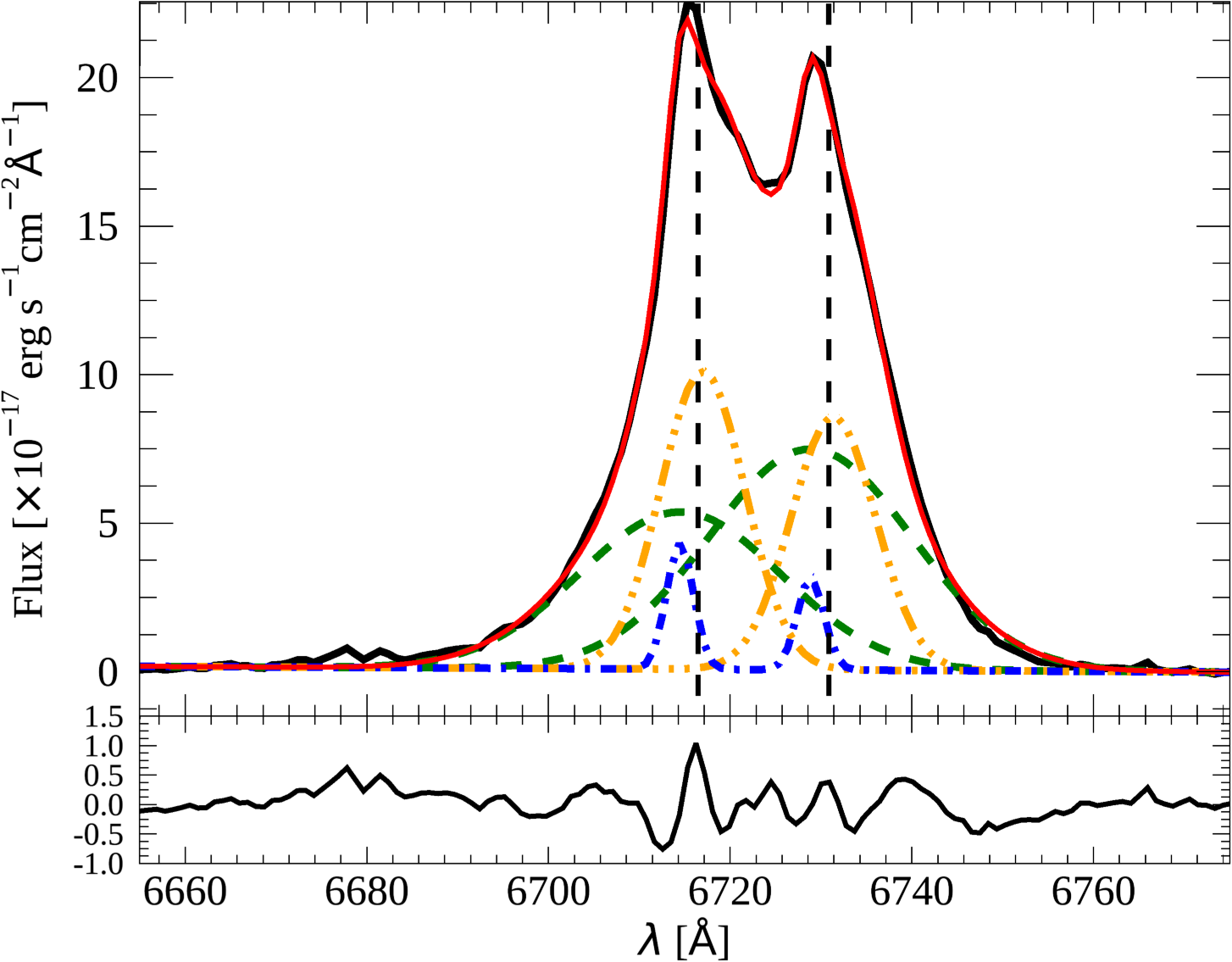}
\end{minipage}
\begin{minipage}[t]{0.4\textwidth}
\includegraphics[width=\textwidth, height=0.21\textheight]{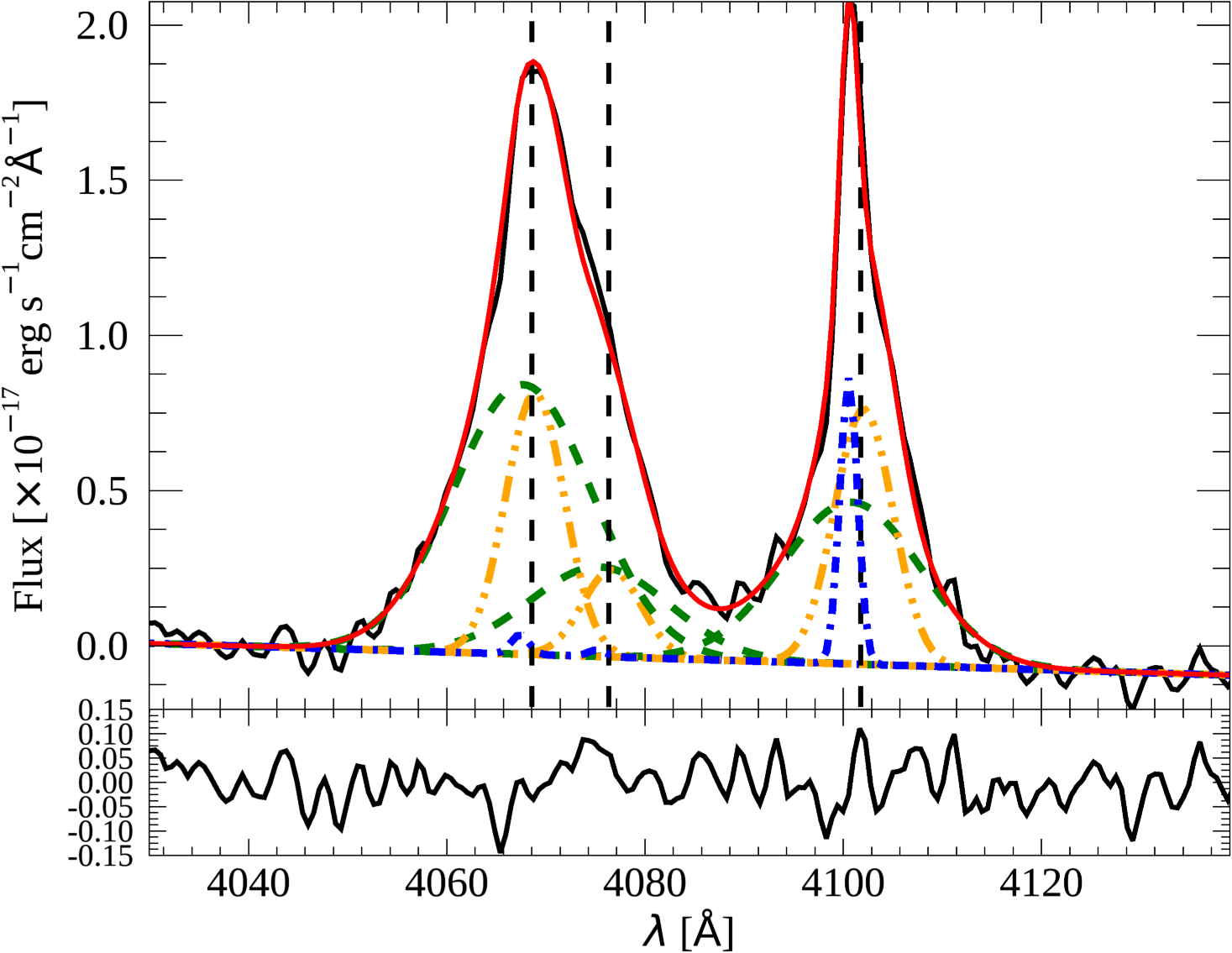}
\end{minipage}

\begin{minipage}[t]{0.4\textwidth}
\includegraphics[width=\textwidth, height=0.21\textheight]{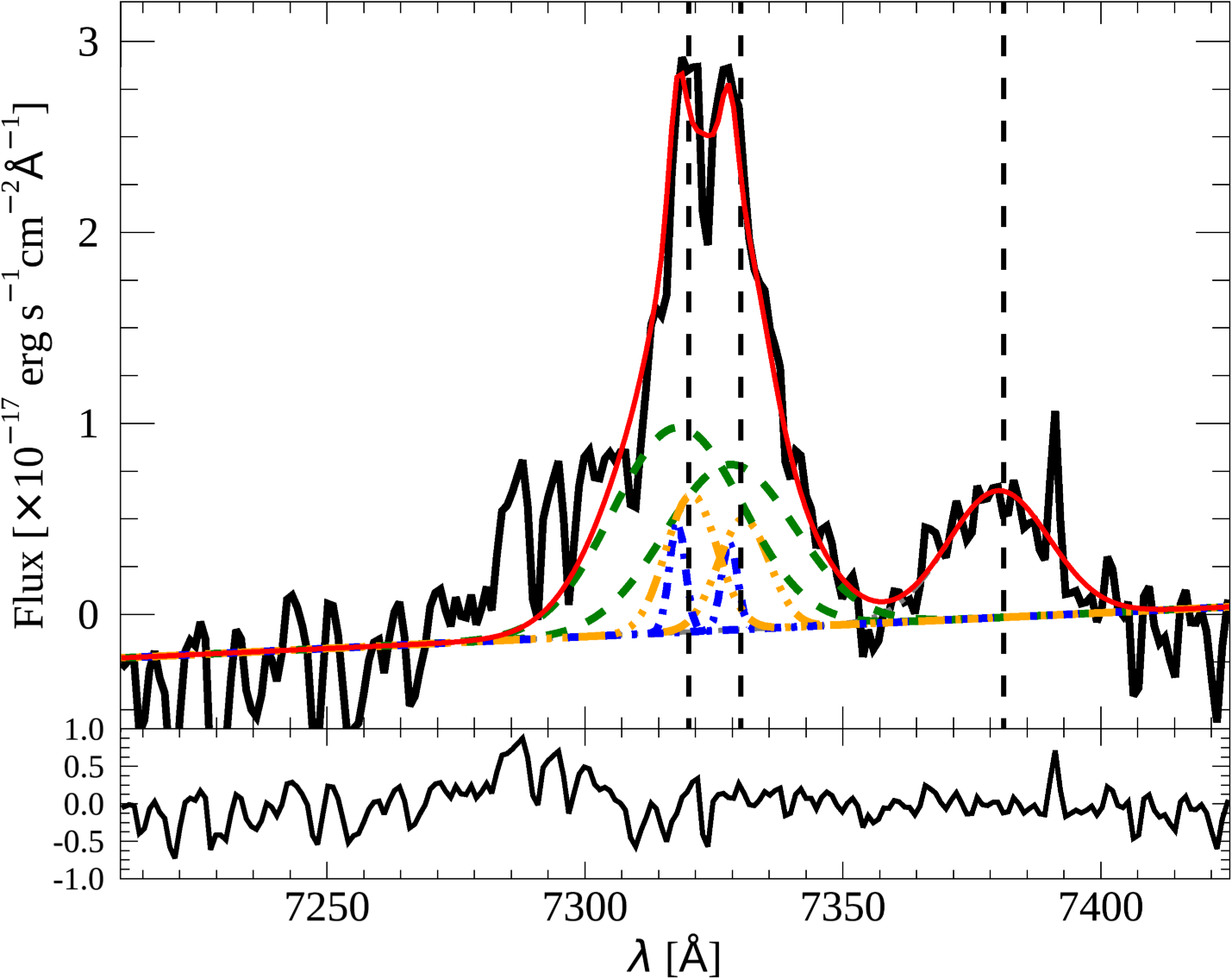}
\end{minipage}
\begin{minipage}[t]{0.4\textwidth}
\includegraphics[width=\textwidth, height=0.21\textheight]{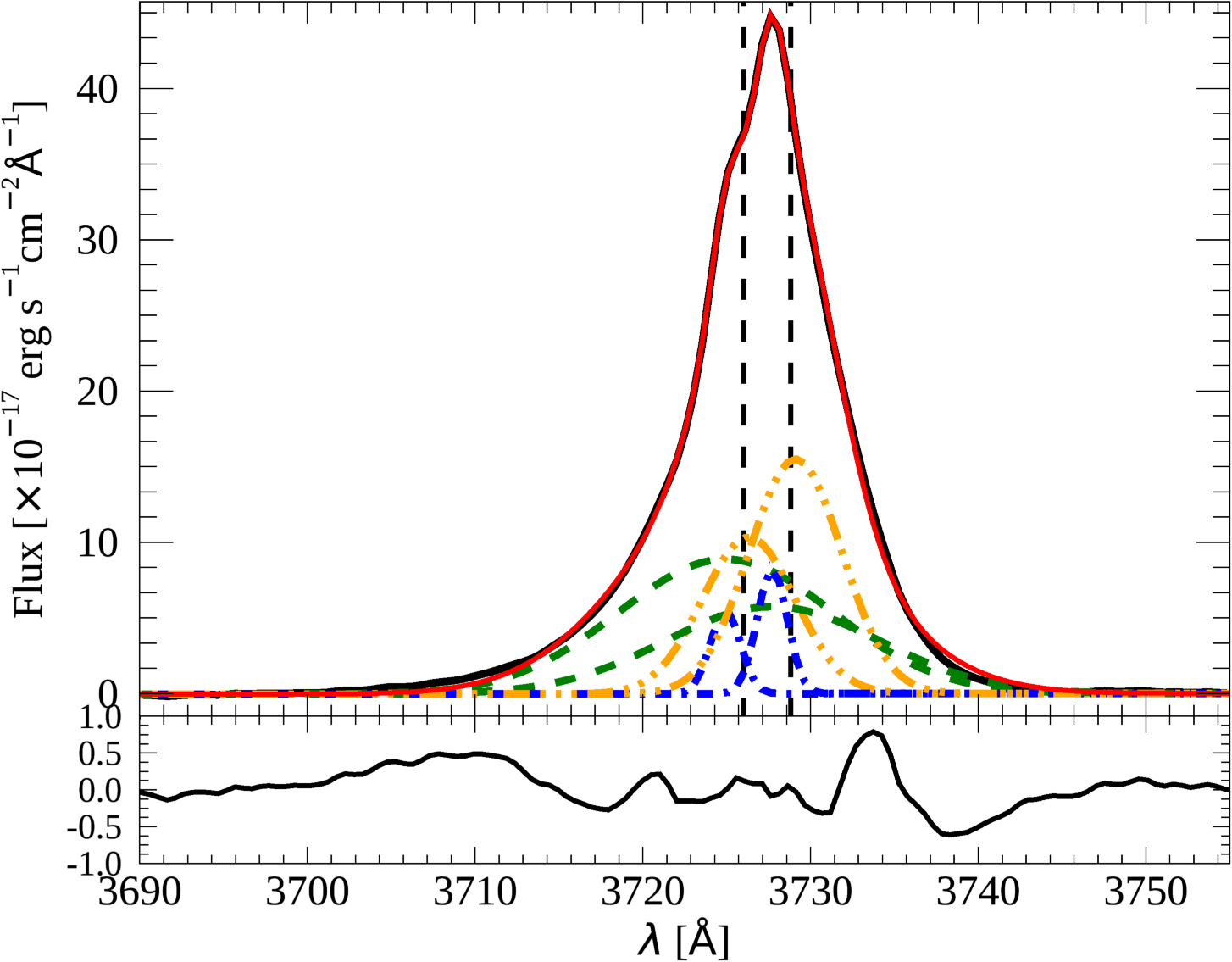}
\end{minipage}
 
 \par}

\caption{ Emission-line profiles for PKS~0023--26: \textit{First row:} \OIII$\lambdaup\lambdaup$4958,5007\AA\ (left panel) and H$\beta$ line (right panel) fits. \textit{Second row:} H$\alpha$+\NII$\lambdaup\lambdaup$6548,84\AA\ line fits. 
\textit{Third row:} \SII$\lambdaup\lambdaup$6717,31\AA\ (left panel) and \SII$\lambdaup\lambdaup$4069,76\AA\ (right panel) trans-auroral line fits. The latter fit includes also the H$\delta$ line. 
\textit{Fourth row:} \OII$\lambdaup\lambdaup$7319,30\AA\ (left panel) and \OII$\lambdaup\lambdaup$3726,29\AA\ (right panel) trans-auroral line fits. The former fit also includes the \OII$\lambdaup$7381\AA\ line which has been modelled with a single Gaussian component. In each of the sub-figures the upper panel shows the best fit (red solid line) of the observed spectrum (black solid line) while the lower panel shows the residuals of the fit. The different kinematic components used for the fit of each emission line are showed with different colors and line styles. In the case of doublets where flux ratios have been fixed (i.e. the \OIII$\lambdaup\lambdaup$4958,5007\AA\ and the \NII$\lambdaup\lambdaup$6548,84\AA\ ) we show the total profile of each doublet kinematic component. The vertical dashed lines marks the rest-frame wavelength of the fitted emission lines. Wavelengths are plotted in \AA,\ and the flux scale is given in units of $10^{-17} \rm{erg~s^{-1}cm^{-2}\AA^{-1}}$. 
} 
\label{Emission lines fits}
\end{figure*}

After subtracting the starlight features from the nuclear spectra of our targets we obtained pure emission-line spectra. In Fig.\ref{Emission lines fits} we show the profiles of some of the main emission lines observed in PKS~0023–26, while analogue plots are shown for the other objects in Appendix~\ref{AppendixFits}. As expected, all targets show complex line profiles with broad wings, confirming the complex kinematics of the gas and the presence of outflows. In order to derive the parameters needed for our analysis, we  performed the modelling of the emission lines by using Gaussian functions and custom-made IDL routines based on the MPFIT \citep{2009ASPC..411..251M} fitting routine. 

For our purposes, we need to recover the fluxes of a significant number of emission lines, many of which are blended with other emission lines and/or have low signal to noise (S/N) ratios. We thus built a reference model by fitting high S/N emission lines that are less affected by blending.
This reference model gives us an indication of the number of kinematic components needed to model the warm ionised gas emission lines, and of the velocity centroid and FWHM of each of these components. 
We fit the emission lines that are needed for our study by performing seven separate fits, more specifically we fit the \OII$\lambdaup\lambdaup$3726,29\AA, the \SII$\lambdaup\lambdaup$4069,76\AA\ plus the H$\delta$, the H$\beta$, the \OIII$\lambdaup\lambdaup$4958,5007\AA , the \NII$\lambdaup\lambdaup$6548,84\AA\ plus the H$\alpha$, the \SII$\lambdaup\lambdaup$6717,31\AA\ and the \OII$\lambdaup\lambdaup$7319,30\AA\ (plus the \OII$\lambdaup$7381\AA\ line, modelled with a single Gausssian function when needed) lines. Constraints on the line separation, width and relative intensities within each group of emission lines have been set according to atomic physics following the approach described in \cite{Santoro2018}. 

To build our reference model we fit the \OIII$\lambdaup\lambdaup$4958,5007\AA\ doublet, whose emission lines are only mildly affected by blending and have very high S/N in the nuclear spectra of our targets, with up to four kinematic components. Each kinematic component consists of two bounded Gaussian functions with the same width, fixed separation (47.9\AA ) and fixed relative fluxes (\OIII$\lambdaup$5007\AA $=$ 2.98$\times$\OIII$\lambdaup$4958\AA) according to atomic physics. Based on $\chi^2$ statistics and residual minimisation we selected as the best fit model the one that minimised the number of kinematic components needed to fit the observed line profiles, and refer to this as the `\OIII\ reference model'. It should be noted that for PKS~1549--79 and PKS~2314+03 we fit the \OIII$\lambdaup\lambdaup$4958,5007\AA\ and the H$\beta$ lines together, due to the difficulties of fitting the H$\beta$ line alone using the model derived  from the \OIII\ line, and used this as our reference model for the remaining lines.

\begin{table*}
\centering 
\resizebox{\textwidth}{!}{
\begin{tabular} {  l  cc  cc cc  cc } 
\toprule 
\multirow{2}{*}{Object} & \multicolumn{2}{c}{Component 1} & \multicolumn{2}{c}{Component 2} & \multicolumn{2}{c}{Component 3} &\multicolumn{2}{c}{Component 4}\\
     &   v$_1$\ & FWHM$_1$  &   v$_2$ & FWHM$_2$ &   v$_3$ & FWHM$_3$  &   v$_4$ & FWHM$_4$    \\  
     & [\kms] & [\kms] & [\kms] & [\kms] & [\kms] & [\kms] & [\kms] & [\kms] \\
            
\toprule
PKS~0023--26 & $-98\pm12$ & $145\pm6$   & $13\pm12$ & $496\pm7$  & $-77\pm13$ &  $1185\pm14$ &   &  \\
\rowcolor{Gray}
PKS~0252--71& $-209\pm12$ & $249\pm7$ &	 $126\pm2$ &  $446\pm9$ & $12\pm13$ & $1228\pm6$ &    &  \\
PKS~1151--34 $_{\rm NLR}$	&$58\pm33$ &  $341\pm7$ &  $82\pm33$ & $786\pm10$ &   $-898\pm33$ & $3613\pm125$ &  &  	 \\
PKS~1151--34$_{\rm BLR}$ &   $-4004\pm55$ &  $4397\pm240$ & $1006\pm54$ & $10504\pm72$ &  &  &  &  	 \\
\rowcolor{Gray}
PKS~1306--09 &  $-437\pm12$ & $130\pm7$ & $220\pm12$ & $221\pm7$ &  $-534\pm12$ & $349\pm7$ &	 $-153\pm13$ &  $1380\pm11$   \\
PKS~1549--79$_{\OIII}$ 	& $-364\pm14$ 	&   $596\pm8$   &  $-990\pm15$   &  $1103\pm10$   &   $-1055\pm16$   &   $2052\pm6$   &     &   \\
PKS~1549--79$_{\OII}$ 	&  $-172\pm14$ 	&  $488\pm9$   &  $-371\pm14$    &   $1667\pm6$   &      &     &   & \\
\rowcolor{Gray}
PKS~1814--63& $198\pm8$   &  $158\pm6$   &   $-19\pm9$  &  $262\pm10$ & $202\pm14$   &  $546\pm17$   &  $-56\pm21$  &   $1583\pm43$      \\
PKS~1934--63&  $-80\pm35$   &  $104\pm4$    & $99\pm35$ &   $128\pm5$   & $25\pm38$   &  $709\pm75$   &    $-302\pm112$  &  $2035\pm207$  \\
\rowcolor{Gray}
 PKS~2135--209& $89\pm14$  &  $805\pm7$   & $-37\pm16$  &  $1768\pm24$   &       &     &    &    \\
PKS~2314+03 & 	$-6.6\pm15$ & $530\pm7$  & $-553\pm17$ & $624\pm21$ & $-542\pm21$ & $1968\pm34$ &  &    \\
\bottomrule
\end{tabular}
}
\caption{Velocity shifts (v) and the FWHM (in \kms) of the `\OIII\ reference model' kinematic components (indicated with a progressive number from 1 to 4 from lower to higher FWHM). For PKS~1151--34 we report the kinematic parameters of both the NLR and the BLR reference model, the velocity of the BLR components are calculated with respect to the H$\beta$ rest-frame wavelength. For PKS~1549--79 we report the kinematic parameters of both the \OIII\ and the \OII\ reference models. The FWHM values have been corrected for instrumental broadening.}
\label{Table_kinmodels}
\end{table*}

The \OIII\ reference model was then used to provide constraints on the fit to the remaining emission lines in the nuclear spectra. It is worth mentioning that the successful fitting of an emission line obtained by using a reference model does not always imply that all the kinematic components of this model are actually detected. This reflects the fact that, while high S/N emission lines such as \OIII\ can give us a reliable model for the gas kinematics, the relative fluxes of the different kinematic components of an emission line depend on the physical properties and ionisation source of the gas.
More generally, when it was not possible to fit an emission line doublet with a reference model because of low S/N (i.e. the \OII$\lambdaup\lambdaup$7319,30\AA\ and/or \SII$\lambdaup\lambdaup$4069,76\AA\ emission lines in   PKS~0252--71, PKS~1306--09, PKS~1814--63 and PKS~2314+03) we recover the total lines flux by using a simple model with a single kinematic component per emission line. 

For two of our galaxies, namely PKS~1549--79 and PKS~1151--34, we needed to adjust the emission line fitting strategy. 
In the case of PKS~1549--79 the \OIII\ reference model did not allow us to recover properly the flux of some of the emission lines \citep[see also the work by][]{Tadhunter2001,Holt2006}. Mismatching kinematics between different emission lines in the nuclear spectrum of a target is expected due to the fact that different emission lines trace gas with different levels of ionisation and/or physical conditions. In this case, we obtained an alternative reference model (that we refer to as the `\OII\  reference model') by fitting the \OII$\lambdaup\lambdaup$3726,29\AA\ doublet with the same procedure described above for the `\OIII ' reference model. This is motivated by the fact that one of our main goals is to study the trans-auroral lines and, among these, the \OII$\lambdaup\lambdaup$3726,29\AA\ lines usually have higher S/N and are less subject to blending with emission lines from other elements.
The \OII\ reference model was used to fit the \NII$\lambdaup\lambdaup$6548,84\AA\ + H$\alpha$, the  \OII$\lambdaup\lambdaup$3726,29\AA\ and the the \SII$\lambdaup\lambdaup$6717,31\AA, while all the other lines have been fitted using the \OIII\ reference model.

In the case of PKS~1151--34, modelling the emission lines is  challenging due to the presence of direct BLR and continuum emission from the AGN, and also subject to higher uncertainties due to our inability in subtracting the contribution of the starlight continuum from its nuclear spectrum. The strategy we adopted to build a reference model for the emission line fit has been aimed at getting some constrains to properly model the H$\beta$ BLR emission from the brighter H$\alpha$ BLR emission, as described in Appendix~\ref{AppendixFits}.

In Fig.~\ref{Emission lines fits} we show the results of the modelling for the warm ionised gas emission lines of PKS~0023--26, while analogue plots are shown for the remaining galaxies of our sample in the Appendix~\ref{AppendixFits}.
In Table~\ref{Table_kinmodels} we report the kinematic properties (i.e. centroid velocity and FWHM of the different kinematic components) of the reference models for all the galaxies in our sample. Errors have been estimated taking into account both the instrumental and the model (fit) uncertainties, and the FWHM have been corrected for instrumental broadening.

\section{Basic results}\label{resuts}

One of the main aims of our study is to derive the mass outflow rates $\dot{M}$, kinetic powers $\dot{E}$, and  AGN feedback efficiencies $F$ for the warm outflows of the targets in our sample (Sec. \ref{outProperties}).
These quantities mainly rely on the estimates of the more basic outflows properties such as their kinematics, electron densities, dust extinction and spatial extent, which are described in the current section. Here we stress the importance of deriving gas electron densities from diagnostics that are able to probe the high gas density regime (Sec.\ref{densities}), and rely on estimates of the gas ionisation parameter (Sec.\ref{ionisationparam}), and photoionisation models (described in detail in Appendix~\ref{appendixModels}).

\subsection{Gas kinematics}\label{gaskinematics}

The results of the emission line modelling presented in Sec. \ref{Data analysis} clearly show that all our targets have complex line profiles with broad wings that require multiple kinematic components to be properly modelled.
We label as `Broad' any kinematic components in the \OIII\ reference model with FWHM$> 500$\kms\ and/or velocity shift v$<-500$\kms\, and associate these components with the outflowing gas.
Confirming previous results \citep[see][and references therein]{Holt2008}, we find all our targets have at least one broad component and thus show signs of hosting an outflow. Remarkably, some of the kinematic components of our reference models show broadening up to FWHM$\sim$2000~\kms\ and blueshifts up to about $-1000$\kms. 
According to our criterion, for PKS~1549--79, PKS~2135--209 and PKS~2314+03 all the \OIII\ kinematic components can be considered as being associated with outflowing gas. This is a clear sign that the outflowing gas comprises a large fraction of the total warm ISM sampled by the spectroscopic slit. Therefore, for these sources we use the total emission of the components detected in \OIII\ when determining the outflow properties.

\begin{table}
\centering 
\begin{tabular} {lccc } 
\toprule         
 Object &   v  & FWHM &  v$_{max}$ \\ 
 & [\kms] & [\kms] & [\kms] \\
\toprule
PKS~0023--26   &   -78$\pm$13  & 1185$\pm$14 &      -892$\pm$13 \\
PKS~0252--71    &   12$\pm$  12 &   1229$\pm$16   &    -848$\pm$12 \\
PKS~1151--34   &    83$\pm$33     &    787$\pm$10    &    590$\pm$33 \\
PKS~1306--09   &   -299$\pm$21    &    1002$\pm$34     &  -967$\pm$13 \\
PKS~1549--79   &   -880$\pm$16  &    1219$\pm$25    &   -1903$\pm$234 \\
PKS~1814--63   &    97$\pm$16     &    972$\pm$45    &   -839$\pm$47 \\
PKS~1934--63	 &   -130$\pm$58	& 	  1339$\pm$120 & 	  -1367$\pm$220 \\
PKS~2135--209  &    46$\pm$12   &      1138$\pm$33  &    -867 $\pm$26 \\
PKS~2314+03    &  	-252$\pm$11    &   1099$\pm$19 &    -1503$\pm$41 \\
\bottomrule
\end{tabular}
\caption{Kinematic parameters for the outflows: flux weighted velocity (col2) and FWHM (col3), and maximum outflow velocity (col4).  All quantities are given in \kms\ and have been determined by making use of the \OIII\ reference model kinematic properties as described in the text. 
}
\label{Table_velocities}
\end{table}

In Table~\ref{Table_velocities} we summarise the kinematic properties of the outflowing gas for each galaxy in our sample.
To quantify the velocity and the FWHM of the outflowing gas, we calculated the flux weighted average velocity (i.e. $v$) and FWHM of the \OIII\ reference model broad components. In addition, we derived an estimate of the maximum velocity (i.e. $v_{max}$) that the outflowing gas can reach by calculating the velocity at which the cumulative flux of the \OIII\ reference model broad components (integrated from low to high velocities in the velocity space) equals  $5\%$, following the approach of \citet{Rose2018}.
Due to the overall redshifted \OIII\ line profile for PKS~1151--34, the $v_{max}$ has been estimated as the velocity at which the cumulative flux of the \OIII\ reference model broad components is $95\%$. 
The errors on the $v_{max}$ have been calculated using the errors on the broad components velocity from the fitting procedure. 

\subsection{The spatial extents of the outflows}\label{radii}

\begin{table}
\centering
\begin{threeparttable}
\begin{tabular} { lccc } 
\toprule
Object         & r$_{optical}$   & r$_{radio}$  &Best\\      
   			   &  [kpc]          & [kpc]        &Estimate \\
\toprule
PKS~0023--26	 &  1.3$\pm$0.1    &	1.53    &HST  \\
PKS~0252--71	 & 0.23$\pm$0.02   &  0.47  &SA    \\
PKS~1151--34	 & $\leq$0.40       &  0.18     &RAD \\
PKS~1306--09	 & 1.9$\pm$0.2     &  1.11     &HST \\
PKS~1549--79   & 0.19$\pm$0.02     &  0.35   &HST   \\
PKS~1814--63	 & $\leq$0.21       &  0.15      &RAD\\
PKS~1934--63	 & 0.059$\pm$0.012*	 &  0.064 &SA   \\
PKS~2135--209  & 0.56$\pm$0.02   &  0.58     &SA \\
PKS~2314+03        &  $\leq$0.46      &  0.36    &RAD  \\
\bottomrule
\end{tabular}

\begin{tablenotes}
       \footnotesize
        \item[*] Radius estimated using the spectro-astrometry technique taken from \cite{Santoro2018}.
\end{tablenotes}
\end{threeparttable}

\caption{Table reporting the radii of the warm ionised gas outflows as estimated from the  optical data (HST or X-shooter) (col~2), and the radio source radii taken from the literature (col~3). In most cases, the latter represent half the radio source diameter values presented in the final column of Table 1, under the assumption that the radio lobes are symmetric about the nucleus. However, in the case of the core-jet source PKS~1549--79, we have taken the maximum extent of the radio emission on the east side of the nucleus. The final column indicates the method used
to determine the best outflow radius estimate for use when calculating general outflow
properties (HST: HST narrow-band imaging;
RAD: radial extent of radio source; SA: X-shooter spectro-astrometry).}
\label{Table_radius}
\end{table}

\begin{table*}[t]
\centering

\begin{threeparttable}
\begin{tabular} { lccc } 
\toprule
Object         & $\Delta \lambdaup_{blue}$   & $\Delta \lambdaup_{red}$ & $\Delta v_{out}$  \\ 
   			   &  [\AA ]          & [\AA ]   &  [\kms ]     \\
\toprule

PKS~0252--71	 & 4890$\leq \lambdaup \leq$4923   & 5040$\leq \lambdaup \leq$5073 &      \\
 PKS~2135--209 & 4890$\leq \lambdaup \leq$4923   & 5040$\leq \lambdaup \leq$5073 &      \\
\bottomrule
PKS~2313+03       &  4756 $\leq \lambdaup \leq$ 4806 & 5040$\leq \lambdaup \leq$5073   &  -2000$\leq$ v $\leq$-800     \\
PKS~1151--34*	 &  4831$\leq \lambdaup \leq$4839       &  4881$\leq \lambdaup \leq$4889  &  -529$\leq$ v $\leq$-279   \\
PKS~1814--63	 &  4890$\leq \lambdaup \leq$4923       &  5040$\leq \lambdaup \leq$5073  &  400$\leq$ v $\leq$1000    \\
\bottomrule
\end{tabular}

\begin{tablenotes}
       \footnotesize
        \item[*] the value reported refer to the H$\beta$ emission line which has been used for this target.
\end{tablenotes}
\end{threeparttable}

\caption{Table reporting the spectral windows used to extract spatial profiles of the stellar continuum light (col~2 and col~3) and of the outflowing gas (col~4) from the X-shooter slit spectra. For PKS~0252--71	and  PKS~2135--209 the indicated spectral windows are used to derive the average continuum light spatial profile in the framework of the spectro-astrometry technique employed to extract the outflow spatial extents.  For PKS~1814--63 and PKS~2314+03 the indicated spectral windows are used to study the outflow spatial extents following the approach described in \citet{Rose2018} and \cite{Spence2018}. The boundaries of the spectral windows are reported in \AA\ for the  stellar continuum and as velocity shifts with respect to the rest-frame velocity of the \OIII$\lambdaup$5007\AA\  emission line for the outflowing gas.
For PKS~1151--34 the spatial study has been carried out using the emission of the H$\beta$ and the bands probing the stellar continuum also contain emission of the BLR, to allow subtraction of the BLR emission as well as the contiunuum emission.}

\label{Table_spectralwindows}
\end{table*}

\begin{figure*}[]
\includegraphics[width=\textwidth]{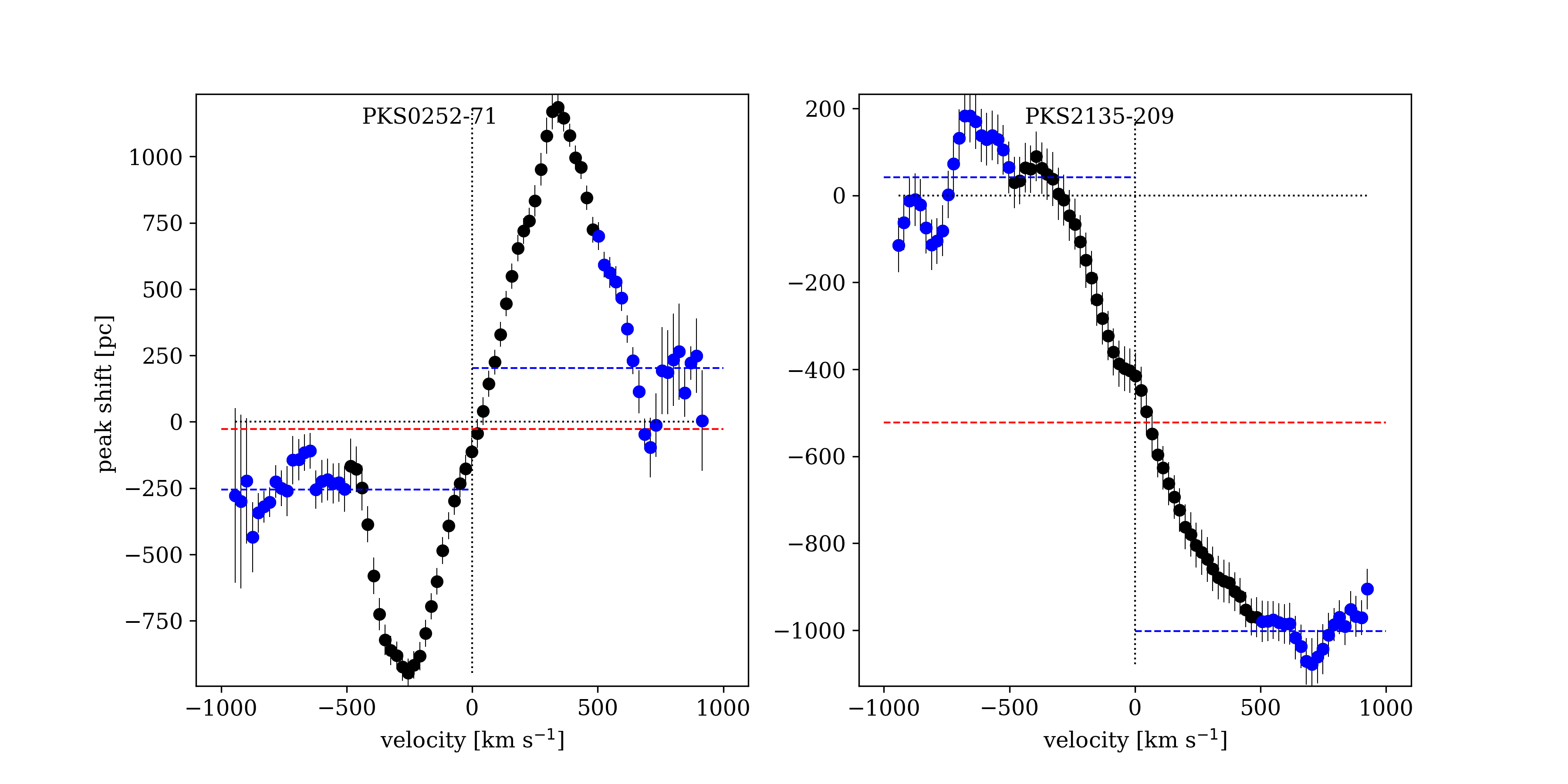}
\caption{Position-velocity diagrams obtained by applying the spectro-astrometry technique to the \OIII$\lambdaup$5007\AA\ line in the slit spectra of PKS~0252--71 and  PKS~2135--209. The diagrams show the position of the fitted centroids of the \OIII$\lambdaup$5007\AA\ emission line spatial profiles, expressed as  offsets in parsecs from the host galaxy centre, as function of velocity measured with respect to the \OIII$\lambdaup$5007\AA\ rest-frame velocity. The black dotted vertical and horizontal lines mark the zero point of the two axes. The blue dashed lines mark the error weighted mean position of the \OIII\ spatial profiles offsets at v<500\kms and at v>500\kms (shown in blue), while the red dashed line marks their average.} \label{pvpot}
\end{figure*}

The radial extent of the outflowing gas is one of the key parameters for estimating the outflow properties, but also
one of the hardest to measure \citep[see][and references therein]{Harrison2018}. Here we adopt the approach of first attempting to estimate the outflow radii using optical HST imaging and X-shooter spectroscopy observations.

Potentially, the most direct estimates of the outflow radii are provided by our HST \OIII\ images, which are available for PKS~0023--26, PKS~1306--09 and PKS~1549--79 (see section 2.2.2, Fig 1, \citealp{Batcheldor2007}  and \citealp{Oosterloo2019}). The main assumption here is that the \OIII\ emission detected in the HST images is dominated by the outflows. 
It should be noted that while this assumption fully holds for PKS~1549--79, whose \OIII\ emission line profile is completely dominated by the outflow (see Sec. \ref{sec:kinmodels}, also \citealp{Oosterloo2019}), it might lead us to overestimate the outflow spatial extents in the other two galaxies if there is a contribution from kinematically-quiescent emission-line gas that lies at larger radial distances from the nuclei than the radio sources 
However, in both PKS~0023--26 and PKS~1306--09 the brightest off-nuclear emission regions are situated along the direction of the radio jets. This agrees with results for other CSS/GPS sources \citep[e.g.][]{deVries1997,Axon2000} and, more generally, for many radio galaxies which show kinematically disturbed \OIII-emitting gas at the location of the radio jets \citep[e.g.][]{Clark1998,VillarMartin1999,Morganti1997,Villar-Mart2017}. We thus consider our initial assumption to be reasonable for our type of source. 

In each of the three objects with HST imaging, we estimate the outflow radius as the distance between the continuum nucleus of the galaxy and the position of the maximum in the flux of the off-nuclear emission in the continuum-subtracted \OIII\ images. These HST estimates for the warm outflow radii are within a factor of two of the radio source radii (see Table~\ref{Table_radius}). 
However, it is perhaps surprising that the brightest off nuclear emission-line region in PKS~1306--09 is apparently situated well beyond the radio source, despite its close alignment with the radio source axis suggesting a jet-cloud interaction. One possible explanation for this is that, rather than the two bright radio components detected in the VLBI observations of PKS~1306--09 being symmetrically placed on either side of the nucleus, as we have assumed, one such ``lobe'' is centred on the nucleus (i.e. the source is highly asymmetric). In that case, the radio source extent (2.2\,kpc) would be similar to the \OIII\ extent (1.9\,kpc). We note  that \cite{2002A&A...392..841T} find that only 30\% of the total radio emission at 2.29\,GHz in PKS~1306--09 is recovered
in their VLBI observations. This leaves open the possibility that there is substantial diffuse radio emission that is resolved out at VLBI resolution; some of this diffuse emission may be situated further to the NW than the ''lobe'' detected in their image.

An alternative to direct imaging for measuring the outflows extent is to use the spectro-astrometry technique \citep[see][]{Santoro2018}. This has the advantage that it can be used to isolate the broad wings associated with the outflowing gas, and measure their spatial extents in the direction of the spectroscopic slit. However, since it is likely that the warm outflows in CSS/GPS sources are closely aligned with the radio axes (see above), this can only be reliably done for objects in which the  X-shooter slit PA is reasonably well aligned with the radio axis (i.e. within 20 degrees, see Sec. \ref{sec:observations}). Three objects in our sample fulfil this criterion: PKS~0252--71, PKS~1934--63 and PKS~2135-20.

Spectro-astrometry  measurements for PKS~1934--63 have already been presented in \citet{Santoro2018}. Following the approach described in that paper, we built the position-velocity diagrams shown in Fig.~\ref{pvpot} for PKS~0252--71 and PKS~2125-20 by using the region of the slit spectrum around the high S/N \OIII$\lambdaup$5007\AA\ emission line. In each case, we extracted spatial slices from the long-slit spectra at different rest-frame velocities across the  \OIII$\lambdaup$5007\AA\ profile, subtracted an appropriately scaled continuum slice, then fitted the resulting \OIII\ spatial profiles with  Gaussians in order to determine the spatial centroids. To increase the S/N of the spatial profiles, especially in the case of PKS~0252--71, for every velocity we extracted the spatial slice over 5 pixels (corresponding to 2.5\AA\ or $\sim$90\kms\ in the wavelength direction). The continuum slices were extracted  on the blue and red sides of the \OIII$\lambdaup\lambdaup$4958,5007\AA\  doublet over the wavelength intervals indicated in Table~\ref{Table_spectralwindows}.
The position-velocity diagrams in Fig.~\ref{pvpot} show the offsets of the fitted centroids of \OIII\ spatial profiles relative to the galaxy continuum centroid as a function of the rest-frame velocity, where the latter was determined using the redshift derived from the stellar population fitting (see Table~\ref{Table_redshift}).

As can be seen in Fig.~\ref{pvpot}, both galaxies show an S-shaped profile that appears symmetric with respect to the systemic velocity, similar to what has been found for PKS~1934--63 \citep{Santoro2018}. There is a clear velocity gradient around the systemic velocity that likely reflects the gas rotation within the galaxies due to gravitational motions, or alternatively a low velocity bipolar outflow. However, at larger velocities, where we clearly probe the outflowing gas, the profile flattens out. 
For  PKS~2135--209 the overall curve is spatially shifted with respect to the zero point along the y axis (i.e. the putative centre of the galaxy). This is likely to be due to dust obscuration, which can potentially shift the position of the galaxy spatial profile peak relative to the position of the AGN.
We use the error-weighted mean positions of the gas at v<500\kms and at v>500\kms (blue dashed lines in Fig.\ref{pvpot}) as an indicators of the approaching and receding positions of the outflowing gas. In both objects we observe a significant offset between these two values, supporting the idea of a bi-polar geometry for the outflows. Under the assumption of a bi-polar outflow we use the latter error-weighted mean positions to estimate the true AGN nucleus position (red dashed line in Fig.\ref{pvpot}) as their average value, and the radial extent of the outflow as half of their separation. 

The outflow radii that we find using the spectro-astrometry method are reported in Table~\ref{Table_radius}. While for  PKS~2135--209 the outflow radius agrees well with the radio source radius, similar to the results found for PKS~1934--63 in \citet{Santoro2018}, in the case of PKS~0252--71 the estimated outflow radius is approximately half the radial extent of the radio source. The latter result can be explained if the outflow maintains a roughly constant surface brightness as a function of radius out to the full extent of the radio source, rather than being concentrated at the edges of the radio lobes.

Finally, for the remaining three galaxies -- PKS~1151--34 and PKS~1814--63, PKS~2314+03 -- we followed the method of \citet{Rose2018} and \cite{Spence2018} and compared the measured FWHM of the spatial profiles of the broad wings of \OIII\ or H$\beta$ emission lines with the seeing FWHM from Table~\ref{Table_obs} (see Sec. \ref{sec:X-shooter-observations}), in order to determine whether these profiles are spatially resolved. In this case, the
emission-line and continuum slices were extracted from the long-slit spectra over the  rest-frame velocity/wavelength ranges given in Table~\ref{Table_spectralwindows}. The continuum slices were scaled to take into account the different window widths, and then subtracted from the  broad-wing spatial profiles of the emission lines before fitting them with Gaussians.
We found for all three objects the outflows are spatially unresolved in the direction of
the X-shooter slit, in the sense that the \OIII\ FWHM are within $3\sigma$ of the seeing FWHM. Therefore, we followed \citet{Rose2018} and determined upper limits on the outflow radii
using the following formula:
\begin{equation}
r \leq \frac{1}{2} \sqrt{({\rm FWHM}+3\sigma)^2-{\rm FWHM}^2}
.\end{equation}
\noindent
Reassuringly, the resulting upper limiting radii are all larger than the estimated radio
source radial extents. Therefore, for these three objects we take the radio source radial extents as
the best available estimates of the warm outflow radii. This is justified on the basis of results obtained above and in previous studies
on the similarities between the scales and position angles of the radio sources and extended
\OIII\ emission-line regions.

To summarise, in cases where we can measure the radial extents of the warm outflows in the CSS/GPS objects using HST narrow-band imaging or X-shooter spectro-astrometry, we find that they are relatively compact -- within a factor 2 of the radio source extents -- and fall in the range $0.06 < r < 1.9$\,kpc. Interestingly, this range is similar to that  measured for the warm outflows in nearby ultra-luminous infrared galaxies (ULIRGs) by \citet{Rose2018}, \citet{Spence2018} and \citet{Tadhunter2018}, despite the fact that the outflows in CSS/GPS source are likely to be driven by the radio jets, whereas those in the most ULIRGs are probably driven by hot winds accelerated by radiation pressure close to the AGN. In subsequent calculations of the general properties of the warm outflows we use the best available
outflow radius estimate available for each object, as derived using the
technique indicated in the final column of
Table~\ref{Table_radius}.

\subsection{Gas ionisation mechanism}\label{ionisationparam}

The warm gas ionisation mechanism can potentially provide clues to the outflow acceleration mechanism. For example, if it could be shown that the gas were shock ionised, this would provide unambiguous evidence that the outflow have been accelerated in shocks. In addition, it is important to establish whether the outflow is predominantly ionised by the AGN (photoionisation or shocks) or by the radiation emitted by young stellar populations in the host galaxy, since in the latter case it would be less clear that the outflow is associated with the AGN activity.

\begin{figure*}
    \centering
    \includegraphics[width=0.9\textwidth]{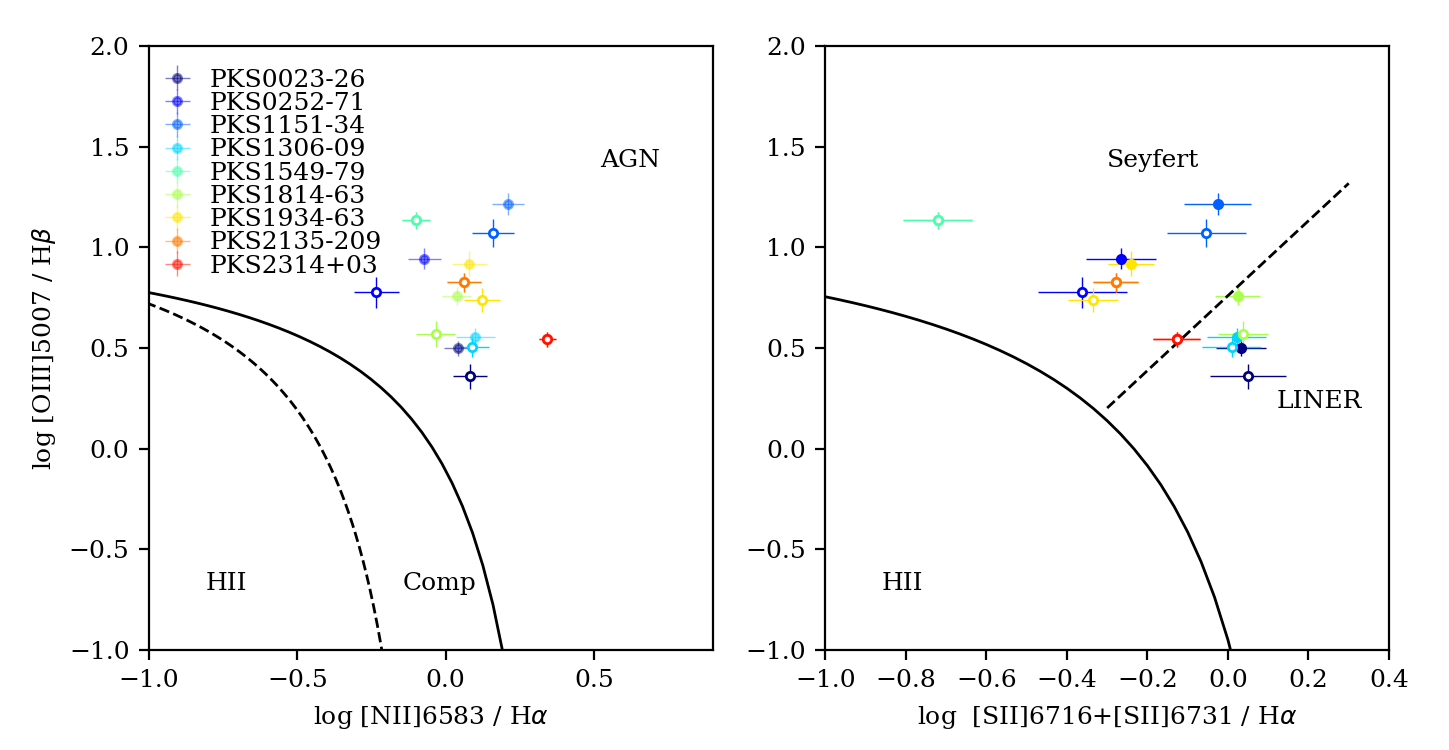}
    \caption{ Locations of our targets in the \OIII/H$\beta$ vs \NII6583/H$\alpha$ and \OIII/H$\beta$ vs \SII6716,31/H$\alpha$ BPT diagrams. Filled circles mark the line ratios obtained from total line fluxes while empty circles are related to the line emission of the broad components only. 
    The solid line in both panels is the \cite{2001ApJS..132...37K} maximum starburst line. The dashed line in the left panel is the semi-empirical \cite{2003MNRAS.346.1055K} line and has bee used together with the \cite{2001ApJS..132...37K} line to separate between line ratios due to photoionisation from stars (HII), AGN or a mixture of both (Comp). 
    The dashed line in the right panel is the \cite{2006MNRAS.372..961K} line separating the AGN between Seyferts and LINERS.
    } 
    \label{fig:BPT}
\end{figure*}

In Fig.~\ref{fig:BPT} we show the location of our targets in two of the classical BPT diagrams \citep{1981PASP...93....5B}. Clearly, not only the outflow component but also the total line emission in our targets shows line ratios that are consistent with AGN ionisation; in none of our sources is photoionisation by the young stellar populations significant. However, based on these diagrams alone, it is not possible in most cases to distinguish between AGN photoionisation  and shocks (e.g. driven by the expanding radio
jets), since the predictions of these two types of ionisation models show strong overlap in the diagrams \citep[e.g][]{RodriguezZaurin2013,Santoro2018}. 

Although some studies have attempted to determine the ionisation mechanism in a more decisive way using fainter diagnostic emission lines such as \OIII$\lambda$4363\AA\ and HeII$\lambda$4646\AA, the results have proved ambiguous \citep[e.g.][]{Holt2009}, apart perhaps from the case of PKS~1934--63 where evidence for shock ionisation of one of the broader emission-line components was found \citep{Santoro2018}. Reasons for the failure to decisively determine the dominant ionisation mechanism using such methods include (a) the low S/N of the faint diagnostic emission lines and their sensitivity to the accuracy of the subtraction of the underlying continuum; (b) the fact that some of the faint lines are in blends
(e.g. \OIII$\lambda$4363\AA), with all the attendant problems of degeneracy and the problems this causes for determining individual line fluxes in the face complex, broad line profiles.

Given the issues surrounding the diagnostic line ratio approach to determining the dominant ionisation mechanism, we adopted the alternative method described in \cite{Baron2017}, which is based on determining whether the H$\beta$ emission-line {\it luminosity} can be reproduced by shock models. Under the assumption that shocks are the dominant ionisation mechanism,  we used the location of our galaxies in the two BPT diagrams shown in Fig.~\ref{fig:BPT} to isolate the shock models which produce line ratios within 0.3 dex of those measured for the outflowing gas. We considered pre-computed shock model grids (with and without precursor) taken from MAPPING~III and spanning different shock velocities, magnetic parameters and pre-shock gas densities \citep[see][and references therein for details on the models]{Baron2017}.
We then extracted, for each galaxy, the H$\beta$ surface brightness of the selected shock models, and predicted the emitting area that the shocked gas should have assuming that it is uniformly distributed in a thin spherical shell with a filling factor of 100\%, and has an H$\beta$ luminosity equal to that measured for the outflowing gas. By comparing the areas and hence radii of the outflows predicted in this way to our observationally-determined estimates of the outflow radii (see Sec.\ref{radii}) it was then possible to test whether shock ionisation is feasible.  

We found that, if the gas were solely ionised by shocks, we would need to observe outflows extending on scales which are larger then the observed ones by a factor between 1.5 and 5 when looking at the entire sample. This means that for all our targets the observed luminosities are far too high (by a factor between 2 and 25) to be produced purely by shocks alone, and that the dominant ionisation mechanism for the warm outflowing gas is most likely to be AGN photoionisation. We note that this argument is conservative in the sense that the gas is likely to be highly clumped rather than uniformly distributed in the putative shocked shell (i.e. filling factor $<<$100\%), and that it is also unlikely that any shocked regions are spherical, 
given the often highly collimated structures visible in emission-line
images of CSS/GPS sources (e.g. Fig. 1). 

Having established that the ionisation of the outflowing gas is likely to be dominated by AGN photoionisation, we can then estimate the ionisation parameter $U$ -- the ratio of the flux density of ionising photons at the face of the ionised cloud to the electron density, normalised by the speed of light. In the following,  estimates of the gas ionisation parameter will be used to isolate the fiducial photoionisation models that are required to determine the electron densities and reddening, as shown in Sec.\ref{densities}.
Using the calibration reported in \citet{Baron19b} (their equation 2) we  estimated the ionisation parameter for the total and the outflowing gas by using the \OIII/H$\beta$ and the \NII6583/H$\alpha$ line ratios.
The derived ionisation parameters are reported in Table~\ref{Table_ionparam}, while the aforementioned line rations are reported in Table~\ref{Table:BPTratios}. 

Finally, we emphasise that, while the outflowing gas is likely to be predominantly AGN photoionised rather than shock ionised, this does not rule out shocks as an acceleration mechanism for the gas, since it is plausible that any shock accelerated gas will be photoionised by the AGN as it cools behind the shock front. 

\begin{table}
\centering
\begin{tabular}{lcc}
\toprule 
Object & Log$U_{T}$ & Log$U_{out} $ \\
\toprule 
PKS~0023--26 & -3.49 $\pm$ 0.04 & -3.62 $\pm$ 0.05 \\
PKS~0252--71 & -2.87 $\pm $0.09 & -3.07 $\pm$ 0.11 \\
PKS~1151--34 & -2.42 $\pm $0.11 & -2.70 $\pm$ 0.13 \\
PKS~1306--09 & -3.44$ \pm$ 0.05 & -3.49$ \pm$ 0.05 \\
PKS~1549--79 & -2.52 $\pm$ 0.08 & -2.52$ \pm $0.08 \\
PKS~1814--63 & -3.18$ \pm $0.07 & -3.40 $\pm$ 0.07 \\
PKS~1934--63 & -2.95 $\pm$ 0.10 & -3.22$ \pm $0.08 \\
PKS~2135--209 & -3.09 $\pm$ 0.07 & -3.09$ \pm $0.07 \\ 
PKS~2314+03 & -3.48 $\pm$ 0.04 & -3.48 $\pm $0.04 \\
\bottomrule
\end{tabular}
\caption{Table reporting the ionisation parameters determined for the total  (col2) and outflowing gas (col3) emission, as determined using the method discussed in \citet{Baron19b}.}
\label{Table_ionparam}
\end{table}

\begin{table*}
\begin{tabular}{lcccccc}
\toprule 
Object & log \(\frac{[OIII]5007}{H\beta} _{T}\) & log \(\frac{[OIII]5007}{H\beta} _{out}\)  & log \(\frac{[NII]6583}{H\alpha}_{T}\) & log\(\frac{[NII]6583}{H\alpha}_{out}\)  & log \(\frac{[SII]6716+6731}{H\alpha}_{T}\) & log \(\frac{[SII]6716+6731}{H\alpha}_{out}\)  \\ [1ex]
\toprule 
PKS~0023--26 & 0.5$\pm$0.04 & 0.36$\pm$0.06 & 0.04$\pm$0.05 & 0.08$\pm$0.06 & 0.03$\pm$0.06 & 0.05$\pm$0.09 \\
PKS~0252--71 & 0.94$\pm$0.05 & 0.78$\pm$0.08 & -0.07$\pm$0.06 & -0.23$\pm$0.07 & -0.27$\pm$0.09 & -0.36$\pm$0.11 \\
PKS~1151--34 & 1.22$\pm$0.05 & 1.07$\pm$0.07 & 0.21$\pm$0.05 & 0.16$\pm$0.07 & -0.03$\pm$0.08 & -0.05$\pm$0.1 \\
PKS~1306--09 & 0.56$\pm$0.04 & 0.5$\pm$0.05 & 0.1$\pm$0.07 & 0.09$\pm$0.06 & 0.02$\pm$0.07 & 0.01$\pm$0.07 \\
PKS~1549--79 & 1.13$\pm$0.04 & 1.13$\pm$0.04 & -0.1$\pm$0.05 & -0.1$\pm$0.05 & -0.72$\pm$0.08 & -0.72$\pm$0.08 \\
PKS~1814--63 & 0.76$\pm$0.05 & 0.57$\pm$0.06 & 0.04$\pm$0.05 & -0.03$\pm$0.07 & 0.02$\pm$0.06 & 0.04$\pm$0.06 \\
PKS~1934--63 & 0.92$\pm$0.06 & 0.74$\pm$0.06 & 0.08$\pm$0.06 & 0.12$\pm$0.06 & -0.24$\pm$0.06 & -0.34$\pm$0.06 \\
PKS~2135--209 & 0.83$\pm$0.05 & 0.83$\pm$0.05 & 0.06$\pm$0.06 & 0.06$\pm$0.06 & -0.28$\pm$0.06 & -0.28$\pm$0.06 \\
PKS~2314+03 & 0.54$\pm$0.04 & 0.54$\pm$0.04 & 0.34$\pm$0.03 & 0.34$\pm$0.03 & -0.13$\pm$0.06 & -0.13$\pm$0.06 \\
\bottomrule 
\end{tabular}
\caption{Table reporting the logarithmic values of the \OIII/H$\beta$, \NII6583/H$\alpha$ and \SII6716+6731/H$\alpha$ line ratios of the total (col2, col4, co6) and  outflowing gas (col3, col5, col7) emission for the sources in our sample.}
\label{Table:BPTratios}
\end{table*}

\subsection{The density and reddening of the outflows}\label{densities}

\begin{table*}
\centering
\begin{tabular}{l|cccc|cccc}
\toprule 
Object & tr \OII$_{T}$ & tr \SII$_{T}$ & log(n$_{e}$)$_{T}$ & E(B-V)$_{T}$ & tr \OII$_{out}$ & tr \SII$_{out}$ & log(n$_{e}$)$_{out}$ & E(B-V)$_{out}$ \\ [0.5ex]
\toprule
PKS~0023--26 & 0.74$\pm$0.11 & -1.32$\pm$0.08 & 3.35$^{+0.06 }_{-0.49 }$ & 0.49$^{+0.10 }_{-0.08 }$ & >0.58 & <-1.24 & <3.53 & >0.53 \\
PKS~0252--71 & 0.84$\pm$0.12 & -0.51$\pm$0.09 & 3.84$^{+0.12 }_{-0.18 }$ & 0.00$^{+0.04 }_{-0.00 }$ & 0.59$\pm$0.15 & -0.35$\pm$0.11 & 4.14$^{+0.18 }_{-0.18 }$ & 0.00$^{+0.10 }_{-0.00 }$ \\
PKS~1151--34 & 0.74$\pm$0.14 & -1.05$\pm$0.19 & 3.53$^{+0.43 }_{-0.12 }$ & 0.27$^{+0.18 }_{-0.18 }$ & >0.41 & <-0.72 & <3.96 & >0.27 \\
PKS~1306--09 & 1.14$\pm$0.13 & -1.19$\pm$0.10 & 3.16$^{+0.80 }_{-0.49 }$ & 0.22$^{+0.12 }_{-0.12 }$ & >1.11 & <-1.16 & <3.22 & >0.20 \\
PKS~1549--79 & 0.19$\pm$0.11 & -0.98$\pm$0.12 & 3.90$^{+0.18 }_{-0.12 }$ & 0.49$^{+0.12 }_{-0.12 }$ & 0.19$\pm$0.11 & -0.98$\pm$0.12 & 3.90$^{+0.18 }_{-0.12 }$ & 0.49$^{+0.12 }_{-0.12 }$ \\
PKS~1814--63 & 0.32$\pm$0.14 & -1.61$\pm$0.13 & 3.35$^{+0.73 }_{-0.43 }$ & 0.88$^{+0.12 }_{-0.14 }$ & >0.20 & <-1.48 & <3.53 & >0.86 \\
PKS~1934--63 & 0.08$\pm$0.06 & -0.44$\pm$0.06 & 4.39$^{+0.06 }_{-0.06 }$ & 0.24$^{+0.08 }_{-0.06 }$ & -0.20$\pm$0.06 & -0.17$\pm$0.06 & 4.76$^{+0.06 }_{-0.06 }$ & 0.20$^{+0.06 }_{-0.06 }$ \\
PKS~2135--209 & 0.81$\pm$0.12 & -1.12$\pm$0.08 & 3.41$^{+0.18 }_{-0.18 }$ & 0.33$^{+0.10 }_{-0.12 }$ & 0.81$\pm$0.12 & -1.12$\pm$0.08 & 3.41$^{+0.18 }_{-0.18 }$ & 0.33$^{+0.10 }_{-0.12 }$ \\
PKS~2314+03 & 1.11$\pm$0.14 & -1.29$\pm$0.09 & 3.10$^{+0.18 }_{-0.25 }$ & 0.29$^{+0.14 }_{-0.12 }$ & 1.11$\pm$0.14 & -1.29$\pm$0.09 & 3.10$^{+0.18 }_{-0.25 }$ & 0.29$^{+0.14 }_{-0.12 }$ \\
\bottomrule
\end{tabular}
\caption{Table reporting the trans-auroral \OII\ and \SII\ ratios, together with the derived density $n_{e}$ and E(B-V) values for the total (col 2, col 3, col 4 and col 5) and  the outflowing (col 6, col 7, col 8 and col 9) gas emission, as determined from the DDD using total line fluxes and broad-component line fluxes respectively.}
\label{Table_densities}
\end{table*}

\begin{figure*}[]
\begin{center}
\includegraphics[width=0.9\textwidth]{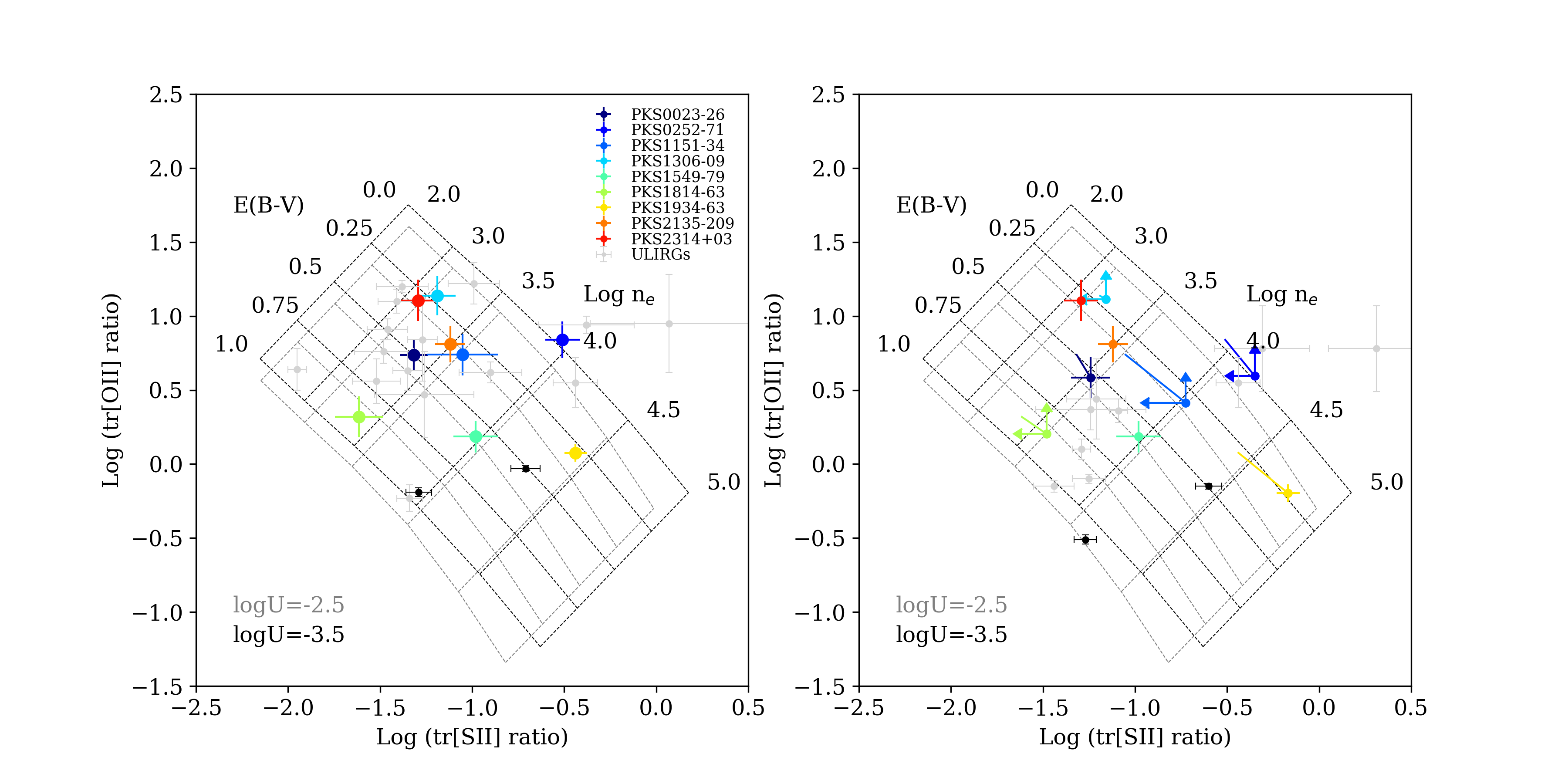}
\caption{ Figure showing the location of our targets in the density diagnostic diagram (DDD) using the logarithm of the tr\OII ={\rm[O}\,{\rm \scriptsize II}{\rm]} (3727+3729)/(7318+7319+ 7330+7331) and of the tr\SII={\rm[S}\,{\rm \scriptsize II}{\rm]} (4068+4076)/(6716+6731) line ratios. Error bars have been estimated by propagating the line flux errors while upper/lower limits have been assigned as described in the text; different sources in our sample are marked with different colours as indicated by the legend in the upper right corner of the left panel. For comparison, the DDD also includes the results obtained for the ULIRGs studied by \cite{Spence2018} and \cite{Rose2018} (thin light grey points), two of which also host a compact radio source (thin black points).
AGN photoionisation grid models for gas with solar metallicity and two ionisation parameters --- log$U$=-2.5 (light grey color) and log$U$=-3.5 (black color) --- are shown with dotted lines. The model grids have been created by fixing the gas metallicity and ionisation parameter while varying the electron density in the interval $n_{\rm e}=100-10^5~\rm{cm^{-3}}$ (from top left to bottom right) and the reddening in the interval E(B-V)=0-1~mag (from top right to bottom left).  \textit{Left Panel}. DDD showing the line ratios based on the total line fluxes. \textit{Right Panel}. DDD showing the outflows line ratios determined from the integrated fluxes of broad components of the emission lines. For each galaxy in our sample a solid line connects the line ratios of the total (as shown in the left panel) and outflowing gas emission.} \label{DDD}
\end{center}
\end{figure*}

Along with the radius, the electron density is a key parameter for determining the properties of the warm outflowing gas. Here we use the density diagnostic diagram (DDD heareafter) approach, first introduced by \cite{Holt2011}, which is sensitive to high electron densities and able to overcome the main limitations of the classical \OII\ and/or \SII\  line ratios \citep[see][for a discussion]{Rose2018, Baron19b}. 
Our X-shooter data are especially well-suited for this purpose, since, by covering a wide wavelength range, they have allowed us to detect and model all the emission lines needed to measure the trans-auroral \OII\ and \SII\ line ratios, as defined in the Introduction.
By comparing the measured line ratios with those of  fiducial photoionisation model grids in the DDD, this method provides estimates of the electron density and the reddening for both the total warm gas emission (i.e. total line fluxes, shown in the left panel of Fig.~\ref{DDD}) and, in some cases, for that of the outflowing gas alone (i.e. broad component integrated line fluxes, shown in the right panel of Fig.~\ref{DDD}). The trans-auroral line fluxes for the total and outflowing gas emission are reported in Table~\ref{Table_transflux_1} of Appendix~\ref{AppendixFits}. 

Some caution is required when using the DDD approach, since the \SII$\lambda\lambda$4069,4076\AA\ and \OII$\lambda\lambda$7319,7330\AA\ lines involved in the trans-auroral line ratios arise from transitions with upper energy levels that have higher energies than those involved in the classical \SII\ and \OII\ ratios. Therefore, trans-auroral ratios are potentially sensitive to the electron temperature of the emitting gas, which in the case of AGN photoionisation, depends on the ionisation parameter $U$, the ionising continuum shape, and the metallicity. To test the sensitivity of the DDD technique to these parameters, we have run a new set of models using {\sc cloudy} version 17.00 \citep{Ferland17}. The results of these models are presented in detail in Appendix~\ref{appendixModels} together with description of the model set-up, assumptions and range of parameters considered. 

Overall, we find that the effect of varying the ionisation parameter, ionising continuum shape and metallicity on the density values determined using the DDD technique is typically at the level of 0.3\,dex (factor $\sim$2) or less. This is illustrated in Fig.\ref{DDD}, where, along with the measured trans-auroral ratios for the CSS/GPS sources, we show two grids of models, one calculated
for a high (i.e. log$U$=-2.5) and the other for a low (log$U$=-3.5) ionisation parameter and solar metallicity. Both grids were created by varying the electron densities in the interval $n_{\rm {e}}=100-10^5~\rm{cm^{-3}}$ and the  reddening ($E(B-V)$)  between zero and one (adopting the \citealt{Cardelli1989} extinction law). The shape of the AGN ionising continuum SED has been chosen by adopting standard assumptions for an AGN and a mean ionising photon energy of 2.56 Ryd (see Appendix~\ref{appendixModels} for further details on this choice). It is important to stress that the changes in the positions of DDD grids induced by varying the model parameters are typically within the observational errors estimated for the tr \OII\ and \SII\ line ratios.

While significant, we emphasise that the level of uncertainty due to varying the model parameters in the DDD approach is  far lower than the orders of magnitude uncertainty associated with using the classical \SII\ and \OII\ ratios  or by {\it assuming} a low electron density ($\sim$100 -- 200\,cm$^{-3}$), as has been done in some studies.

In order to determine the best possible density and reddening estimates for the outflows, we selected a specific model grid for each object (see Appendix~\ref{appendixModels} for details). In the absence of a robust metallicity calibration for AGN-photoionised gas, we chose to use model grids with solar metallicity. On the other hand, we were able to select specific reference model grids (one grid for the total and one for the outflowing gas emission) for each object,  based on the ionisation parameter $U$ estimates determined from the line ratios in Sec.~\ref{ionisationparam}. 
The electron density and reddening estimates derived by comparing the measured trans-auroral ratios with the object-specific DDD grids are presented in Table~\ref{Table_densities}, together with the respective trans-auroral \OII\ and \SII\ line ratio measurements.

By comparing the observed line ratios to the models in Fig.~\ref{DDD}, it is clear that most of our targets show high gas electron densities that are sometimes well outside the regime that can be probed by classical density diagnostics. 
For PKS~1934--63 we find that the gas density of the outflow is higher than that measured using the total line fluxes, in line with the trend seen in \cite{Santoro2018}. The same trend is seen also for PKS~0023-26 even though the increase in density is less pronounced and the density values are compatible within the errors. The upper limits and values obtained for the densities of the outflow components (right panel of Fig.~\ref{DDD}) suggest a similar behaviour for the remaining sources; however, the sensitivity of our observations does not allow us to confirm this trend in a definitive way.
We find no significant difference in the reddening between the total emission and the outflows.

It should be noted that for PKS~0023--26, PKS~1549--79, PKS~1934--63 and PKS~2135--209  we can isolate the emission from the outflowing gas (i.e. the broad kinematic components) in all the trans-auroral lines and thus have a precise estimate of its density. The choice of using only total line fluxes for PKS~1549--79, PKS~2135--209 and PKS~2314+03 gives us a single estimate of the gas density and dust attenuation which can be considered representative of the outflowing gas whose emission prevails in the line profiles (see Sec.\ref{gaskinematics}).

As previously mentioned, broad kinematic components are often faint and difficult to detect in the  \OII$\lambdaup\lambdaup$7319,30\AA\ and \SII$\lambdaup\lambdaup$4069,76\AA\ doublets.
For the targets where this happens we estimate a tr\OII\ lower limit and a  tr\SII\ upper limit by considering the total fluxes of the \OII$\lambdaup\lambdaup$7319,30\AA\ and \SII$\lambdaup\lambdaup$4069,76\AA\ lines respectively as upper limits on the fluxes of the outflowing gas. By construction of the DDD, for these cases we can then extract an upper limit on the density and a lower limit on the reddening of the outflowing gas.

In Fig.~\ref{DDD} we also include sources from the sample of nearby ULIRGs presented in \cite{Rose2018} and \cite{Spence2018},  which have been studied using a similar DDD technique. Two of these sources --- F~23389+0303N  and  PKS~1345+12 --- are highlighted in the DDD because they are are known to host high power ($P_{1.4GHz} > 10^{25}$\,W Hz$^{-1}$), compact ($D_L <1$\,kpc) radio sources associated with their AGN activity, and are therefore similar in their radio properties to the CSS/GPS sources considered in this paper. Clearly, the warm outflows in ULIRGs show a similar range of electron density ($3.45 < log(n_e~\rm{cm}^{-3}) < 4.75$) and reddening ($0 < E(B-V) < 1$\,mag) to those in the CSS/GPS objects. It should be noted that also for F~23389+0303N and  PKS~1345+12 the density of the outflowing components are higher then the densities of the whole ISM estimated using total line fluxes.

In Appendix~\ref{appendixComparison} we discuss the comparison between our measurements and i) the dust attenuation estimated via the classical H$\alpha$/H$\beta$ Balmer decrement, ii) the gas densities estimated using an alternative method proposed by \cite{Baron19b}. 

\subsection{AGN bolometric luminosties}\label{subsec:Lbolometric}

\begin{table*}
\centering
\begin{tabular}{lccccc}
\toprule
Object & F\OIII$_{T}$ & L\OIII$_{obs}$ & L\OIII$_{intr}$ & L$_{BOL~L09}$ & L$_{BOL~H04}$ \\
\toprule
PKS~0023--26 & $\left(3.3 \pm 0.2\right) \times 10^{-15}$ & $\left(1.1 \pm 0.1\right) \times 10^{42}$ & $\left(5.4 \pm 1.8\right) \times 10^{42}$ & $\left(2.5 \pm 0.8\right) \times 10^{45}$ & $\left(4.0 \pm 0.2\right) \times 10^{45}$ \\
PKS~0252--71 & $\left(3.1 \pm 0.2\right) \times 10^{-15}$ & $\left(4.1 \pm 0.2\right) \times 10^{42}$ & $\left(4.1 \pm 0.6\right) \times 10^{42}$ & $\left(1.8 \pm 0.3\right) \times 10^{45}$ & $\left(1.4 \pm 0.1\right) \times 10^{46}$ \\
PKS~1151--34 & $\left(1.2 \pm 0.1\right) \times 10^{-14}$ & $\left(2.5 \pm 0.2\right) \times 10^{42}$ & $\left(5.8 \pm 3.4\right) \times 10^{42}$ & $\left(2.6 \pm 1.6\right) \times 10^{45}$ & $\left(8.7 \pm 0.6\right) \times 10^{45}$ \\
PKS~1306--09 & $\left(1.5 \pm 0.1\right) \times 10^{-15}$ & $\left(1.3 \pm 0.1\right) \times 10^{42}$ & $\left(2.6 \pm 1.0\right) \times 10^{42}$ & $\left(1.2 \pm 0.5\right) \times 10^{45}$ & $\left(4.5 \pm 0.3\right) \times 10^{45}$ \\
PKS~1549--79 & $\left(4.3 \pm 0.3\right) \times 10^{-14}$ & $\left(2.7 \pm 0.2\right) \times 10^{42}$ & $\left(1.3 \pm 0.5\right) \times 10^{43}$ & $\left(6.0 \pm 2.4\right) \times 10^{45}$ & $\left(9.6 \pm 0.6\right) \times 10^{45}$ \\
PKS~1814--63 & $\left(3.5 \pm 0.2\right) \times 10^{-15}$ & $\left(3.4 \pm 0.2\right) \times 10^{40}$ & $\left(5.7 \pm 2.3\right) \times 10^{41}$ & $\left(8.1 \pm 3.2\right) \times 10^{43}$ & $\left(1.2 \pm 0.1\right) \times 10^{44}$ \\
PKS~1934--63 & $\left(1.0 \pm 0.1\right) \times 10^{-14}$ & $\left(9.5 \pm 0.9\right) \times 10^{41}$ & $\left(2.1 \pm 0.6\right) \times 10^{42}$ & $\left(9.4 \pm 2.6\right) \times 10^{44}$ & $\left(3.3 \pm 0.3\right) \times 10^{45}$ \\
PKS~2135--209 & $\left(5.9 \pm 0.4\right) \times 10^{-15}$ & $\left(1.0 \pm 0.1\right) \times 10^{43}$ & $\left(2.9 \pm 1.0\right) \times 10^{43}$ & $\left(1.3 \pm 0.4\right) \times 10^{46}$ & $\left(3.6 \pm 0.3\right) \times 10^{46}$ \\
PKS~2314+03 & $\left(4.7 \pm 0.1\right) \times 10^{-15}$ & $\left(6.8 \pm 0.1\right) \times 10^{41}$ & $\left(1.7 \pm 0.8\right) \times 10^{42}$ & $\left(7.7 \pm 3.5\right) \times 10^{44}$ & $\left(2.4 \pm 0.04\right) \times 10^{45}$ \\

\bottomrule
\end{tabular}
\caption{ Table reporting the total \OIII5007\AA\ line fluxes (col 2), the \OIII\ observed (col3) and dust corrected (col4) luminosities and the AGN bolometric luminosities calculated following the \cite{2009A&A...504...73L} (col5) and the \cite{Heckman2004} (col6) calibrations. Fluxes are given in units of $\rm{erg~s^{-1}cm^{-2}\AA^{-1}}$ while luminosities are in $\rm erg~s^{-1}$.
The distances of the targets were estimated using the redshifts reported in Table~\ref{Table_redshift}, and to apply the dust extinction correction we used the DDD-derived E(B-V) values for total line fluxes reported in column 5 of Table~\ref{Table_densities}.}
\label{Table_OIII}
\end{table*}

The bolometric luminosity ($L_{BOL}$) provides a key indication of the overall level of radiative AGN activity, and in the following will be compared with the outflow kinetic power, in order to derive the feedback efficiency -- an important parameter for comparison with AGN feedback models. Determinations of $L_{BOL}$ using optical continuum measurements are challenging for most of the objects in our sample, which are Type 2 AGN, because
the direct AGN continuum is blocked out by circum-nuclear dust. Even in wavelength regions that are less strongly affected by dust extinction, the direct emission from the energy-generating regions close to the AGN of the CSS/GPS sources is potentially contaminated by non-thermal emission from the radio sources \citep[mid-to-far infrared and X-ray wavelength regions:][Dicken et al. in prep.]{Dicken2008,Ming2014} and/or emission from star formation regions \cite[particularly mid-to-far infrared:][]{Dicken2008,Dicken2012}. Therefore, we must rely on emission-line indicators of $L_{BOL}$.

We have considered two emission-line based bolometric
luminosity indicators. The first uses the total \OIII$\lambdaup$5007\AA\ luminosity ($L_{[OIII]}$) along with the $L_{[OIII]}$-to-$L_{BOL}$ correction factor of 3500 derived by
\citet{Heckman2004}. Note that this method uses $L_{[OIII]}$ values 
that have not been corrected for dust extinction. In contrast, the second indicator -- developed by \citet{2009A&A...504...73L} -- uses $L_{[OIII]}$ values corrected for dust extinction, as well as luminosity-dependent correction factors. When applying this latter method to our CSS/GPS sample, the extinction correction was performed using the E(B-V) values measured as part of the DDD analysis described above. The results are
shown in Table~\ref{Table_OIII}, from which it is clear that the bolometric luminosities derived using the \citet{Heckman2004} method are systematically higher, by typically a factor of a few but up to a factor of ten, than those derived using the \citet{2009A&A...504...73L} method.

Both the above methods assume that the covering factor of the \OIII-emitting gas is the same for all AGN
of a given $L_{BOL}$. However, this is not necessarily the case, and indeed there are reasons to believe
that covering factor may in reality be higher in CSS/GPS than in typical AGN: first, because these are young 
radio sources in an early stage of evolution, the circum-nuclear regions may not yet have been swept clear
by AGN-induced outflows \citep{Tadhunter1999}, thus leading to larger amounts of gas in the narrow-line region (NLR)
\citep[see discussion in][]{Tadhunter2001}; second, 
there is evidence from mid-to-far infrared and optical observations that CSS/GPS sources have higher
rates of star formation than more typical, extended radio galaxies \citep{Tadhunter2011,Dicken2014}, 
thus suggesting more gas-rich near-nuclear
environments that could, potentially, also be associated with high NLR covering factors.

In this context, it is interesting to note that if we attempt to estimate the gas electron densities following the technique of \citet{Baron19b}, based on measurements of ionisation parameter $U$, AGN bolometric luminosity $L_{BOL}$,
and radius $r$ (N.B. $n_e \propto L_{BOL}/(r^2 U)$), the derived values are systematically higher
than those we derived using the DDD technique above (see Appendix~\ref{appendixComparison} for a detailed discussion). This applies to densities derived using both the \citet{Heckman2004} and \citet{2009A&A...504...73L} $L_{BOL}$ estimates, but the discrepancy is higher in \citet{Heckman2004} case.
Part of the reason for this apparent discrepancy might be that, in making this calculation, 
we have not corrected our radius estimates
for line-of-sight projection effect. However, assuming that the radio sources are oriented at random, the 
mean angle ($\theta$) of the the radio axis relative to the line of sight is expected to be $\theta \sim 57.3$\,degrees. The
implied mean correction factor is then 1.8 for $r$, or 3.4 for $r^2$, which is not sufficient alone to explain the
density discrepancy. One way to bring the density estimates fully into agreement would
be to decrease the $L_{BOL}$ values; this provides indirect evidence that the $L_{BOL}$ values derived using both the emission-line methods may be over-estimated by a factor 2 or more. As we discuss in detail in Appendix C, alternative explanations for the discrepancy in terms of $U$ being overestimated from the line ratios, or the trans-auroral \SII\ and \OII\ lines being emitted by lower density clouds situated at larger radii than those emitting \OIII\ and H$\beta$ lines, are less plausible.

In the following sections we will use the lower $L_{BOL}$ values derived via the \citet{2009A&A...504...73L}  approach, since they take into account dust extinction, and we feel that they are more likely to reflect  the true $L_{BOL}$ values, given the above discussion. However, it is important to bear in mind that even these lower $L_{BOL}$ values may represent over-estimates if the covering factor of the \OIII-emitting gas is substantially higher in CSS/GPS sources than in the general population of AGN used to derive the calibration. 

\section{General outflow properties}\label{outProperties}

\subsection{Warm gas masses}\label{subsec:Outf_Masses}

\begin{table*}
\resizebox{\textwidth}{!}{
\begin{tabular}{lcccccc}
\toprule
Object & F H$\beta_{T}$ & L H$\beta_{T~obs}$ & L H$\beta_{T~intr}$ & F H$\beta_{out}$ & L H$\beta_{out~obs}$ & L H$\beta_{out~intr}$ \\
\toprule
PKS~0023--26 & $\left(1.1 \pm 0.1\right) \times 10^{-15}$ & $\left(3.7 \pm 0.2\right) \times 10^{41}$ & $\left(1.9 \pm 0.6\right) \times 10^{42}$ & $\left(5.2 \pm 0.5\right) \times 10^{-16}$ & $\left(1.8 \pm 0.2\right) \times 10^{41}$ & $\left(1.1 \pm 0.5\right) \times 10^{42}$ \\
PKS~0252--71 & $\left(3.5 \pm 0.4\right) \times 10^{-16}$ & $\left(4.7 \pm 0.5\right) \times 10^{41}$ & $\left(4.7 \pm 0.8\right) \times 10^{41}$ & $\left(2.3 \pm 0.3\right) \times 10^{-16}$ & $\left(3.0 \pm 0.4\right) \times 10^{41}$ & $\left(3.0 \pm 1.1\right) \times 10^{41}$ \\
PKS~1151--34 & $\left(7.3 \pm 0.8\right) \times 10^{-16}$ & $\left(1.5 \pm 0.2\right) \times 10^{41}$ & $\left(3.7 \pm 2.3\right) \times 10^{41}$ & $\left(4.2 \pm 0.5\right) \times 10^{-16}$ & $\left(8.9 \pm 1.1\right) \times 10^{40}$ & $\left(2.1 \pm 1.3\right) \times 10^{41}$ \\
PKS~1306--09 & $\left(4.3 \pm 0.3\right) \times 10^{-16}$ & $\left(3.6 \pm 0.3\right) \times 10^{41}$ & $\left(7.6 \pm 3.2\right) \times 10^{41}$ & $\left(4.0 \pm 0.3\right) \times 10^{-16}$ & $\left(3.4 \pm 0.3\right) \times 10^{41}$ & $\left(6.7 \pm 3.2\right) \times 10^{41}$ \\
PKS~1549--79 & $\left(3.2 \pm 0.2\right) \times 10^{-15}$ & $\left(2.1 \pm 0.1\right) \times 10^{41}$ & $\left(1.0 \pm 0.4\right) \times 10^{42}$ & $\left(3.2 \pm 0.2\right) \times 10^{-15}$ & $\left(2.1 \pm 0.1\right) \times 10^{41}$ & $\left(1.0 \pm 0.4\right) \times 10^{42}$ \\
PKS~1814--63 & $\left(6.1 \pm 0.6\right) \times 10^{-16}$ & $\left(6.1 \pm 0.6\right) \times 10^{39}$ & $\left(1.1 \pm 0.5\right) \times 10^{41}$ & $\left(4.2 \pm 0.5\right) \times 10^{-16}$ & $\left(4.2 \pm 0.5\right) \times 10^{39}$ & $\left(7.3 \pm 3.6\right) \times 10^{40}$ \\
PKS~1934--63 & $\left(1.2 \pm 0.1\right) \times 10^{-15}$ & $\left(1.2 \pm 0.1\right) \times 10^{41}$ & $\left(2.6 \pm 0.8\right) \times 10^{41}$ & $\left(7.6 \pm 0.8\right) \times 10^{-16}$ & $\left(7.2 \pm 0.7\right) \times 10^{40}$ & $\left(1.4 \pm 0.3\right) \times 10^{41}$ \\
PKS~2135--209 & $\left(8.8 \pm 0.7\right) \times 10^{-16}$ & $\left(1.6 \pm 0.1\right) \times 10^{42}$ & $\left(4.6 \pm 1.6\right) \times 10^{42}$ & $\left(8.8 \pm 0.7\right) \times 10^{-16}$ & $\left(1.6 \pm 0.1\right) \times 10^{42}$ & $\left(4.6 \pm 1.6\right) \times 10^{42}$ \\
PKS~2314+03 & $\left(1.4 \pm 0.1\right) \times 10^{-15}$ & $\left(2.0 \pm 0.2\right) \times 10^{41}$ & $\left(5.1 \pm 2.5\right) \times 10^{41}$ & $\left(1.4 \pm 0.1\right) \times 10^{-15}$ & $\left(2.0 \pm 0.2\right) \times 10^{41}$ & $\left(5.1 \pm 1.8\right) \times 10^{41}$ \\
\bottomrule
\end{tabular}
}
\caption{ Table reporting the H$\beta$ flux, the observed and intrinsic H$\beta$ luminosity of the total (col 2-3-4) and the outflowing gas (col 5-6-7). Fluxes are given in units of $\rm{erg~s^{-1}cm^{-2}\AA^{-1}}$ while luminosities in $\rm erg~s^{-1}$
The distances of the targets have been estimated using the redshifts reported in Table~\ref{Table_redshift}, to apply the dust correction we use the DDD-derived E(B-V) values reported in Table~\ref{Table_densities}.
}\label{Table_Hbeta}

\end{table*}

\begin{table}
\resizebox{\columnwidth}{!}{
\begin{tabular}{lccc}
\toprule
Object & $M_{T} [M_\odot]$ & $M_{out} [M_\odot]$ & $M_{out}/M_{T} [\%]$ \\
\toprule
PKS~0023--26 & $(5.7\pm 3.0)\times10^{6}$ & $(2.1\pm 1.4)\times10^{6}$ & $37\pm 31$ \\
PKS~0252--71 & $4.7\pm 1.76\times10^{5}$ & $>1.5\times10^{5}$ & $>32$ \\
PKS~1151--34 & $7.4\pm 6.57\times10^{5}$ & $>1.6\times10^{5}$ & $>22$ \\
PKS~1306--09 & $3.5\pm 2.13\times10^{6}$ & $>2.7\times10^{6}$ & $>77$ \\
PKS~1549--79 & $(9.0\pm 5.1)\times10^{5}$ & $(9.0\pm 5.1)\times10^{5}$ & $100$ \\
PKS~1814--63 & $3.5\pm 2.20\times10^{5}$ & $>1.5\times10^{5}$ & $>42$ \\
PKS~1934--63 & $(7.3\pm 2.4)\times10^{4}$ & $(1.7\pm 0.5)\times10^{4}$ & $23\pm 10$ \\
PKS~2135--209 & $(1.2\pm 0.7)\times10^{7}$ & $(1.2\pm 0.7)\times10^{7}$ & $100$ \\
PKS~2314+03 & $(2.7\pm 1.9)\times10^{6}$ & $(2.7\pm 1.6)\times10^{6}$ & $100$ \\
\bottomrule
\end{tabular}
}
\caption{Table reporting the masses of the gas emitting the total ($M_T$) and  outflow ($M_{out}$) emission in units of solar mass, and the fraction of the masses associated with the outflowing gas expressed as a percentage of the total gas mass.} 
\label{Table_masses}

\end{table}

Having derived the main basic properties of the outflows in our galaxies (e.g. their electron densities, extents and kinematic features), we now characterise the outflows in terms of their masses and kinetic powers, in order to asses the effect that the AGN feedback has on the host galaxies. 

The warm ionised gas masses associated with both the total emission and the outflows have been calculated using the following equation:

\begin{equation}
M_{\rm gas}= \frac{L({\rm H}\beta)m_{\rm p}}{n_{\rm e} \alpha^{\rm eff}_{\rm H\beta} h \nu_{\rm H\beta}}
,\end{equation}
where $L({\rm H\beta})$ is the intrinsic $\rm{H\beta}$ luminosity (i.e. corrected for dust extinction), $m_{\rm p}$ is the proton mass, $n_{\rm e}$ is the gas electron density, $\alpha^{\rm eff}_{\rm H\beta}$ is the effective $\rm{H\beta}$ recombination coefficient \citep[taken as 3.03 $\times10^{-14}~\rm{cm^3 s^{-1}}$ for case B, T$_e$=10$^4$~K, n$_e$=10$^2$~cm$^{-3}$ ;][]{2006agna.book.....O},  $\nu_{\rm H\beta}$ is the frequency of the $\rm{H\beta}$, and $h$ is the Planck constant.
To obtain these estimates we used the intrinsic H$\beta$ luminosities ($L H\beta_{intr}$) reported in Table~\ref{Table_Hbeta} and the DDD-derived electron densities reported in Table~\ref{Table_densities}. When calculating the intrinsic H$\beta$ luminosities we used the redshifts extracted from the stellar population fitting (see Table~\ref{Table_redshift}) to estimate the target distances, and the DDD-derived E(B-V) values (see Table~\ref{Table_densities}) to correct for the dust.

As can be seen in Table~\ref{Table_masses} we find that the fraction of the total warm gas mass in the outflow  varies considerably from galaxy to galaxy, ranging from $\sim$20$\%$ up to $\sim$70$\%$. However, for three of our targets --  PKS~1549--79, PKS~2135--209 and PKS~2314+03  -- the outflowing gas fraction is 100$\%$ due to the fact that the emission line profiles are dominated by the outflowing gas components, and all the kinematic components used to model the line emission are considered broad according to our criteria.

\subsection{Mass outflow rates,  kinetic powers and AGN feedback efficiencies}\label{subsec:Outf_numbers}

\begin{table*}
\resizebox{\textwidth}{!}{
\begin{tabular}{lcccccccc}
\toprule
Object & $\dot{M}_{class} [M_\odot\ yr^{-1}]$ & $\dot{E}_{class} [erg s^{-1}]$ & $F_{classic}$ & $\dot{M}_{vmax} [M_\odot\ yr^{-1}]$ & $\dot{E}_{vmax} [erg s^{-1}]$ & $F_{vmax}$  \\
\toprule
PKS~0023--26 & $0.13\pm 0.09$ & $(3.1\pm 2.1)\times10^{40}$ & $(1\pm 1)\times10^{-5}$ & $1.47\pm 0.09$ & $(3.7\pm 0.2)\times10^{41}$ & $(1\pm 1)\times10^{-4}$ \\
PKS~0252--71 & $>0.008$ & $>2.1\times10^{39}$ & $>1\times10^{-6}$ & $>0.6$ & $>1.3\times10^{41}$ & $>7\times10^{-5}$ \\
PKS~1151--34 & $>0.07$ & $>8.1\times10^{39}$ & $>3\times10^{-6}$ & $>0.5$ & $>5.9\times10^{40}$ & $>2\times10^{-5}$ \\
PKS~1306--09 & $>0.4$ & $>8.8\times10^{40}$ & $>7\times10^{-5}$ & $>1.4$ & $>4.2\times10^{41}$ & $>4\times10^{-4}$ \\
PKS~1549--79 & $4.3\pm 2.5$ & $(2.1\pm 1.2)\times10^{42}$ & $(4\pm 2)\times10^{-4}$ & $9.2\pm 2.5$ & $(1.1\pm 0.3)\times10^{43}$ & $(2\pm 1)\times10^{-3}$ \\
PKS~1814--63 & $>0.09$ & $>1.6\times10^{40}$ & $>2\times10^{-4}$ & $>0.8$ & $>1.8\times10^{41}$ & $>2\times10^{-3}$ \\
PKS~1934--63 & $0.04\pm 0.02$ & $(1.2\pm 0.7)\times10^{40}$ & $(1\pm 1)\times10^{-5}$ & $0.4\pm 0.01$ & $(2.4\pm 0.1)\times10^{41}$ & $(3\pm 1)\times10^{-4}$ \\
PKS~2135--209 & $1.0\pm 0.6$ & $(2.3\pm 1.4)\times10^{41}$ & $(2\pm 1)\times10^{-5}$ & $19.3\pm 0.5$ & $(4.6\pm 0.1)\times10^{42}$ & $(3\pm 1)\times10^{-4}$ \\
PKS~2314+03 & $1.8\pm 1.2$ & $(4.5\pm 2.7)\times10^{41}$ & $(6\pm4)\times10^{-4}$ & $12\pm 1.2$ & $(8.3\pm 0.8)\times10^{42}$ & $(1\pm 1)\times10^{-2}$ \\
\bottomrule
\end{tabular}
}
\caption{ Table reporting the mass outflow rates $\dot{M}$, the outflow kinetic powers $\dot{E}$ and the AGN feedback efficiencies $F$ for the targets of our sample, as estimated by following the classical (col 2-3-4) and the v$_{max}$ (col 5-6-7) approaches.}
\label{outflows_properties}
\end{table*}

In this section we use the outflow masses, radii and kinematic properties, along with the AGN bolometric luminosities to finally determine the mass outflow rates $\dot{M}$, outflow kinetic powers $\dot{E}$ and  AGN feedback efficiencies $F$. Following \cite{Rose2018} we use the equations listed below:

\begin{equation}
\dot{M}= M_{gas}\frac{v_{\rm out}}{r}
\label{Mdot}
,\end{equation}  

\begin{equation}
\dot{E}=\frac{\dot{M}}{2} (v^2_{\rm out}+ 3\sigma^2), ~~{\rm and}
\label{Edot}  
\end{equation}

\begin{equation}
F=\dot{E}/L_{BOL}
\label{F}  
,\end{equation}
where $r$ is the radius of the outflow and $L_{BOL}$ is the AGN blometric luminosity.

In order to assign values to the $v_{out}$ and the $\sigma$ terms we use the kinematic properties of our outflows reported in Table~\ref{Table_kinmodels} and adopt two different approaches.
The so-called `classical approach' assumes that the centroid velocity of an emission associated with the outflowing gas represents the true velocity of the outflow, and its broadening is due to the intrinsic velocity dispersion in each part
of the outflow. In this case, we set $v_{out}=v$ and $\sigma={\rm FWHM}/2.355$. This classical
approach does not explicitly account for line-of-sight projection effects. Therefore, we also consider the less conservative approach (hereafter referred to as the `$v_{max}$ approach')
discussed in \citet{Rose2018}.
This attempts to account for projection effects by setting $v_{\rm out}=v_{\rm max}$ and $\sigma=0$, and assumes that the broadening of the emission lines is entirely due to the different projections of the velocity vectors of the outflowing gas, rather than localised velocity dispersion.   

Table~\ref{outflows_properties} reports the $\dot{M}$, $\dot{E}$ and $F$ we obtain by using the equations above and following the two approaches. 
It appears clear that by using the $v_{max}$ approach the kinetic powers and the AGN feedback efficiencies, which are proportional to the mass outflow rates $\dot{M}$,  typically increase by one order of magnitude with respect to the classical approach. The target that is most affected is PKS~0252--71, with a change of two orders of magnitude for both these quantities, mainly due to the large difference between $v$ and $v_{max}$.  

In the following and in Sec. \ref{discussion} we will discuss only the estimates that have been obtained with the $v_{max}$ approach. As already noted, this approach is thought to overcome some of the limitations due to projection effects and therefore gives a more realistic characterisation of the outflows properties. Moreover, since similar approaches have been widely used in the literature, it allows us to make a fair comparison with previous outflow studies.   

For our sample of compact radio galaxies we find mass outflow rates in the range $0.4 < \dot{M} <  20$\,M$_{\odot}$ yr$^{-1}$. The highest values are found for PKS~1549--79,  PKS~2135--209 and PKS~2314+03 where we consider that all the line-emitting gas is outflowing.
The AGN feedback efficiencies are remarkably small and in general lower then 1$\%$, ranging from a minimum of 0.002$\%$ to a maximum of 1$\%$\footnote{Note that some of these $F$ values represent lower limits due to the upper limits on the n$_e$.} in agreement with what has been found by some of the more recent studies on outflows that, similarly to ours, adopted density diagnostics sensitive to high densities, thus overcoming the limitations of the classical \OII\ and \SII\ line ratios \citep[e.g.][]{Spence2018,Rose2018,Baron19b,Davies2020}. In particular, we  highlight that the general warm outflow properties of the CSS/GPS sources are comparable with those derived by \citet{Rose2018} and \citet{Spence2018} using similar techniques for nearby ULIRGs with optical AGN nuclei ($0.07 < \dot{M} <  20$\,M$_{\odot}$ yr$^{-1}$; $ 0.03 < F  < 2.5$\%).

\section{Discussion}\label{discussion}

\begin{figure}
\includegraphics[width=0.5\textwidth]{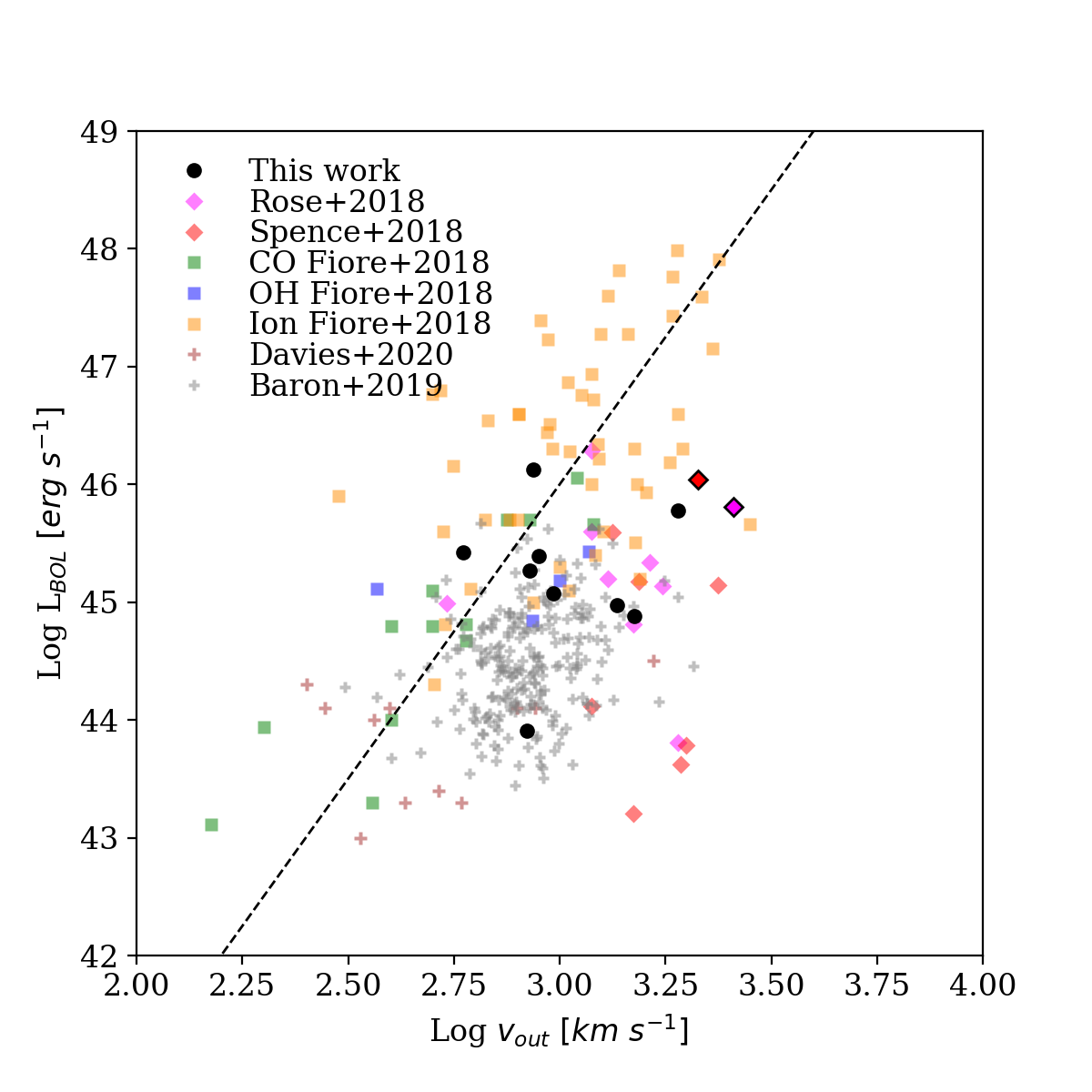}
\caption{AGN bolometric luminosity as a function of the outflow velocity, $v_{out}$ for our sample (back filled points) and the comparison samples (see legend in the upper left). To distinguish them from the rest of the comparison sample, the markers of the two ULIRGs hosting a compact radio source have a black edge colour. The black dashed line marks a $v_{out}^5$ scaling, as shown in Fig.~2 of \cite{Fiore2017}, and can be used as a reference for the $v_{out}^{4.6\pm1.8}$ scaling found in their paper considering molecular and warm ionised gas outflows only.}
\label{Lbol_vmax}
\end{figure}

\begin{figure*}
\includegraphics[width=\textwidth]{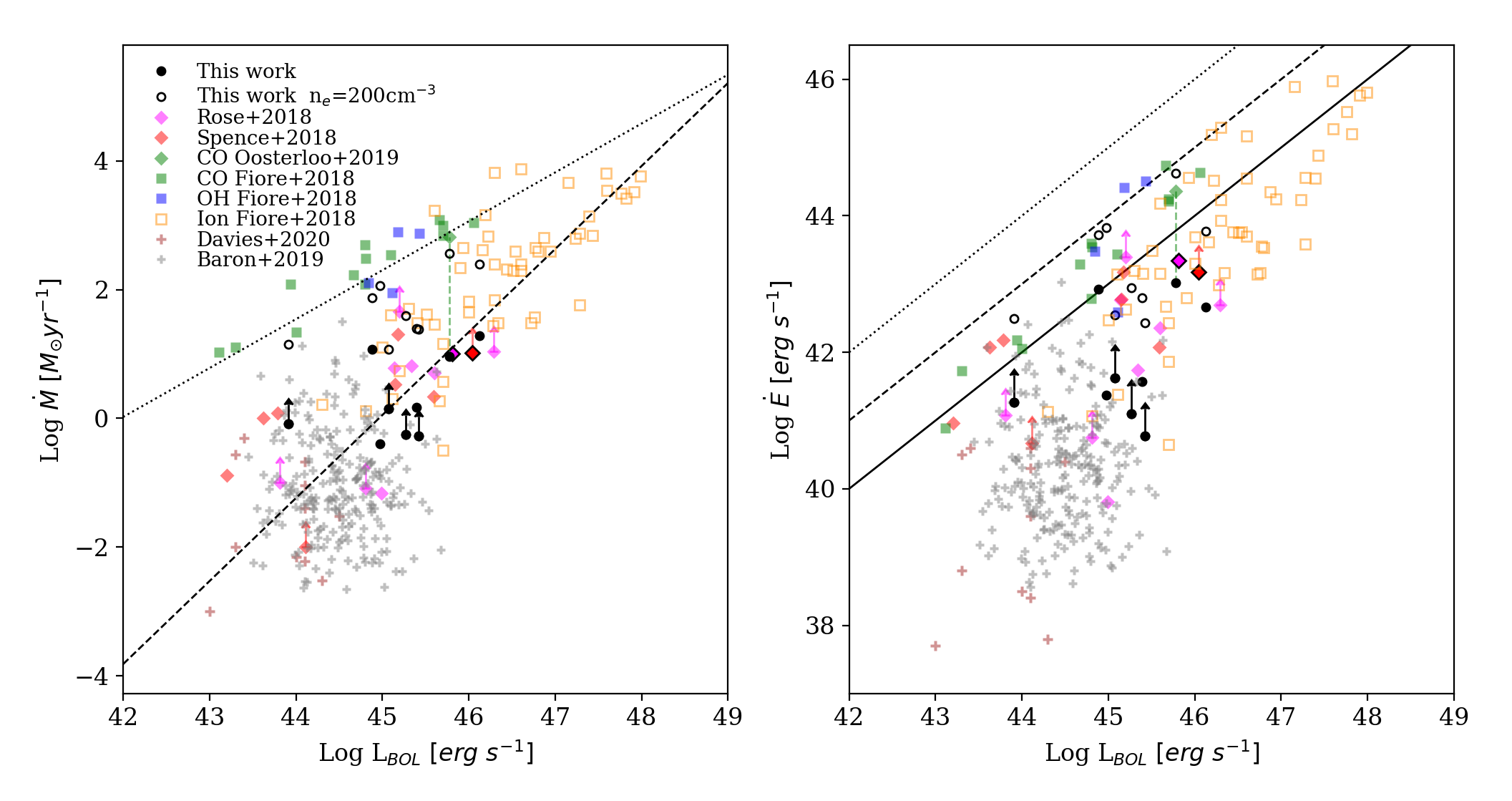}
\caption{Mass outflow rate (left panel) and kinetic power of the outflow (right panel) as a function of the AGN bolometric luminosity for our sample (black points) and the comparison samples (see legend in the upper left of the left panel). To distinguish them from the rest of the comparison sample, the markers of the two ULIRGs hosting a compact radio source have a black edge colour. To enable a fair comparison with the ionised gas outflows of \cite{Fiore2017} we have also assume an outflow gas density of $n_e=200~cm^{-3}$ for all our targets (black empty circles). The CO outflow detected in PKS~1549--79 by \cite{Oosterloo2019} has been also included in this plot using our estimate of the source $L_{BOL}$, a green dashed vertical line connects the warm ionised and the molecular outflow $\dot{M}$ and $\dot{E}$ values for this object to highlight the differences between the outflows properties in the two gas phases.  The dotted and dashed black lines in the left panel represent the best fit relations found by \cite{Fiore2017} for the molecular gas and ionised gas outflows respectively. The dotted, dashed and solid black lines in the right panel mark the correlations $\dot{E}=1,0.1,0.01 L_{bol}$. }
\label{sampleComparison}
\end{figure*}

In this section we compare the properties of the outflows detected in our sample of compact radio sources with those of outflows in other AGN samples collected from the literature. These samples are described below to highlight the effect that alternative electron density estimators can have on the final numbers quantifying the mass outflow rates, and more in general, the AGN feedback efficiency.

The mass outflow rates and kinetic energies we collected for the comparison samples have been calculated under the assumption that the true velocity of the outflowing gas is the maximum velocity detected in the wings of the emission line profiles -- similar to what we call the `$v_{max}$ approach' in Sec. \ref{outProperties}.
It should be noted, however, that the way $v_{max}$ has been calculated differs in detail from sample to sample, and more details are given below. We will compare our results with those for the outflows included in the following studies.

\begin{itemize}
  \item The \cite{Fiore2017} study including molecular (CO and OH) and warm ionised outflow results for AGN covering a wide redshift range 0.003 < z < 3.5.
\vspace{2.5mm}

  This study concerns an heterogeneous collection of outflows from the literature that were selected to have a good estimate (or a robust limit) of the spatial extent of the high velocity gas involved in the outflow. For the ionised gas outflows the authors re-computed the outflow properties from the original data, homogenizing the estimates of the outflow velocities and densities. In particular, following \cite{Rupke2013}, they defined $v_{max} = v_{broad} + 2\sigma _{broad}$, where $v_{broad}$ and $\sigma _{broad}$ are respectively the centroid velocity (with respect to the systemic velocity) and the $\sigma$ of the broad Gaussian component used to model the outflow. The electron densities of the outflowinig material were assumed to be $n_e=200$ cm$^{-3}$ in the absence of more accurate measurements.

\vspace{2.5mm} 
  \item The \cite{Rose2018} and \cite{Spence2018} study of the warm outflows in a 90\% complete sample of 15 local ($z < 0.175$) ULIRGs with nuclear AGN activity detected at optical wavelengths.
\vspace{2.5mm}

  This study considers rapidly evolving galaxies in the local universe that are hosting and AGN and are, in most cases, involved in a merger event. It is the most directly comparable study to the current one, since the methods used to determine outflow radii and densities are similar to those we have
  used for the CSS/GPS sources. In particular, the DDD approach was used to determine the electron densities of the outflows. As already noted above, two of these ULIRGs --- F~23389+0303N and PKS~1345+12 ---  have radio properties that are similar to the CSS/GPS sources 
  in our sample.
\vspace{2.5mm}

  \item The \cite{Baron19b} and the \cite{Davies2020} studies of warm ionised outflows of low-to-moderate luminosity AGN in the local 
  universe ($z < 0.15$). 
  \vspace{2.5mm}
  
  The  outflows considered in these studies are mainly associated with type II AGN in the local universe with low-to-moderate luminosities.
  The \cite{Baron19b} sample includes 234 outflows for which $v_{out}=(v_{shift}^2 + \sigma^2)^{1/2}$, where $v_{shift}$ is the velocity shift of the broad emission line centroid with respect to the narrow lines,  and $\sigma$ is the velocity dispersion of the broad emission lines \citep[following][]{Karouzos2016}, and the gas density has been estimated using a novel method based on the ionisation parameters and sensitive to high densities. 
  The \cite{Davies2020} sample includes 11 outflows for which the outflows velocity is set equal to $v_{98}$, the absolute value of the velocity above (or below) which 98$\%$ of the line flux is contained, and outflow gas densities have been obtained following the \cite{Baron19b} prescriptions.
  
\end{itemize}

\noindent
From the description of the selected comparison samples is clear that we are here comparing AGN at different redshifts and hosted by galaxies at different evolutionary stages. In addition, we are not making a distinction between the AGN mode (jet mode vs quasar mode) that is driving the outflows. Most of the outflows in the comparison samples are claimed to be driven by the central AGN radiation pressure (so-called AGN winds), while for our sample we have indications that the radio jets can be the main driver. However, with the available data it is often not possible to draw a clear demarcation line: some of the galaxies in the comparison samples host radio sources (e.g. the sources of the \citealp{Nesvadba2008} study included in the \citealp{Fiore2017} sample) that are known/might contribute to driving the outflows, even at relatively moderate radio
powers (e.g. IC~5063 \citealp{Tadhunter2014,Morganti2015}) and, on the other hand, we cannot rule out radiation pressure playing a role in driving the outflows in the CSS/GPS sources in our sample.
Multi-wavelength high spatial resolution data allowing detailed comparison of the radio and warm gas morphologies are usually the best way to discriminate between the AGN feedback modes, but unfortunately they are available for only a handful of sources.
The comparison we present in this section is thus meant to show the overall effect that the AGN feedback (operated by both radio jets and radiation pressure) can have on the ISM of the host galaxies.
Therefore, it is important to bear in mind that some of the differences between the outflow properties that we are going to discuss can be reasonably attributed to the different nature of the AGN we are comparing.

Fig.~\ref{Lbol_vmax} shows the outflow velocity  as a function of the AGN bolometric luminosity for our sample and the comparison samples.
It is notable that, although CSS/GPS and ULIRG samples  have intermediate bolometric luminosities, they both  include objects whose extreme velocities are characteristic of the outflows hosted in the most luminous AGN in the \cite{Fiore2017} sample.
By considering their molecular and ionised gas outflows, \cite{Fiore2017} claim to find a trend of increasing outflow velocity with increasing AGN bolometric luminosity (i.e. the dashed line in  Fig.\ref{Lbol_vmax} can be used as a reference). However, considering
the \cite{Fiore2017} sample along with the GPS/CSS and other comparison samples included in the plot, the trend appears less clear.

In Fig.~\ref{sampleComparison} we show the mass outflow rate (left panel) and kinetic power (right panel) as a function of the AGN bolometric luminosity for our sample and the comparison samples, reproducing Fig.1 presented in \cite{Fiore2017}.
It is important to emphasise that the way the density of the outflowing gas has been estimated plays a crucial role in the interpretation of this figure, as also stressed in \cite{Spence2018} and \cite{Davies2020}. Apart from the ionised gas outflows in the sample of \cite{Fiore2017}, for which the electron density has been assumed to be $n_e=200$ cm$^{-3}$, the electron densities of all the outflows presented in the Fig.~\ref{sampleComparison} have been estimated using diagnostics that are sensitive to the high density regime. Despite this, we choose to show the best fit relation found for the ionised gas outflows by \cite{Fiore2017} (dashed black line) for its importance as a reference point in the current literature, and to stress the differences that a more precise electron density estimate can make. 
In fact, in Fig.~\ref{sampleComparison} is also shown how assuming the same gas density of $n_e=200$ cm$^{-3}$ (black empty circles) as \cite{Fiore2017}  increases the $\dot{M}$ and the $\dot{E}$ estimates for our warm outflows by one or two orders of magnitude. This would place our warm ionised outflows among those with the highest mass outflow rates and kinetic energies at a given AGN bolometric luminosity, and in some cases make them similar to the massive molecular gas outflows in the sample of \cite{Fiore2017}. 
An analogous behaviour is expected for the outflows of the remaining comparison samples, as noted in the original studies.

Conversely, we argue that more precise estimates of the outflow densities would potentially bring the warm ionised outflows of \cite{Fiore2017} down to lower $\dot{M}$ and $\dot{E}$ values  \citep[see][who first tried to quantify the expected shift]{Davies2020}.
Considering this effect, we predict that the current compilation of sources shown in Fig.~\ref{sampleComparison} would result in a correlation between $\dot{M}$ and L$_{BOL}$ with a flatter slope than that found by \cite{Fiore2017} (i.e. the dashed black line in Fig.\ref{sampleComparison}). Similarly, most of the \cite{Fiore2017} warm ionised outflows in the right panel of Fig.~\ref{sampleComparison} would then lie below the solid line marking an AGN feedback efficiency of $F$=1$\%$, consistent with the results for our sample and the remaining comparison samples, but well below the $F\sim$5 -- 10\% required by some AGN feedback models \citep[e.g.][]{Silk1998,Fabian1999,DiMatteo2005}.

Interestingly, the CSS/GPS and ULIRG objects also show evidence for higher mass outflow rates, kinetic powers and feedback efficiencies in their warm outflows than objects of similar AGN bolometric luminosity in the \citet{Baron19b} and \citet{Davies2020} samples. This may reflect the fact that the AGN in both CSS/GPS and the ULIRG objects have been triggered in unusually gas-rich environments, so that the drivers of their warm outflows, whether relativistic radio jets (CSS/GPS and some ULIRGs) or hot, radiation-driven  winds (most ULIRGs), are able to couple more effectively with the cooler phases of the ISM.

It is important to emphasise that the scatter of the points in these plots and/or the lack of correlations for single samples such as ours, spanning only few orders of magnitudes in $L_{BOL}$, is not surprising. As stressed by \cite{Zubovas2018} the evolutionary times of outflows are longer then the AGN activity event (at most 10$^6$~yr for young jets as we consider here) that inflates them \citep[typically by one order of magnitude, see][]{King2011}, and for this reason a significant scatter is expected between quantities measuring the properties of outflows and the level of the current AGN activity in the nuclear regions of the host galaxies.

Finally, we stress that we have only considered the warm ionised outflows in the CSS/GPS sources, and that significantly higher mass outflow rates and kinetic powers may be associated with the neutral \citep[e.g][]{Morganti2005,Rupke2005} and molecular \citep{2014A&A...562A..21C,Morganti2015} gas phases in the outflows (e.g. see the green points representing the molecular outflows taken from the \citealt{Fiore2017} in Fig.~\ref{sampleComparison}).  Therefore, the total feedback efficiencies may be higher that those we have calculated for the warm outflowing gas phase alone. This is illustrated by the case of one of the CSS/GPS objects in our sample -- PKS~1549--79 -- which has recently been observed using ALMA at high resolution in the CO(1-0) and CO(2-3) molecular lines \citep{Oosterloo2019}. The observations detect a compact, high velocity
molecular outflow whose mass outflow rate ($\dot{M}_{mol} = 650$\,M$_{\odot}$yr$^{-1}$), kinetic power ($\dot{E}_{mol}= 2.3\times10^{44}$\,\ergs) and feedback efficiency ($F_{mol} \sim 3$\%) are more than an order of magnitude higher than those we measure for the warm outflow in the 
same object ($\dot{M}_{warm} = 9$\,M$_{\odot}$yr$^{-1}$, $\dot{E}_{warm} = 1.1\times10^{43}$\,\ergs; $F_{warm} \sim 0.2$\% -- see Fig.~\ref{sampleComparison}).


\section{Conclusions}\label{conclusion}

We have used VLT/X-shooter and HST/ACS data to investigate the warm, AGN-driven outflows in a sample of 9 CSS/GPS objects whose AGN have intermediate bolometric luminosities (in the interval $10^{44-46}$\ergs). This class of galaxies is known to host young and powerful radio sources whose jets are expanding through the surrounding ISM and interacting with it. The signs of this interaction are detected in the form of gas outflows in different gas phases, and help  shed  light on the AGN feedback process and some of the still open question surrounding this topic (e.g. the typical mass content of outflows and the overall AGN feedback efficiency).  Our main results are as follows.
\begin{itemize}
    
    \item {\bf Kinematics.} The emission line profiles clearly indicate very disturbed gas kinematics and, when modelled, show the presence of outflows reaching velocities ranging  from  $\sim$600 \kms\ up to 2000 \kms\ (i.e. $v_{max}$ in Table~\ref{Table_velocities}) for all the galaxies in our sample. Many of the outflows we detect have extreme kinematic features compared to the outflows found in more extended radio sources with comparable radio powers and redshifts,
    or radio-quiet AGN of similar bolometric luminosities, confirming what has been found in the literature \citep{Holt2008}.
    \item {\bf Spatial extents.} The spatial extents of the warm ionised outflows ($0.06 < r < 1.9$\,kpc) agree well with those of the radio sources, as measured using the separation of the radio lobes in high resolution VLBI radio images (see Table~\ref{Table_radius}); in some cases the emission-line structures are also closely aligned with the radio axes. This supports the idea that the so-called `jet mode' feedback is in action in compact radio sources, and hence the expansion of the radio jets is likely to be the dominant driver of the outflows we observe, similar to what has been found in other studies \citep[e.g.][]{Tadhunter2014}.
    \item {\bf Densities.} The warm  outflows include dense (log$(n_e$~cm$^{-3})\sim3-4.8$) gas components that are often denser than the gas of the host galaxy ISM, possibly indicating gas compression operated, at some level, by the AGN feedback. 
    \item {\bf Gas masses.} The outflows comprise 20 -- 70\% of the total host galaxy warm ISM mass sampled by our spectroscopic slits, while the mass outflow rates cover the range   $0.4 < \dot{M} <  20$\,M$_{\odot}$yr$^{-1}$.
    \item {\bf Feedback efficiency.} The AGN feedback efficiencies $F$ for the warm outflows of our sample are relatively low: usually below 1\%, and covering the range $0.002 < F < 1$\%. 
\end{itemize}

To put our study in a broader context, we compared our results for the CSS/GPS sources with those for an heterogeneous collection of warm outflows (claimed to be mainly driven by the AGN quasar mode) taken from the literature. The results of this comparison further emphasise the sensitivity of the derived outflow properties (e.g. masses, energetics, feedback efficiency) to the precision with which the electron densities are measured. 

In all aspects, the properties of warm outflows we detect in the CSS/GPS sources are strikingly similar to those measured using the same techniques for  nearby ULIRGs with nuclear AGN activity, despite the fact that the dominant outflow driving mechanisms are likely to be different in the two types of objects. This suggests that the feedback effect of the relativistic jets on kpc-scales can be as effective as that of the hot winds driven by the radiation of the AGN, in line with the predictions of recent numerical simulations \citep[e.g.][]{Wagner2013,Mukherjee2016,Mukherjee2018}.

Although we are aware of the caveats to consider when comparing the AGN feedback efficiency derived from observations to the prescription of cosmological models \citep[see][]{Harrison2018}, we note here that all the latest results from studies attempting to improve the measurements of the warm outflow gas density indicate that the AGN feedback efficiencies are far below the 5-10$\%$ expected by classical AGN feedback models \citep[such as][]{Fabian1999,DiMatteo2005}, and in general lower then those that have been typically estimated for warm ionised gas outflows in the literature \citep{Fiore2017}.

This puts the accent on the need for further precise measurements of the basic outflows parameters such as the gas density, as well as more sophisticated AGN feedback models for both the jet-mode and the quasar-mode \citep[see e.g.][]{Cielo2017,Weinberger2017}, in order to fully understand the role of AGN-driven outflows in the galaxy evolution context.

\begin{acknowledgements}
The authors thank the anonymous referee for their useful comments that improved the clarity of the paper.
FS and CT acknowledge support from STFC.
The research leading to these results has received funding from the European Research Council under the European Union's Seventh Framework Programme (FP/2007-2013) / ERC Advanced Grant RADIOLIFE-320745. Based on observations collected at the European Organisation for Astronomical Research in the Southern Hemisphere under ESO programmes 060.A-9412(A) and 87.B-0614A. 
Based on observations taken with the NASA/ESA Hubble Space Telescope, obtained at the Space Telescope Science Institute (STScI), which is operated by AURA, Inc. for NASA under contract NAS5-26555.
This research has made use of the NASA/IPAC Extragalactic Database (NED) which is operated by the Jet Propulsion Laboratory, California Institute of Technology, under contract with the National Aeronautics and Space Administration. The python packages \url{https://pypi.org/project/PyNeb/}{{\it PyNeb}} \citep{Pyneb2013} and \url{https://uncertainties-python-package.readthedocs.io/en/latest/index.html}{{\it uncertainties}} have been used throughout this work to respectively derive dust corrections and propagate the errors.

\end{acknowledgements}

\bibliographystyle{aa}
\bibliography{biblio.bib}

\newpage

\begin{appendix}
\section{Starlight and line emission fitting}\label{AppendixFits}

In this section we provide supporting material and additional details related to the data analysis. 

In the figures going from Fig.\ref{3C459Stellar_continuum} to Fig.\ref{PKS1151Emission lines fits} we present the results of the stellar population and emission-line modelling.
In Table~\ref{Table_transflux_1} we report the total and outflowing gas trans-auroral line fluxes that have been used to obtain the DDD diagram presented in Fig.\ref{DDD}. Similarly, Table~\ref{TableHalphaNIIfluxes} reports the H$\alpha$ and \NII6584\AA\ fluxes of the total and outflowing warm ionised gas.

As mentioned in Sec.\ref{sec:kinmodels}, the emission line modelling for PKS~1151--34 was performed following a different strategy with respect to the remaining sources because of the presence of emission from the BLR of the AGN. The first step was to obtain constraints on the kinematical components needed to model the H$\beta$ BLR emission using the more prominent H$\alpha$ BLR emission line. 
To do so we obtained a first-guess kinematic model for the emission lines of the NLR by fitting the \OIII$\lambdaup\lambdaup$4958,5007\AA\ and the H$\beta$ lines together with an additional very broad component to take into account the H$\beta$ BLR emission. The NLR components of this model were then used as a first, guess together with two additional free-to-vary Gaussian components (i.e. to recover the BLR flux) to fit the \OI$\lambdaup\lambdaup$6300,63\AA + \NII$\lambdaup\lambdaup$6548,84\AA\ + H$\alpha$ +  \SII$\lambdaup\lambdaup$6717,31\AA\ lines. 
The kinematic properties of the two Gaussian components used to model the BLR were then used to model the \OIII$\lambdaup\lambdaup$4958,5007\AA\ + H$\beta$ blend again, in order to obtain our final \OIII\ reference model. 
We then used all the components of the \OIII\ reference model to perform the final fit of the \OI$\lambdaup\lambdaup$6300,63\AA + \NII$\lambdaup\lambdaup$6548,84\AA\ + H$\alpha$ +  \SII$\lambdaup\lambdaup$6717,31\AA\ lines. We used only the \OIII\  kinematic components to fit the \SII$\lambdaup\lambdaup$4069,76\AA\ + H$\delta$, the \OII$\lambdaup\lambdaup$7319,30\AA\ and \OII$\lambdaup\lambdaup$3726,29\AA\ emission lines, due to the absence of the BLR contamination of these lines.
It should be noted that the \OIII\ reference model includes three kinematic components for the NLR emission, however the \OIII$\lambdaup\lambdaup$4958,5007\AA\ lines are the only emission lines that need three kinematic components to be properly modelled --  all the remaining emission lines are well recovered by using only two of the kinematic components of the `\OIII' reference model.
As can be seen in Fig.~\ref{PKS1151Emission lines fits}, the emission related to the third, broader [OIII] component is at a similar level to the H$\beta$ BLR emission, and is a very small fraction of the total lines flux. This broader [OIII] component might represent some residual emission from the H$\beta$ BLR emission that is not properly modelled. Therefore, we exclude the broadest \OIII\ reference model component (in terms of its flux and kinematical properties) from the analysis presented in this paper.

\begin{figure*}[h]
\begin{minipage}[t]{0.5\textwidth}
\includegraphics[angle=90]{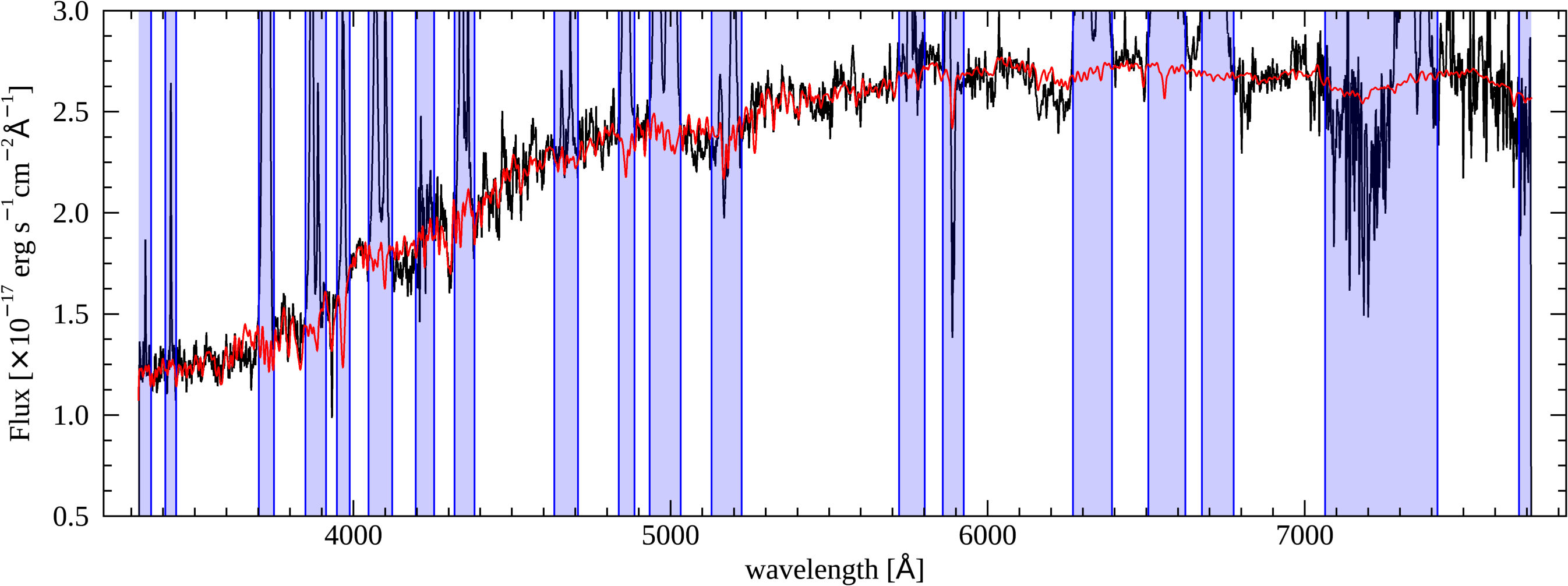}
\end{minipage}
\begin{minipage}[t]{0.5\textwidth}
\includegraphics[ angle=90]{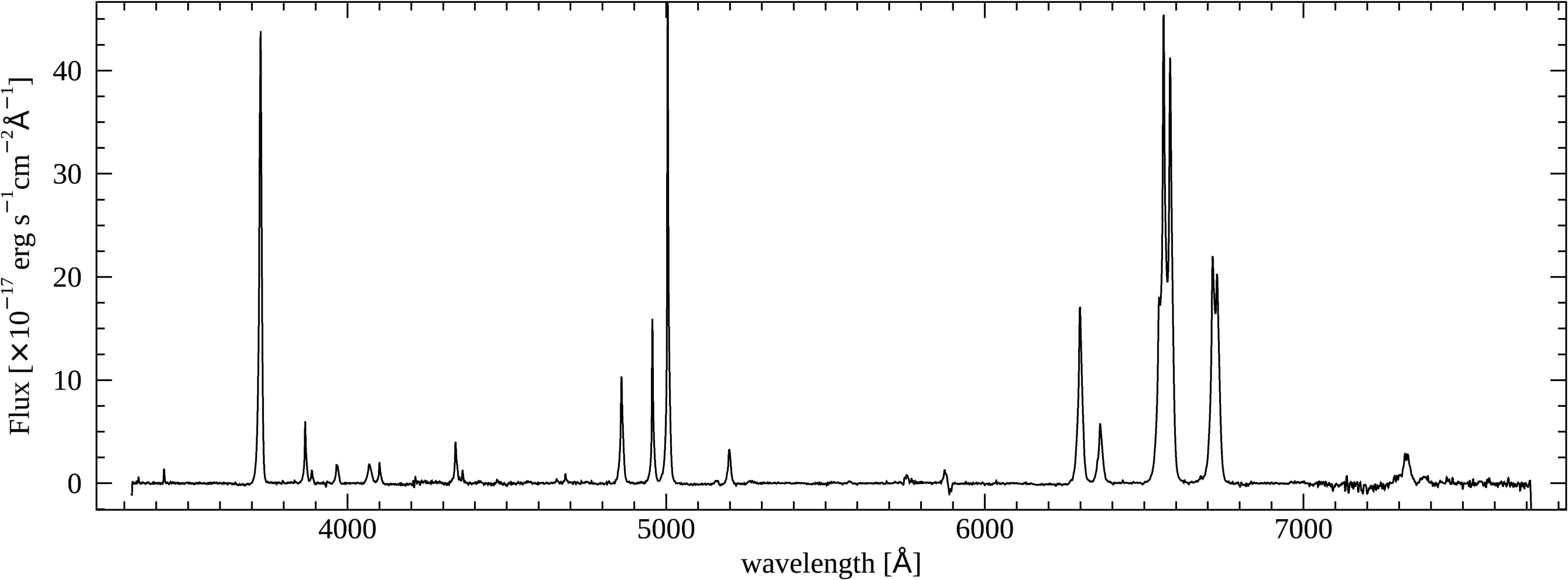}
\end{minipage}
\caption{ PKS~0023--26: \textit{Upper panel :} UVB+VIS Nuclear spectrum (black solid line) and best-fit model (red solid line) for the continuum emission, regions of the spectrum corresponding to masked emission lines are indicated in blue. \textit{Lower panel :} Residual spectrum after subtraction of the stellar continuum showing the emission lines associated with the warm ionised gas. Wavelengths are plotted in \AA, and the flux scale is given in units of $10^{17} \rm{erg~s^{-1}cm^{-2}\AA^{-1}}$. Note that plots with detailed fits to the emission lines in this source are presented in the main text.} 
\label{Stellar_continuum}
\end{figure*}

\begin{figure*}[h]
\begin{minipage}[t]{0.5\textwidth}
\includegraphics[angle=90]{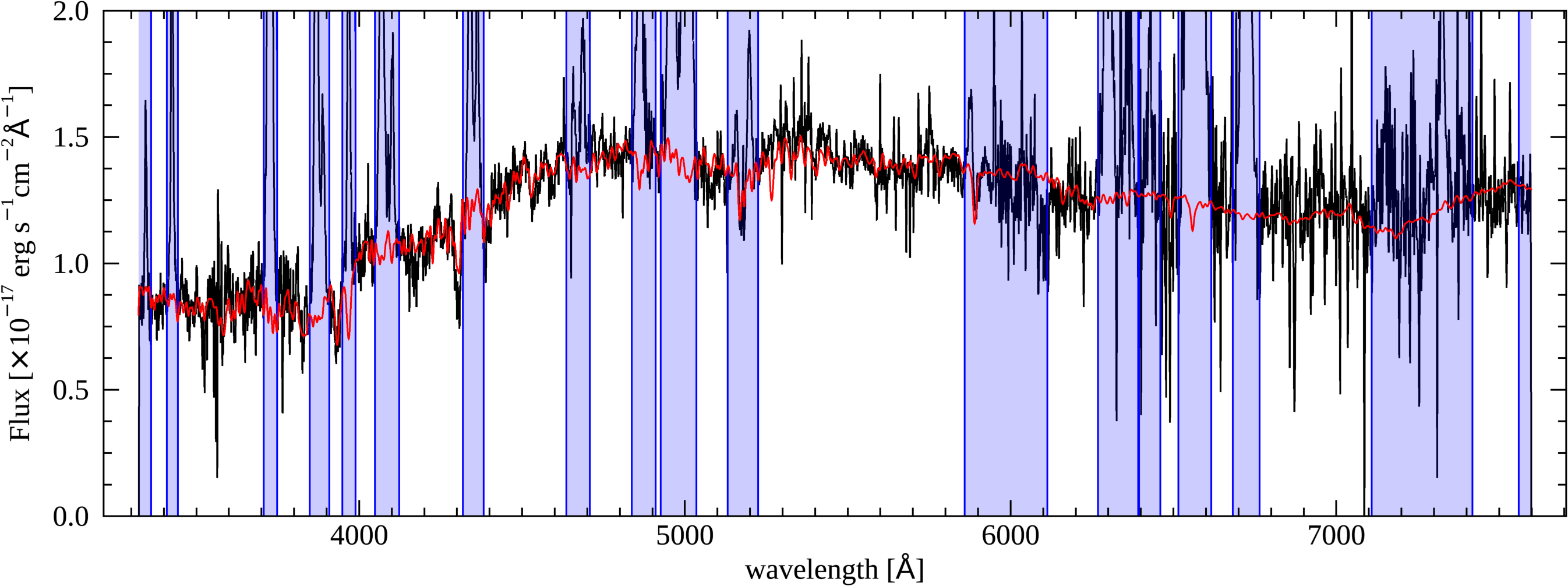}
\end{minipage}
\begin{minipage}[t]{0.5\textwidth}
\includegraphics[ angle=90]{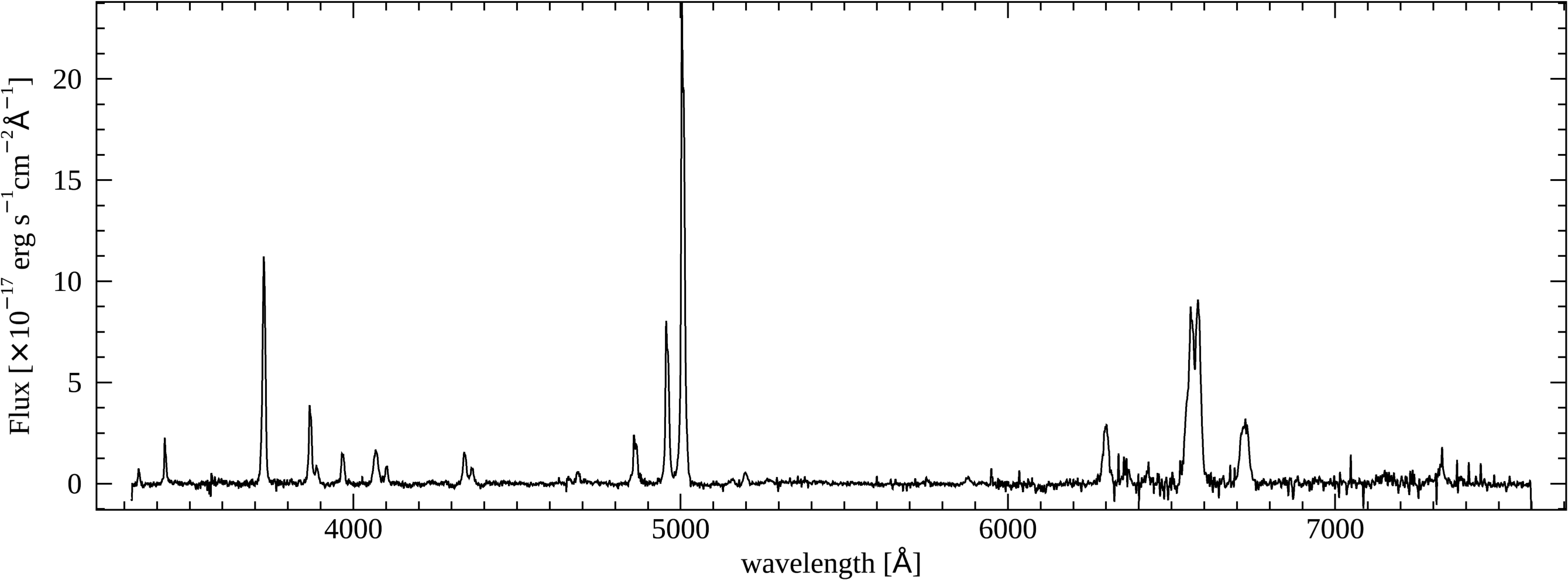}
\end{minipage}
\caption{ PKS0252--71: As in Fig.\ref{Stellar_continuum}} 
\label{PKS0252Stellar_continuum}
\end{figure*}

\begin{figure*}[h]
\begin{minipage}[t]{0.5\textwidth}
\includegraphics[width=\textwidth, height=0.21\textheight]{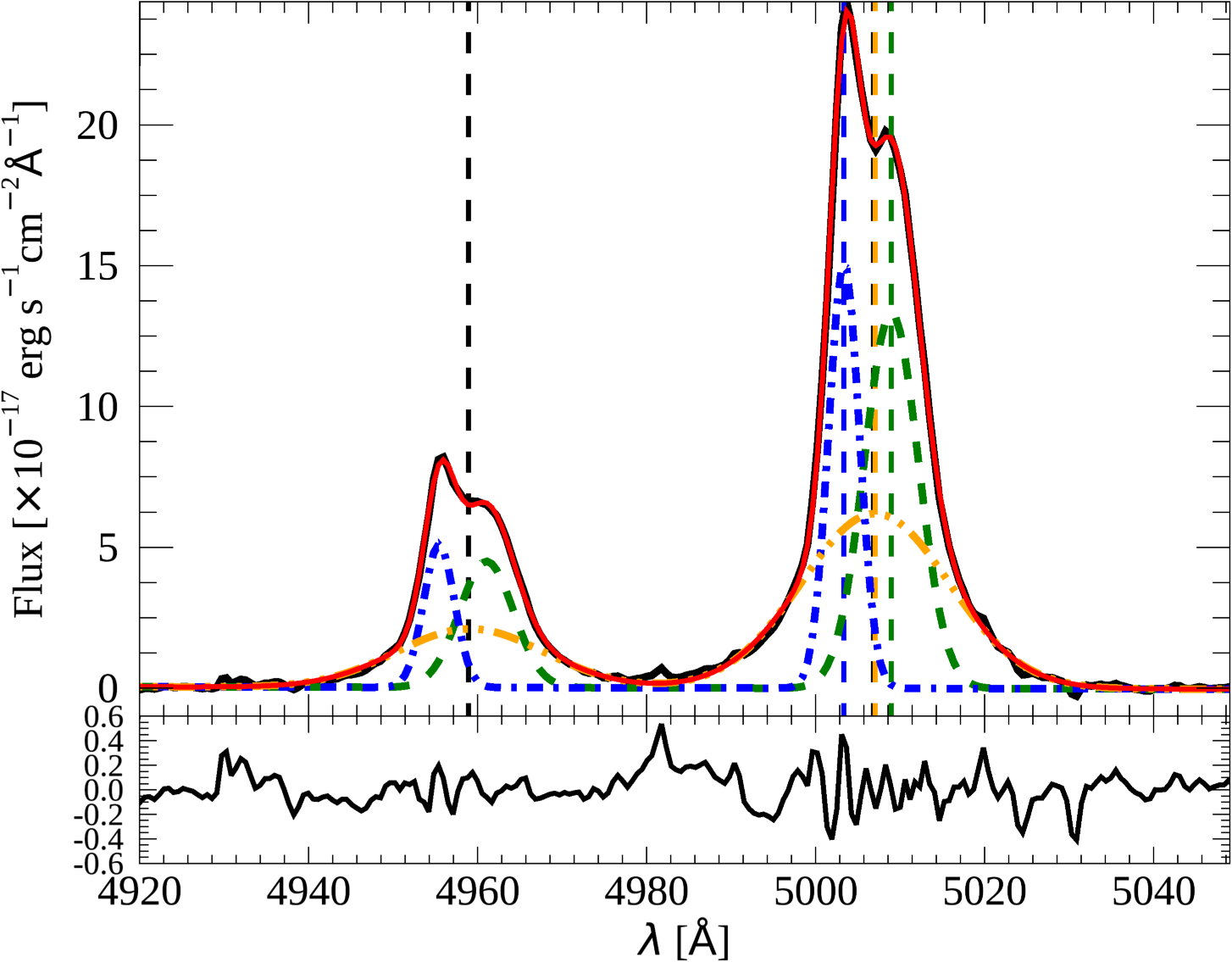}
\end{minipage}
\begin{minipage}[t]{0.5\textwidth}
\includegraphics[width=\textwidth, height=0.21\textheight]{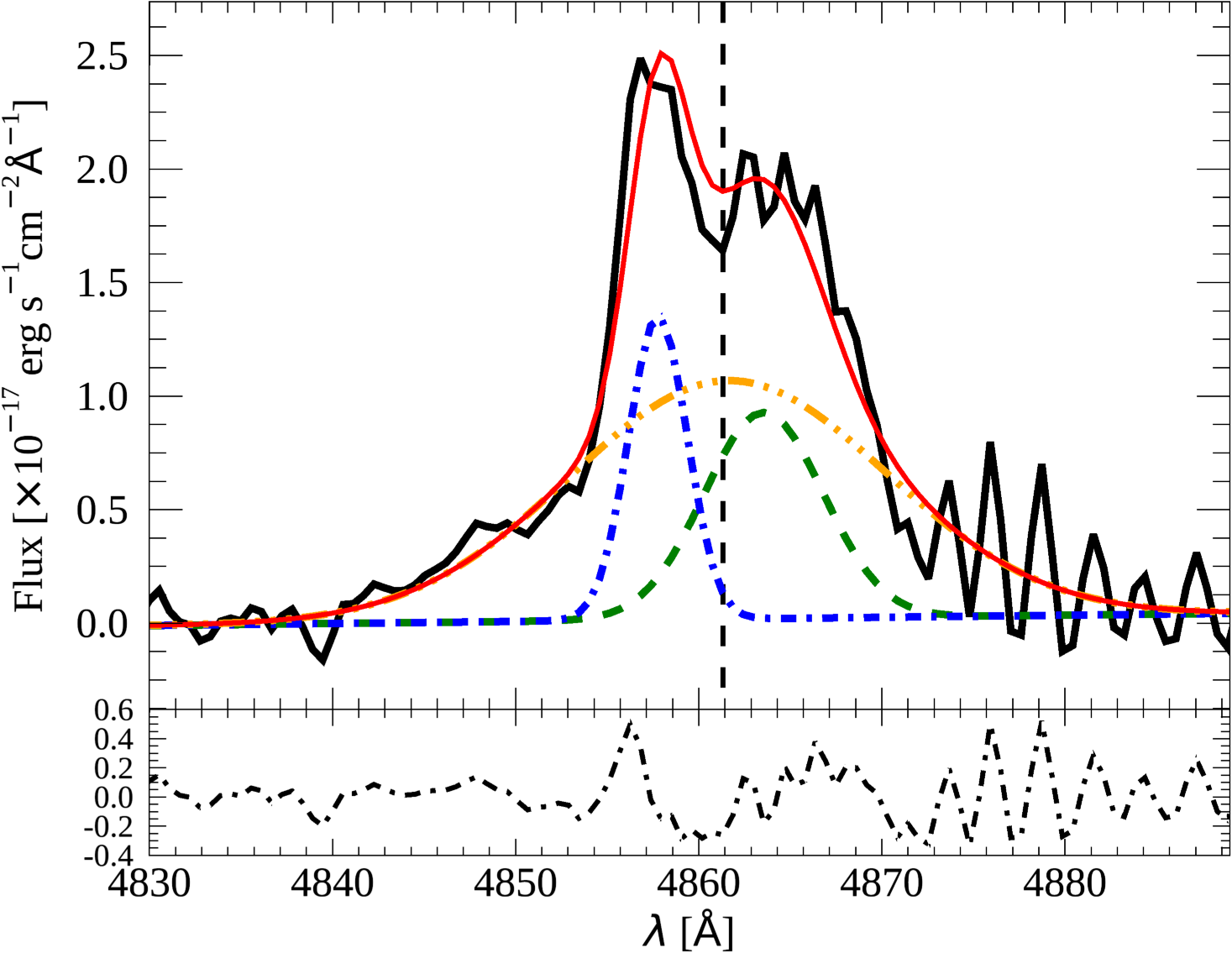}
\end{minipage}

{\centering
\begin{minipage}[b]{0.5\textwidth}
\includegraphics[width=\textwidth, height=0.21\textheight]{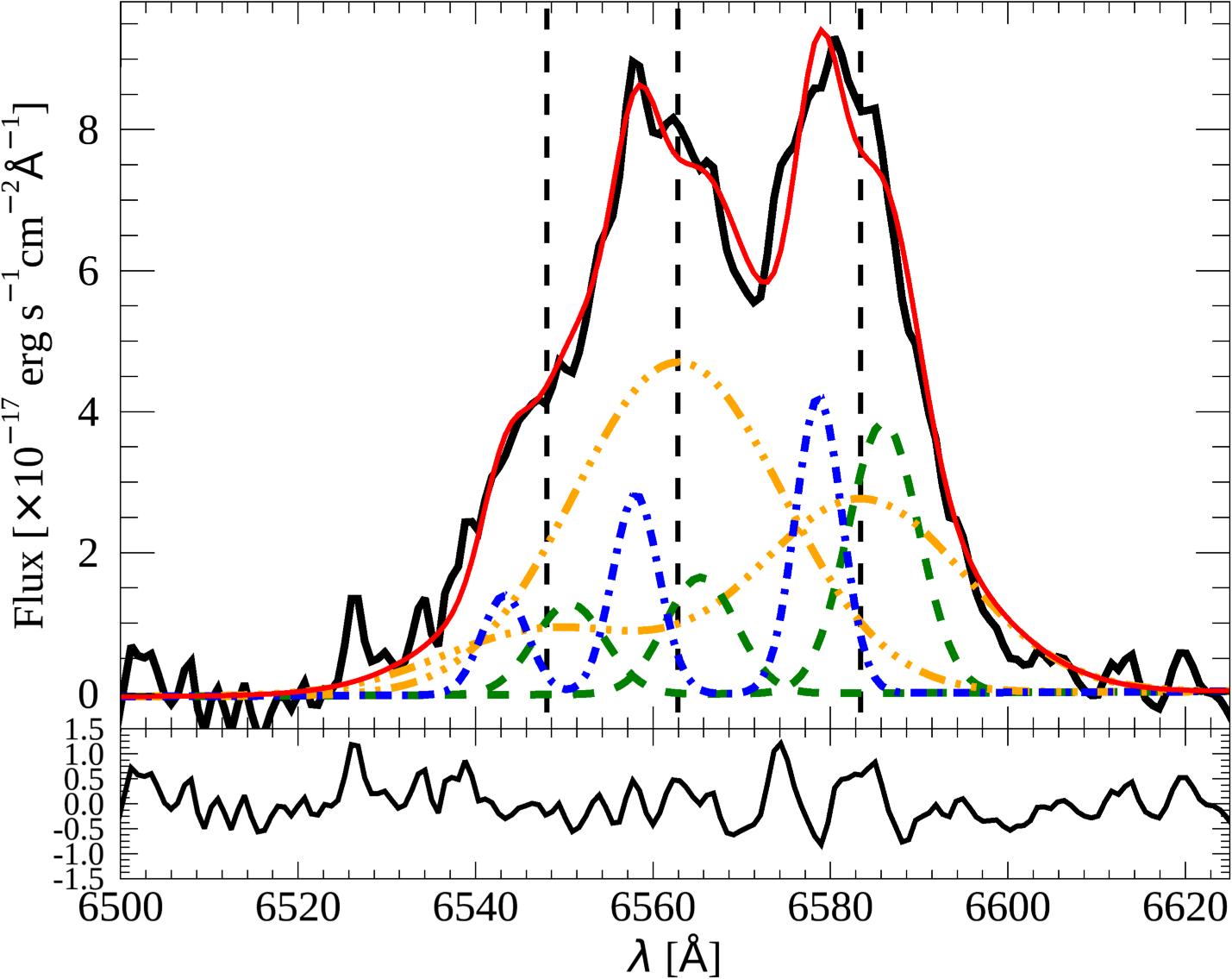}
\end{minipage}   \par}

\begin{minipage}[t]{0.5\textwidth}
\includegraphics[width=\textwidth, height=0.21\textheight]{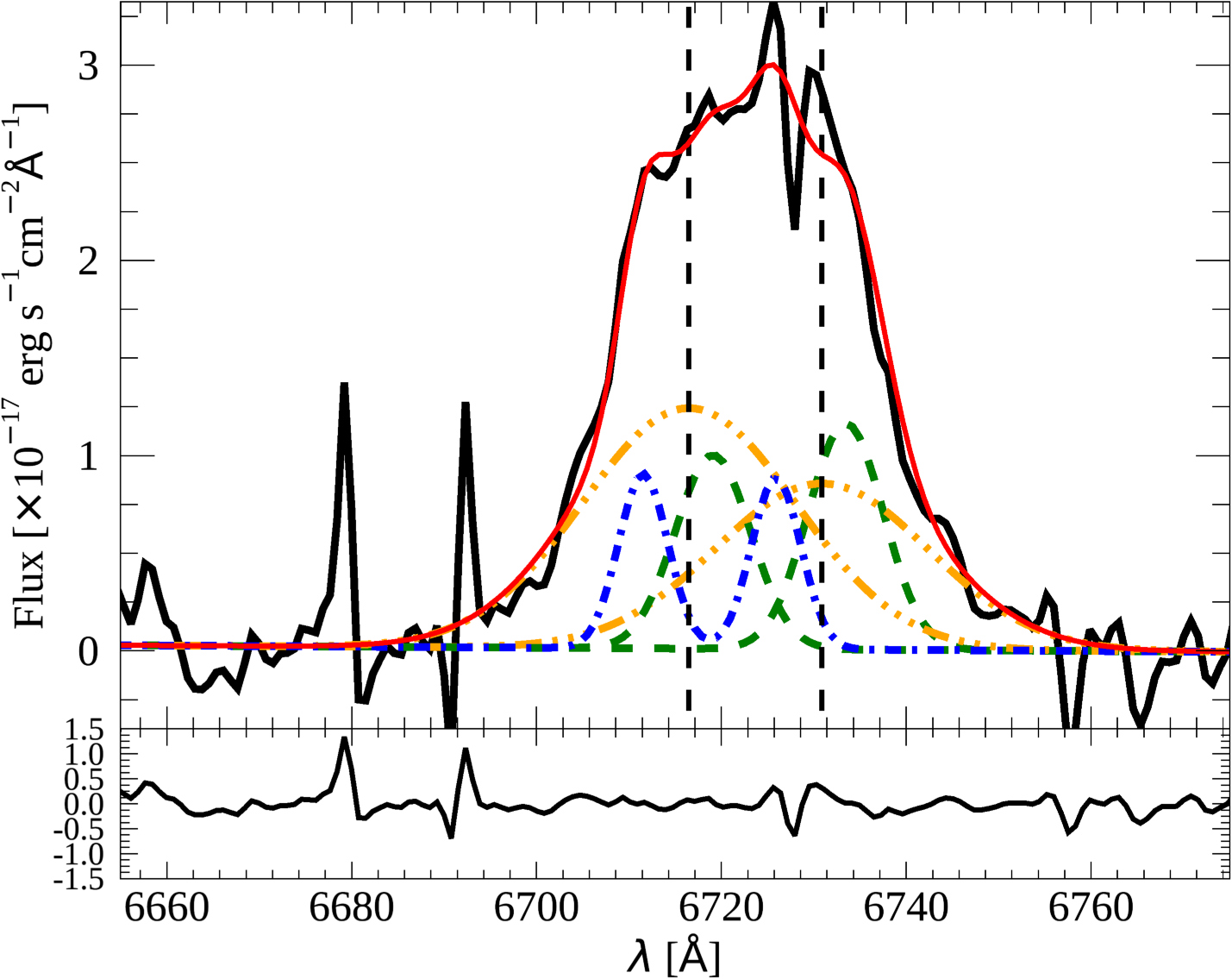}
\end{minipage}
\begin{minipage}[t]{0.5\textwidth}
\includegraphics[width=\textwidth, height=0.21\textheight]{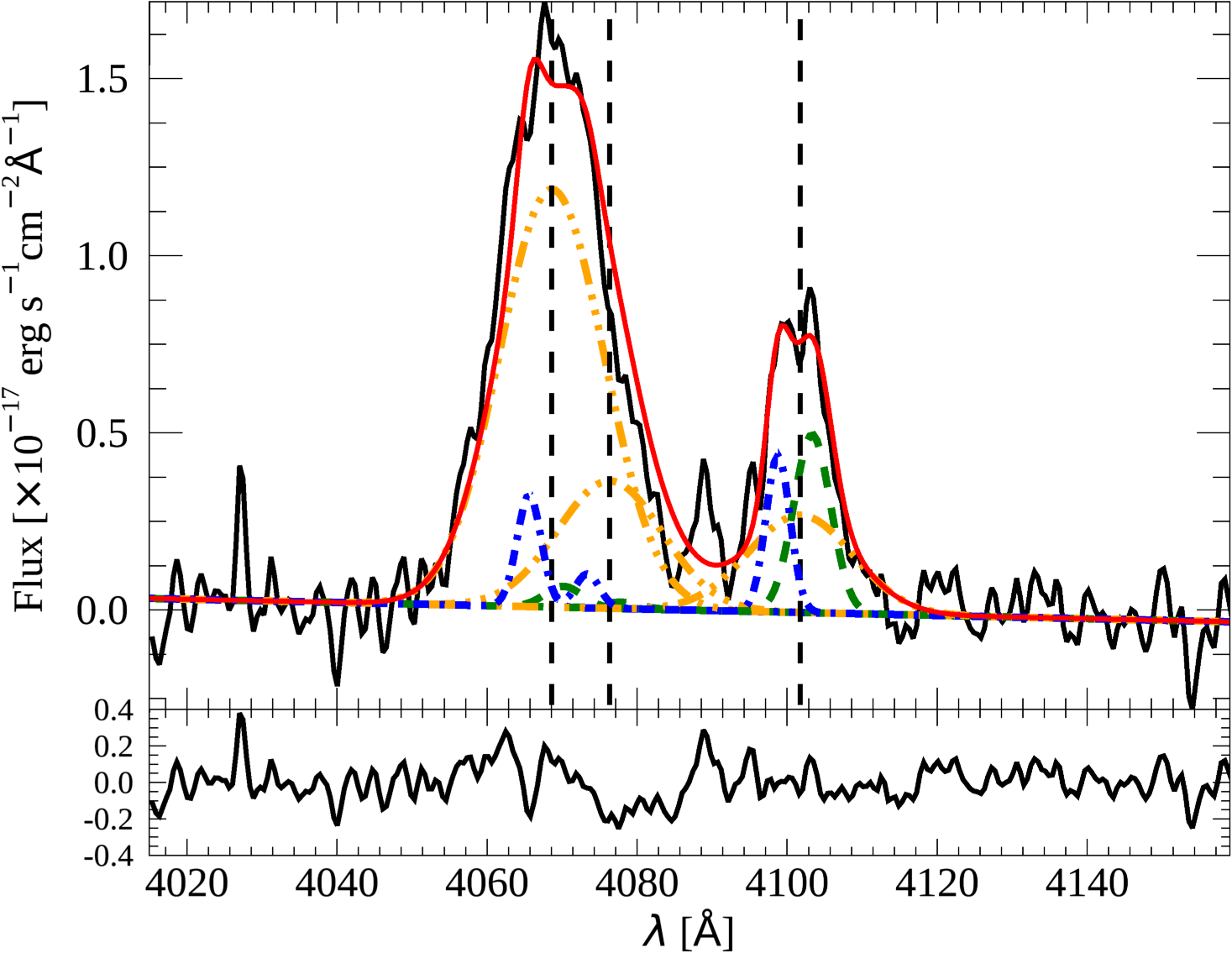}
\end{minipage}

\begin{minipage}[t]{0.5\textwidth}
\includegraphics[width=\textwidth, height=0.21\textheight]{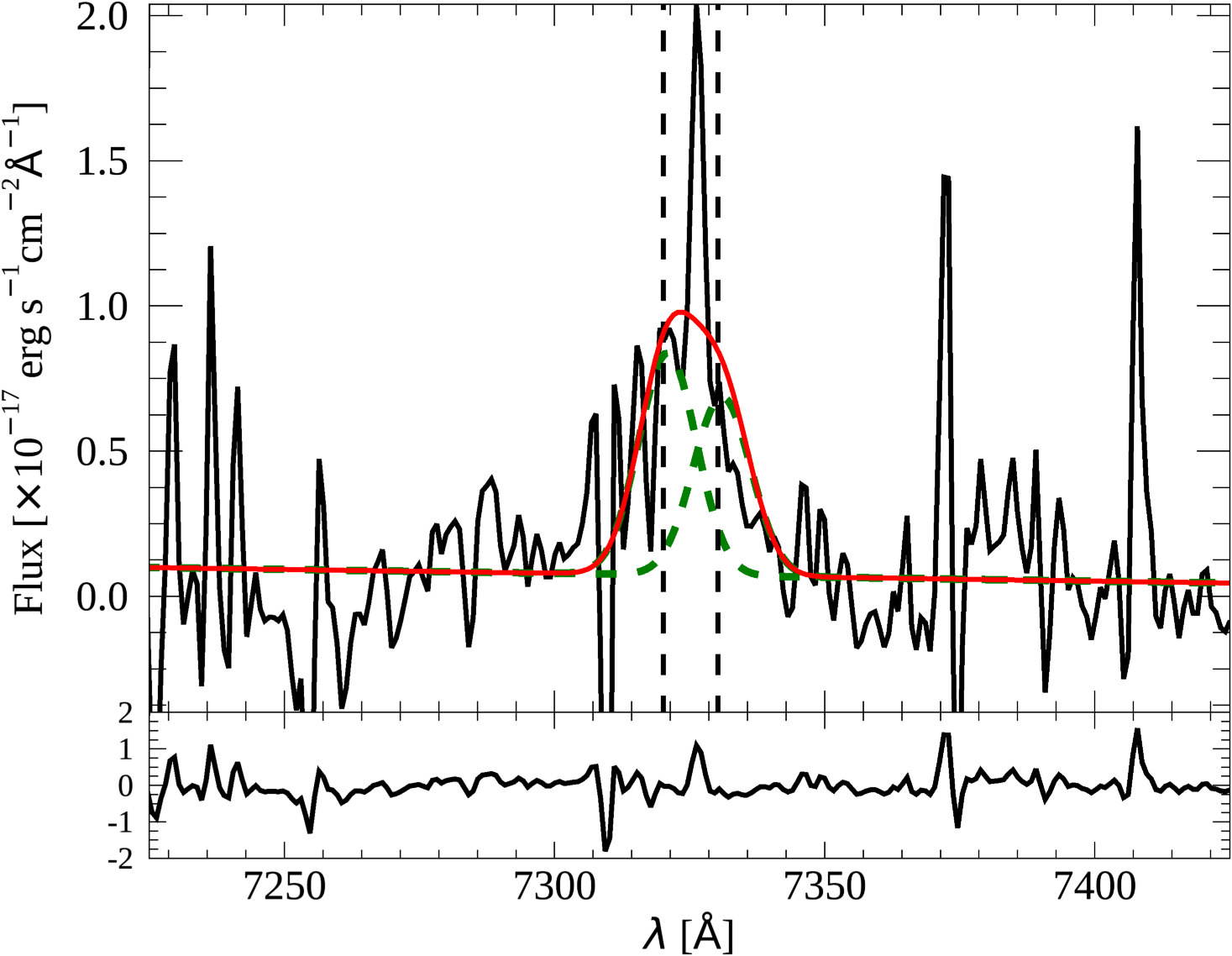}
\end{minipage}
\begin{minipage}[t]{0.5\textwidth}
\includegraphics[width=\textwidth, height=0.21\textheight]{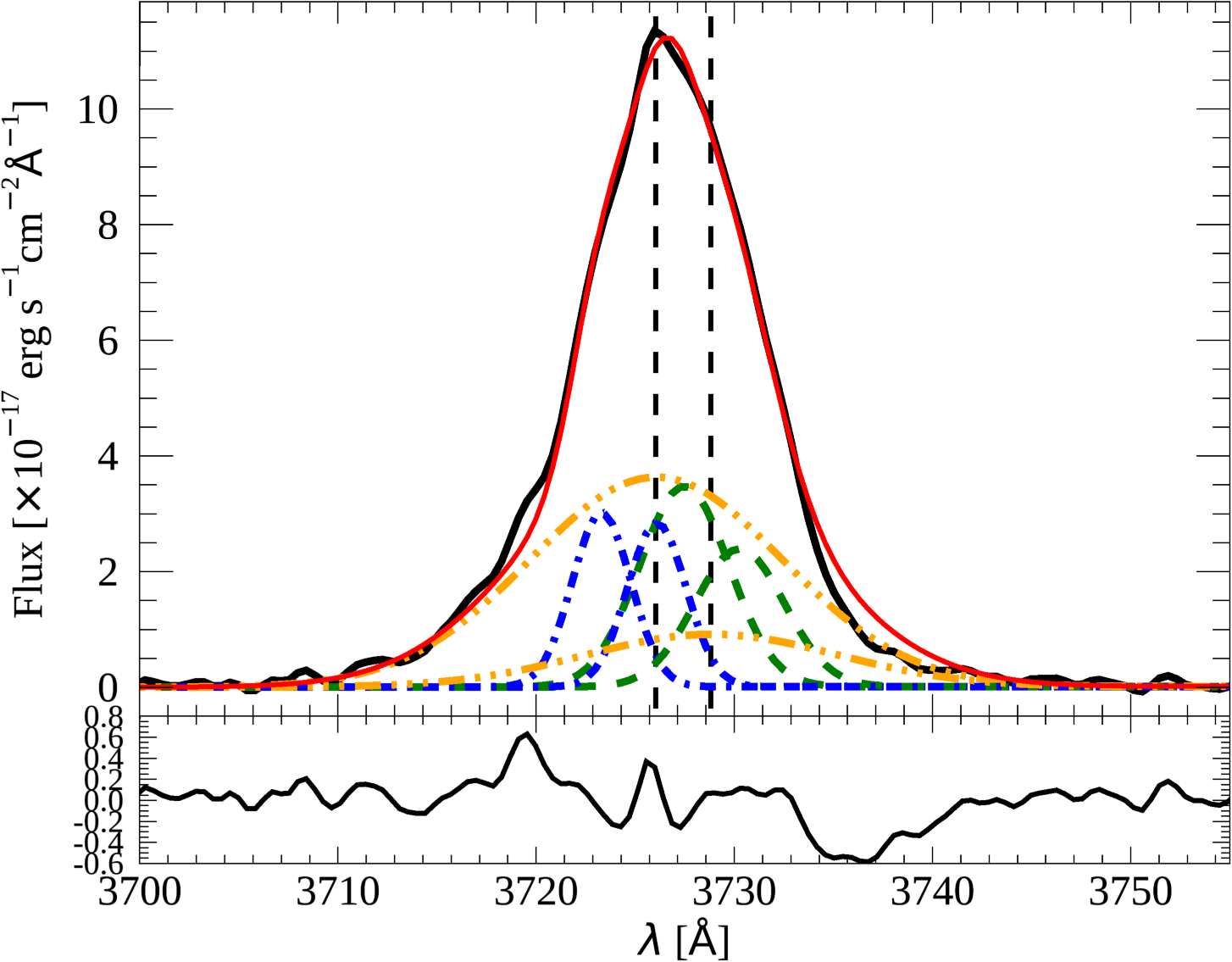}
\end{minipage}

\caption{ PKS0252--71: \textit{First row:} \OIII$\lambdaup\lambdaup$4958,5007\AA\ (left panel) and H$\beta$ line (right panel) fits. \textit{Second row:} H$\alpha$+\NII$\lambdaup\lambdaup$6548,84\AA\ line fits. 
\textit{Third row:} \SII$\lambdaup\lambdaup$6717,31\AA\ (left panel) and \SII$\lambdaup\lambdaup$4069,76\AA\ (right panel) trans-auroral line fits. The latter fit includes also the H$\delta$ line. 
\textit{Fourth row:} \OII$\lambdaup\lambdaup$7319,30\AA\ (left panel) and \OII$\lambdaup\lambdaup$3726,29\AA\ (right panel) trans-auroral line fits. When needed, the former fit includes also the \OII$\lambdaup$7381\AA\ line which has been modelled with a single Gaussian component. In each of the figures the upper panel shows the best fit (red solid line) of the observed spectrum (black solid line) while the lower panel shows the residuals of the fit. The different kinematic components used for the fit of each emission line are showed with different colors and line styles, in the case of doublets where flux ratios have been fixed (i.e. the \OIII$\lambdaup\lambdaup$4958,5007\AA\ and the \NII$\lambdaup\lambdaup$6548,84\AA\ ), we show the total profile of each doublet kinematic component. The vertical dashed lines marks the rest-frame wavelength of the fitted emission lines. Wavelengths are plotted in \AA,\ and the flux scale is given in units of $10^{-17} \rm{erg~s^{-1}cm^{-2}\AA^{-1}}$. } 
\label{PKS0252Emission lines fits}
\end{figure*}

\begin{figure*}[h]
\begin{minipage}[t]{0.54\textwidth}
\includegraphics[angle=90]{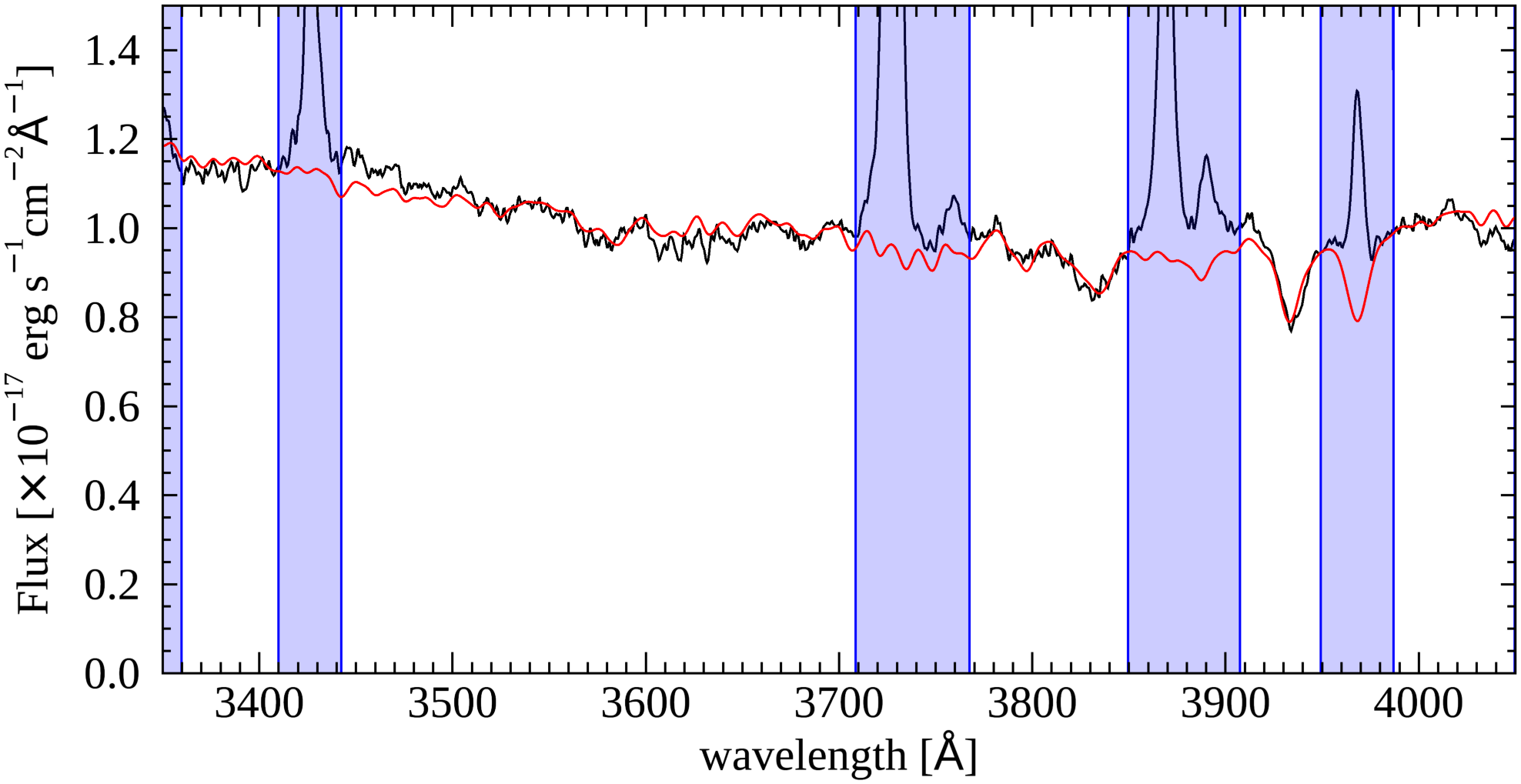}
\end{minipage}
\begin{minipage}[t]{0.5\textwidth}
\includegraphics[ angle=90]{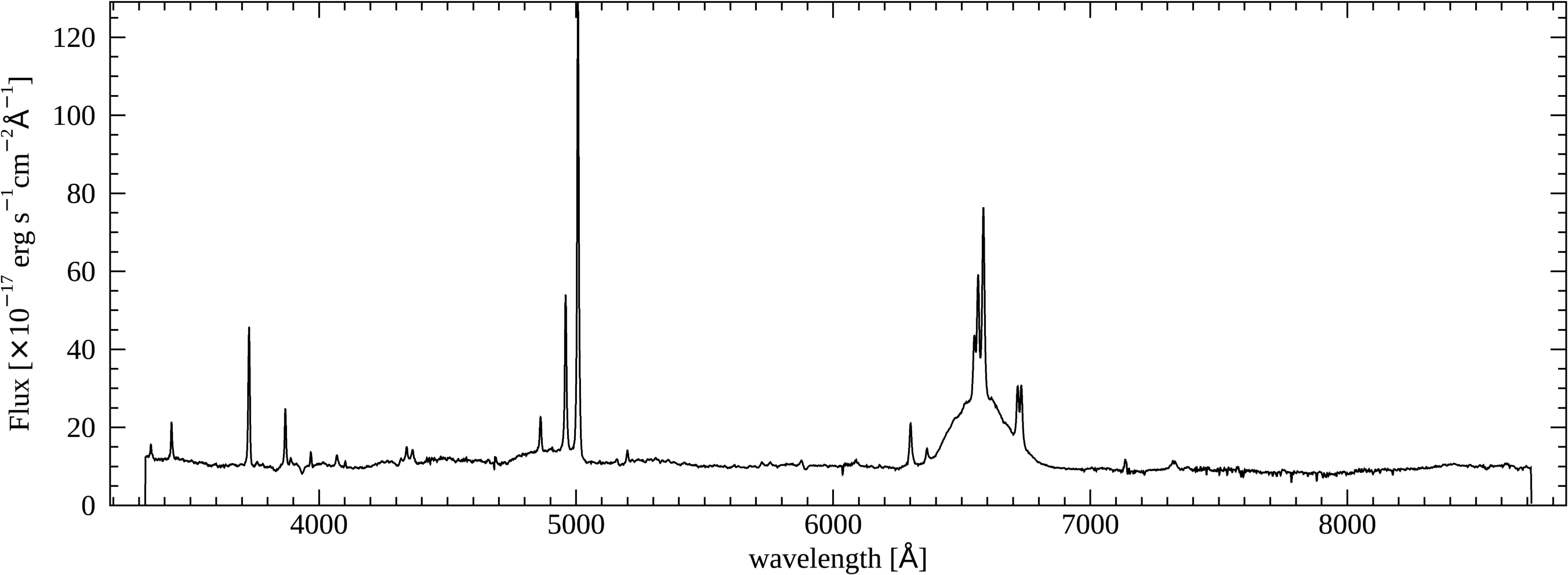}
\end{minipage}
\caption{ PKS~1151--34: \textit{Upper panel :} UVB+VIS Nuclear spectrum (black solid line) and best-fit model (red solid line) for the continuum emission in the wavelength range 3350-4450\AA\, regions of the spectrum corresponding to masked emission lines are indicated in blue. \textit{Lower panel :} Nuclear spectrum (with no subtraction of the stellar continuum) showing the emission lines associated with the warm ionised gas an the BLR emission. Wavelengths are plotted in \AA, and the flux scale is given in units of $10^{17} \rm{erg~s^{-1}cm^{-2}\AA^{-1}}$.} 
\label{PKS1151Stellar_continuum}
\end{figure*}

\begin{figure*}[h]
\begin{center}

\begin{minipage}[t]{0.4\textwidth}
\includegraphics[width=\textwidth, height=0.2\textheight]{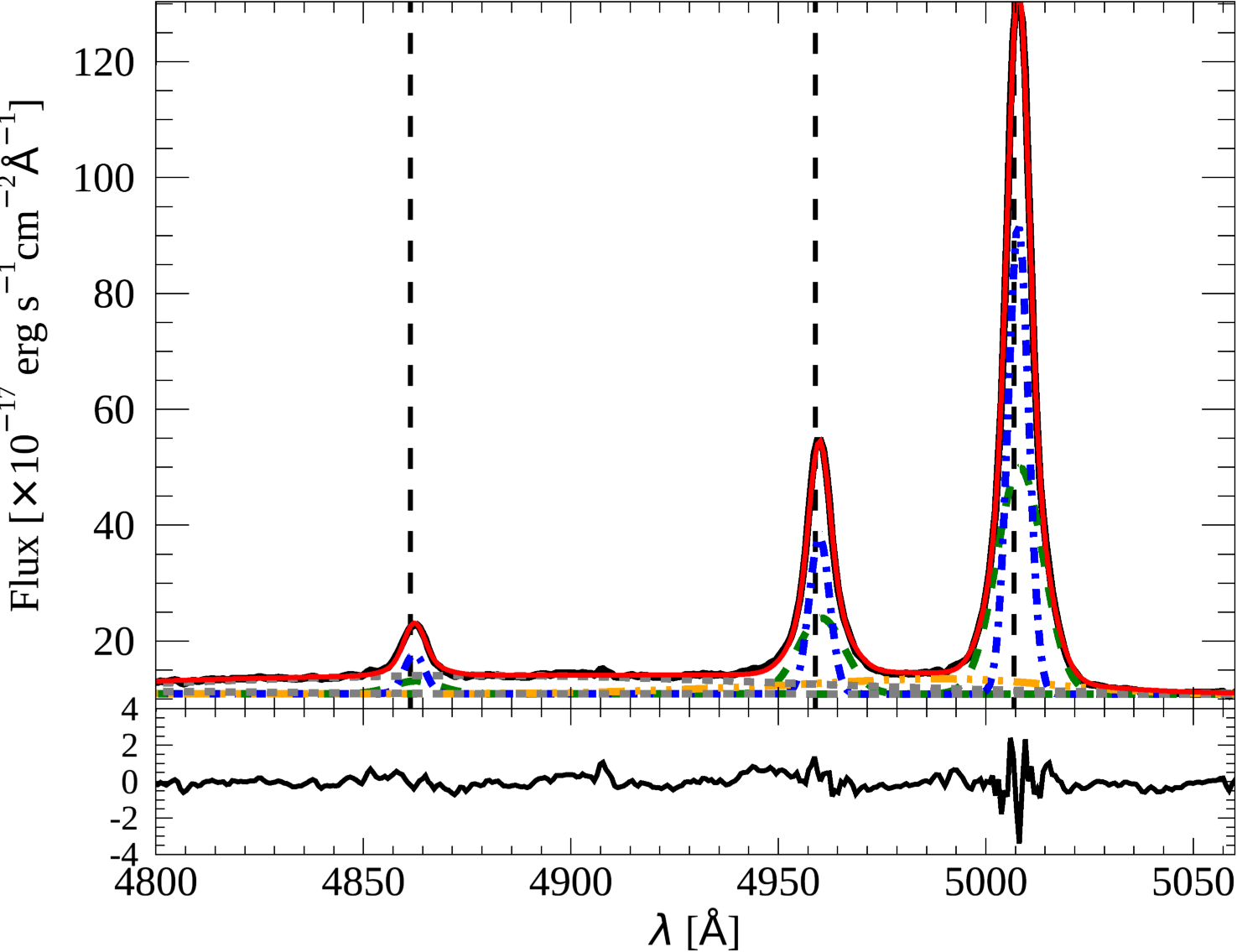}
\end{minipage}
\begin{minipage}[t]{0.4\textwidth}
\includegraphics[width=\textwidth, height=0.2\textheight]{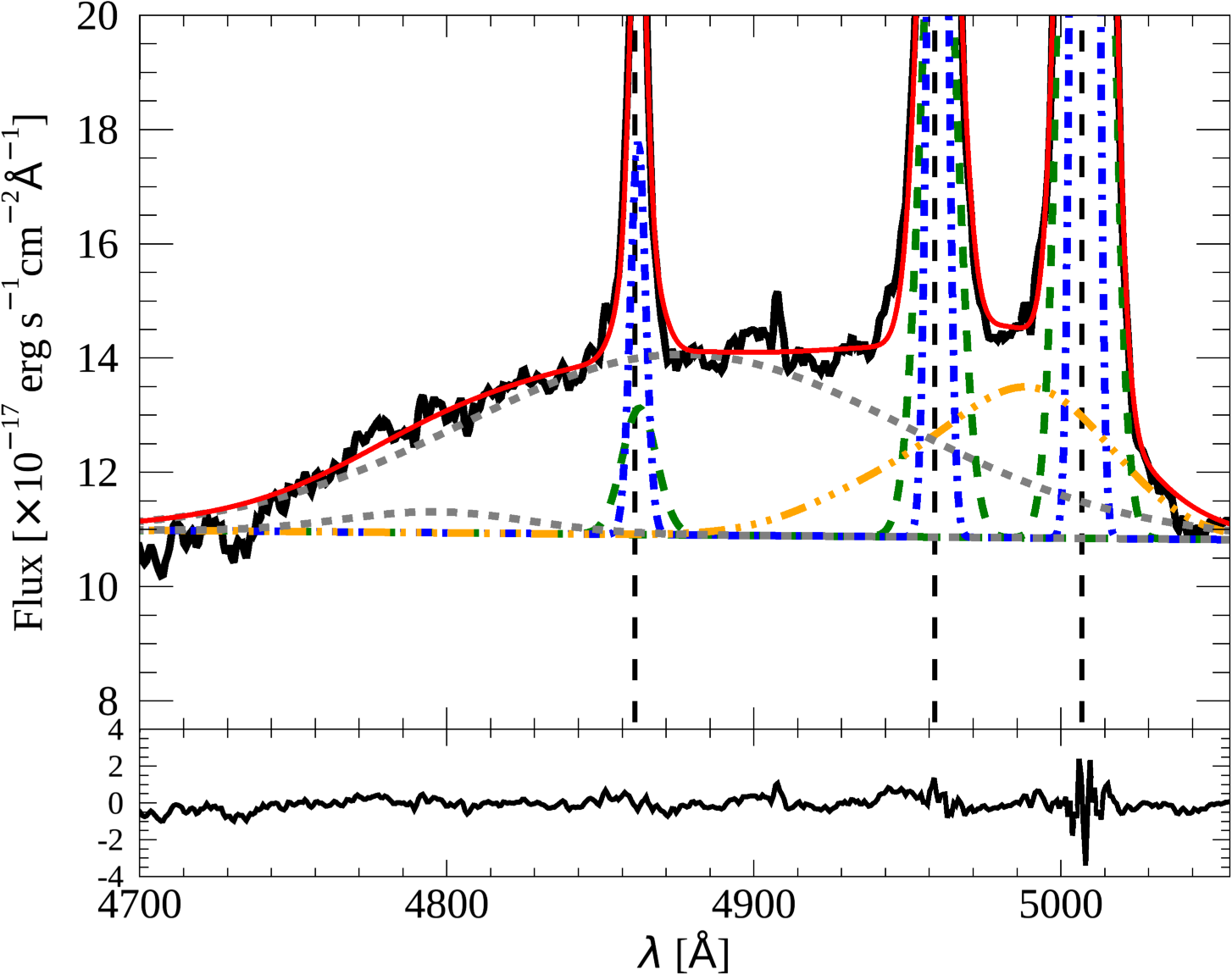}
\end{minipage}

\begin{minipage}[b]{0.8\textwidth}
\includegraphics[width=\textwidth, height=0.2\textheight]{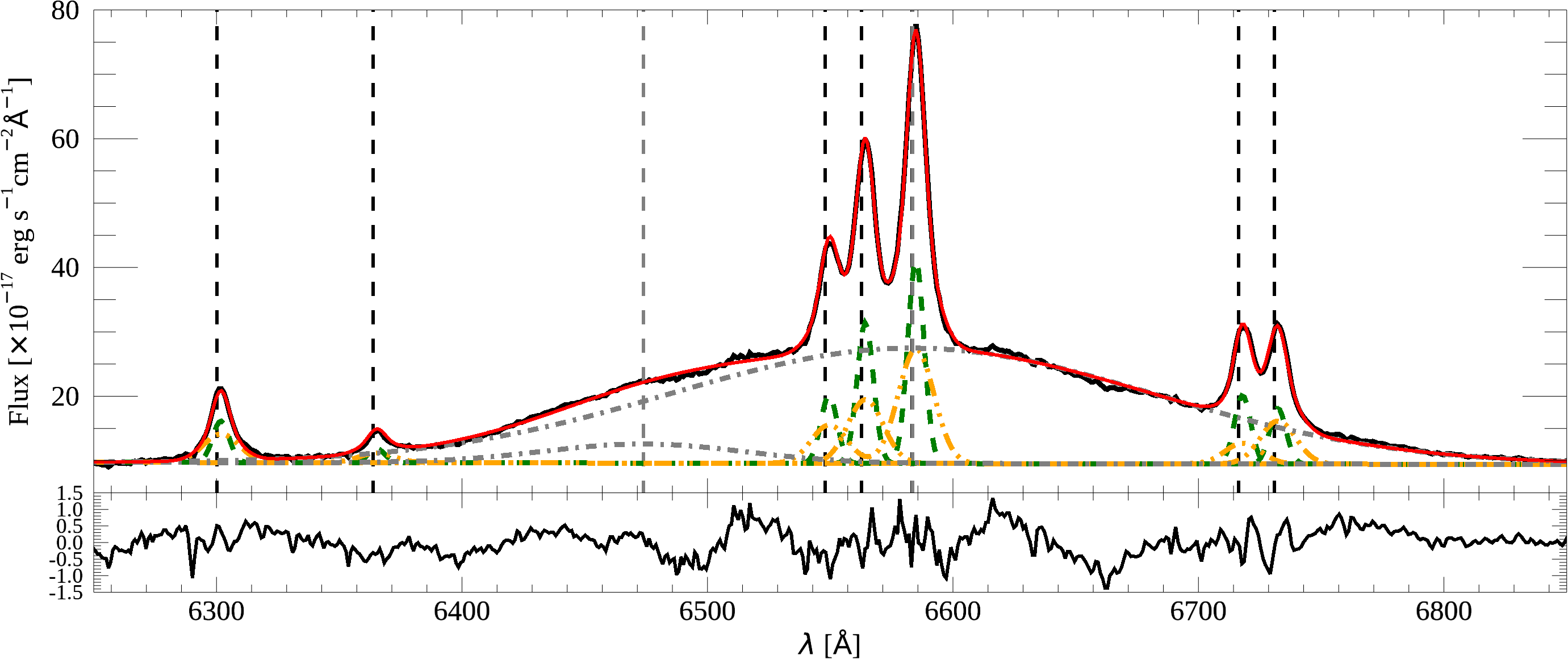}
\end{minipage}

\begin{minipage}[t]{0.4\textwidth}
\includegraphics[width=\textwidth, height=0.2\textheight]{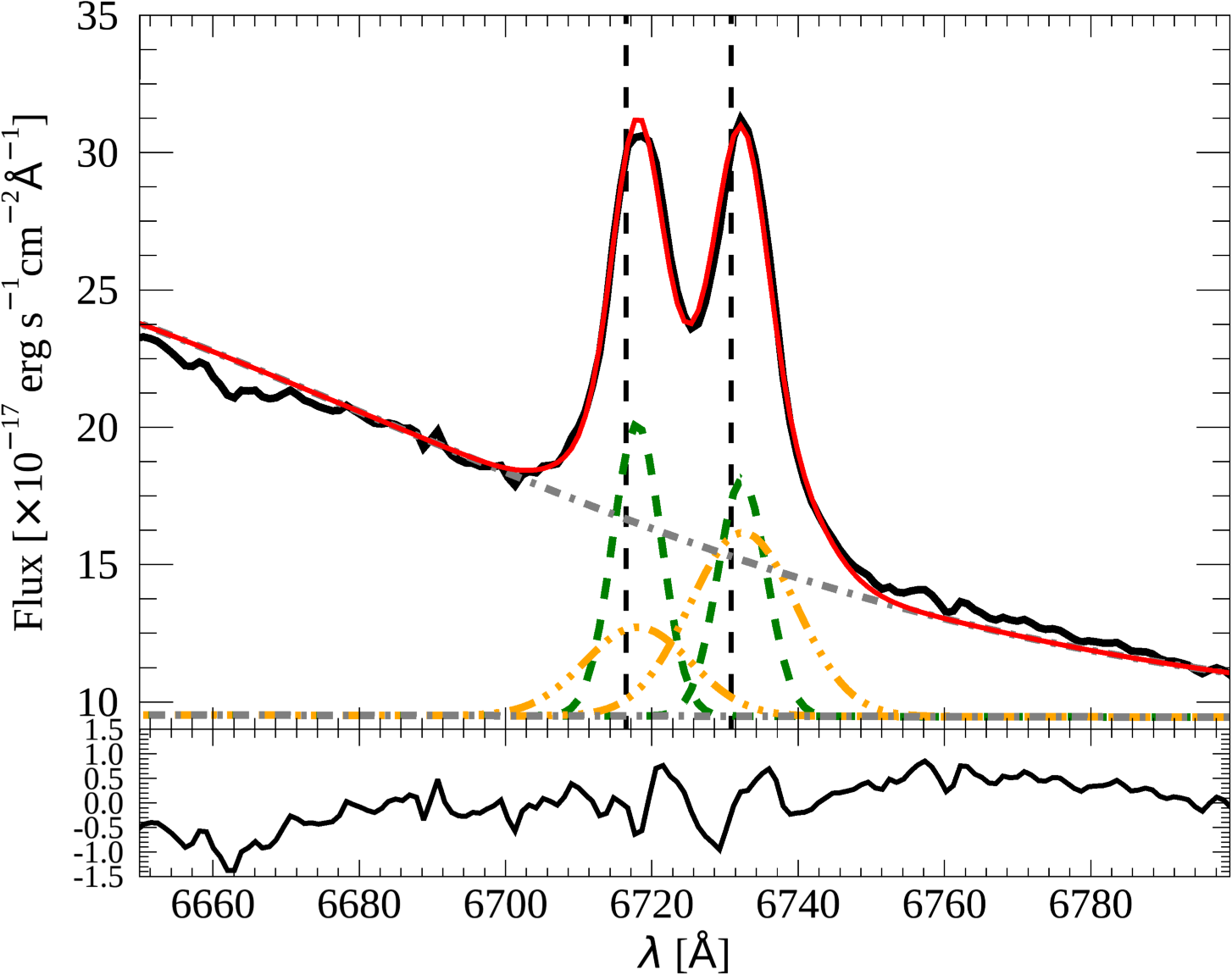}
\end{minipage}
\begin{minipage}[t]{0.4\textwidth}
\includegraphics[width=\textwidth, height=0.2\textheight]{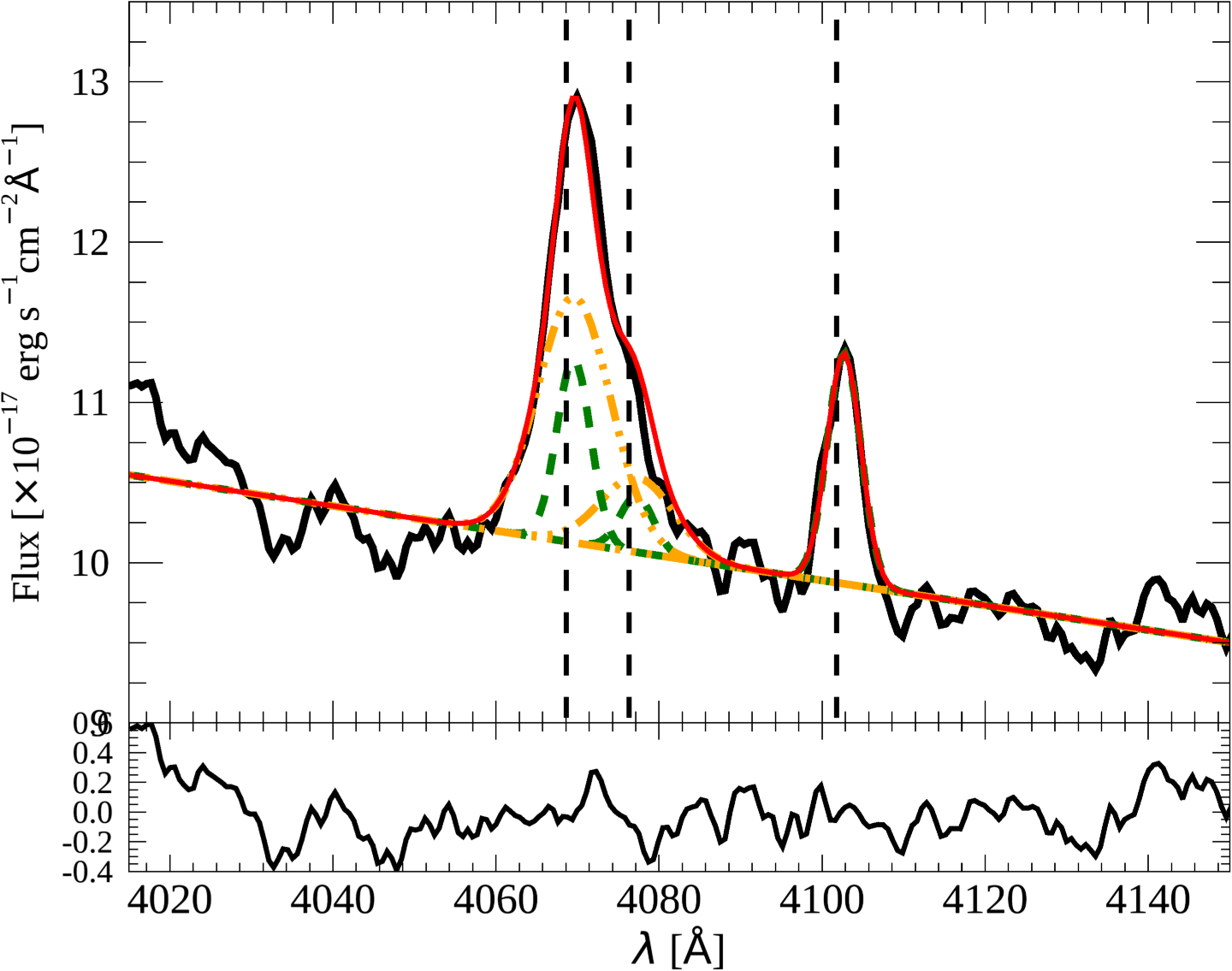}
\end{minipage}

\begin{minipage}[t]{0.4\textwidth}
\includegraphics[width=\textwidth, height=0.2\textheight]{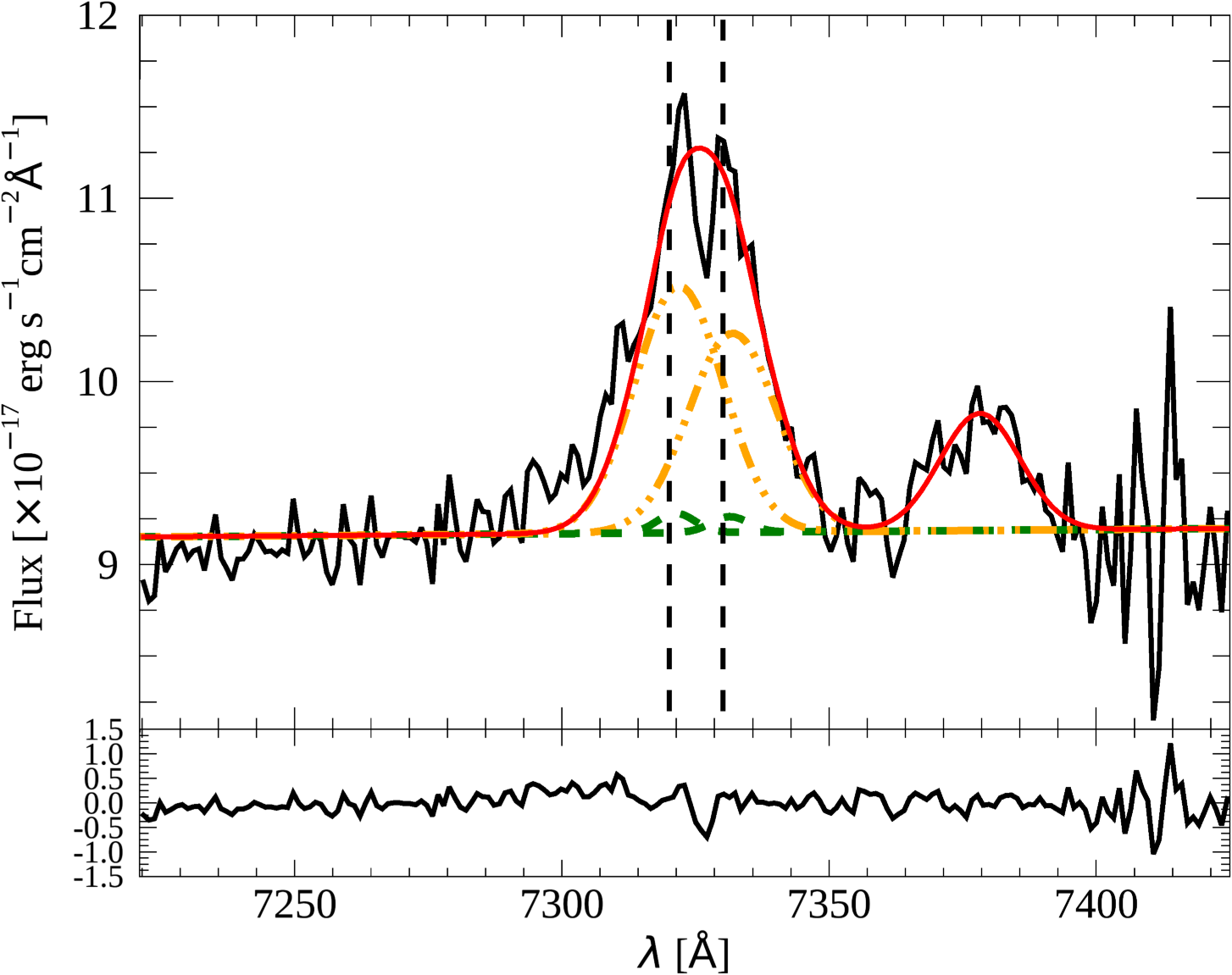}
\end{minipage}
\begin{minipage}[t]{0.4\textwidth}
\includegraphics[width=\textwidth, height=0.2\textheight]{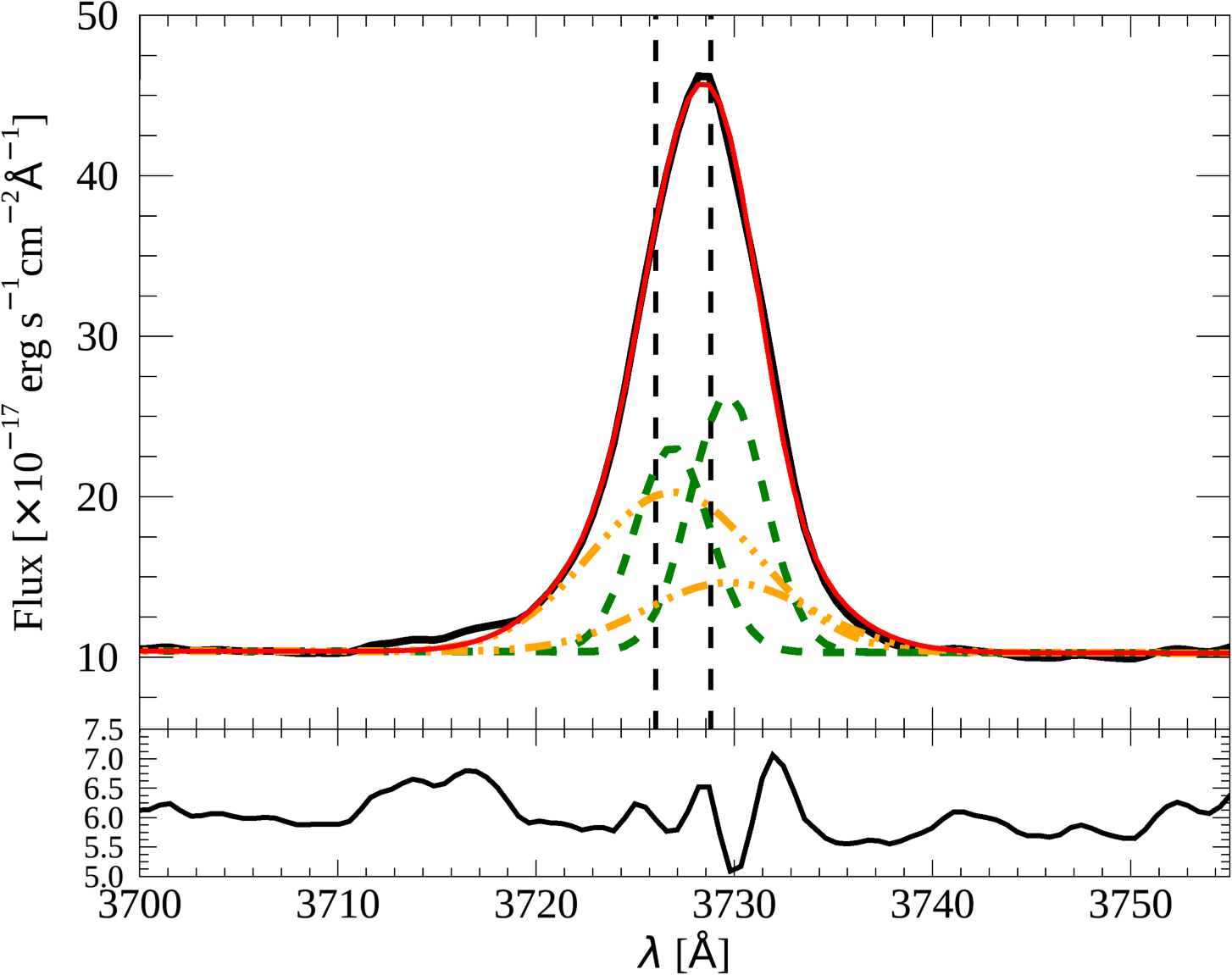}
\end{minipage}
\end{center}

\caption{ PKS~1151--34: \textit{First row:} \OIII$\lambdaup\lambdaup$4958,5007\AA\ + H$\beta$ line fit (left panel) and a zoom-in (right panel) highlighting the presence of the emission from the AGN BLR.  \textit{Second row:} \OI$\lambdaup\lambdaup$6300,6363\AA\ + H$\alpha$+\NII$\lambdaup\lambdaup$6548,84\AA\ +\SII$\lambdaup\lambdaup$6717,31\AA\ line fits. 
\textit{Third row:} \SII$\lambdaup\lambdaup$6717,31\AA\ (left panel) and \SII$\lambdaup\lambdaup$4069,76\AA\ (right panel) trans-auroral line fits. The latter fit includes also the H$\delta$ line.
\textit{Fourth row:} \OII$\lambdaup\lambdaup$7319,30\AA\ (left panel) and \OII$\lambdaup\lambdaup$3726,29\AA\ (right panel) trans-auroral line fits. When needed, the former fit includes also the \OII$\lambdaup$7381\AA\ line which has been modelled with a single Gaussian component. In each of the figures the upper panel shows the best fit (red solid line) of the observed spectrum (black solid line) while the lower panel shows the residuals of the fit. The different kinematic components used for the fit of each emission line are showed with different colors and line styles, in the case of doublets where flux ratios have been fixed (i.e. the \OIII$\lambdaup\lambdaup$4958,5007\AA\ and the \NII$\lambdaup\lambdaup$6548,84\AA\ ), we show the total profile of each doublet kinematic component. The vertical dashed lines marks the rest-frame wavelength of the fitted emission lines. The AGN BLR has been modelled by the components shown with grey color whose centroid velocity is also marked by the grey vertical dashed line.  Wavelengths are plotted in \AA,\ and the flux scale is given in units of $10^{-17} \rm{erg~s^{-1}cm^{-2}\AA^{-1}}$.} 
\label{PKS1151Emission lines fits}
\end{figure*}

\begin{figure*}[h]
\begin{minipage}[t]{0.5\textwidth}
\includegraphics[angle=90]{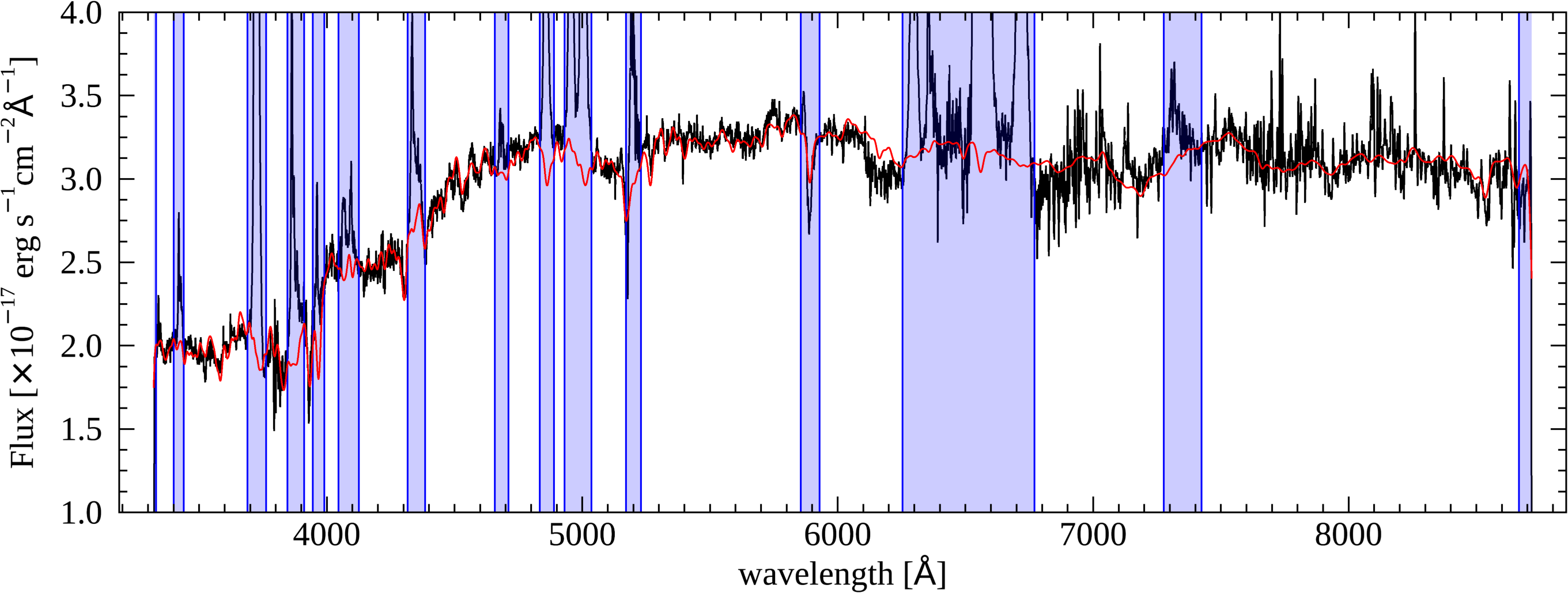}
\end{minipage}
\begin{minipage}[t]{0.5\textwidth}
\includegraphics[ angle=90]{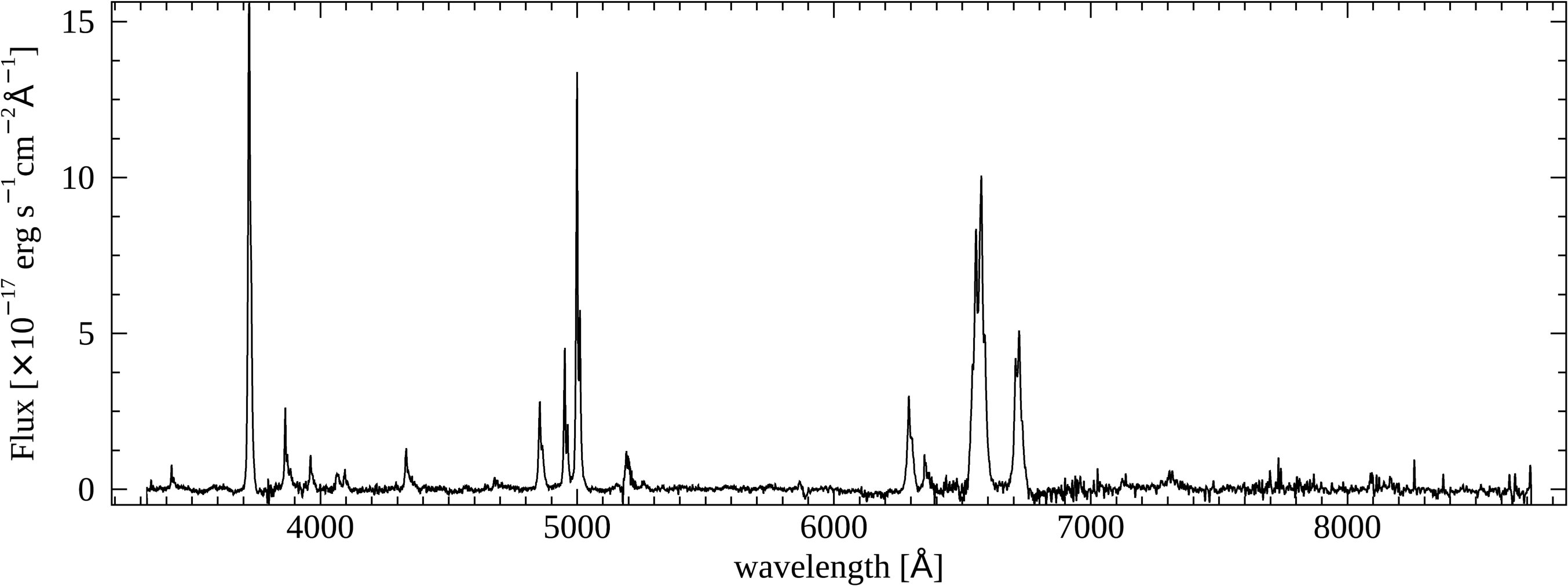}
\end{minipage}
\caption{  PKS1306--09: As in Fig.\ref{Stellar_continuum}.} 
\label{PKS1306Stellar_continuum}
\end{figure*}

\begin{figure*}[h]
\begin{minipage}[t]{0.5\textwidth}
\includegraphics[width=\textwidth, height=0.225\textheight]{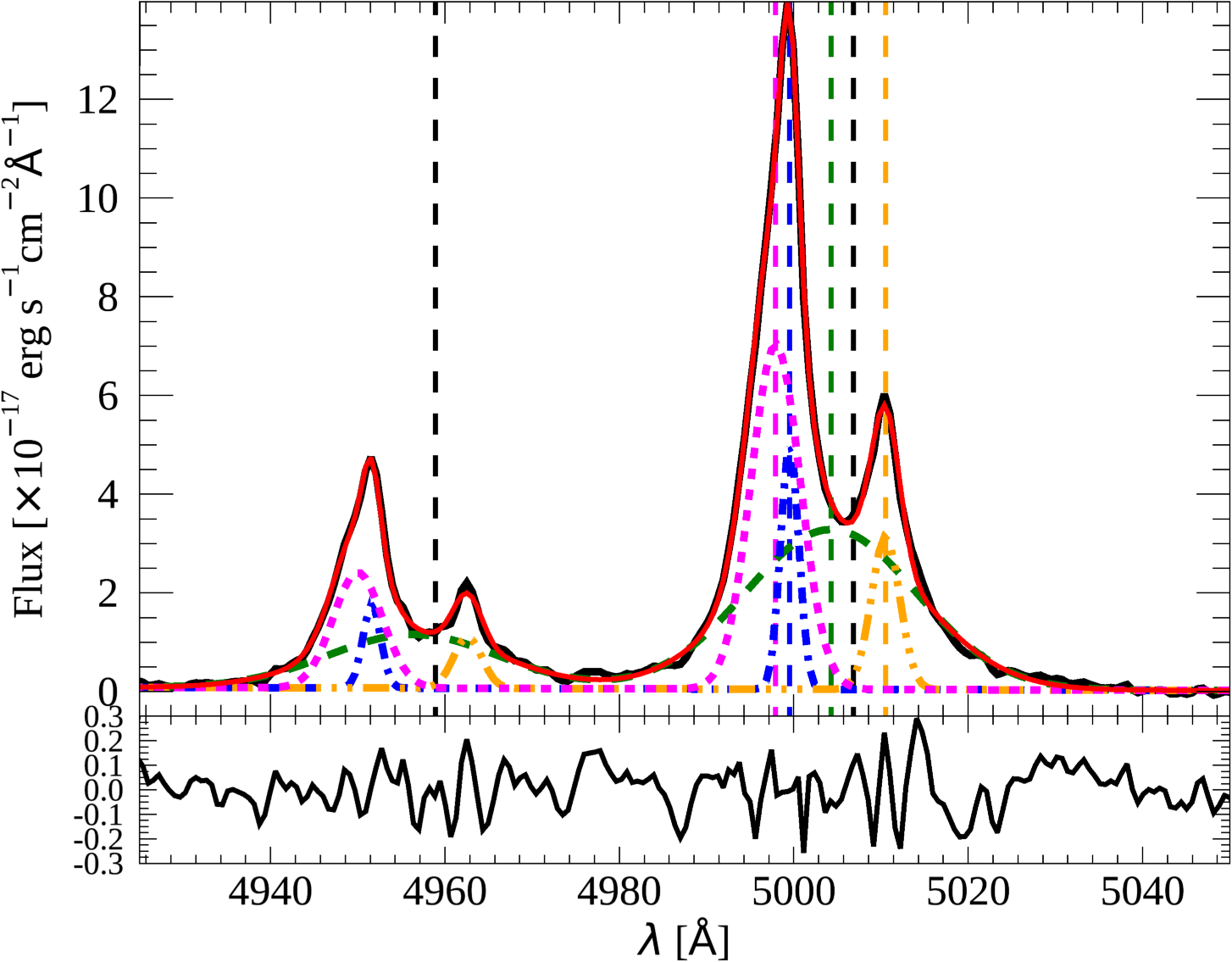}
\end{minipage}
\begin{minipage}[t]{0.5\textwidth}
\includegraphics[width=\textwidth, height=0.225\textheight]{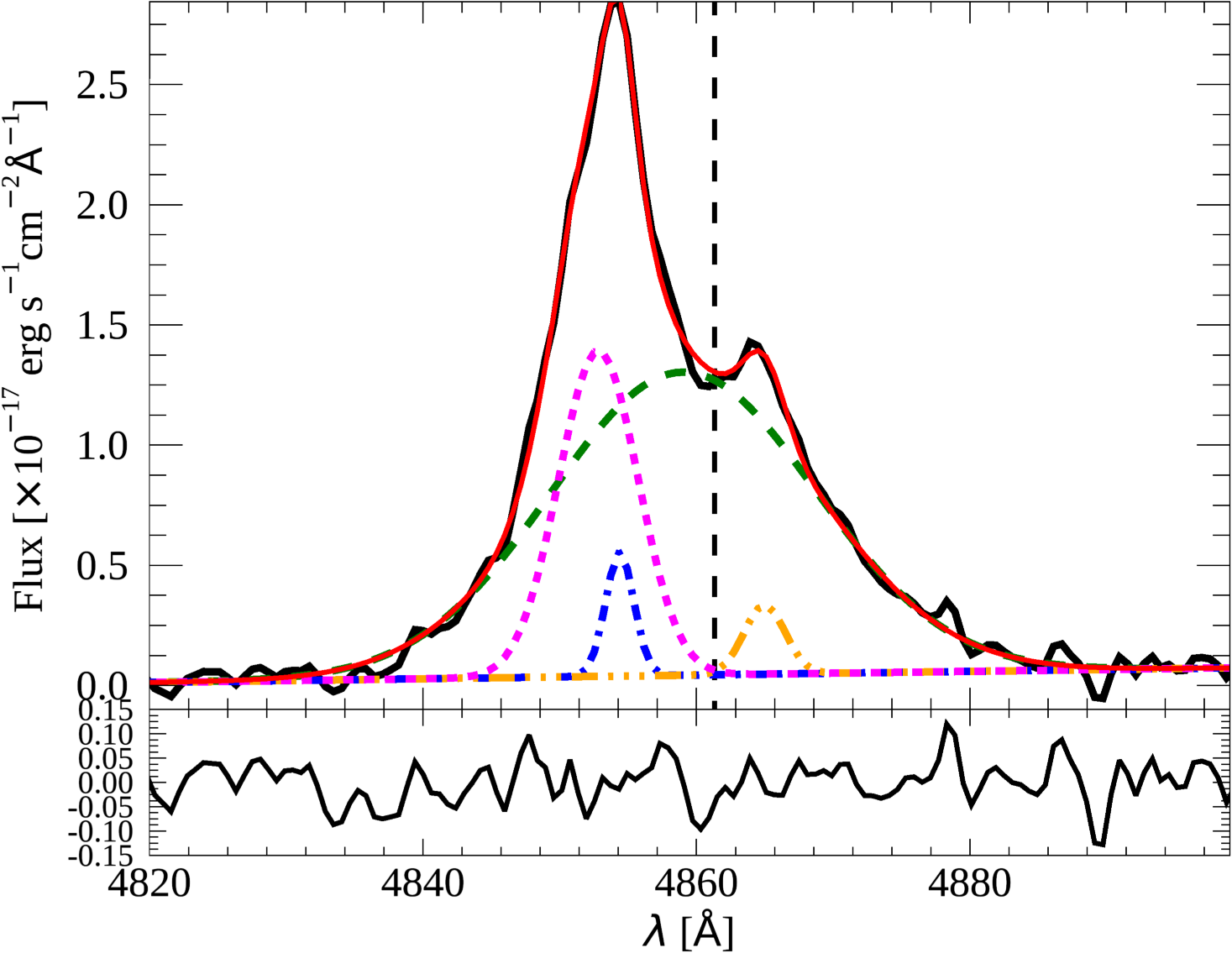}
\end{minipage}

{\centering
\begin{minipage}[b]{0.5\textwidth}
\includegraphics[width=\textwidth, height=0.225\textheight]{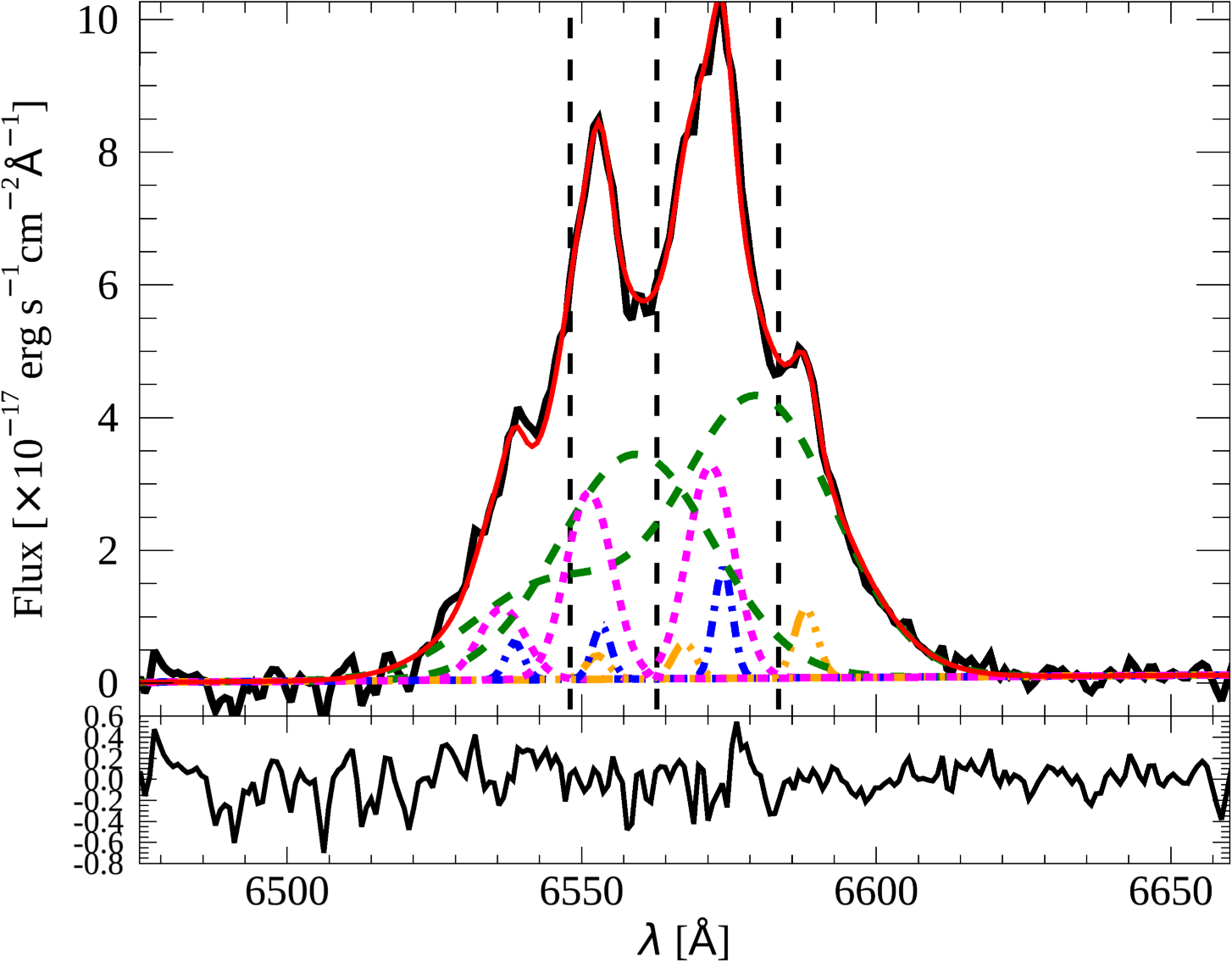}
\end{minipage}   \par}

\begin{minipage}[t]{0.5\textwidth}
\includegraphics[width=\textwidth, height=0.225\textheight]{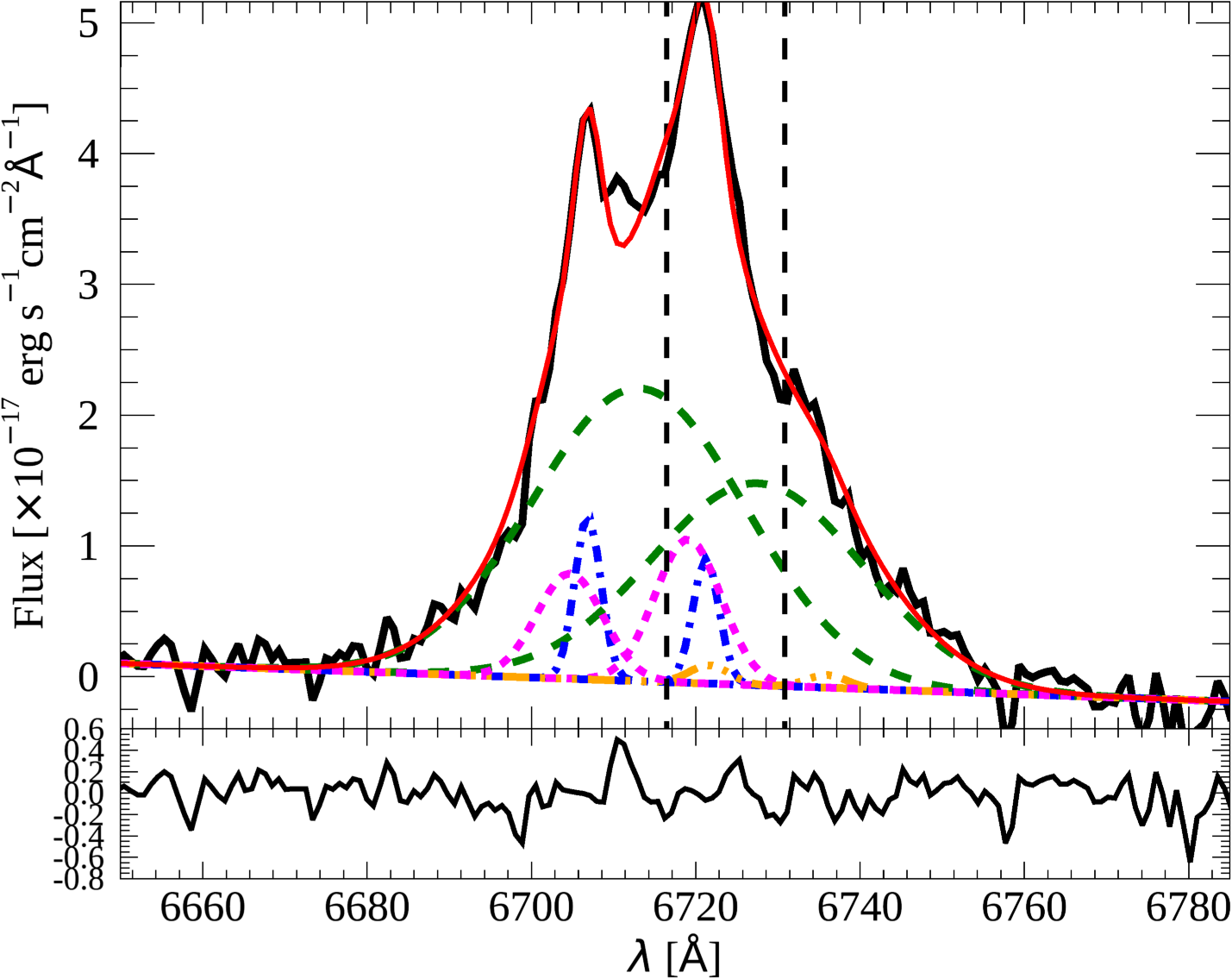}
\end{minipage}
\begin{minipage}[t]{0.5\textwidth}
\includegraphics[width=\textwidth, height=0.225\textheight]{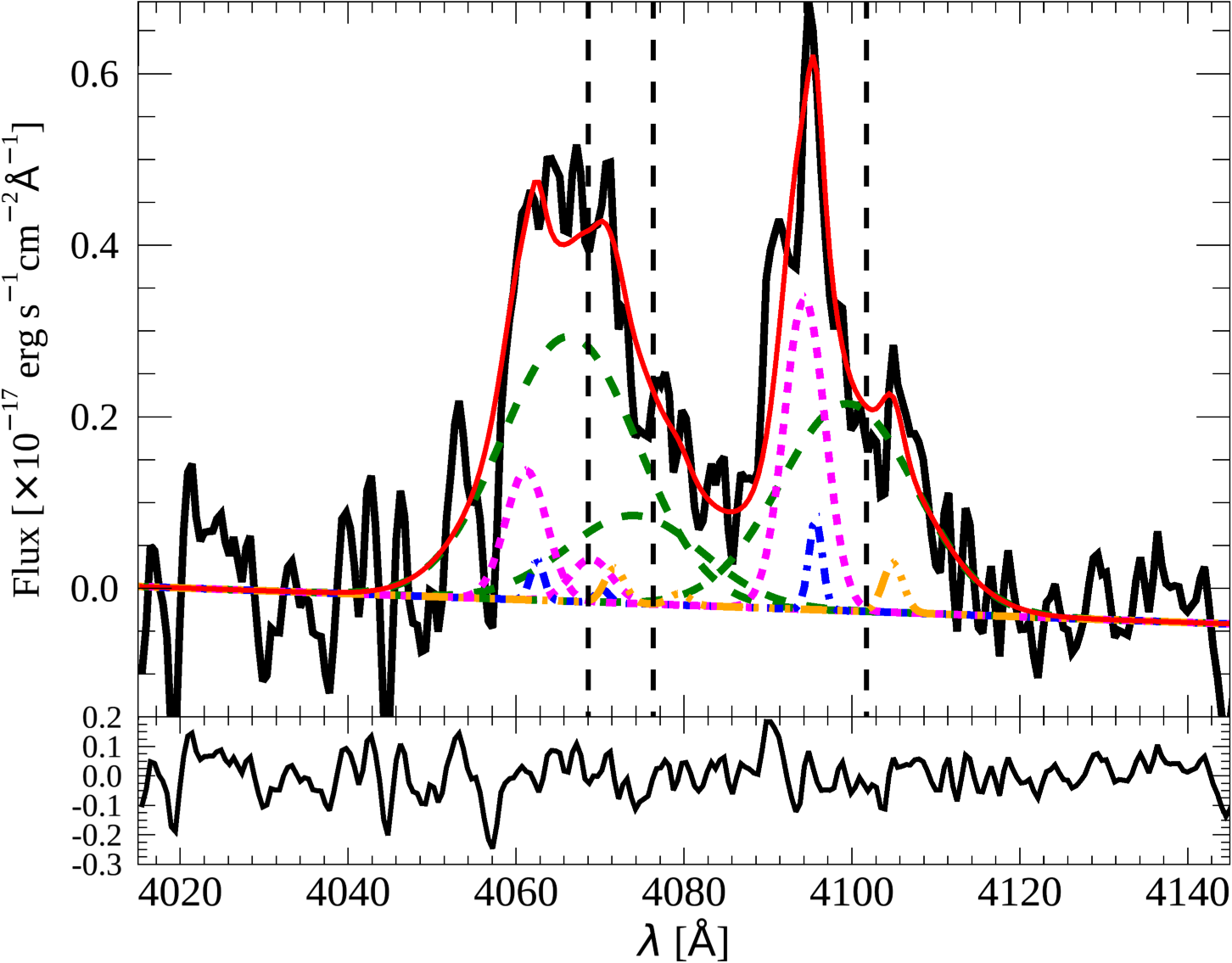}
\end{minipage}

\begin{minipage}[t]{0.5\textwidth}
\includegraphics[width=\textwidth, height=0.225\textheight]{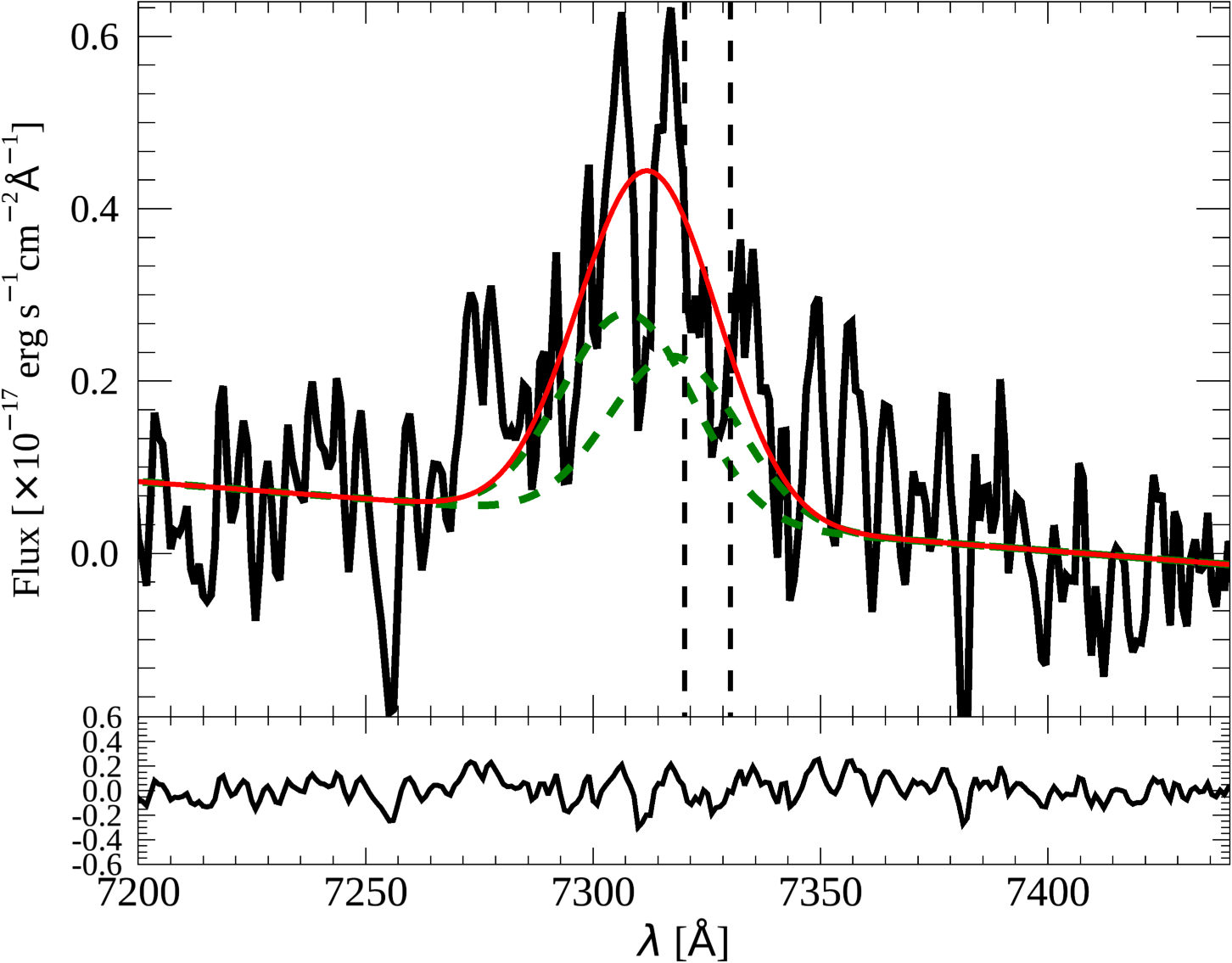}
\end{minipage}
\begin{minipage}[t]{0.5\textwidth}
\includegraphics[width=\textwidth, height=0.225\textheight]{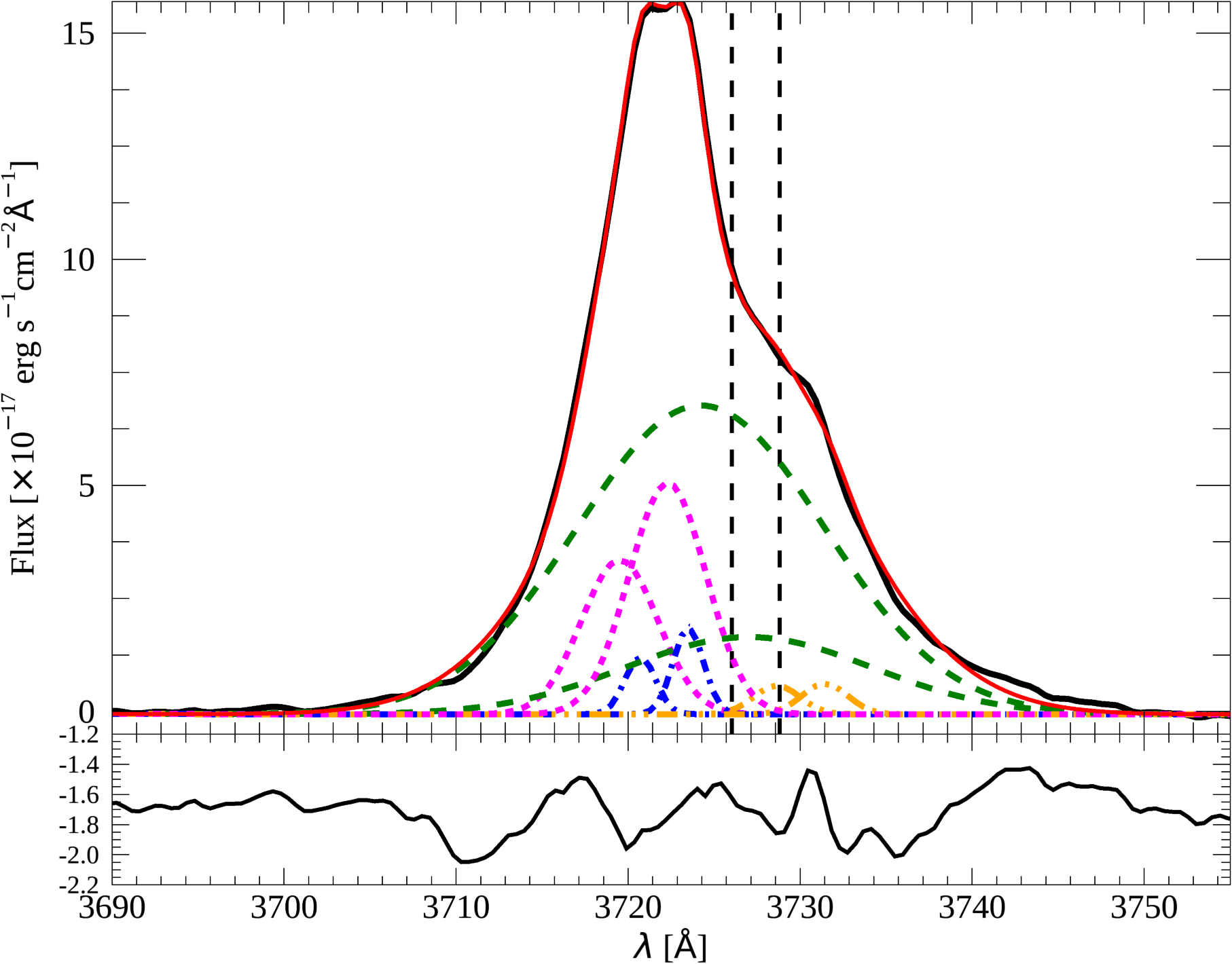}
\end{minipage}

\caption{ PKS~1306--09: As in Fig.\ref{PKS0252Emission lines fits}.} 
\label{PKS1306Emission lines fits}
\end{figure*}

\begin{figure*}[h]
\begin{minipage}[t]{0.5\textwidth}
\includegraphics[angle=90]{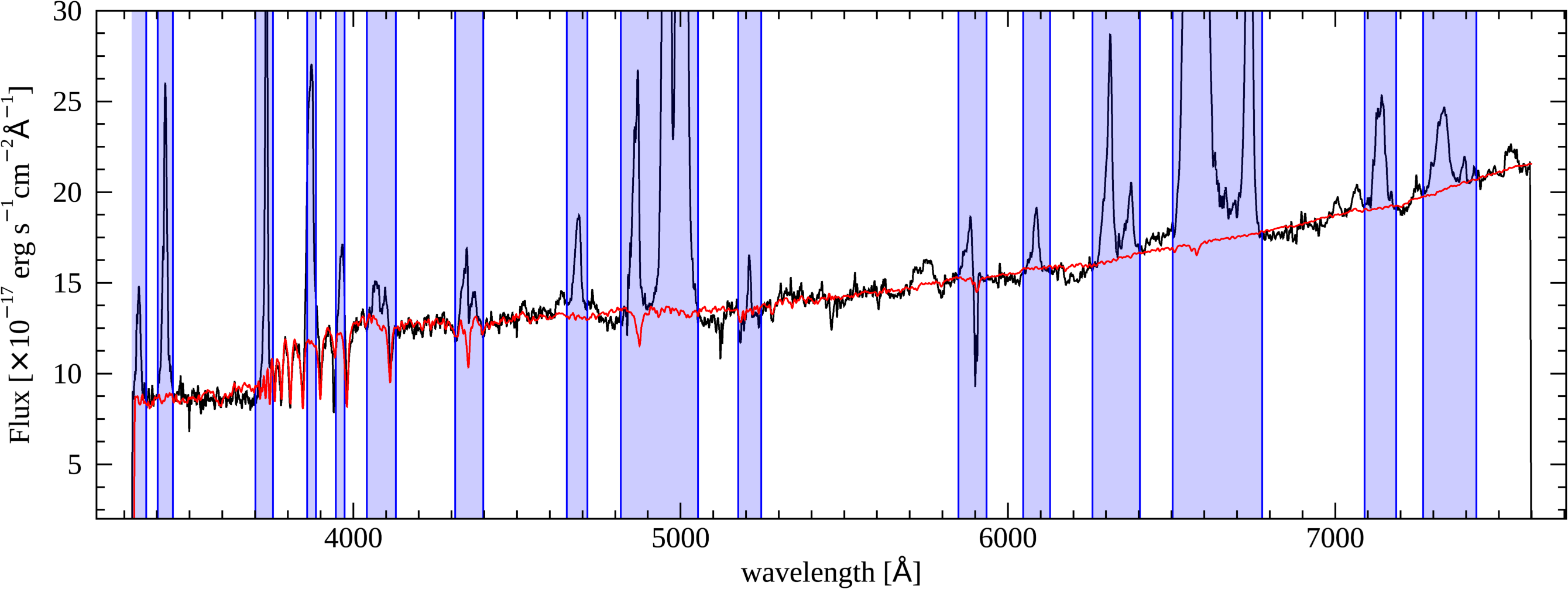}
\end{minipage}
\begin{minipage}[t]{0.5\textwidth}
\includegraphics[ angle=90]{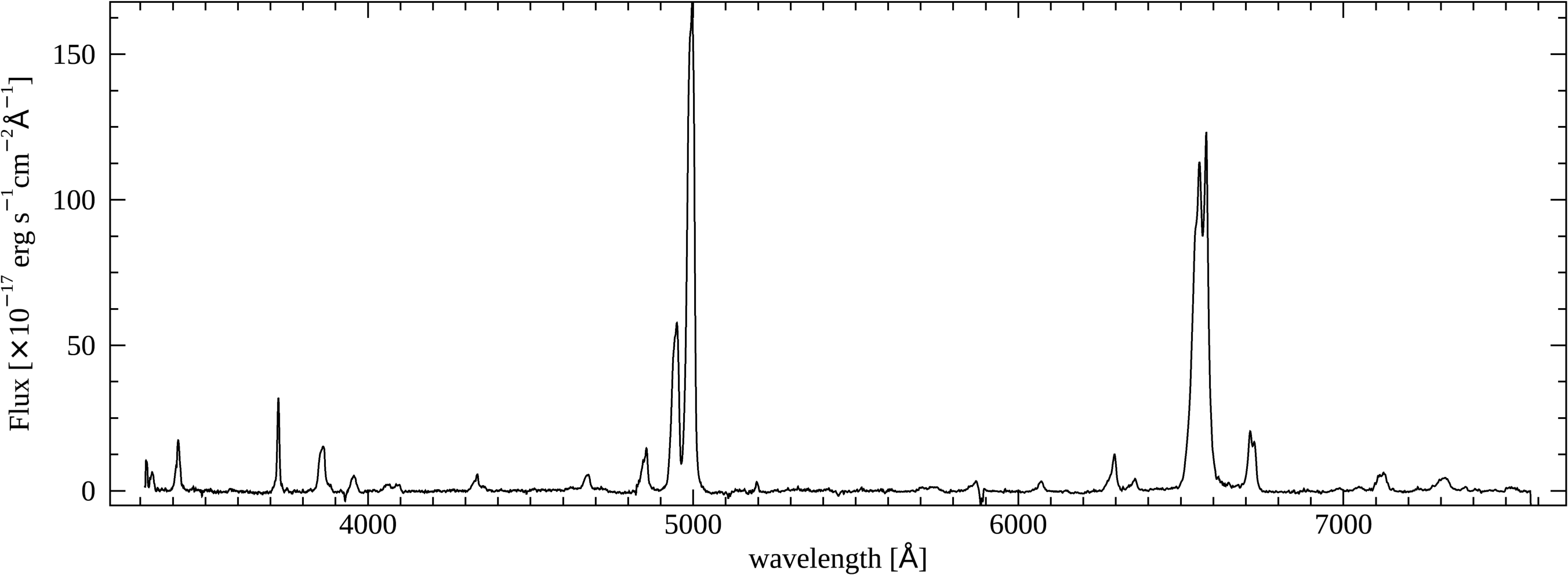}
\end{minipage}
\caption{ PKS~1549--79:As in Fig.\ref{Stellar_continuum}.} 
\label{PKS1549Stellar_continuum}
\end{figure*}

\begin{figure*}[h]
\begin{minipage}[t]{1\textwidth}
\includegraphics[width=\textwidth, height=0.22\textheight]{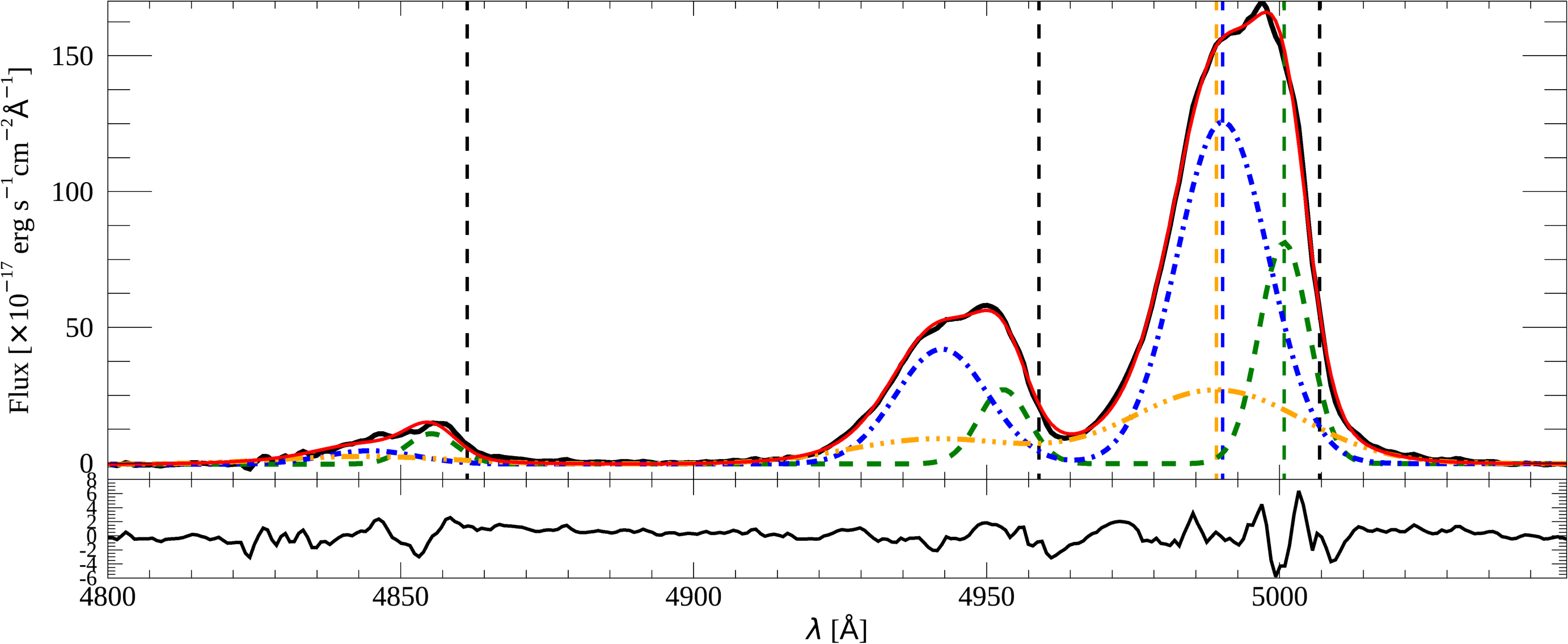}
\end{minipage}

{\centering
\begin{minipage}[b]{0.5\textwidth}
\includegraphics[width=\textwidth, height=0.22\textheight]{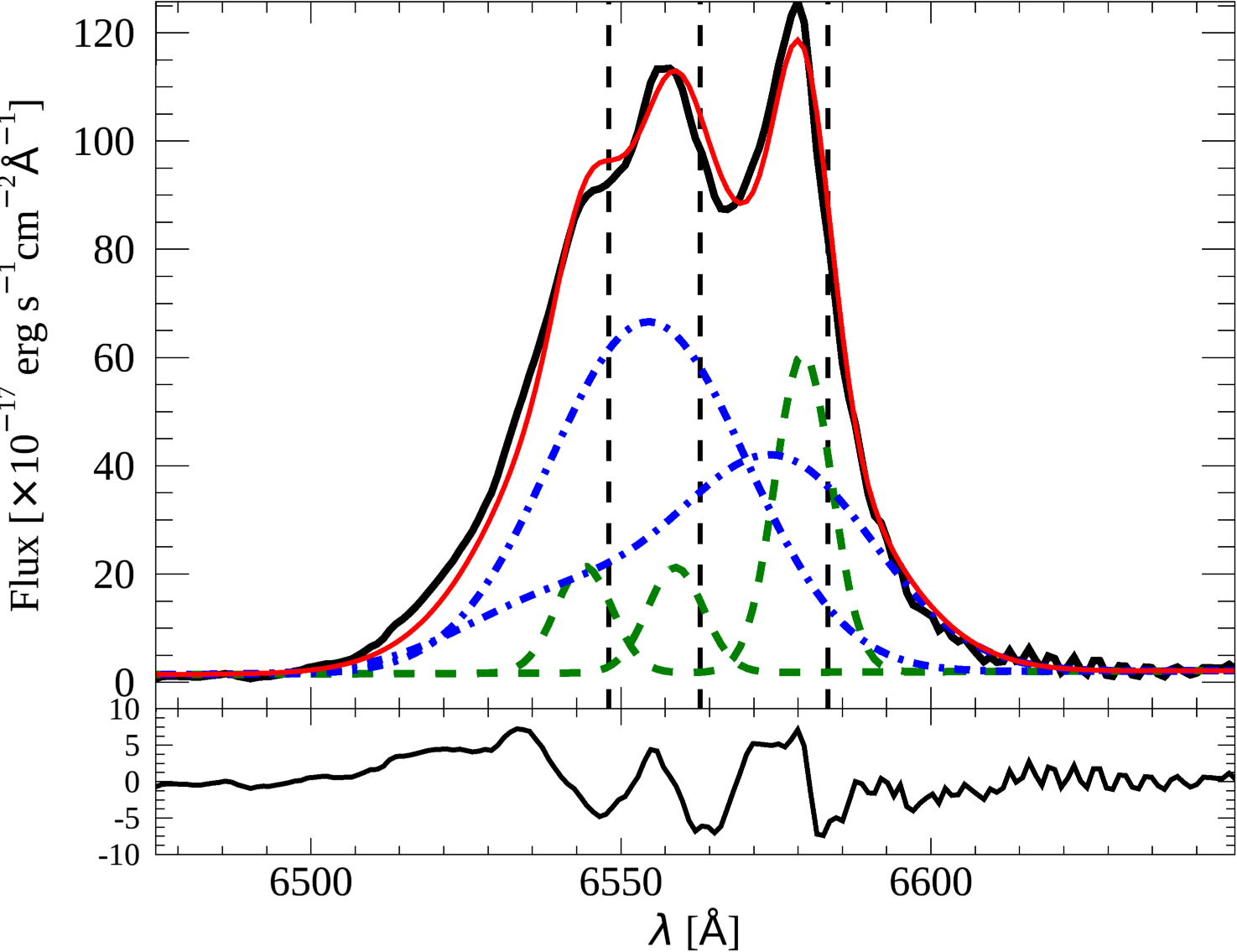}
\end{minipage}   \par}

\begin{minipage}[t]{0.5\textwidth}
\includegraphics[width=\textwidth, height=0.22\textheight]{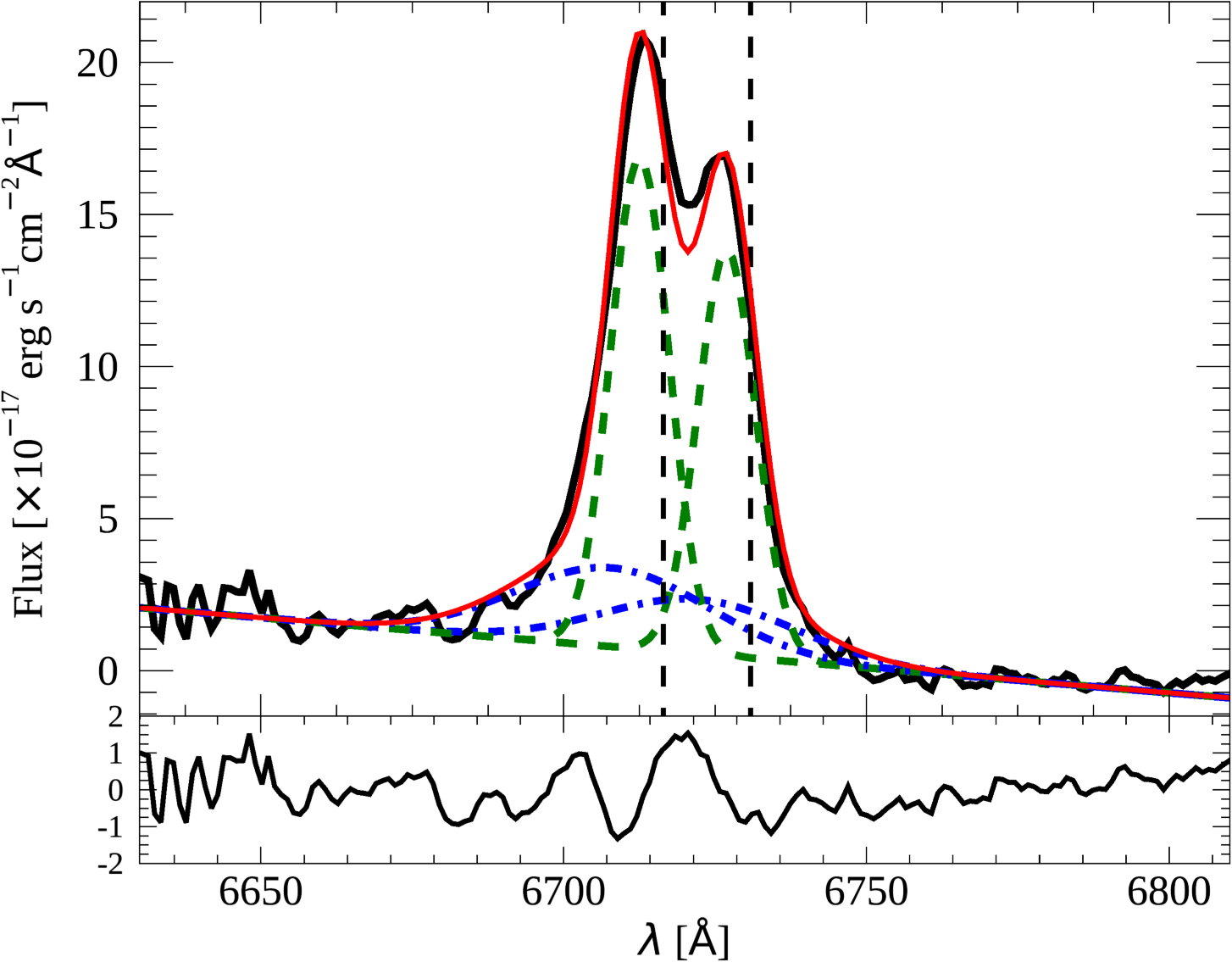}
\end{minipage}
\begin{minipage}[t]{0.5\textwidth}
\includegraphics[width=\textwidth, height=0.22\textheight]{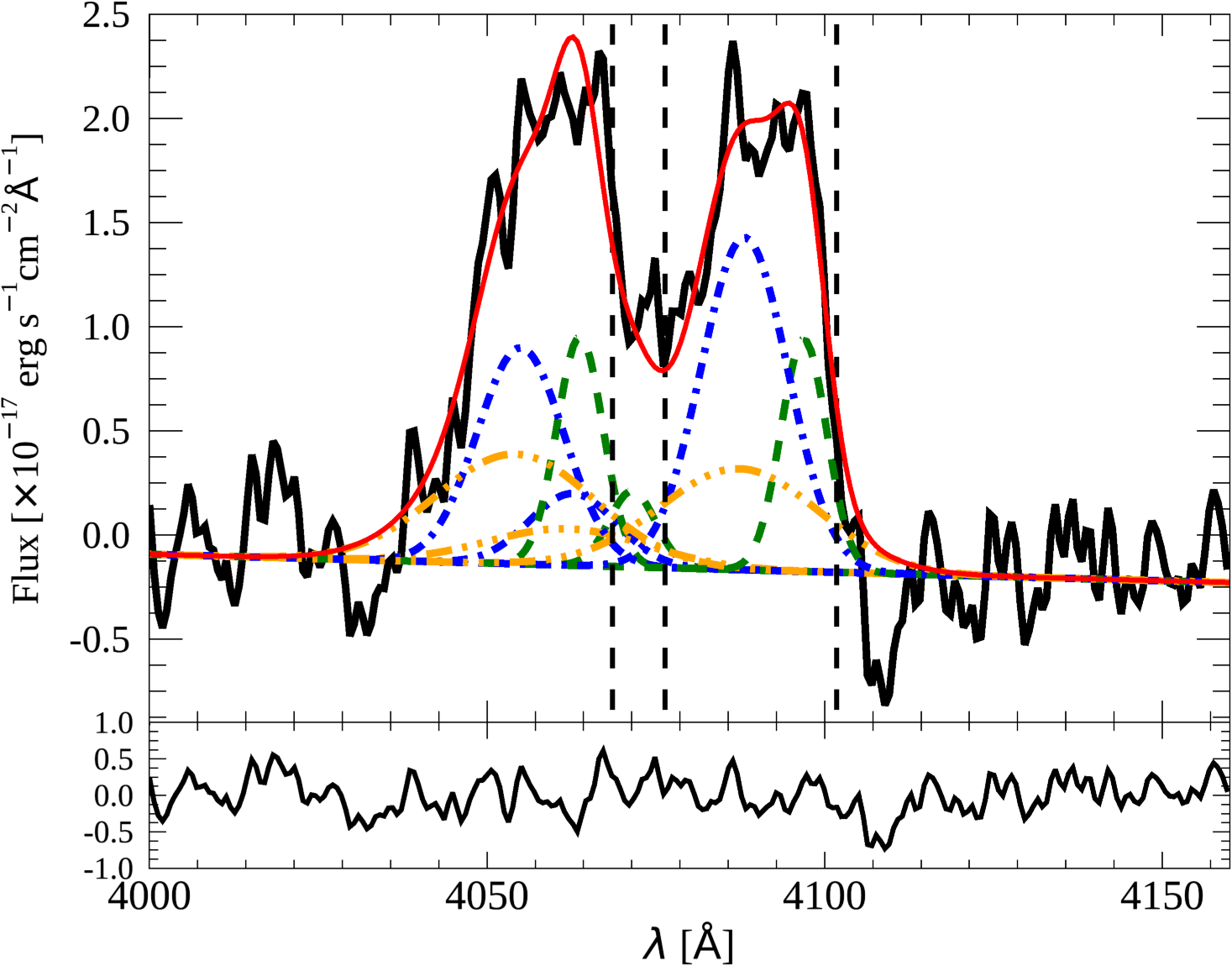}
\end{minipage}

\begin{minipage}[t]{0.5\textwidth}
\includegraphics[width=\textwidth, height=0.22\textheight]{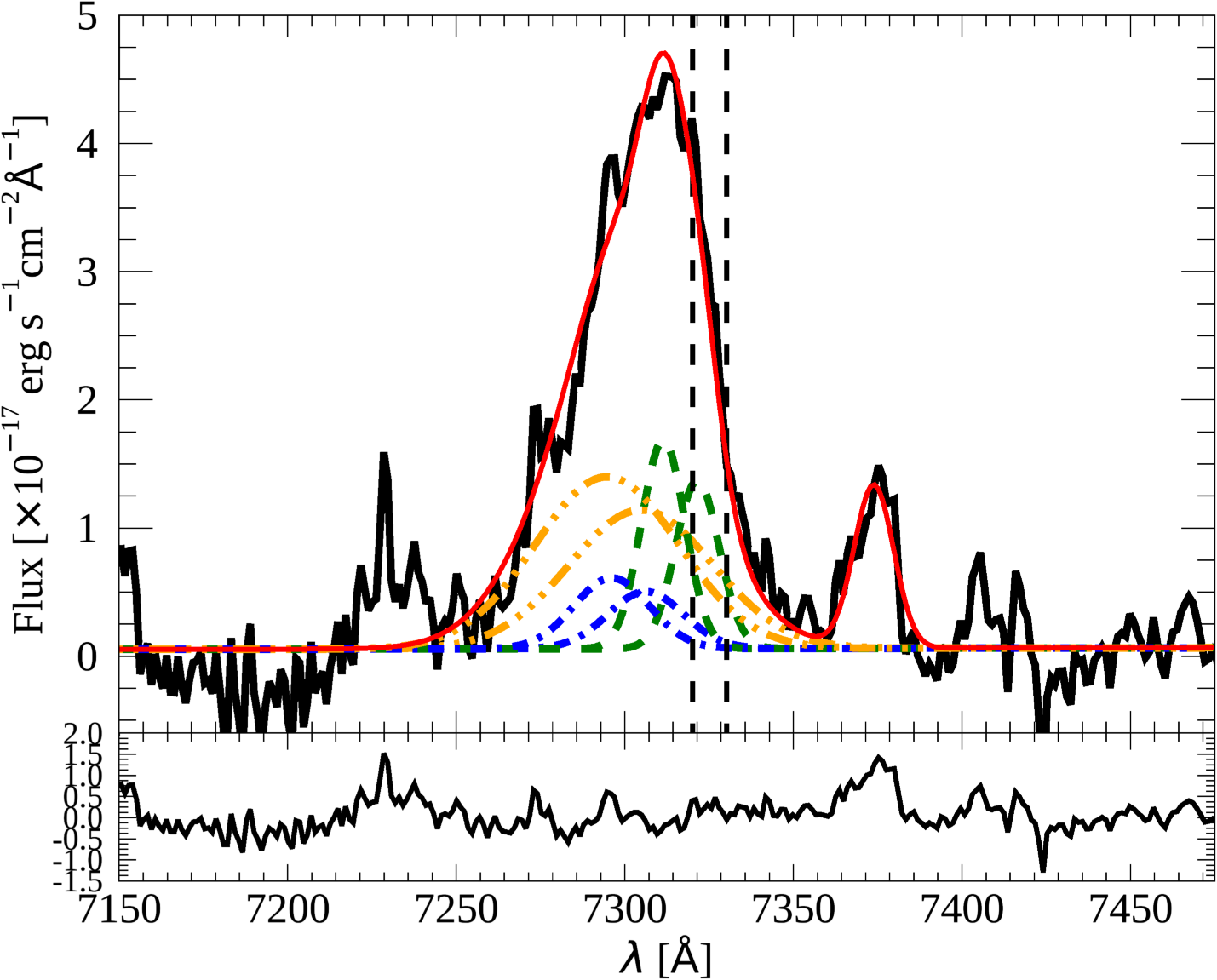}
\end{minipage}
\begin{minipage}[t]{0.5\textwidth}
\includegraphics[width=\textwidth, height=0.22\textheight]{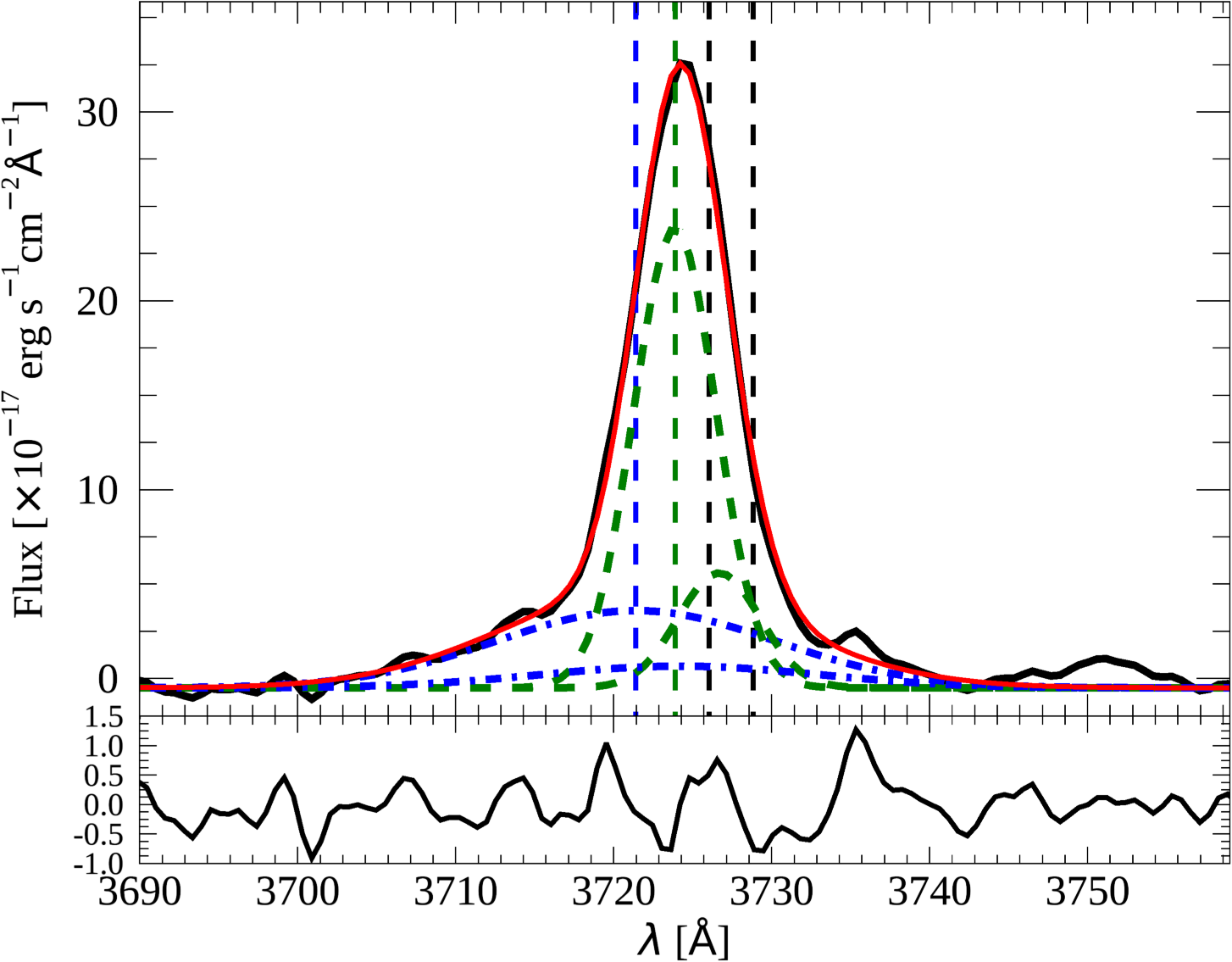}
\end{minipage}

\caption{ PKS~1549--79: \textit{First row:} \OIII$\lambdaup\lambdaup$4958,5007\AA\ + H$\beta$ line fit. \textit{Second row:} H$\alpha$+\NII$\lambdaup\lambdaup$6548,84\AA\ line fits. 
\textit{Third row:} \SII$\lambdaup\lambdaup$6717,31\AA\ (left panel) and \SII$\lambdaup\lambdaup$4069,76\AA\ + H$\delta$ (right panel) trans-auroral line fits.
\textit{Fourth row:} \OII$\lambdaup\lambdaup$7319,30\AA\ (left panel) and \OII$\lambdaup\lambdaup$3726,29\AA\ (right panel) trans-auroral line fits. In each of the figures the upper panel shows the best fit (red solid line) of the observed spectrum (black solid line) while the lower panel shows the residuals of the fit. The different kinematic components used for the fit of each emission line are showed with different colors and line styles, in the case of doublets where flux ratios have been fixed (i.e. the \OIII$\lambdaup\lambdaup$4958,5007\AA\ and the \NII$\lambdaup\lambdaup$6548,84\AA\ ), we show the total profile of each doublet kinematic component. The vertical dashed lines marks the rest-frame wavelength of the fitted emission lines. Wavelengths are plotted in \AA,\ and the flux scale is given in units of $10^{-17} \rm{erg~s^{-1}cm^{-2}\AA^{-1}}$.} 
\label{PKS1549Emission lines fits}
\end{figure*}

\begin{figure*}[h]
\begin{minipage}[t]{0.5\textwidth}
\includegraphics[angle=90]{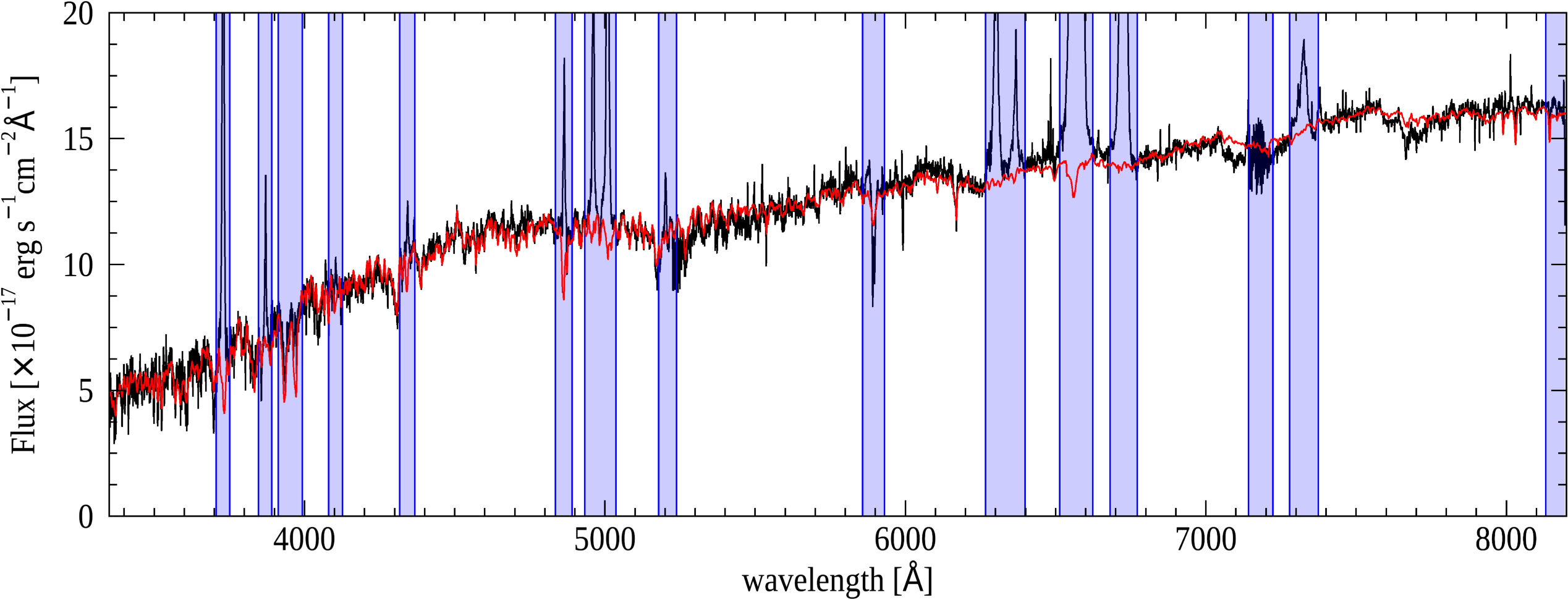}
\end{minipage}
\begin{minipage}[t]{0.5\textwidth}
\includegraphics[ angle=90]{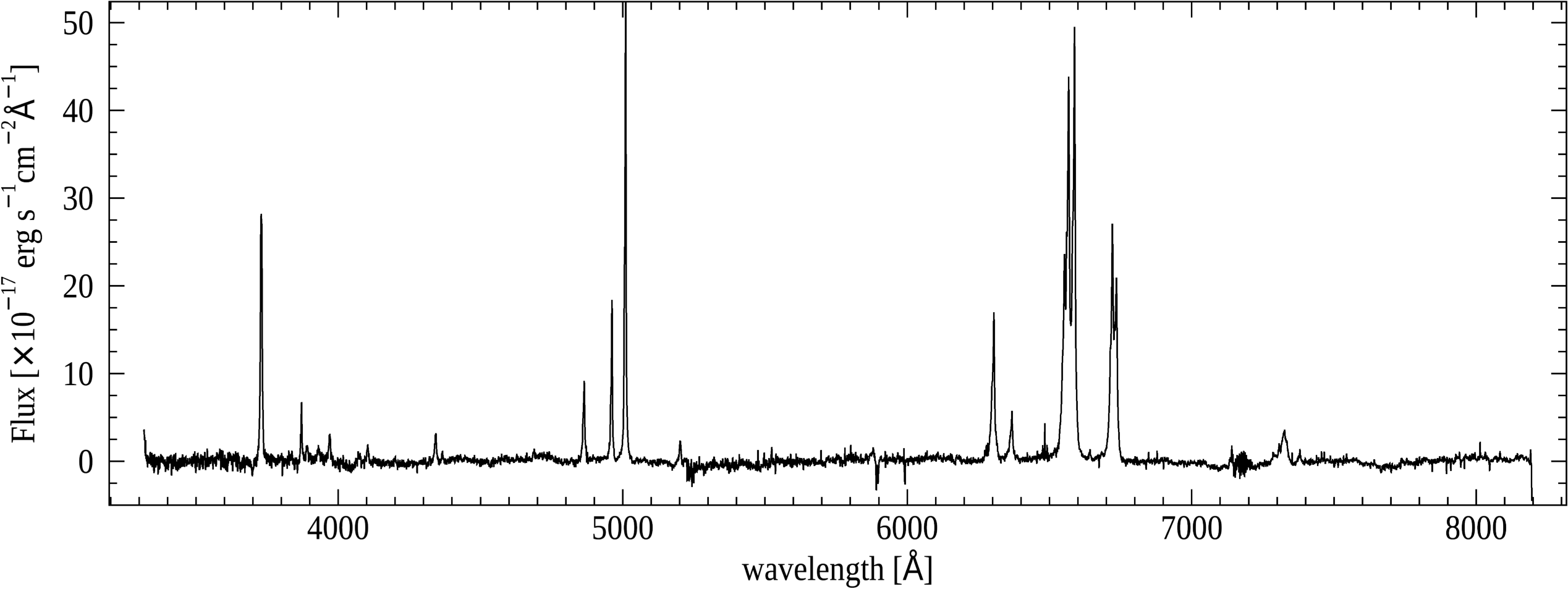}
\end{minipage}
\caption{  PKS~1814--63: As in Fig.\ref{Stellar_continuum}. } 
\label{PKS1814Stellar_continuum}
\end{figure*}

\begin{figure*}[h]
\begin{minipage}[t]{0.5\textwidth}
\includegraphics[width=\textwidth, height=0.225\textheight]{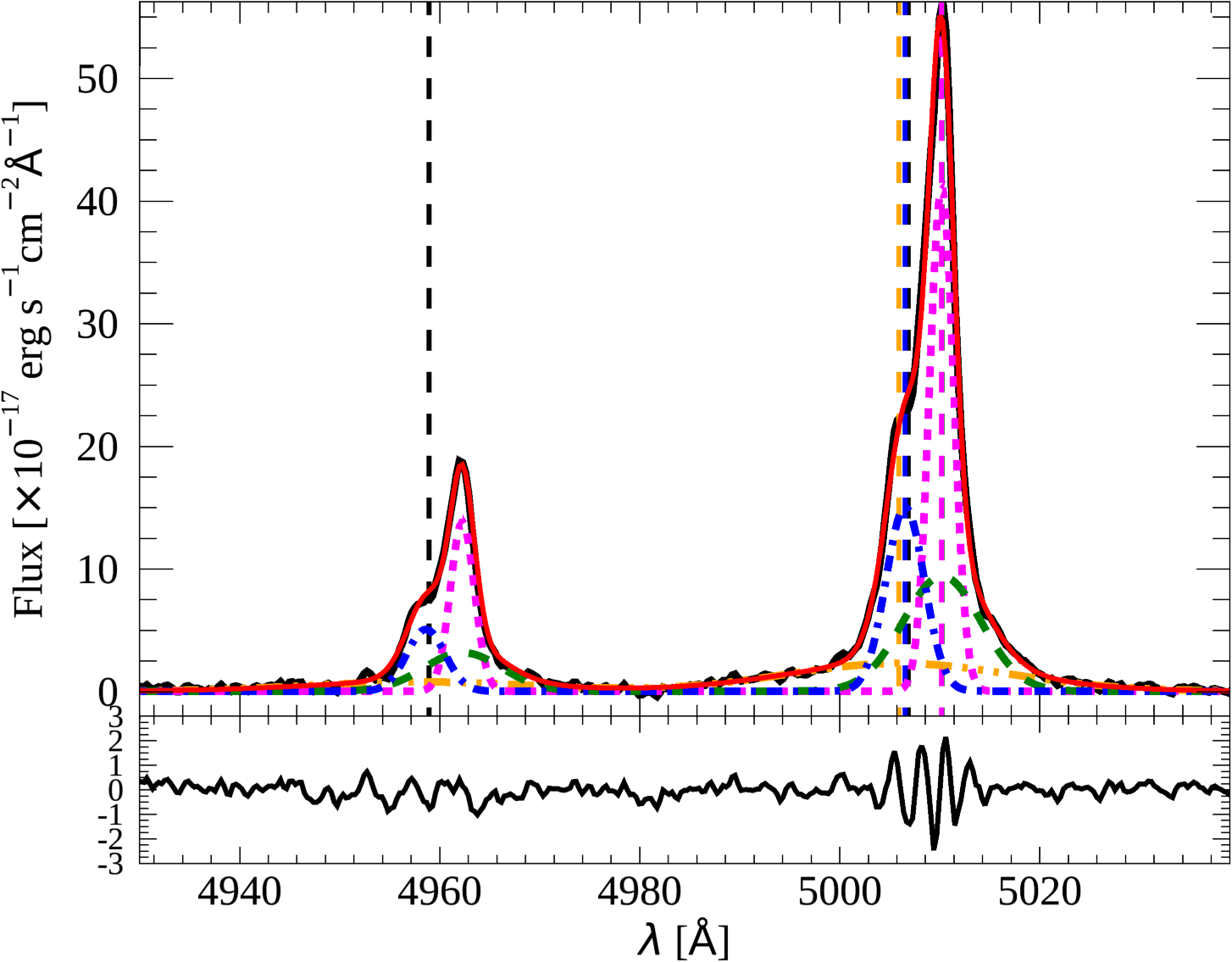}
\end{minipage}
\begin{minipage}[t]{0.5\textwidth}
\includegraphics[width=\textwidth, height=0.225\textheight]{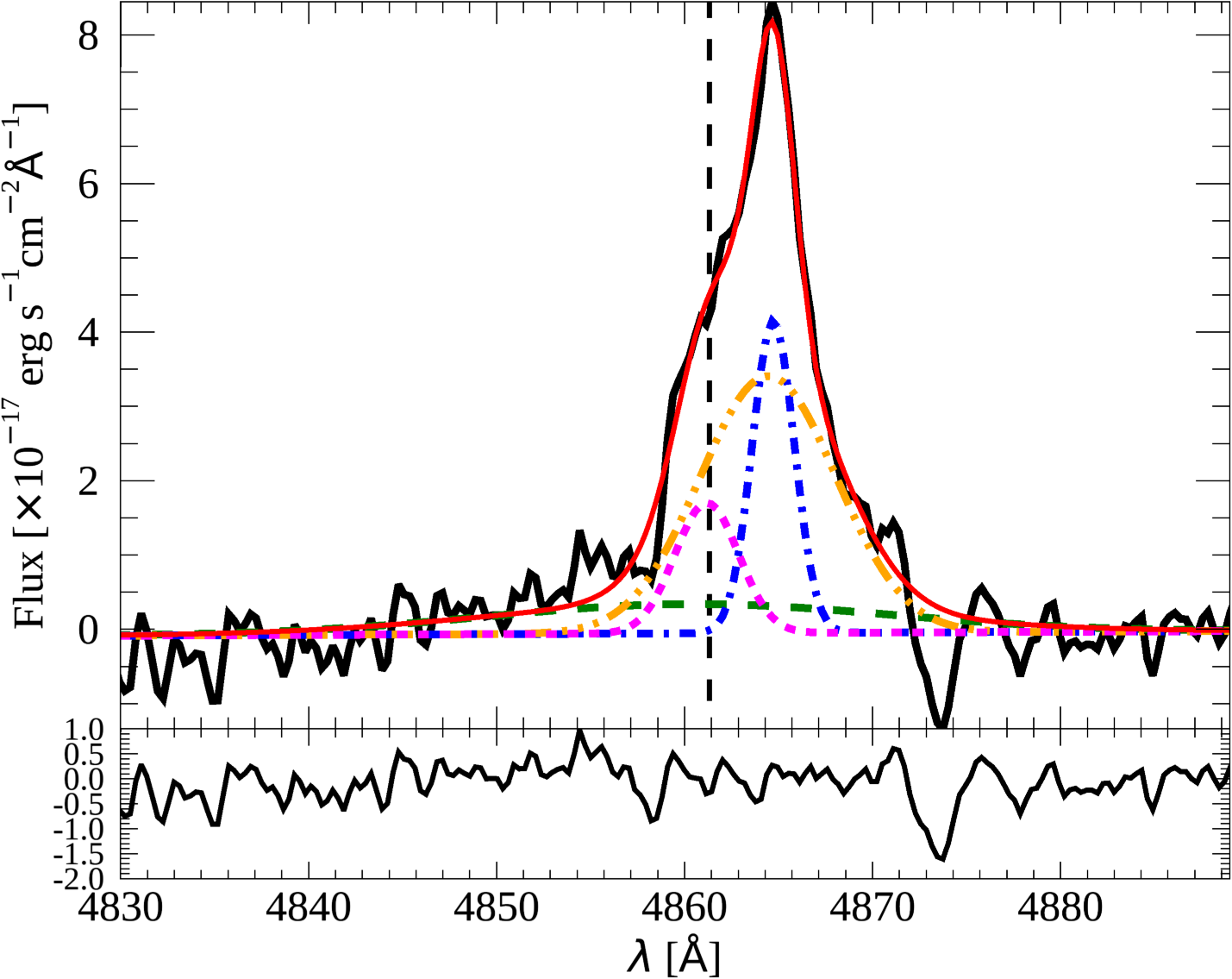}
\end{minipage}

{\centering
\begin{minipage}[b]{0.5\textwidth}
\includegraphics[width=\textwidth, height=0.225\textheight]{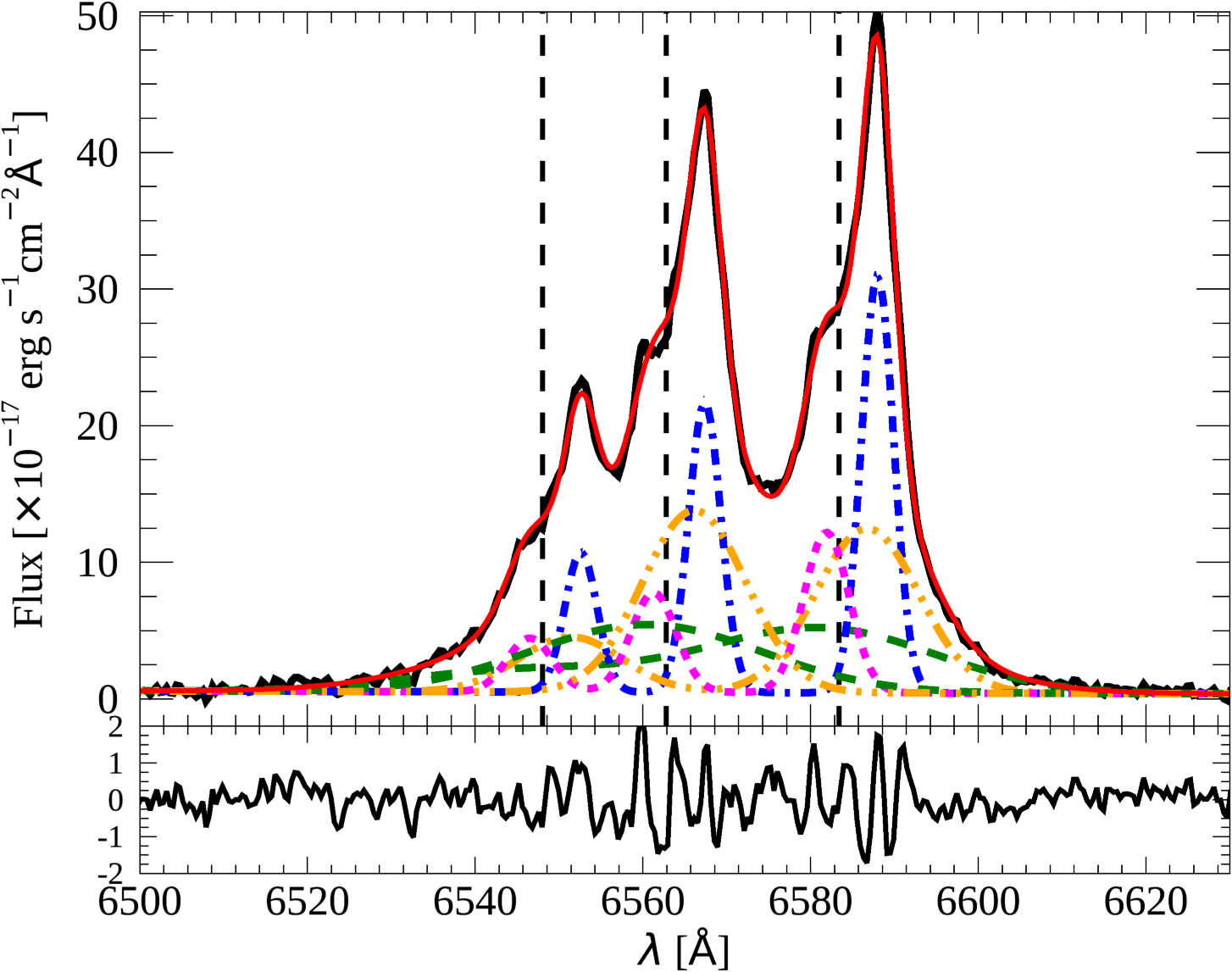}
\end{minipage}   \par}

\begin{minipage}[t]{0.5\textwidth}
\includegraphics[width=\textwidth, height=0.225\textheight]{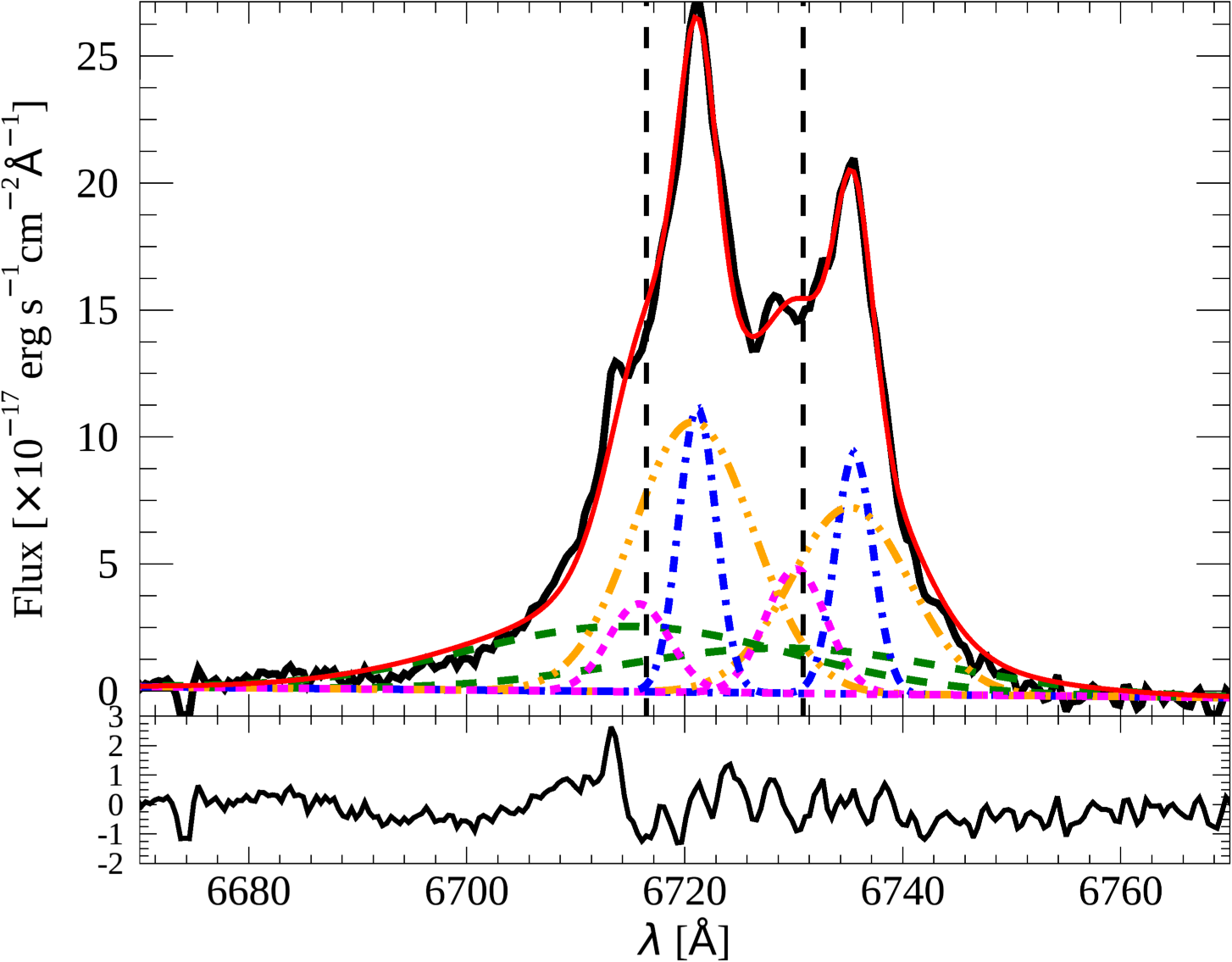}
\end{minipage}
\begin{minipage}[t]{0.5\textwidth}
\includegraphics[width=\textwidth, height=0.225\textheight]{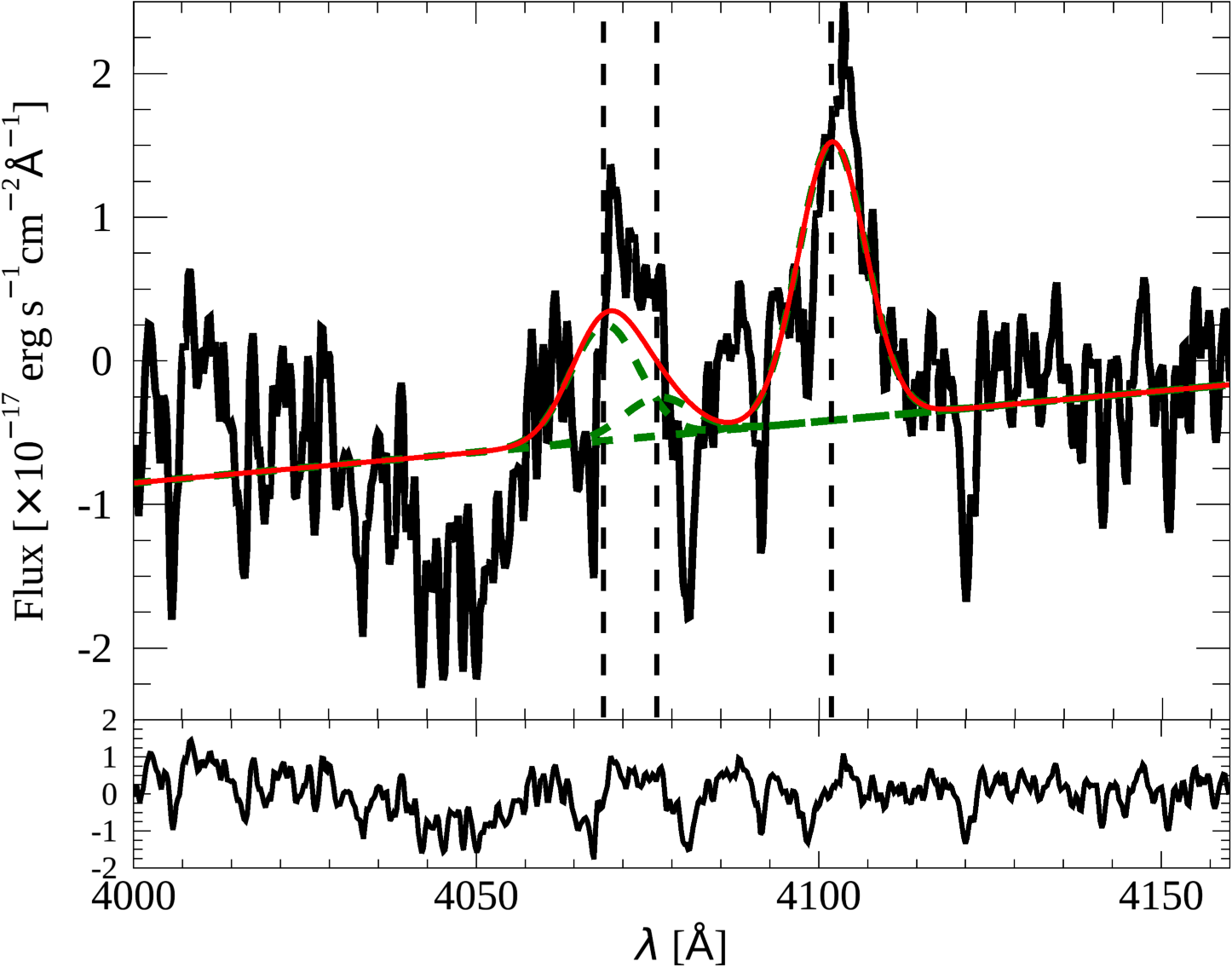}
\end{minipage}

\begin{minipage}[t]{0.5\textwidth}
\includegraphics[width=\textwidth, height=0.225\textheight]{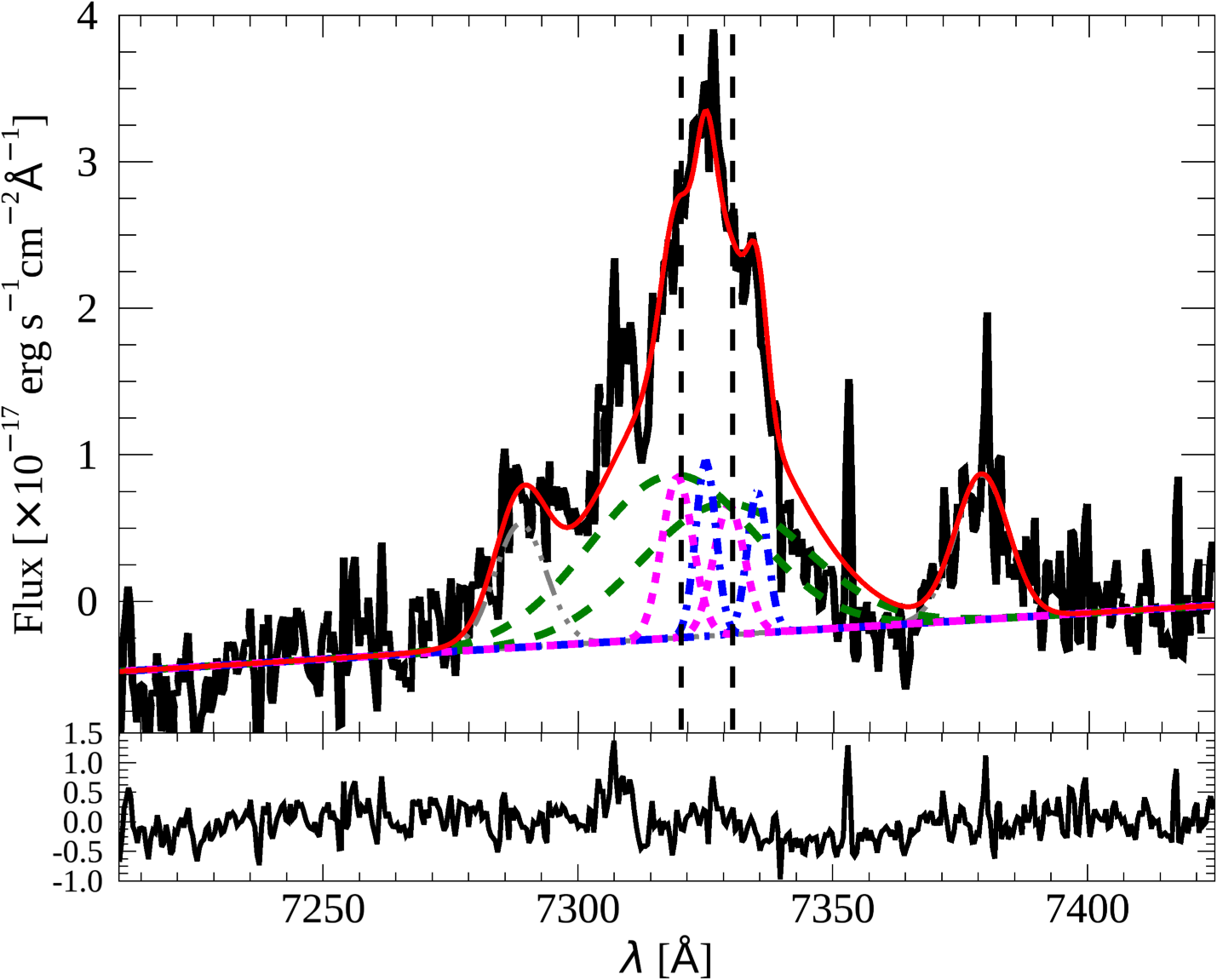}
\end{minipage}
\begin{minipage}[t]{0.5\textwidth}
\includegraphics[width=\textwidth, height=0.225\textheight]{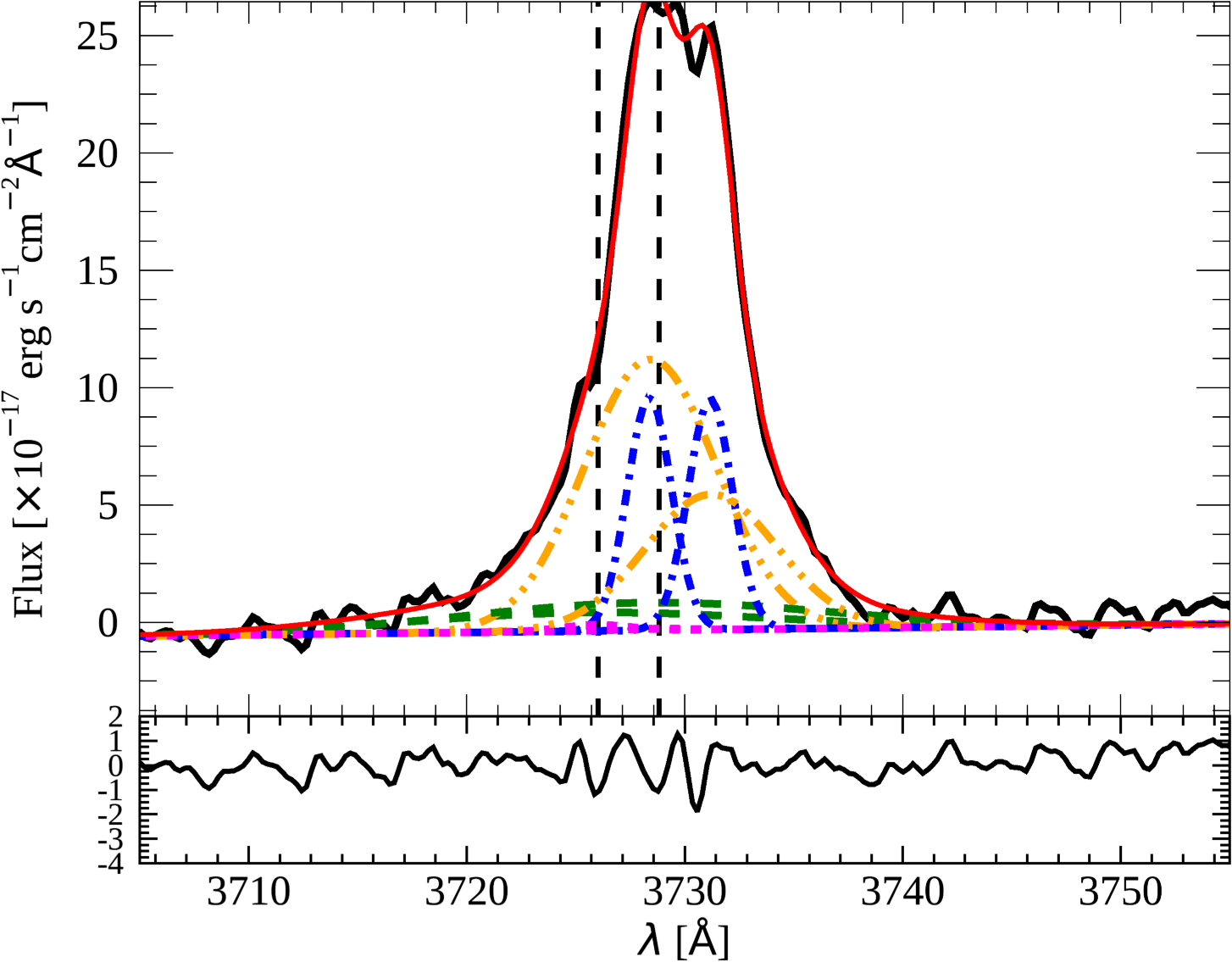}
\end{minipage}

\caption{ PKS~1814--63: As in Fig.\ref{PKS0252Emission lines fits}.} 
\label{PKS1814Emission lines fits}
\end{figure*}

\begin{figure*}[h]
\begin{minipage}[t]{0.5\textwidth}
\includegraphics[angle=90]{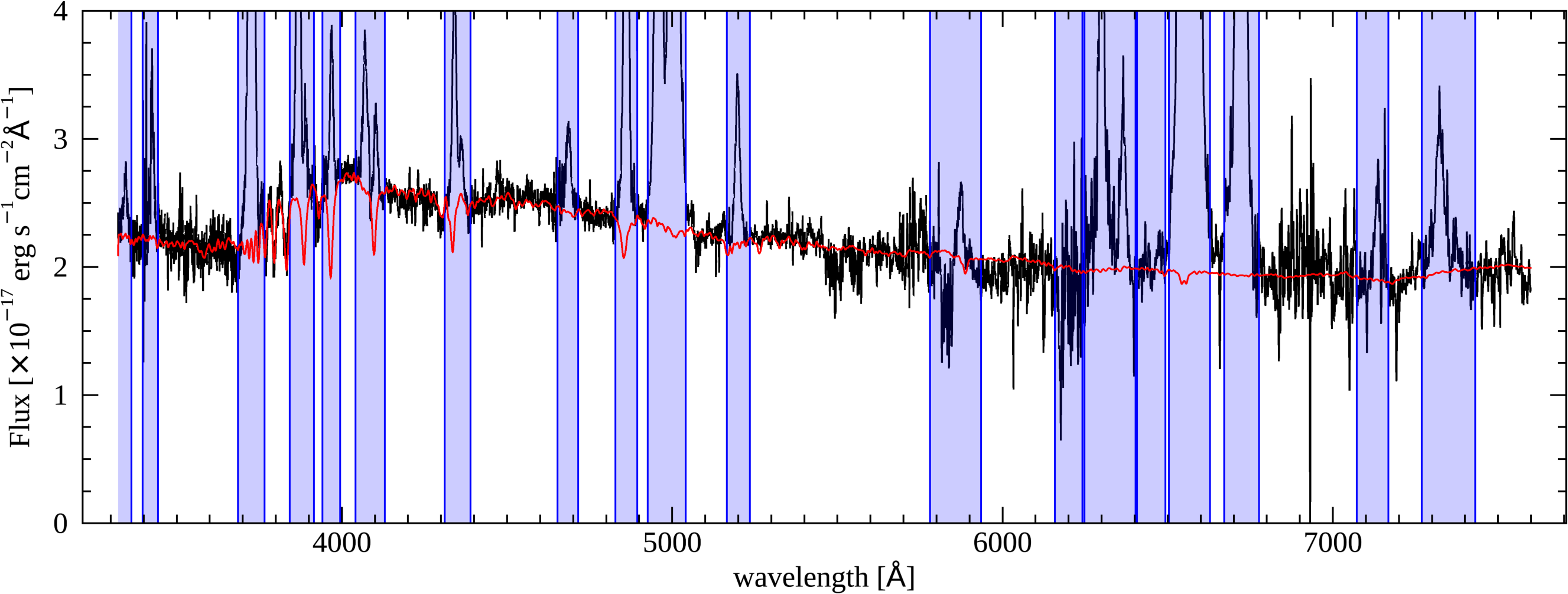}
\end{minipage}
\begin{minipage}[t]{0.5\textwidth}
\includegraphics[ angle=90]{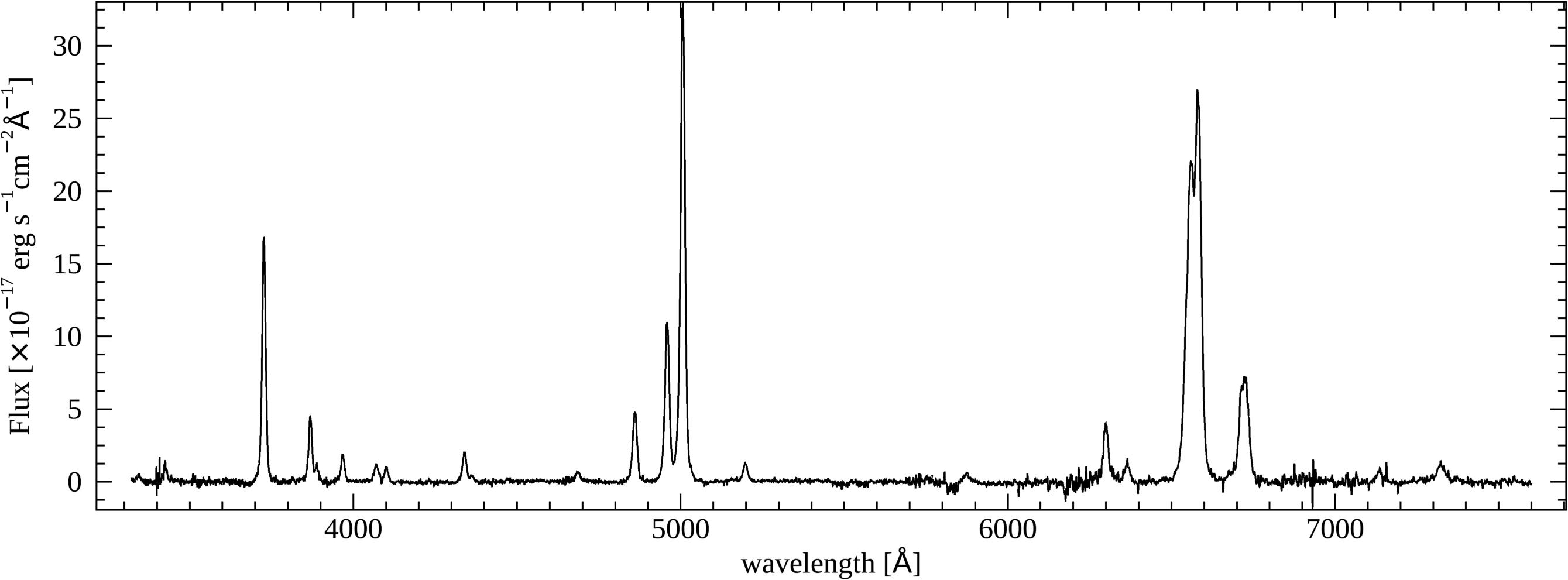}
\end{minipage}
\caption{ PKS~2135--209: As in Fig.\ref{Stellar_continuum}.} 
\label{PKS2135Stellar_continuum}
\end{figure*}

\begin{figure*}[h]
\begin{minipage}[t]{0.5\textwidth}
\includegraphics[width=\textwidth, height=0.225\textheight]{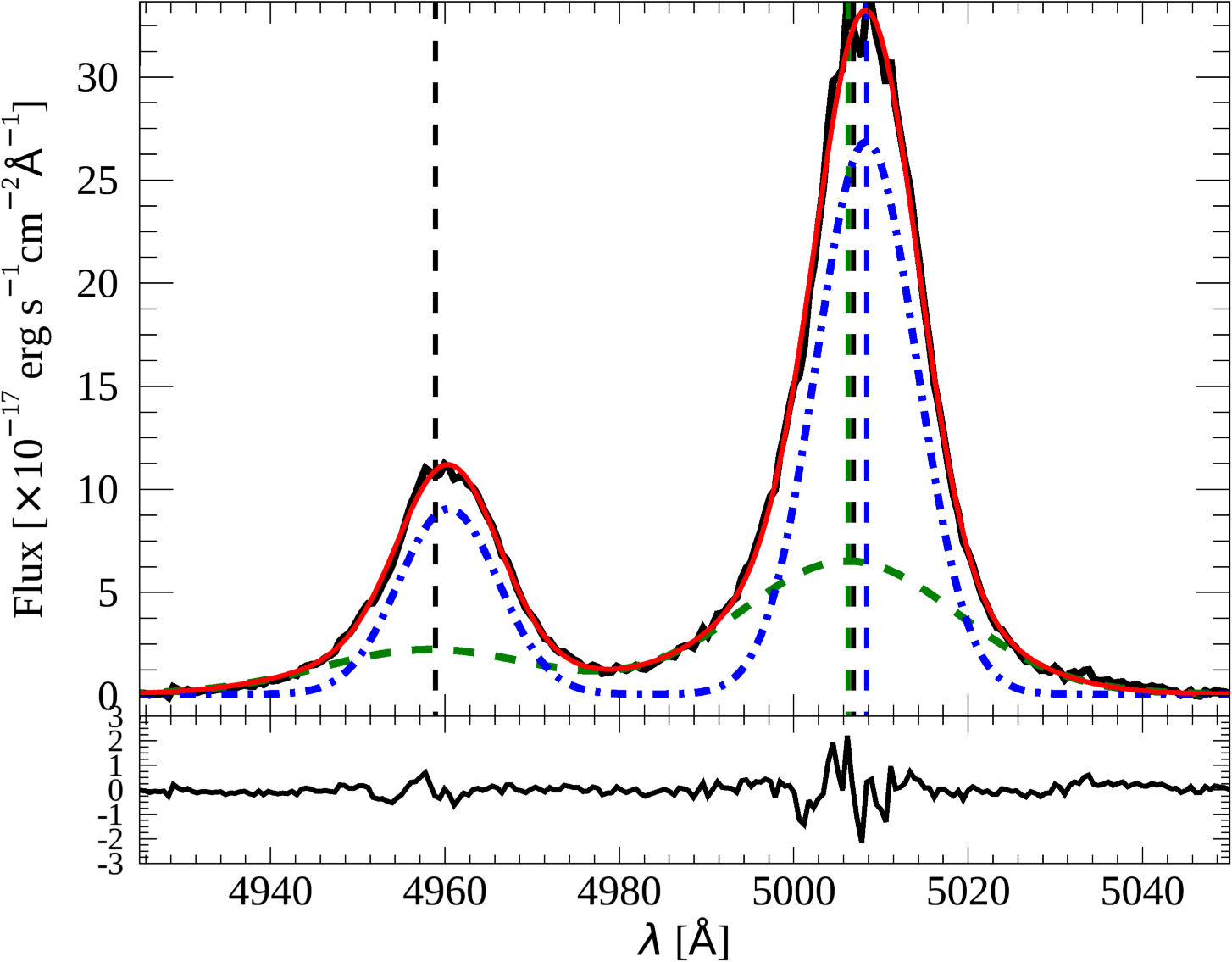}
\end{minipage}
\begin{minipage}[t]{0.5\textwidth}
\includegraphics[width=\textwidth, height=0.225\textheight]{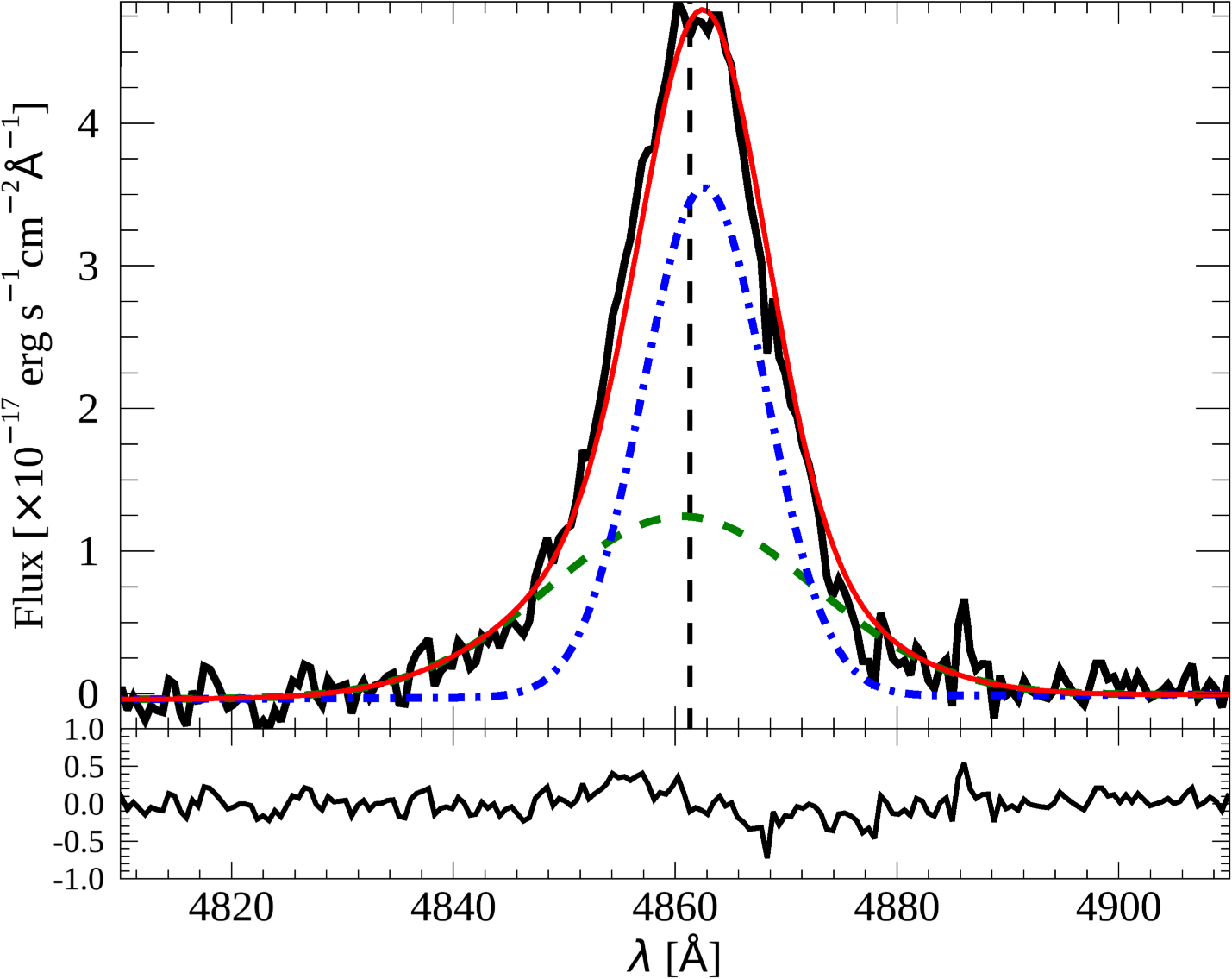}
\end{minipage}

{\centering
\begin{minipage}[b]{0.5\textwidth}
\includegraphics[width=\textwidth, height=0.225\textheight]{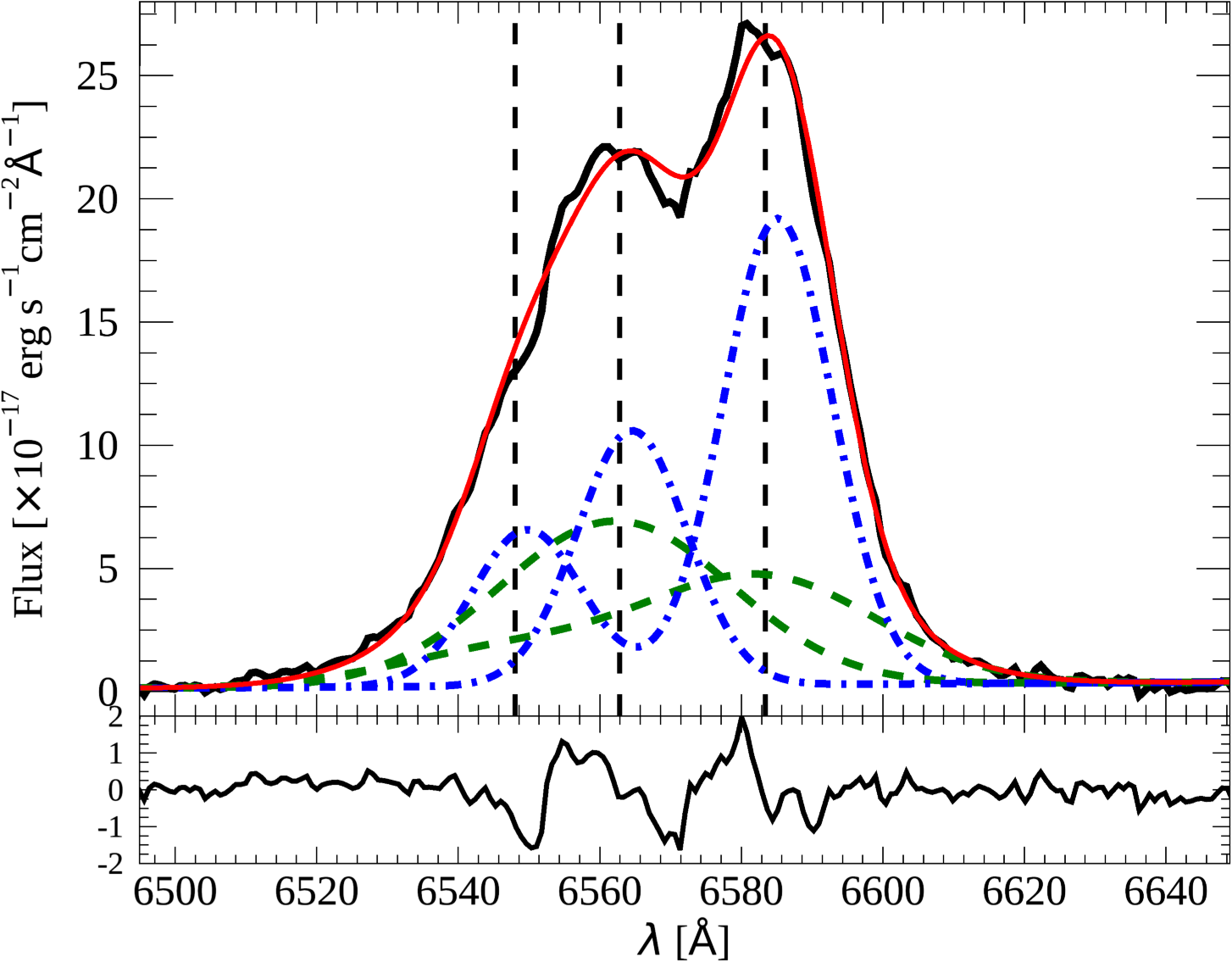}
\end{minipage}   \par}

\begin{minipage}[t]{0.5\textwidth}
\includegraphics[width=\textwidth, height=0.225\textheight]{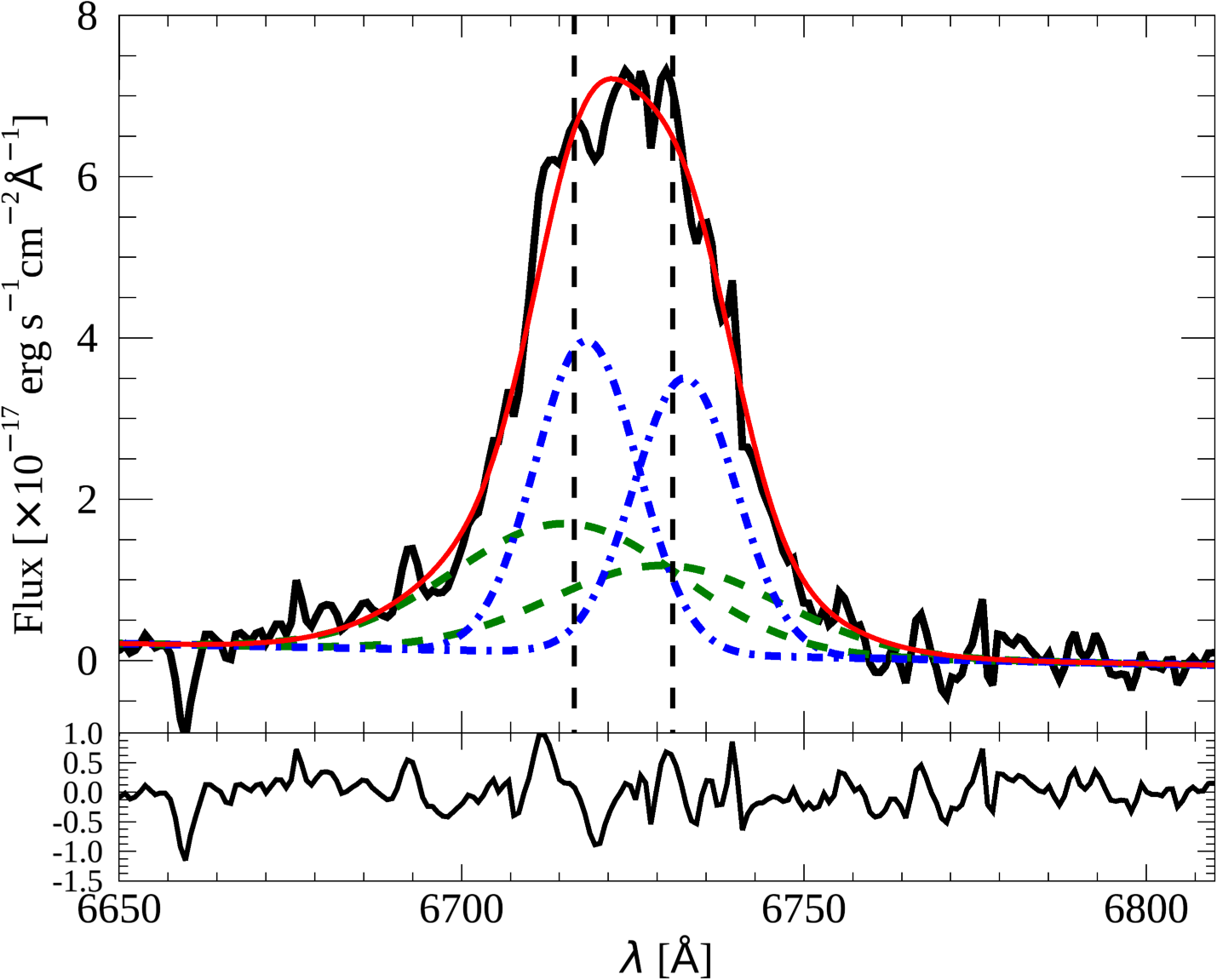}
\end{minipage}
\begin{minipage}[t]{0.5\textwidth}
\includegraphics[width=\textwidth, height=0.225\textheight]{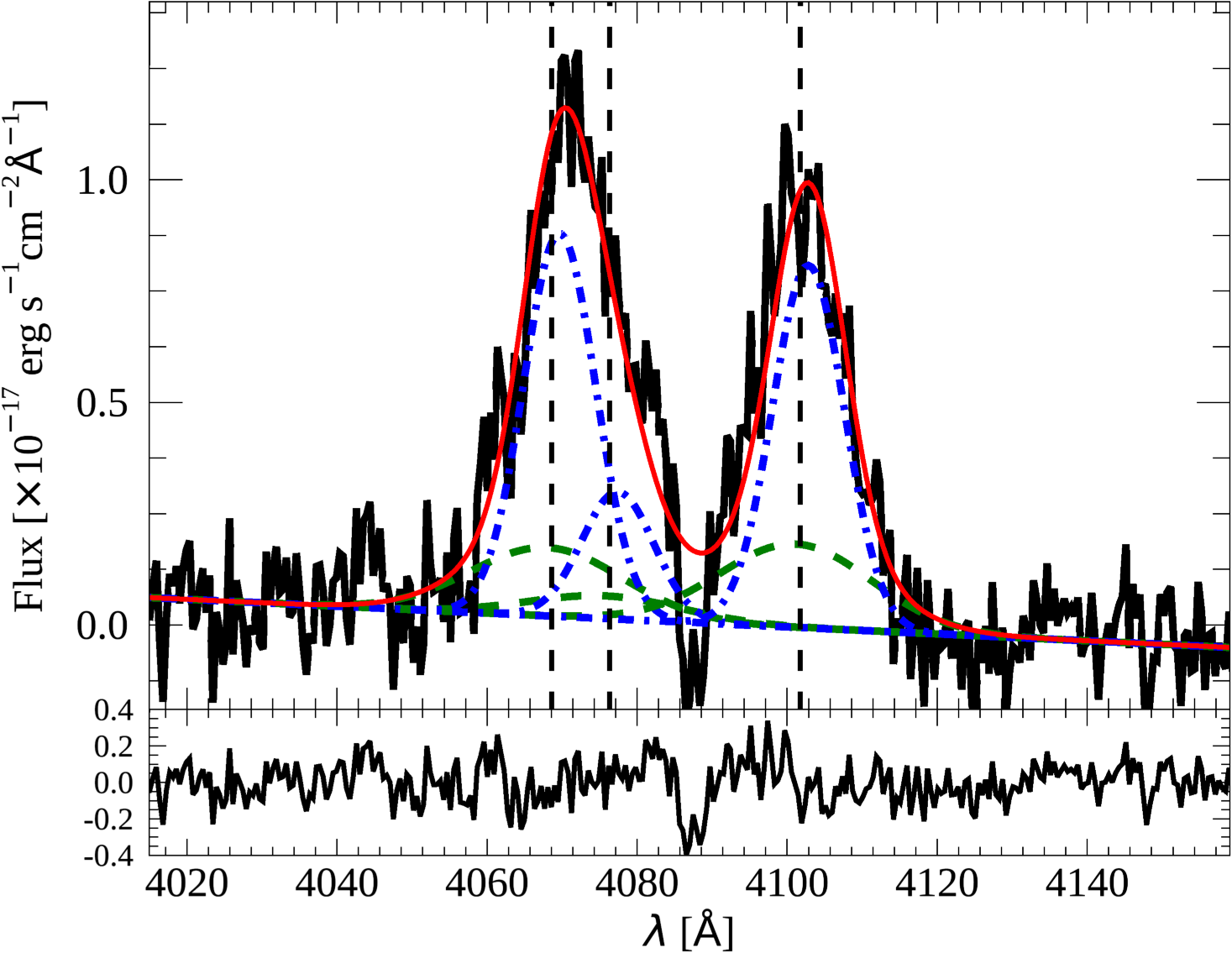}
\end{minipage}

\begin{minipage}[t]{0.5\textwidth}
\includegraphics[width=\textwidth, height=0.225\textheight]{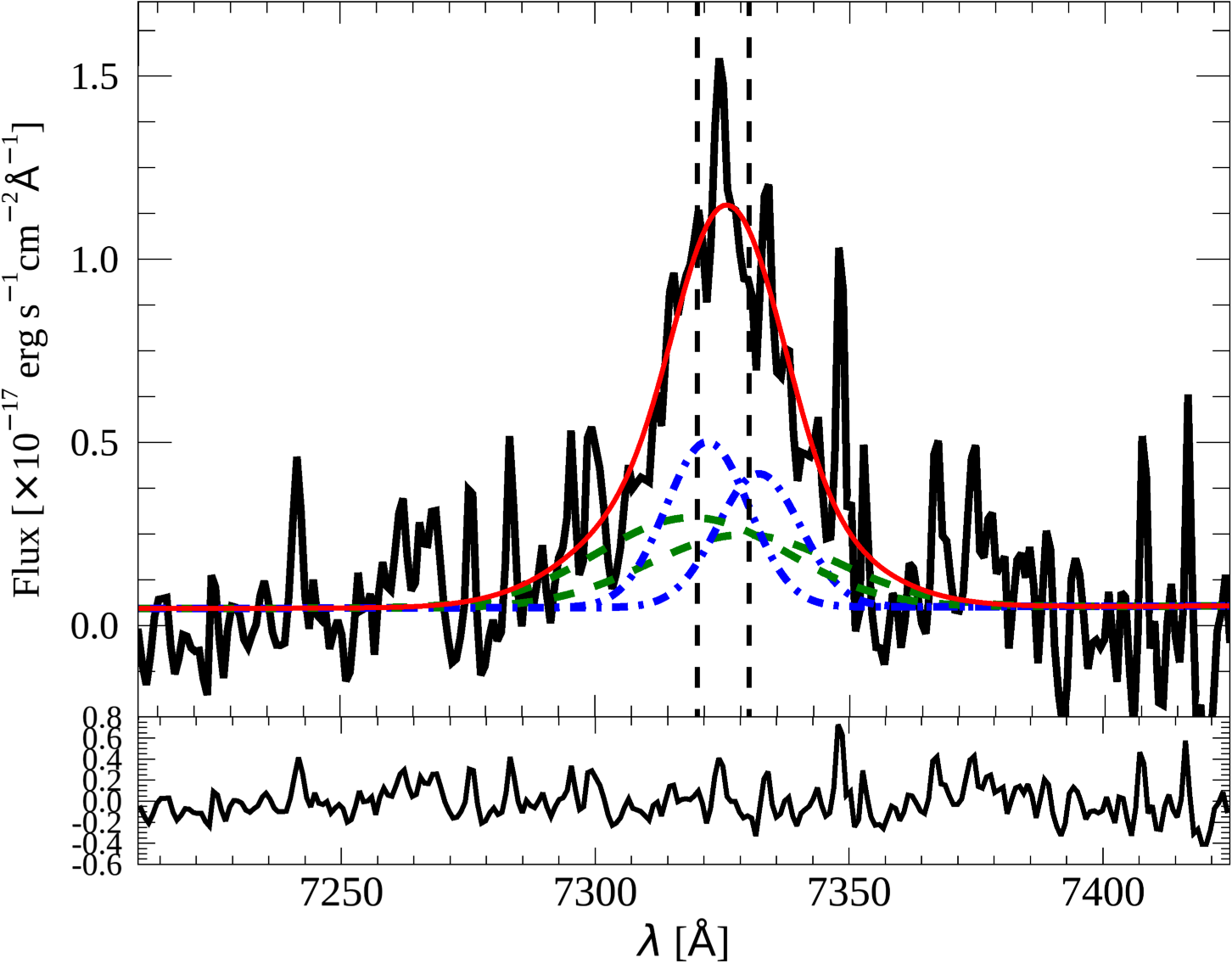}
\end{minipage}
\begin{minipage}[t]{0.5\textwidth}
\includegraphics[width=\textwidth, height=0.225\textheight]{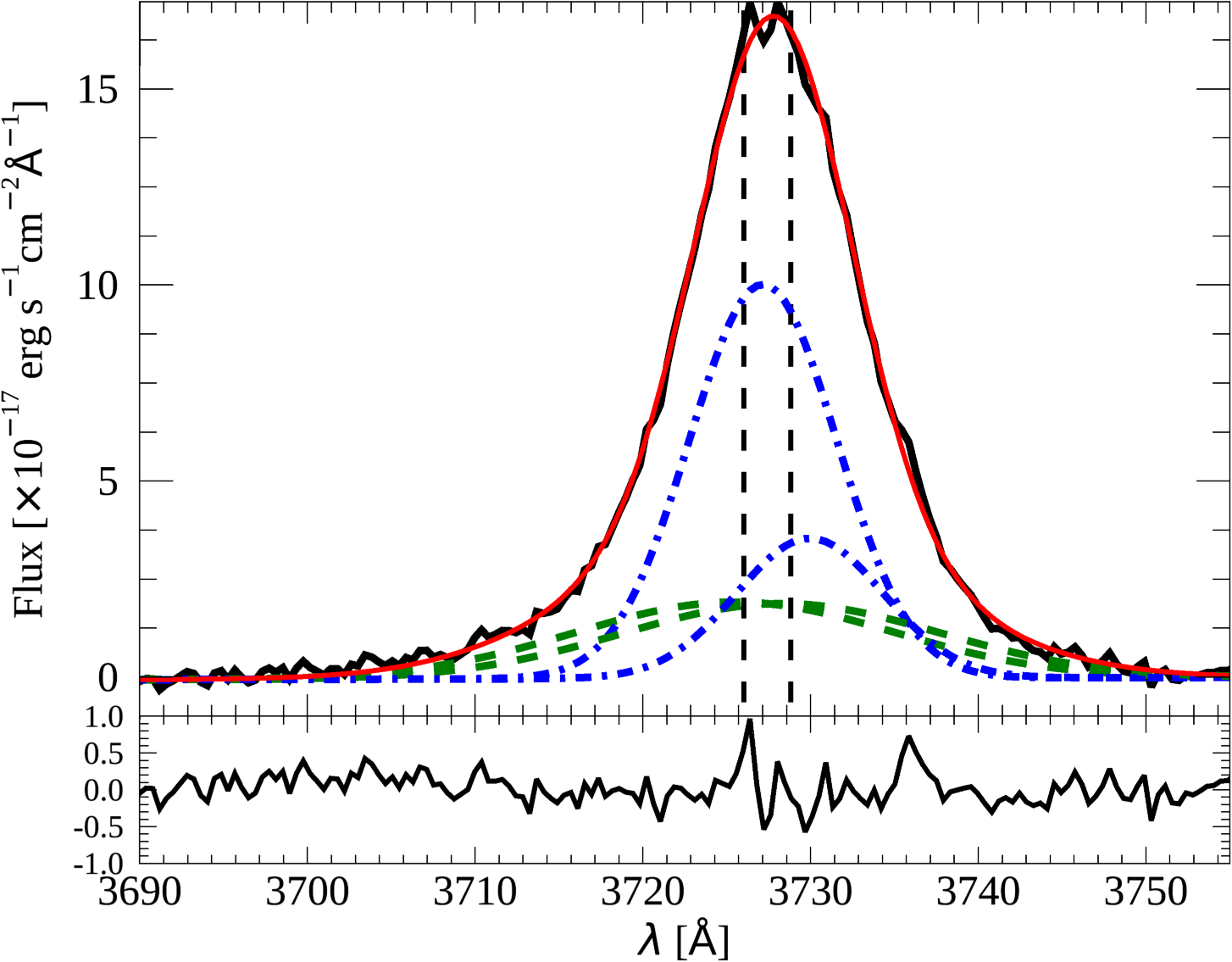}
\end{minipage}

\caption{ PKS~2135--209: As in Fig.\ref{PKS0252Emission lines fits}.} 
\label{PKS2135Emission lines fits}
\end{figure*}

\begin{figure*}[h]
\begin{minipage}[t]{0.5\textwidth}
\includegraphics[angle=90]{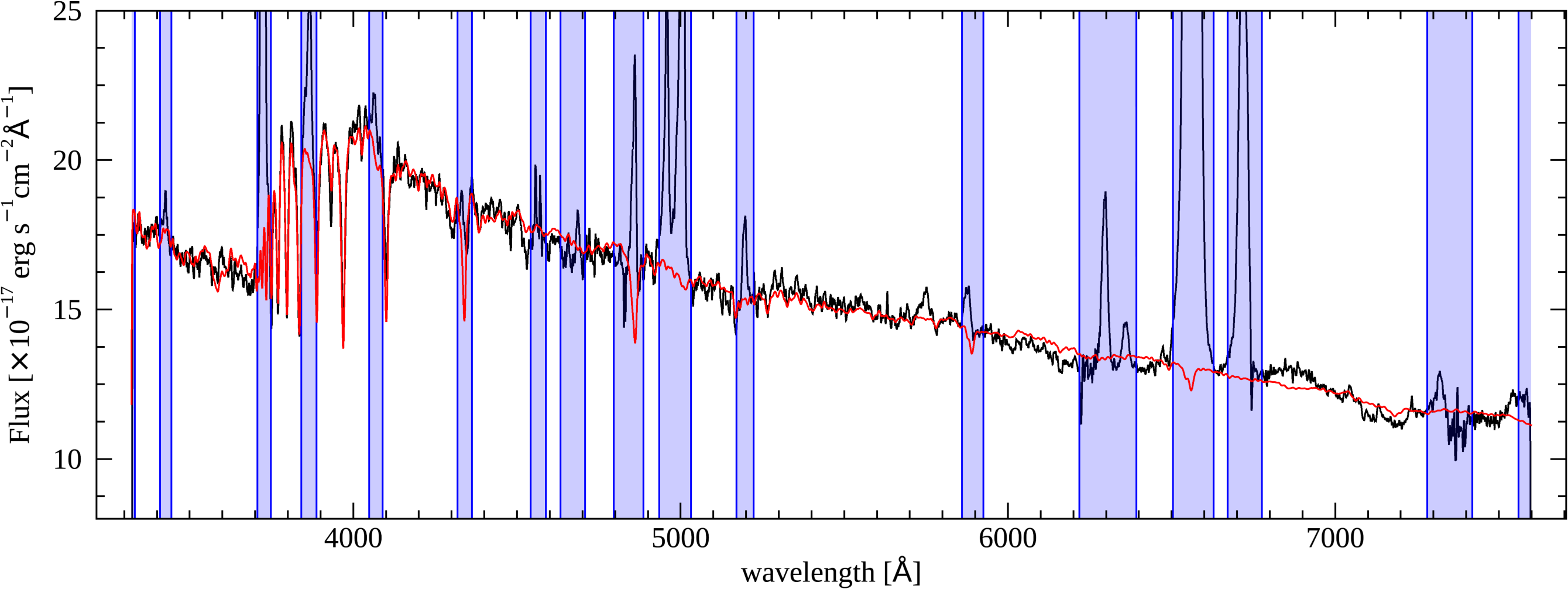}
\end{minipage}
\begin{minipage}[t]{0.5\textwidth}
\includegraphics[ angle=90]{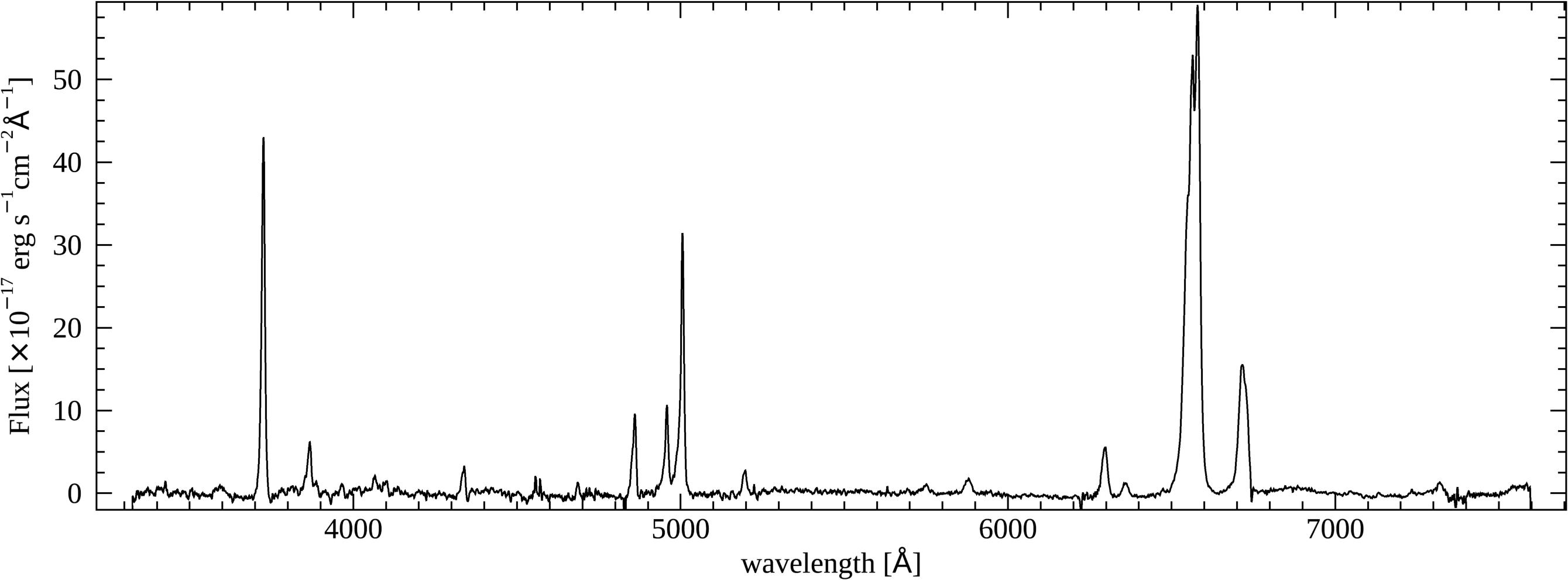}
\end{minipage}
\caption{PKS2314+03: As in Fig.\ref{Stellar_continuum}.} 
\label{3C459Stellar_continuum}
\end{figure*}

\begin{figure*}[h]
\begin{minipage}[t]{1\textwidth}
\includegraphics[width=\textwidth, height=0.225\textheight]{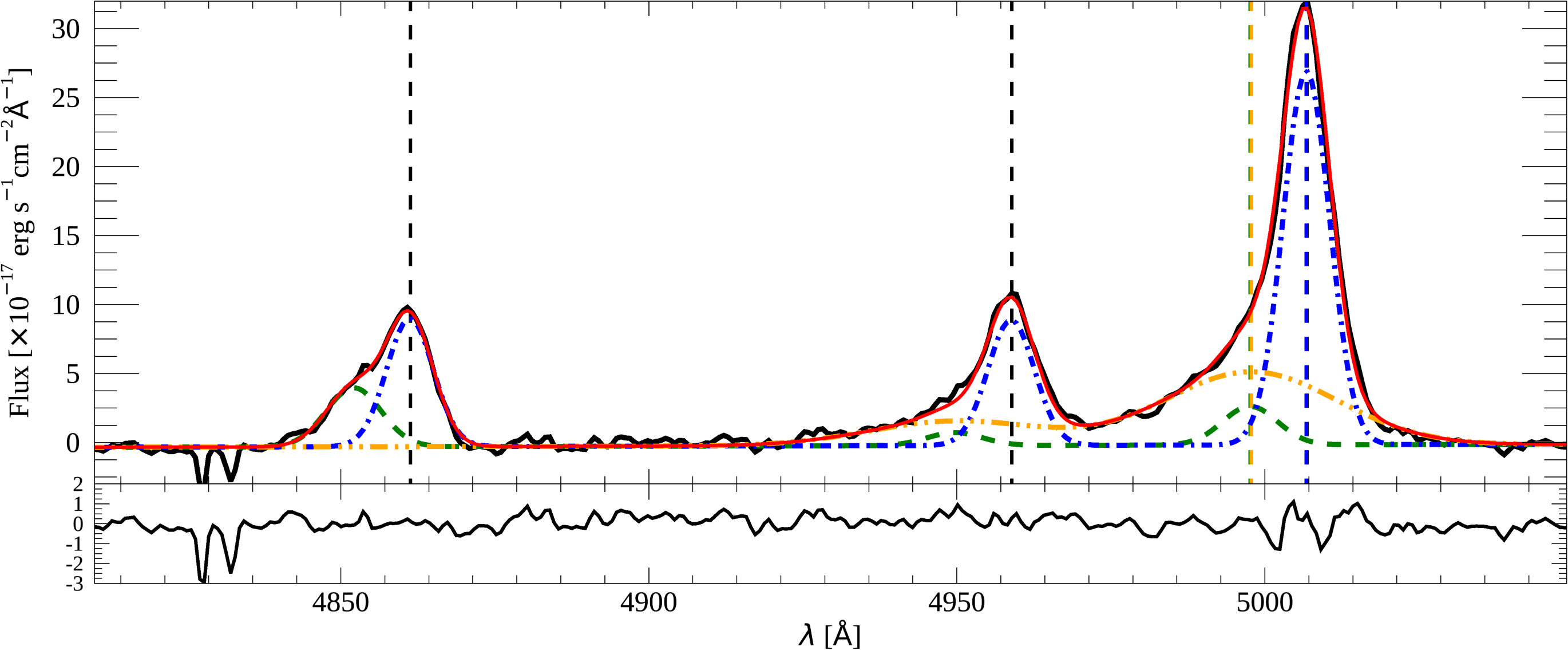}
\end{minipage}

{\centering
\begin{minipage}[b]{0.5\textwidth}
\includegraphics[width=\textwidth, height=0.225\textheight]{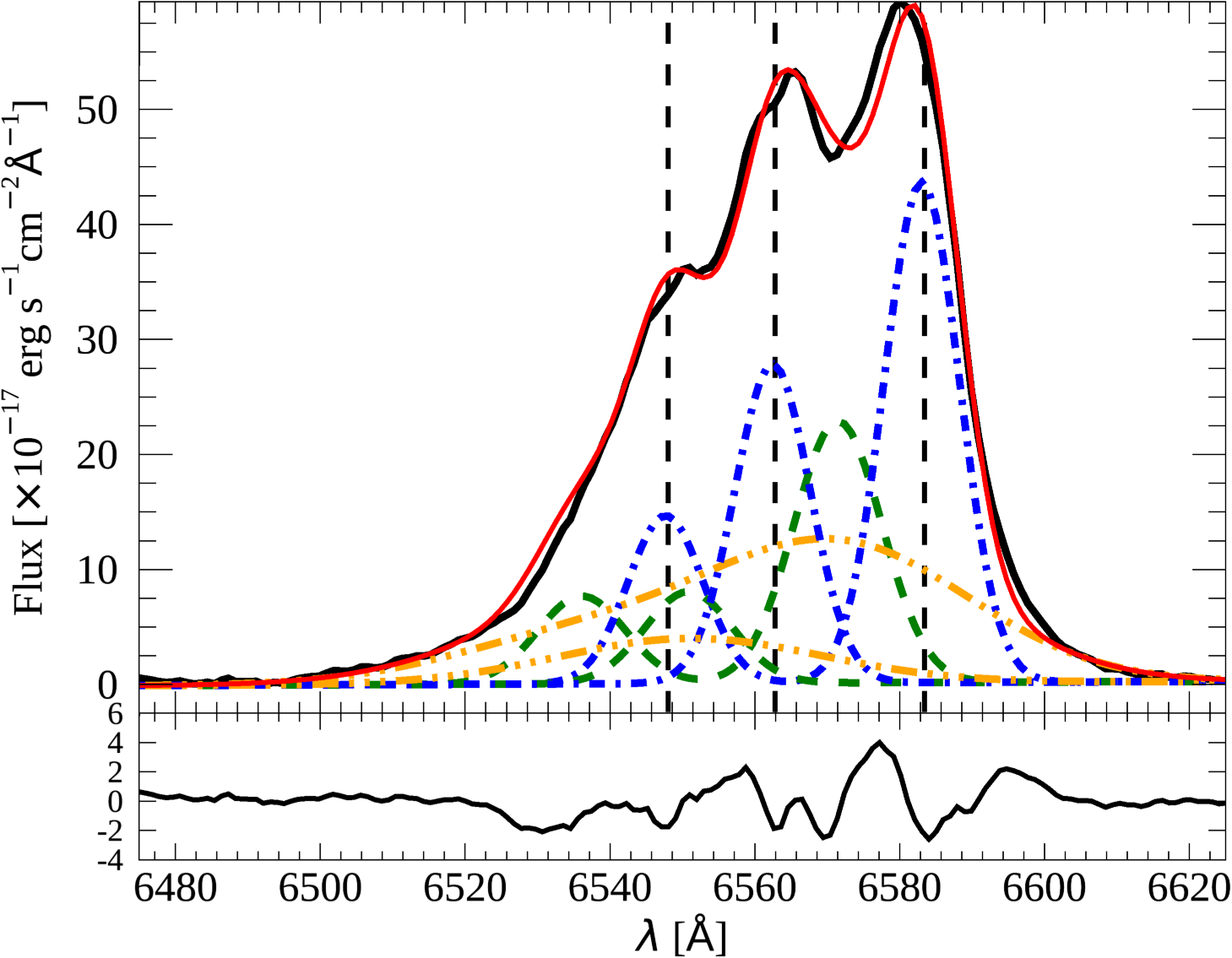}
\end{minipage}   \par}

\begin{minipage}[t]{0.5\textwidth}
\includegraphics[width=\textwidth, height=0.225\textheight]{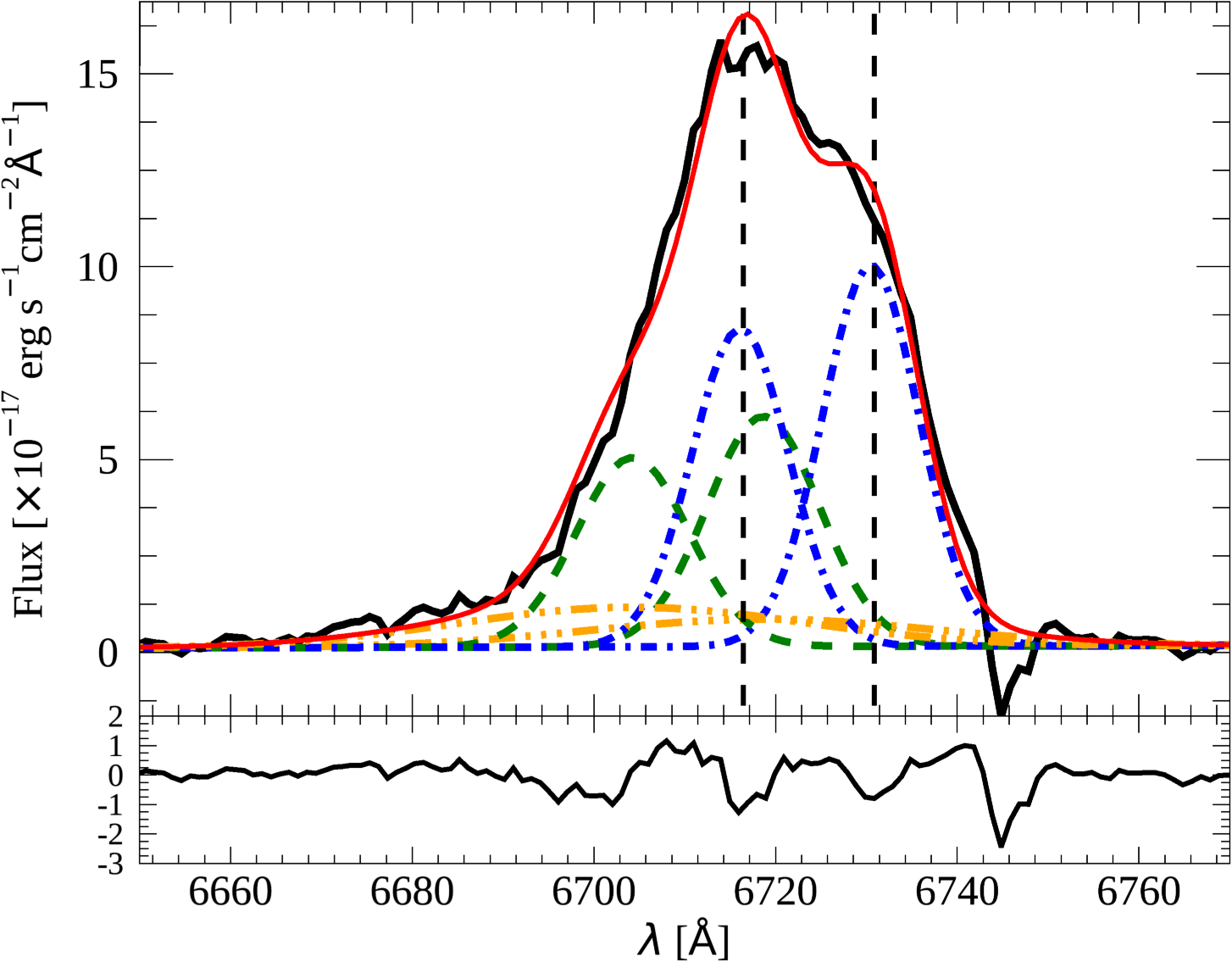}
\end{minipage}
\begin{minipage}[t]{0.5\textwidth}
\includegraphics[width=\textwidth, height=0.225\textheight]{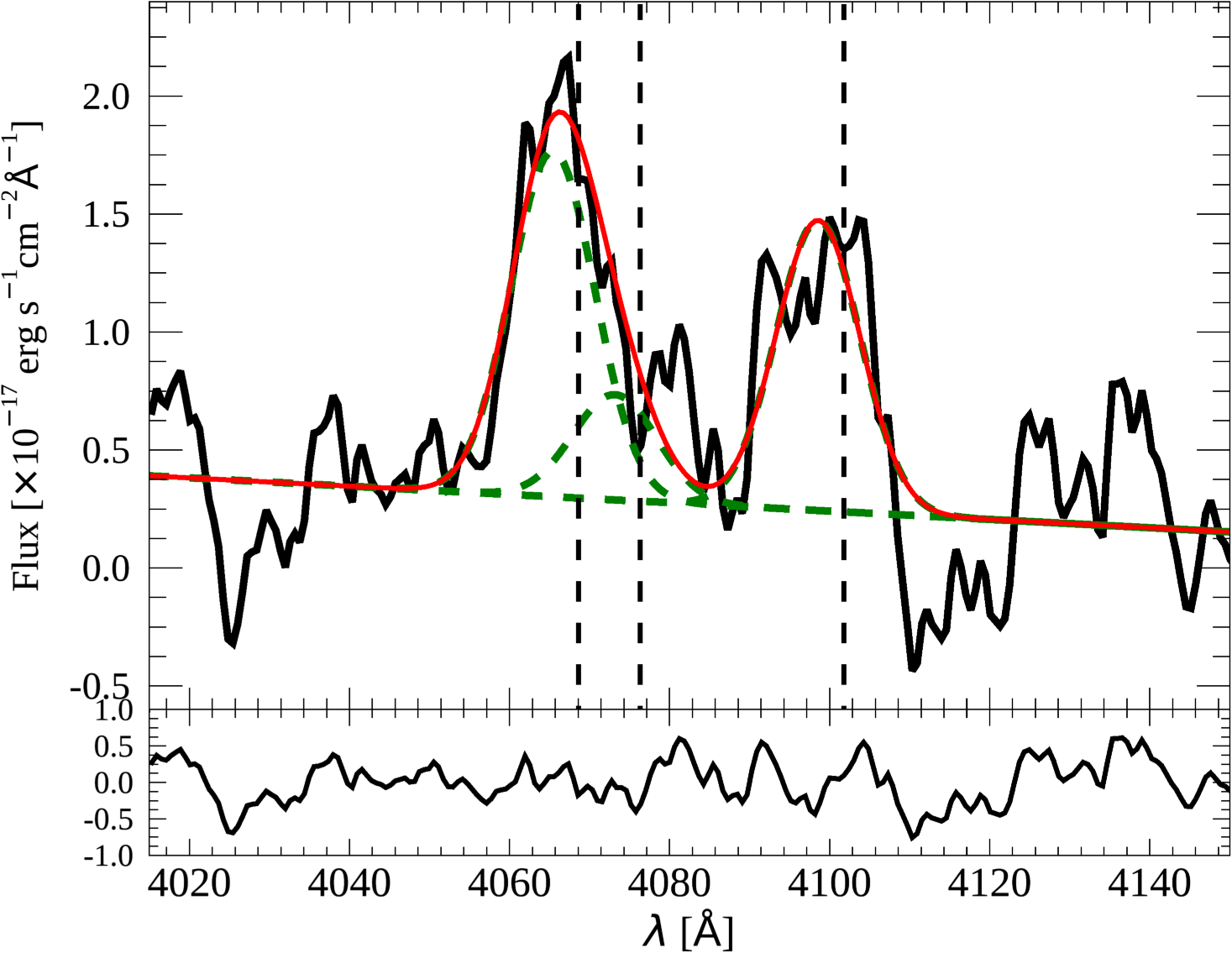}
\end{minipage}

\begin{minipage}[t]{0.5\textwidth}
\includegraphics[width=\textwidth, height=0.225\textheight]{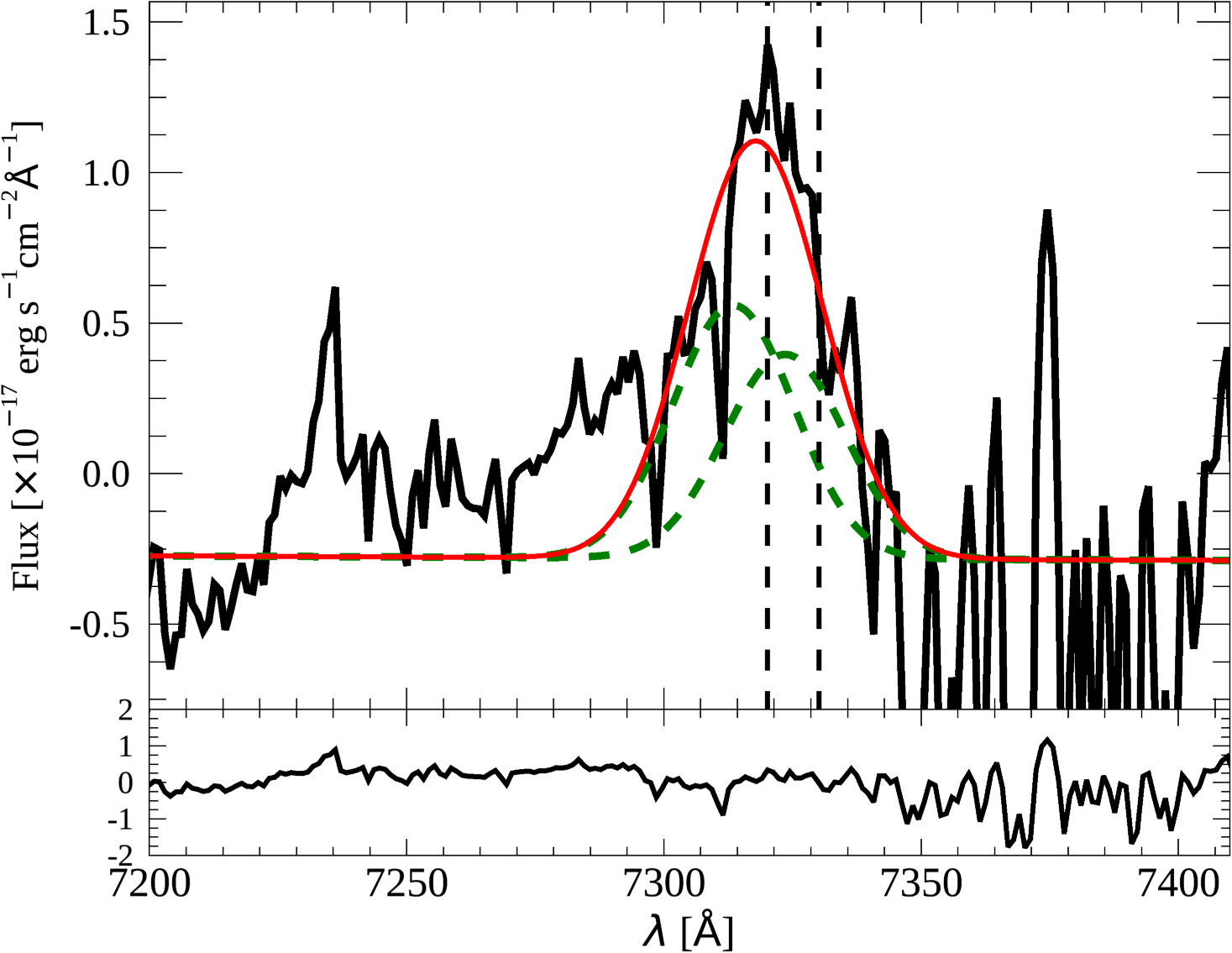}
\end{minipage}
\begin{minipage}[t]{0.5\textwidth}
\includegraphics[width=\textwidth, height=0.225\textheight]{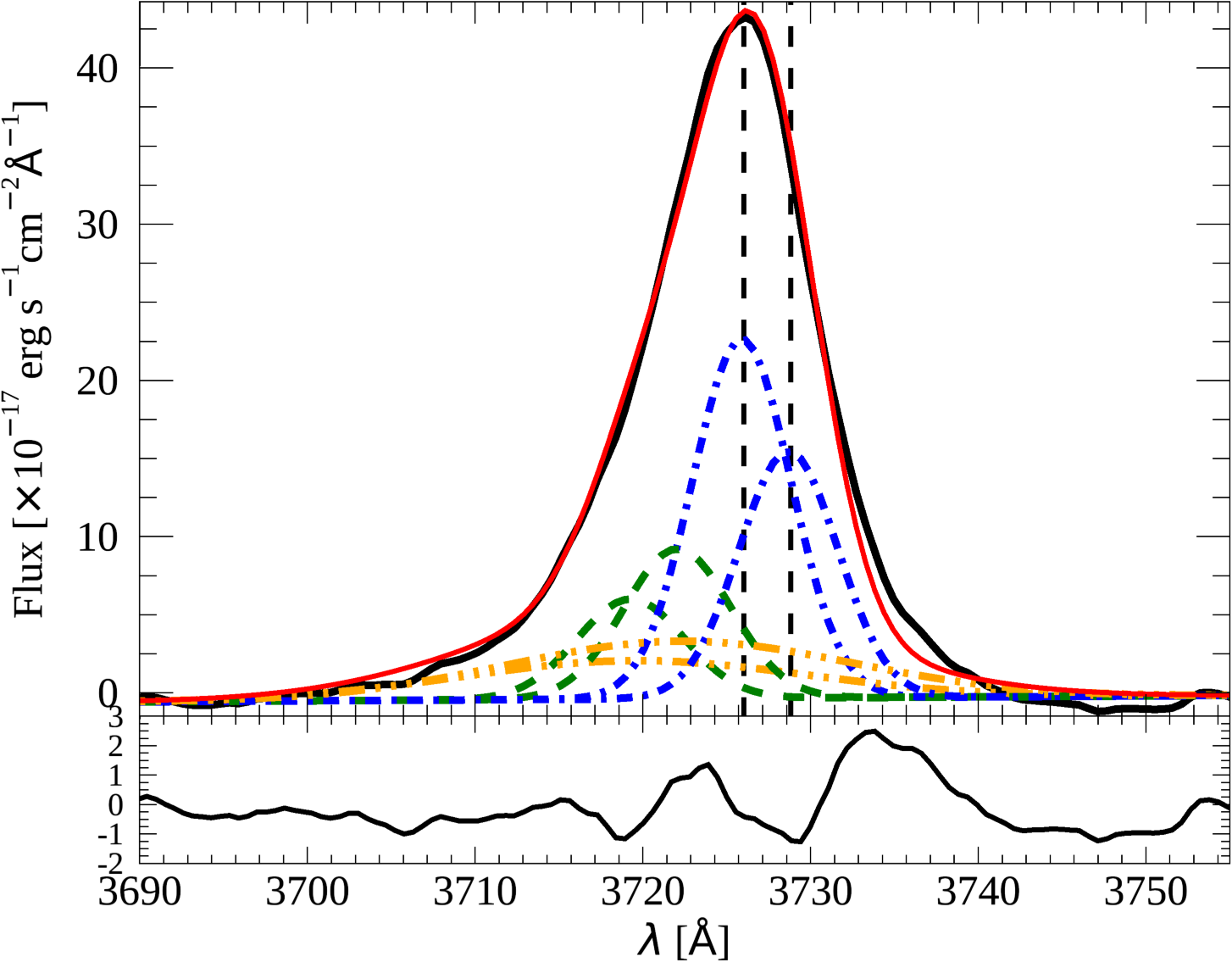}
\end{minipage}

\caption{PKS2314+03: As in Fig.~\ref{PKS1549Emission lines fits}. } 
\label{3C459Emission lines fits}
\end{figure*}

\begin{table*}[h]
\centering 
\begin{tabular}{p{3cm}cccc}
\toprule 
Object & \OII 3726,29  & \OII 7319,30  & \SII 4069,76  & \SII 6717,31  \\ [0.5ex]  
\toprule
PKS$~$0023--26 $-T$ & (4.4$\pm$0.5)$\times$10$^{-15}$ & (8.1 $\pm$1.7)$\times$ 10$^{-16}$ & (2.9$\pm$0.3)$\times$10$^{-16}$ & (6.0$\pm$0.8)$\times$10$^{-15}$ \\
PKS$~$0023--26 $-out$ & $\left(2.3 \pm 0.5\right) \times 10^{-15}$ & $\left(6.1 \pm 1.5\right) \times 10^{-16}$ & $\left(2.0 \pm 0.3\right) \times 10^{-16}$ & $\left(3.5 \pm 0.7\right) \times 10^{-15}$ \\
\rowcolor{Gray}
PKS~0252--71 $-T$ & $\left(1.3 \pm 0.2\right) \times 10^{-15}$ & $\left(1.8 \pm 0.4\right) \times 10^{-16}$ & $\left(2.9 \pm 0.4\right) \times 10^{-16}$ & $\left(9.4 \pm 1.6\right) \times 10^{-16}$ \\
\rowcolor{Gray}
PKS~0252--71 $-out$ & $\left(7.2 \pm 1.8\right) \times 10^{-16}$ & $\left(1.8 \pm 0.4\right) \times 10^{-16}$ & $\left(2.7 \pm 0.4\right) \times 10^{-16}$ & $\left(6.0 \pm 1.3\right) \times 10^{-16}$ \\
PKS~1151--34 $-T$ & $\left(2.9 \pm 0.4\right) \times 10^{-15}$ & $\left(5.2 \pm 1.5\right) \times 10^{-16}$ & $\left(3.0 \pm 1.2\right) \times 10^{-16}$ & $\left(3.4 \pm 0.6\right) \times 10^{-15}$ \\
PKS~1151--34 $-out$ & $\left(1.4 \pm 0.2\right) \times 10^{-15}$ & $\left(5.2 \pm 1.5\right) \times 10^{-16}$ & $\left(3.0 \pm 1.2\right) \times 10^{-16}$ & $\left(1.6 \pm 0.3\right) \times 10^{-15}$ \\
\rowcolor{Gray}
PKS~1306--09 $-T$ & $\left(2.1 \pm 0.4\right) \times 10^{-15}$ & $\left(1.6 \pm 0.4\right) \times 10^{-16}$ & $\left(9.8 \pm 1.8\right) \times 10^{-17}$ & $\left(1.5 \pm 0.2\right) \times 10^{-15}$ \\
\rowcolor{Gray}
PKS~1306--09 $-out$ & $\left(2.0 \pm 0.4\right) \times 10^{-15}$ & $\left(1.6 \pm 0.4\right) \times 10^{-16}$ & $\left(9.8 \pm 1.8\right) \times 10^{-17}$ & $\left(1.4 \pm 0.2\right) \times 10^{-15}$ \\
PKS~1549--79 $-T$ & $\left(3.1 \pm 0.6\right) \times 10^{-15}$ & $\left(2.0 \pm 0.3\right) \times 10^{-15}$ & $\left(5.5 \pm 1.2\right) \times 10^{-16}$ & $\left(5.3 \pm 0.9\right) \times 10^{-15}$ \\
PKS~1549--79 $-out$ & $\left(3.1 \pm 0.6\right) \times 10^{-15}$ & $\left(2.0 \pm 0.3\right) \times 10^{-15}$ & $\left(5.5 \pm 1.2\right) \times 10^{-16}$ & $\left(5.3 \pm 0.9\right) \times 10^{-15}$ \\
\rowcolor{Gray}
PKS~1814--63 $-T$ & $\left(2.2 \pm 0.6\right) \times 10^{-15}$ & $\left(1.1 \pm 0.2\right) \times 10^{-15}$ & $\left(1.4 \pm 0.4\right) \times 10^{-16}$ & $\left(5.6 \pm 0.6\right) \times 10^{-15}$ \\
\rowcolor{Gray}
PKS~1814--63 $-out$ & $\left(1.7 \pm 0.5\right) \times 10^{-15}$ & $\left(1.1 \pm 0.2\right) \times 10^{-15}$ & $\left(1.4 \pm 0.4\right) \times 10^{-16}$ & $\left(4.1 \pm 0.4\right) \times 10^{-15}$ \\
PKS~1934--63 $-T$ & $\left(3.3 \pm 0.3\right) \times 10^{-15}$ & $\left(2.8 \pm 0.3\right) \times 10^{-15}$ & $\left(1.2 \pm 0.1\right) \times 10^{-15}$ & $\left(3.3 \pm 0.3\right) \times 10^{-15}$ \\
PKS~1934--63 $-out$ & $\left(1.7 \pm 0.2\right) \times 10^{-15}$ & $\left(2.6 \pm 0.3\right) \times 10^{-15}$ & $\left(1.1 \pm 0.1\right) \times 10^{-15}$ & $\left(1.7 \pm 0.2\right) \times 10^{-15}$ \\
\rowcolor{Gray}
PKS~2135--209 $-T$ & $\left(2.4 \pm 0.6\right) \times 10^{-15}$ & $\left(3.7 \pm 0.5\right) \times 10^{-16}$ & $\left(1.9 \pm 0.3\right) \times 10^{-16}$ & $\left(2.5 \pm 0.2\right) \times 10^{-15}$ \\
\rowcolor{Gray}
PKS~2135--209 $-out$ & $\left(2.4 \pm 0.6\right) \times 10^{-15}$ & $\left(3.7 \pm 0.5\right) \times 10^{-16}$ & $\left(1.9 \pm 0.3\right) \times 10^{-16}$ & $\left(2.5 \pm 0.2\right) \times 10^{-15}$ \\

PKS~2314+03 $-T$ & $\left(5.7 \pm 1.3\right) \times 10^{-15}$ & $\left(4.5 \pm 1.0\right) \times 10^{-16}$ & $\left(2.5 \pm 0.4\right) \times 10^{-16}$ & $\left(4.9 \pm 0.6\right) \times 10^{-15}$ \\
\bottomrule
\end{tabular}

\caption{Table reporting the total ($T$) and outflowing ($out$) gas flux of the \OII$\lambdaup\lambdaup$3726,29\AA\ (col 2), \OII$\lambdaup\lambdaup$7319,30\AA\ (col 3), \SII$\lambdaup\lambdaup$4069,76\AA\ (col 4) and \SII$\lambdaup\lambdaup$6717,31\AA\ (col 5) trans-auroral lines. The fluxes are given in units of $\rm{erg~s^{-1}cm^{-2}\AA^{-1}}$. The total flux is assumed as an upper limit on the outflowing gas flux for those lines where is not possible to detect broad components during the fitting procedure.} 
\label{Table_transflux_1}
\end{table*}

\begin{table*}[h]
\centering 
\begin{tabular}{lcccc}
\toprule
Object & F H$\alpha_{T}$ & F H$\alpha_{out}$ & F $\NII_{T}$ & F $\NII_{out}$ \\
\toprule
PKS~0023--26 & $(5.5\pm 0.4)\times10^{-15}$ & $(3.2\pm 0.3)\times10^{-15}$ & $(6.1\pm 0.6)\times10^{-15}$ & $(3.8\pm 0.3)\times10^{-15}$ \\
PKS~0252--71 & $(1.7\pm 0.2)\times10^{-15}$ & $(1.4\pm 0.2)\times10^{-15}$ & $(1.5\pm 0.1)\times10^{-15}$ & $(8.1\pm 0.9)\times10^{-16}$ \\
PKS~1151--34 & $(3.6\pm 0.4)\times10^{-15}$ & $(1.8\pm 0.2)\times10^{-15}$ & $(5.9\pm 0.4)\times10^{-15}$ & $(2.6\pm 0.3)\times10^{-15}$ \\
PKS~1306--09 & $(1.4\pm 0.1)\times10^{-15}$ & $(1.4\pm 0.1)\times10^{-15}$ & $(1.8\pm 0.2)\times10^{-15}$ & $(1.7\pm 0.2)\times10^{-15}$ \\
PKS~1549--79 & $(2.8\pm 0.3)\times10^{-14}$ & $(2.8\pm 0.3)\times10^{-14}$ & $(2.2\pm 0.0)\times10^{-14}$ & $(2.2\pm 0.0)\times10^{-14}$ \\
PKS~1814--63 & $(5.3\pm 0.4)\times10^{-15}$ & $(3.8\pm 0.4)\times10^{-15}$ & $(5.7\pm 0.5)\times10^{-15}$ & $(3.5\pm 0.4)\times10^{-15}$ \\
PKS~1934--63 & $(5.7\pm 0.5)\times10^{-15}$ & $(3.6\pm 0.4)\times10^{-15}$ & $(6.8\pm 0.7)\times10^{-15}$ & $(4.8\pm 0.5)\times10^{-15}$ \\
PKS~2135--209 & $(4.7\pm 0.5)\times10^{-15}$ & $(4.7\pm 0.5)\times10^{-15}$ & $(5.5\pm 0.5)\times10^{-15}$ & $(5.5\pm 0.5)\times10^{-15}$ \\
PKS~2314+03 & $(6.6\pm 0.4)\times10^{-15}$ & $(6.6\pm 0.4)\times10^{-15}$ & $(1.5\pm 0.0)\times10^{-14}$ & $(1.5\pm 0.0)\times10^{-14}$ \\
\bottomrule
\end{tabular}
\caption{Table reporting the H$\alpha$ and the \NII$\lambdaup\lambdaup$6548,84\AA\ total and outflowing gas line fluxes (col 2, col 3 and col 4, col 5 respectively). Fluxes are given in units of $\rm{erg~s^{-1}cm^{-2}\AA^{-1}}$.}
\label{TableHalphaNIIfluxes}
\end{table*}

\section{Models}\label{appendixModels}
The electron density and the reddening derived using the DDD diagram presented in Fig.~\ref{DDD} depends on the chosen fiducial photoionisation model grids. We have run a new set of models using {\sc cloudy} version 17.00 \citep{Ferland17} to test the effect that the different model parameters, and in particular the ionisation parameter $U$ and the metallicity of the gas, have on the final results.
For that, we constructed a grid of photoionisation models with {\sc cloudy}  \citep[version 17.00][]{Ferland17}, covering a large range of gas and AGN properties. The details and assumptions of these models are similar to those presented and discussed by \citet{Baron19b}, and we briefly summarise them here. In particular, we used the `single cloud' approach, where we assumed spherical shells of gas with different densities, column densities, and distances from the central ionising source. The gas was assumed to be optically-thick and geometrically-thin, and we dis not model cases in which the gas is optically-thin and extends over regions that are as large, or even larger, than the distance from the central source (e.g., see \citealt{Baron18}). Such models require information on the gas spatial distribution, which is not available in our case.

\begin{figure*}[]
\begin{center}
\includegraphics[width=0.7\textwidth]{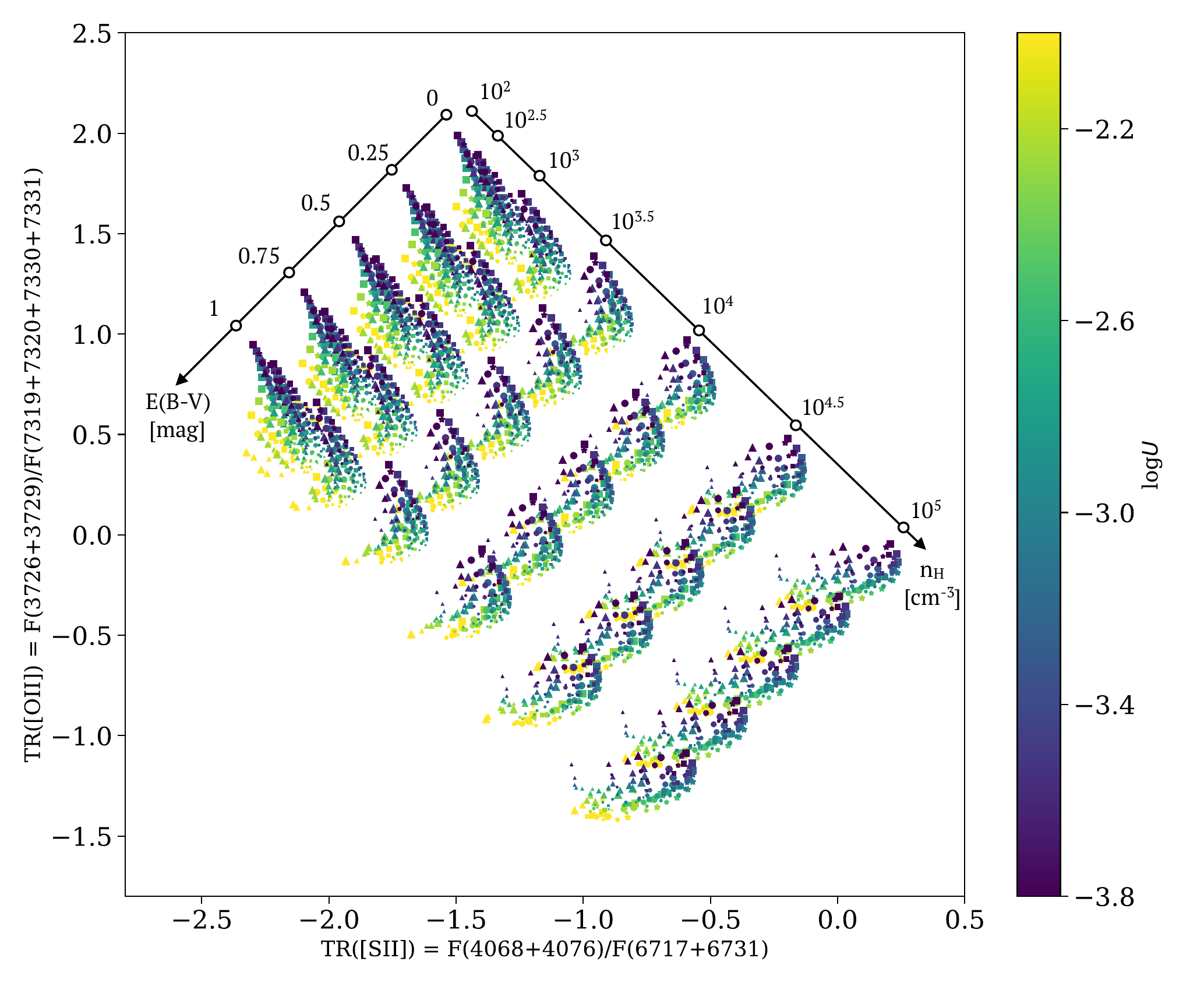}
\caption{Predicted auroral [OII] and trans-auroral [SII] line ratios for the 896 models considered in this work. The hydrogen number density ranges from $10^2\,\mathrm{cm^{-3}}$ to $10^5\,\mathrm{cm^{-3}}$, and the reddening ranges from $\mathrm{E}(B-V)=0$ mag to $\mathrm{E}(B-V)=1$ mag, both of which are indicated with arrows in the diagram. The colour of the markers represents the ionisation parameter, where $\log U=-3.8$ is marked with purple and $\log U=-2$ is marked with yellow. The size of the markers represents the gas metallicity, where the smallest markers correspond to $\mathrm{0.5 Z_{\odot}}$ and the largest correspond to $\mathrm{2 Z_{\odot}}$. The shapes of the markers represent different shapes of the ionising SED, where squares, stars, circles, and triangles correspond to SEDs with a mean energy of ionising photon of 2.56, 2.65, 3.15, and 4.17 Ryd respectively. See text for additional details about the models.
}\label{photoionisation_models}
\end{center}
\end{figure*}  

We modelled the central ionising source using standard assumptions about the spectral energy distribution (SED) of AGN. The SED consisted of an optical-UV continuum emitted by an optically-thick geometrically-thin accretion disk, with an additional X-ray power-law source that extends to 50 keV with a photon index of $\Gamma = 1.9$. We took the normalisation of the UV (2500\AA) to X-ray (2 keV) to be $\mathrm{\alpha_{OX} = 1.37}$. We considered four different cases for the shape of the ionising SED, which correspond to sources with different AGN properties (see table A1 in \citealt{Baron19b}). In particular, the SEDs we considered had different mean energies of ionising photon of 2.56, 2.65, 3.15, and 4.17 Ryd. The typical AGN in our sample is well-described by an SED with a mean energy of an ionising photon of 2.56 Ryd (SED 2 in table A1 in \citealt{Baron19b}), which we take as our fiducial AGN SED. We note that using different SEDs has a negligible effect on our results and conclusions. 

The grid of models consists of a geometrically-thin, optically-thick, shells of dusty gas, with ISM-type grains and the appropriate depletion. We assumed constant density clouds\footnote{In this work we dis not consider constant pressure clouds as they are expected to give similar line ratios and luminosities to constant density clouds, for the ionisation parameters considered here. See \citet{Baron19b} for additional details.}, with hydrogen number density ranging from $\mathrm{n_{H} = 10^{2}\, cm^{-3}}$ to $\mathrm{n_{H} = 10^{5}\, cm^{-3}}$. We considered four different gas metallicities: 0.5, 1, 1.5, and 2$\mathrm{Z_{\odot}}$. Finally, many of the ionised gas properties depend on the ionisation parameter, $U = Q(\mathrm{Lyman})/4 \pi r^{2} n_{\mathrm{H}} c$, where $Q(\mathrm{Lyman})$ is the number of the hydrogen-ionising photons, $r$ is the gas distance from the central ionising source, $n_{\mathrm{H}}$ is the hydrogen number density, and $c$ is the speed of light. We therefore considered a wide range in ionisation parameter, from $\log U = -3.8$ to $\log U = -2$. The distance of the gas from the central source was set according to each combination of hydrogen number density and ionisation parameter.

The resulting model grid consists of 896 models with different gas and AGN properties. For each model, we extracted the predicted auroral \OII\ and trans-auroral \SII\ line luminosities, and applied dust reddening with color excess ranging from $\mathrm{E}(B-V)=$0 mag to  $\mathrm{E}(B-V)=$1 mag, assuming \citet{Cardelli1989} extinction curve. Figure \ref{photoionisation_models} shows the predicted tr \OII\ and \SII\ line ratios as a function of the different gas and AGN properties. The colour of the markers represents the ionisation parameter of the gas, and the size of the markers represents the metallicity, where 0.5$\mathrm{Z_{\odot}}$ corresponds to the smallest marker, while 2$\mathrm{Z_{\odot}}$ corresponds to the largest. The shapes of the markers represent the ionising SED, where the squares, stars, circles, and triangles correspond to SEDs with a mean energy of ionising photon of 2.56, 2.65, 3.15, and 4.17 Ryd respectively. The predicted line ratios  depend primarily on the hydrogen number density and the dust reddening, both of which are indicated with arrows in the diagram. The strong dependence of the line ratios on these two parameters suggests that the line ratios can be used to estimate the density and reddening of the ionised cloud \citep[as initially discussed by][]{Holt2011}.

Inspecting Fig.~\ref{photoionisation_models}, one can see that the ionisation parameter and the gas metallicity have a non-negligible effect on the predicted line ratios, and thus on the derived density and reddening of the ionised cloud. In particular, for $n_{\mathrm{H}} \leq 10^{4} \, \mathrm{cm^{-3}}$, different ionisation parameters can change the derived reddening by 0.1--0.2 mag, and the inferred hydrogen density by 0.2--0.7 dex. For high density clouds, $n_{\mathrm{H}} \sim 10^{5} \, \mathrm{cm^{-3}}$, varying the ionisation parameter has a stronger impact on the derived reddening, but less so on the derived hydrogen density. On the other hand, for lower density clouds, $n_{\mathrm{H}} \leq 10^{4} \, \mathrm{cm^{-3}}$, varying the gas metallicity affects the derived hydrogen density (by 0.1--0.3 dex), while for higher density clouds, it mostly effects the derived reddening (by $\sim$0.2 mag). We note that the shape of the ionising SED can change the tr \OII\ and \SII\ line ratios but its effect on the derived density and reddening is smaller compared to variations in ionisation parameter and/or gas metallicity.

As discussed in the main text, in order to estimate the gas density and reddening due to dust for a given gas component of a given target from the DDD, a fiducal model needs to be assumed. 
In this work we considered only the models with solar metallicity and our fiducial AGN SED and we selected as fiducal the photoionisation model with the ionisation parameter closest to the empirically derived one in Sec.\ref{ionisationparam}.
We then calculated a denser grid of models, with hydrogen density ranging from $n_{\mathrm{H}} = 10^{2} \, \mathrm{cm^{-3}}$ to $n_{\mathrm{H}} = 10^{5} \, \mathrm{cm^{-3}}$ with 0.1 dex jumps, and E(B-V) colour excess ranging from 0 to 1 mag with 0.05 mag jumps and use this finer grid to derive the electron density and the reddening from our observed tr\OII\ and \SII\ line ratios. 

\section{Comparison with the E(B-V) and n${_e}$ measured with alternative methods}\label{appendixComparison}    
 
\begin{figure}
\includegraphics[width=0.5\textwidth]{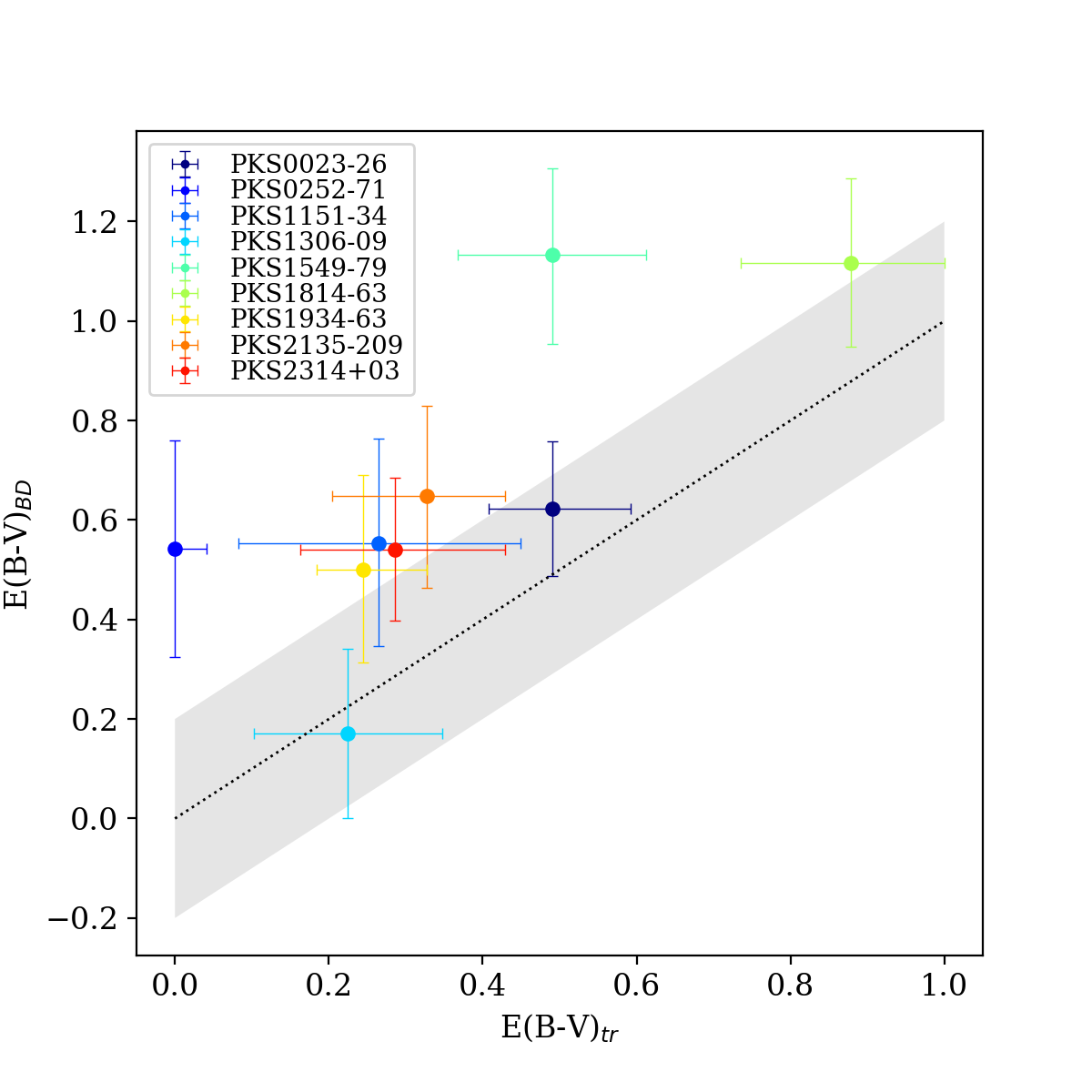}
\caption{Comparison between the gas reddening estimated using the DDD presented in Fig.\ref{DDD} (E(B-V)$_{tr}$) and the  H$\alpha$/H$\beta$ Balmer decrement (E(B-V)$_{BD}$).
Different galaxies have different colours following the convention showed in the legend in upper left part. The dotted line marks the 1:1 relation and the shaded area a scatter of 0.2 in E(B-V).}
\label{Figure_reddeening}
\end{figure}

To check our DDD-derived reddening estimates and densities, we have compared them with reddening values estimated via the classical H$\alpha$/H$\beta$ Balmer decrement, and gas densities determined using the method proposed by \cite{Baron19b}.
For the reddening estimates we used total line fluxes only, which we believe give more robust measurements. Values of E(B-V) were determined using the observed H$\alpha$/H$\beta$ ratio using the \cite{Cardelli1989} reddening law with R$_{V}$=3.1 and a theoretical H$\alpha$/H$\beta$=2.85 consistently with the way reddenings are derived from the models in the DDD. Errors were calculated by taking into account the errors on the fluxes of the emission lines.
From the Fig.~\ref{Figure_reddeening} it is clear that the Balmer decrement estimates are systematically higher for most of the sample. However, taking into account the errors, the values are consistent within 0.2 magnitudes for all the galaxies in our sample excluding PKS~1549--79 and PKS~0252--71. It is most likely that these discrepancies can be ascribed to the degeneracies in the emission line modelling which can particularly affect the H$\alpha$+\NII\ blend due to the extreme kinematics of the ionised gas in our sources, as can be observed from the figures presented in Appendix~\ref{AppendixFits}.

PKS~1549--79 and PKS~0252--71 are the two targets that show the highest discrepancy between the reddenings derived with the two methods, and we believe this might be related to the uncertainties in the stellar population fitting.
In fact, the nuclear spectrum of PKS~0252--71 has a higher noise level compared to the rest of the sample, especially around the H$\alpha$ line, which increases the uncertainties on the flux of this emission line due to the starlight continuum subtraction.
In the case of PKS~1549--79 the difficulty in taking into account the level of the continuum is related to the fact that it noticeably increase toward the red part of the spectrum. Looking at the stellar population fit is seems possible that around the \OIII and H$\beta$ lines our best fit is overestimating the continuum level leading us to underestimate the H$\beta$ flux. An additional factor to consider in the case of PKS~1549--79 is that there may be an underlying BLR component to the H$\alpha$+\NII\ blend that we did not include in our fit because we obtained an adequate fit using NLR components alone. However, \citet{Holt2006} did detect this BLR component, and its presence in our spectrum could potentially boost the NLR H$\alpha$ flux used in our analysis of reddening.

\begin{figure}
\includegraphics[width=0.5\textwidth]{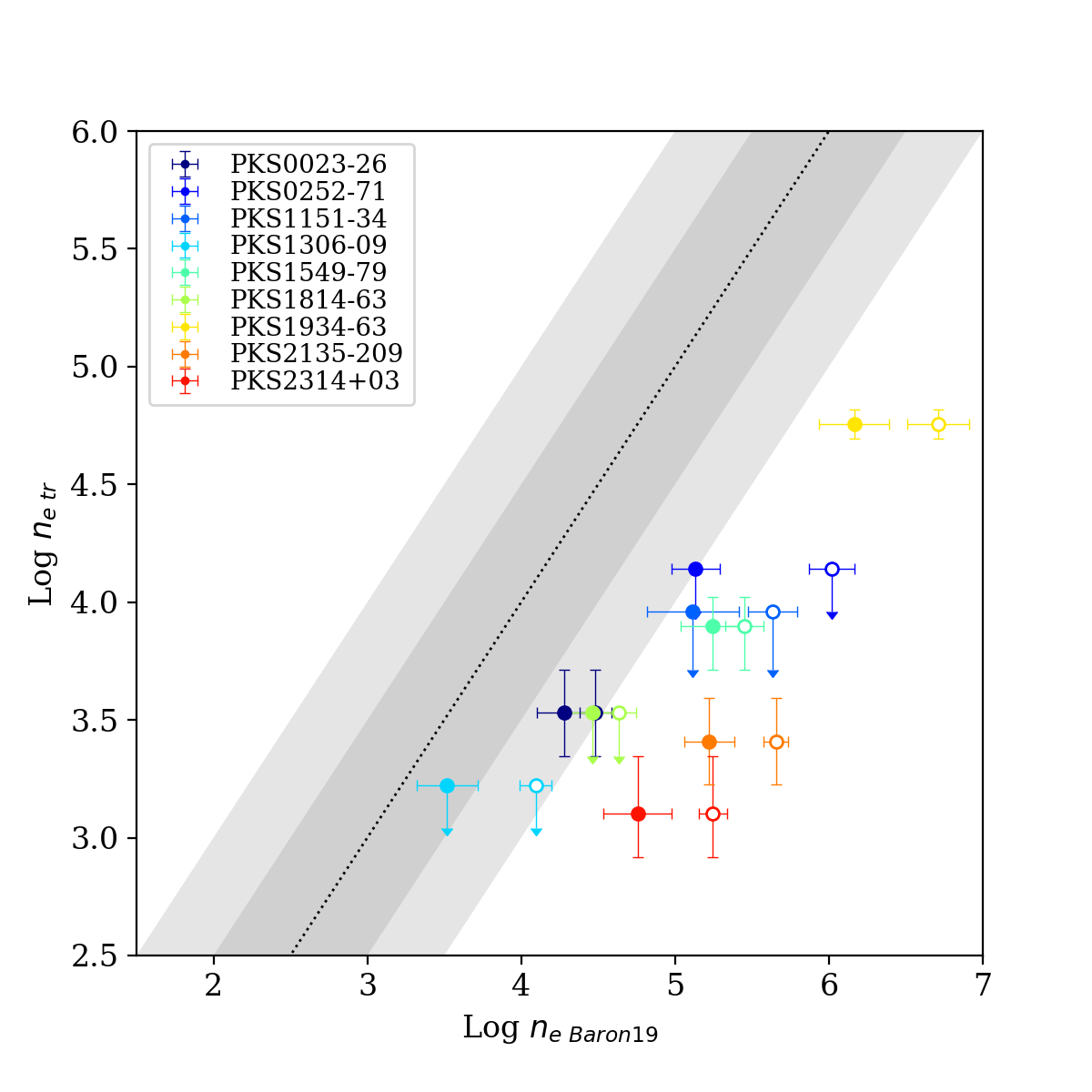}
\caption{Comparison between the outflowing gas electron densities extracted using the DDD presented in Fig.\ref{DDD} (n$_{e~tr}$) and the \cite{Baron19b} calibration (n$_{e~Baron19}$).
The n$_{e~Baron19}$ have been extracted using both the \cite{Heckman2004} (empty circles) and the \cite{2009A&A...504...73L} (filled circles) calibrations for the $L_{BOL}$.
Different galaxies have different colours following the convention showed in the legend in upper left part. 
The dotted line marks the 1:1 relation and the shaded areas a scatter of 0.5 and 1 in log~n$_{e}$. The errors do not take into account the scatter of the $L_{BOL}$ calibration.}
\label{dens_comp}
\end{figure}

In Fig.~\ref{dens_comp} we compare the densities of the outflowing gas estimated using the DDD diagram and the equation
\begin{equation}
    n_e \approx 3.2 \left( \frac{L_{bol}}{10^{45} erg/sec} \right) \left( \frac{r}{1~kpc} \right)^{-2} \left( \frac{1}{U} \right) cm^{-3} 
\end{equation}
taken from \cite{Baron19b}.

In order to estimate the densities with the latter method we made use of our estimates of the outflow radii (Sec.\ref{radii}) and ionisation parameter (Sec.\ref{ionisationparam}), and took the AGN bolometric luminosities derived using both the \cite{Heckman2004} and the \cite{2009A&A...504...73L} calibrations as described in Sec.\ref{subsec:Lbolometric}.
This is the first time that a one-to-one comparison between these two methods has been performed, with the goal of setting the ground to explore the possible biases and limitations of the two methods \citep[see also][]{Davies2020}.
Fig.~\ref{dens_comp} shows that we find discrepancies higher than one order of magnitude between the densities derived in this paper using the trans-auroral lines and those derived using the \cite{Baron19b} method, with the latter giving systematically higher densities for all the galaxies in our sample
As discussed in Sec.~\ref{subsec:Lbolometric}, the \cite{Heckman2004} calibration gives higher $L_{BOL}$ compared to the \cite{2009A&A...504...73L} calibration, thus increasing the discrepancy between the two sets of densities.

It is worth noting here that, by comparing the densities estimated with the two methods, we are actually checking if the quantities we have derived in our work (mainly the outflow densities and radii, and the AGN $L_{BOL}$) are compatible with the ionisation state of the gas (measured via the ionisation parameter $U$) that produces the observed line ratios.

We have estimated $U$ from the observed line ratios following the prescriptions of \cite{Baron19b}. The latter is a robust method that does not strongly depend on the details of the photoionisation models (e.g. AGN SED shape and gas metallicity). In addition, looking at both our sample and other, often larger, AGN samples \citep[e.g.][]{Baron19a,Davies2020}, the derived values of the ionisation parameter typically change by less than an order of magnitude across each sample. For this reason, we believe that the errors on $U$ are playing a minor role, and attribute the discrepancy between the densities mainly to the uncertainties on the derived $L_{BOL}$ and outflow radii. 
The comparison between the densities is thus indicating that we are likely to be overestimating $L_{BOL}$, while also underestimating the outflow radii $r$ due to projection effects (see the discussion in Sec.~\ref{subsec:Lbolometric}).

In Sec.~\ref{subsec:Lbolometric} we have already described the possible uncertainties on the estimated $L_{BOL}$ for our type of sources and our choice to adopt the \cite{2009A&A...504...73L} calibration.
We have also considered different calibrations for the AGN bolometric luminosites, namely the \cite{Netzer2009} calibration using the H$\beta$ fluxes from our observations, the \cite{Runnoe2012} calibration using 24$\mu$m IR fluxes taken from \cite{dicken09}, and finally the bolometric luminosities reported by \cite{Ming2014} using X-ray data, however this did not mitigate the discrepancies between the two sets of electron densities in a significant way.
As discussed in Sec.\ref{ionisationparam} we can exclude the idea that shocks are significantly boosting the \OIII\ emission, leading us to overestimate the AGN bolometric luminosities.
We further emphasise that the X-ray bolometric correction (XBC) adopted in the \cite{2009A&A...504...73L} paper \citep[taken from][]{Marconi2004} is the main source of uncertainty on the $L_{BOL}$ calculated from X-ray luminosities, and can lead up to a factor 2 uncertainty \citep{Hopkins2007}.

The uncertainties on the outflow radii and on the $L_{BOL}$, taken together, make it feasible to lower the densities estimated with the \cite{Baron19b} method by about one order of magnitude and thus reconcile their values with the DDD densities.

As well as a lower $L_{BOL}$ and larger $r$, an alternative explanation might be that the clouds emitting the trans-auroral \SII\ and \OII\ lines have lower densities and are emitted on larger scales than those emitting the \OIII\ and H$\beta$ lines, which are key for determining the ionisation parameter.   However, we believe that this alternative explanation is unlikely for the following reasons.
\begin{itemize}
\item[-] It would be difficult to avoid the higher density clouds emitting substantial \SII\ and \OII\ trans-auroral line emission, since the  latter lines have significantly higher critical densities than \OIII$\lambda5007$. One possibility is that the high density clouds are matter-bounded, such that the partially ionised zones at the backs of the clouds that would normally emit the \SII\ and \OII\ lines are absent. However, any matter-bounded high density clouds would need to have just the right column depth to reduce the trans-auroral line flux, but avoid decreasing the H$\beta$ flux too much, otherwise the \OIII/H$\beta$ ratio and ionisation parameter would be overestimated, and other line ratios such as \HeII(4686)/H$\beta$ would be too high.
\item[-] If the putative lower density clouds were at the same radius as the higher density clouds emitting the [OIII] and H$\beta$, then they would inevitably have a higher ionisation parameter and therefore be expected to emit substantial \OIII\ and H$\beta$ emission. Moreover, the lower density clouds would probably be larger than the higher density ones, increasing their covering factor for the ionising continuum and boosting their \OIII\ and H$\beta$ luminosity further.
\item[-] To avoid the lower density clouds emitting substantial \OIII\ and H$\beta$ emission, they must be situated at much larger radial distances from the nucleus. For example, for them to have the same ionisation parameter as the high density clouds emitting the
\OIII\ and H$\beta$ lines, they would need to be $\sim3\times$ further away; and to achieve an ionisation parameter that is, say, 10x lower (to avoid significant \OIII\ emission)  for the low density clouds than for the high density clouds, they would need to be ~10x further away (several kpc for some sources). We would already be able to detect such large radial extents for the trans-auroral lines in our spectra, but we do not. 
\item[-] The trans-auroral lines are broad and show disturbed kinematics (even if not measured as well as for [OIII] due to blending issues). It seems unlikely that this would be the case if the clouds were situated at large radii. In this context, it is notable that, for the one case in which we have been able to measure precise densities for the narrow, intermediate and broad components separately using the DDD technique -- PKS~1934--63 -- it is the broadest kinematic components that have the highest trans-auroral densities \citep{Santoro2018}.
\end{itemize}

\end{appendix}

\end{document}